\documentclass{amsarte}

\usepackage{latexsym,pifont}
\usepackage{epsfig}
\usepackage{rotating}

\usepackage{ifpdf}
\ifpdf
\usepackage{epstopdf}
\usepackage{hyperref}
\else
\usepackage[hypertex]{hyperref}
\fi

\usepackage{amsfonts,amsmath,amssymb,mathrsfs,extarrows,MnSymbol}
\usepackage[all]{xy}

\usepackage{tikz}
\usetikzlibrary{matrix,arrows}

\newcommand{\alxydim}[2]{\begin{aligned}\xymatrix#1{#2}\end{aligned}}

\newcommand{\brem}{\begin{Rem}}
\newcommand{\erem}{\end{Rem}\medskip}
\newcommand{\beg}{\begin{Eg}}
\newcommand{\eeg}{\end{Eg}}
\newcommand{\bedef}{\begin{Def}}
\newcommand{\exdef}{\begin{flushright}$\diamond$\end{flushright}
\end{Def}\vskip0.1cm}
\newcommand{\berop}{\begin{Prop}}
\newcommand{\eerop}{\end{Prop}}
\newcommand{\belem}{\begin{Lem}}
\newcommand{\elem}{\end{Lem}}
\newcommand{\bethe}{\begin{Thm}}
\newcommand{\ethe}{\end{Thm}}
\newcommand{\becor}{\begin{Cor}}
\newcommand{\ecor}{\end{Cor}}
\newcommand{\beroof}{\noindent\begin{proof}}
\newcommand{\eroof}{\end{proof}}
\newcommand{\becon}{\begin{Conv}}
\newcommand{\econ}{\begin{flushright}$\checkmark$\end{flushright}\end{Conv}}
\newcommand{\befact}{\begin{Fact}}
\newcommand{\efact}{\begin{flushright}$\checkmark$\end{flushright}\end{Fact}}
\newcommand{\bequest}{\begin{Quest}}
\newcommand{\equest}{\end{Quest}}
\newcommand{\brob}{\begin{Prob}}
\newcommand{\erob}{\end{Prob}}

\newcommand{\barr}{\begin{array}}
\newcommand{\earr}{\end{array}}
\newcommand{\ben}{\begin{enumerate}}
\newcommand{\een}{\end{enumerate}}
\newcommand{\bit}{\begin{itemize}}
\newcommand{\eit}{\end{itemize}}

\newcommand{\qq}{\begin{eqnarray}}
\newcommand{\qqq}{\end{eqnarray}}

\newcommand{\nn}{\nonumber}

\newcommand{\ovl}[1]{\overline{#1}}
\newcommand{\unl}[1]{\underline{#1}}

\newcommand{\Reqref}[1]{Eq.\,\eqref{#1}}
\newcommand{\Rcite}[1]{Ref.\,\cite{#1}}
\newcommand{\Rxcite}[2]{Ref.\,\cite[#1]{#2}}

\newcommand\void[1]{}

\newcommand{\tx}[1]{\textrm{#1}} 
\newcommand{\ciut}[1]{\tiny$#1$}

\newcommand{\gt}[1]{\mathfrak{#1}}

\def\cA{\mathcal{A}}

\def\cD{\mathcal{D}}

\def\cG{\mathcal{G}}
\def\ceH{\mathcal{H}}
\def\cI{\mathcal{I}}

\def\cK{\mathcal{K}}

\def\cM{\mathcal{M}}

\def\cO{\mathcal{O}}

\def\cT{\mathcal{T}}
\def\cU{\mathcal{U}}
\def\cV{\mathcal{V}}
\def\cW{\mathcal{W}}

\def\xcC{\mathscr{C}}
\def\xcD{\mathscr{D}}
\def\xcE{\mathscr{E}}
\def\xcF{\mathscr{F}}
\def\xcG{\mathscr{G}}

\def\xcI{\mathscr{I}}
\def\xcJ{\mathscr{J}}

\def\xcL{\mathscr{L}}
\def\xcM{\mathscr{M}}

\def\xcP{\mathscr{P}}
\def\xcQ{\mathscr{Q}}

\def\xcS{\mathscr{S}}
\def\xcT{\mathscr{T}}
\def\xcU{\mathscr{U}}

\def\xcW{\mathscr{W}}
\def\xcX{\mathscr{X}}
\def\xcY{\mathscr{Y}}
\def\xcZ{\mathscr{Z}}

\def\t{\mathbf{t}}

\def\bC{{\mathbb{C}}}
\def\bD{{\mathbb{D}}}

\def\bH{{\mathbb{H}}}
\def\bK{{\mathbb{K}}}

\def\bN{{\mathbb{N}}}

\def\bR{{\mathbb{R}}}
\def\bS{{\mathbb{S}}}

\def\bV{{\mathbb{V}}}
\def\bZ{{\mathbb{Z}}}

\def\a{\alpha}
\def\b{\beta}
\def\g{\gamma}
\def\G{\Gamma}
\def\d{\delta}
\def\D{\Delta}
\def\ep{\epsilon}
\def\vep{\varepsilon}

\def\Th{\Theta}

\def\k{\kappa}

\def\la{\lambda}

\def\om{\omega}
\def\Om{\Omega}

\def\si{\sigma}
\def\Si{\Sigma}

\def\t{\tau}

\def\z{\zeta}

\def\agt{\gt{a}}

\def\Egt{\gt{E}}

\def\ggt{\gt{g}}

\def\hgt{\gt{h}}

\def\igt{\gt{i}}

\def\jgt{\gt{j}}

\def\kgt{\gt{k}}

\def\Pgt{\gt{P}}

\def\Sgt{\gt{S}}
\def\tgt{\gt{t}}

\def\Vgt{\gt{V}}

\def\Xgt{\gt{X}}

\def\zgt{\gt{z}}

\newcommand{\sfa}{{\mathsf a}}

\newcommand{\sfd}{{\mathsf d}}

\newcommand{\sfF}{{\mathsf F}}

\newcommand{\sfi}{{\mathsf i}}

\newcommand{\sfL}{{\mathsf L}}
\newcommand{\sfm}{{\mathsf m}}

\newcommand{\sfP}{{\mathsf P}}

\newcommand{\sfT}{{\mathsf T}}

\newcommand{\sfY}{{\mathsf Y}}

\newcommand{\txa}{{\rm a}}
\newcommand{\txA}{{\rm A}}
\newcommand{\txb}{{\rm b}}
\newcommand{\txB}{{\rm B}}

\newcommand{\txc}{{\rm c}}
\newcommand{\txC}{{\rm C}}

\newcommand{\ee}{{\rm e}}
\newcommand{\txE}{{\rm E}}

\newcommand{\txF}{{\rm F}}
\newcommand{\txg}{{\rm g}}
\newcommand{\txG}{{\rm G}}
\newcommand{\txh}{{\rm h}}
\newcommand{\txH}{{\rm H}}

\newcommand{\txJ}{{\rm J}}

\newcommand{\Lx}{{\rm L}}
\newcommand{\txm}{{\rm m}}

\def\Cv{\v{C}}

\def\vH{\check{H}}

\def\exp{{\rm exp}}
\def\id{{\rm id}}
\newcommand{\pr}{{\rm pr}}
\def\sign{{\rm sign}}

\def\too{\longrightarrow}
\def\ev{{\rm ev}}
\def\Deg{{\rm Deg}}

\def\morf{{\rm Mor}}
\def\1morf{1{\rm -Mor}}
\def\2morf{2{\rm -Mor}}
\def\dim{{\rm dim}}
\def\im{{\rm im}}
\def\ker{{\rm ker}}

\def\End{{\rm End}}

\def\Inv{{\rm Inv}}
\newcommand\Mod{{\rm Mod}}

\def\Vol{{\rm Vol}}

\def\p{\partial}

\def\con{\righthalfcup}
\newcommand{\Diff}{{\rm Diff}}

\def\curv{{\rm curv}}
\def\Hol{{\rm Hol}}

\newcommand{\sG}{\mathcal{sG}}

\def\bd1{{\boldsymbol{1}}}
\def\brd0{{\boldsymbol{0}}}

\def\det{{\rm det}}

\def\diag{\textrm{diag}}

\def\Ad{{\rm Ad}}

\def\Cliff{{\rm Cliff}}

\newcommand{\uj}{{\rm U}(1)}

\def\x{\times}
\def\ox{\otimes}

\def\lx{{\hspace{-0.04cm}\ltimes\hspace{-0.05cm}}}
\def\rx{\rtimes}

\def\ract{\vartriangleleft}
\def\lact{\vartriangleright}
\def\must{\stackrel{!}{=}}

\def\rstr{\mathord{\restriction}}

\newcommand{\corr}[1]{\left\langle #1 \right\rangle}

\newtheorem{Thm}{Theorem}
\newtheorem{Prop}[Thm]{Proposition}
\newtheorem{Lem}[Thm]{Lemma}
\newtheorem{Cor}[Thm]{Corollary}
\theoremstyle{definition}
\newtheorem{Rem}[Thm]{Remark}
\newtheorem{Def}[Thm]{Definition}
\newtheorem{Eg}[Thm]{Example}
\newtheorem{Conv}[Thm]{Convention}
\newtheorem{Fact}[Thm]{Fact}
\newtheorem{Quest}[Thm]{Question}
\newtheorem{Prob}[Thm]{Problem}

\numberwithin{equation}{section} \numberwithin{Thm}{section}

\DeclareMathOperator{\Hom}{Hom}

\newcount\hour\newcount\minute
        \hour=\time \divide\hour by60 \minute=\time
        {\multiply\hour by60 \global\advance\minute by-\hour}
        \edef\militarytime{\number\hour:\ifnum\minute<10 0\fi\number\minute}

\begin{document}

\title{Equivariant Cartan--Eilenberg supergerbes\\ 
for the Green--Schwarz superbranes\\[2pt] I. The super-Minkowskian case}

\author{Rafa\l ~R. ~Suszek}
\address{R.R.S.:\ Katedra Metod Matematycznych Fizyki, Wydzia\l ~Fizyki
Uniwersytetu Warszawskiego, ul.\ Pasteura 5, PL-02-093 Warszawa,
Poland} \email{suszek@fuw.edu.pl}

\begin{abstract}
An explicit gerbe-theoretic description of the super-$\sigma$-models of the Green--Schwarz type is proposed and its fundamental structural properties, such as equivariance with respect to distinguished isometries of the target supermanifold and $\kappa$-symmetry, are studied at length for targets with the structure of a homogeneous space of a Lie supergroup. The programme of (super)geometrisation of the Cartan--Eilenberg super-$(p+2)$-cocycles that determine the topological content of the super-$p$-brane mechanics and ensure its $\kappa$-symmetry, motivated by the successes of and guided by the intuitions provided by its bosonic predecessor, is based on the idea of a supercentral extension of a Lie supergroup in the presence of a nontrivial super-2-cocycle in the Chevalley--Eilenberg cohomology of its Lie superalgebra, the gap between the two cohomologies being bridged by a super-variant of the classic Chevalley--Eilenberg construction. A systematic realisation of the programme is herewith begun with a detailed study of the elementary homogeneous space of the super-Poincar\'e group, the super-Minkowskian spacetime, whose simplicity affords straightforward identification of the supergeometric mechanisms and unobstructed development of formal tools to be employed in more complex circumstances.
\end{abstract}

\void{\date{\today, \militarytime\,(GMT+1)}}

\begin{flushright}
{\it Krzysztofowi Gaw\c{e}dzkiemu, Mistrzowi i Przyjacielowi,\\ z okazji Jego siedemdziesi\c{a}tych urodzin\\ -- z szacunkiem, wdzi\c{e}czno\'sci\c{a} i oddaniem}\vspace{30pt}
\end{flushright}

\maketitle

\tableofcontents

\section{Introduction}

The naturality and adequacy of the language of gerbe theory in the setting of the mechanics of the topologically charged bosonic loop, captured by the two-dimensional non-linear $\si$-model, and the efficiency of its higher-geometric, -cohomological and -categorial methods in the canonical description \cite{Gawedzki:1987ak,Suszek:2011hg}, symmetry analysis \cite{Gawedzki:2007uz,Gawedzki:2008um,Gawedzki:2010rn,Gawedzki:2012fu,Suszek:2012ddg} and constructive geometric quantisation \cite{Gawedzki:1987ak,Gawedzki:1999bq,Gawedzki:2002se,Gawedzki:2003pm,Gawedzki:2004tu,Suszek:2011hg} of field theories from this distinguished class, has, by now, attained the status of a widely documented, albeit clearly insufficiently exploited, fact. Introduced in the disguise of the Deligne--Beilinson hypercohomology in the pioneering works of Alvarez \cite{Alvarez:1984es} and Gaw\c{e}dzki \cite{Gawedzki:1987ak}, the language has found -- since the advent of the geometric formulation of gerbe theory worked out by Murray {\it et al.} in Refs.\,\cite{Murray:1994db,Murray:1999ew,Stevenson:2000wj,Bouwknegt:2001vu,Carey:2002,Carey:2004xt} -- ample structural applications in the study of $\si$-models and the associated conformal field theories and string theories, and in particular in a neat cohomological classification of quantum-mechanically consistent field theories of the type indicated (also in the presence of boundaries and defects in the two-dimensional spacetime \cite{Fuchs:2007fw,Runkel:2008gr,Suszek:2011hg}), in a concrete formulation of a universal Gauge Principle \cite{Gawedzki:2010rn,Gawedzki:2012fu,Suszek:2011,Suszek:2012ddg,Suszek:2013}, going well beyond the na\"ive minimal-coupling scheme, and in the resulting classification of obstructions against the gauging of rigid symmetries (or gauge anomalies) and of inequivalent gaugings, and -- finally -- in a rigorous geometric description of defects and their fusion in the said theories, in which the r\^ole of defects in the modelling of symmetries and dualities between theories has been elucidated and turned into a handy field-theoretic tool \cite{Fuchs:2007fw,Runkel:2008gr,Suszek:2011hg}. The models that afford the farthest insight and the richest pool of formal methods and constructions are those with a high internal symmetry, reflecting -- in consequence of their geometric nature -- a high symmetry of the target of propagation of the loop, to wit, the Wess--Zumino--Witten $\si$-models of loop dynamics on (compact) Lie groups \cite{Witten:1983ar,Gawedzki:1990jc,Gawedzki:1999bq,Gawedzki:2001rm} and their gauged variants \cite{Goddard:1984vk,Gawedzki:1988hq,Gawedzki:1988nj,Karabali:1988au,Hori:1994nc,Gawedzki:2001ye} defining that dynamics on the associated homogeneous spaces. The generating nature of these models in the category of rational conformal field theories in two dimensions and -- not unrelatedly -- their holographic correspondence with the three-dimensional Chern--Simons topological gauge field theory in the presence of Wilson lines, give a measure of the theoretical significance of a good understanding of these models offered by gerbe theory, and simultaneously provide us with numerous and varied means of verification of its field-theoretic predictions. From it, a picture of a coherent and unified higher-geometric and -algebraic description scheme of two-dimensional field theories with a topological charge emerges in which the constructions central to the systematic development of conformal field theory, often beyond the scope of alternative methods, find their manageable geometrisation, {\it e.g.}, a methodical construction of orbifolds and orientifolds of known $\si$-models in terms of gerbes with an equivariant structure resp.\ a Jandl structure \cite{Gawedzki:2003pm,Schreiber:2005mi,Gawedzki:2007uz,Gawedzki:2010G}, extending naturally to the formulation of $\si$-models on spaces of orbits of the action of continuous groups in what can be thought of as a natural generalisation of the concept of a worldsheet orbifold of \Rcite{Frohlich:2009gb} (going back to the seminal papers \cite{Dixon:1985jw,Dixon:1986jc} of Dixon, Harvey, Vafa and Witten) using the gauge-symmetry defects of Refs.\,\cite{Suszek:2011,Suszek:2012ddg,Suszek:2013} determined by the data of the relevant equivariant structure ({\it cp} also \Rcite{Runkel:2008gr} for an early instantiation of the idea); explicit equivariant geometric quantisation \cite{Gawedzki:1987ak,Gawedzki:1999bq,Gawedzki:2002se,Gawedzki:2003pm,Gawedzki:2004tu} in terms of the Cheeger--Simons differential characters provided by gerbe theory, leading to a hands-on realisation of Segal's idea of functorial quantisation advanced in \Rcite{Segal:2002}, and to the discovery of a new species of Dirichlet branes (the so-called non-abelian branes) over fixed points of the action of an orbifold group \cite{Gawedzki:2004tu} (the latter were first noticed by Douglas and Fiol in Refs.\,\cite{Douglas:1998xa,Douglas:1999hq}); and even, somewhat surprisingly, the elucidation of the peculiar structure of the emergent spectral noncommutative geometry of the maximally symmetric D-branes on the target Lie group \cite{Recknagel:2006hp}, determined by the loop-mechanical deformation of the Dirac operator, and so also of the associated differential calculus, given by the superconformal current of the relevant super-WZW $\si$-model in the spirit of \Rcite{Frohlich:1993es}.\smallskip

Among the phenomena and constructions of the loop mechanics {\it not} covered (at least not in all generality) by gerbe theory to date, two stand out as particularly significant and hence pressing: 
\bit
\item[-] a rigorous and exhaustive treatment of purely loop-mechanical dualities, such as T-duality, with view -- among other things -- to the construction, by means of an adaptation of the aforementioned generalised worldsheet orbifolding procedure, of (classical) geometries modelled on riemannian geometries of fixed topology (of a toroidal principal bundle over a given base) only locally, and with the global structure of an `orbifold' with respect to a suitably defined action of -- instead of the standard diffeomorphism group of the model space $\,\bR^{\x n}\,$ -- the T-duality `group' of gerbe-theoretic $\si$-model dualities (with the group law captured by fusion of the corresponding T-duality bi-branes) determining the relevant `gluing' data -- these constructions, known under the name of T-folds \cite{Hull:2004in,Hull:2006qs}, would place us outside the paradigm of riemannian geometry;
\item[-] an extension of the hitherto successful formalism of gerbe theory to models with supersymmetry.
\eit
As for the former issue, we shall not have anything to say in the present work, except for the comment that its proper analysis calls for the application of the methods recently developed in Refs.\,\cite{Gawedzki:2012fu,Suszek:2012ddg} and is a subject of an ongoing research, to be reported shortly. It is the latter point that we want to tackle herein, with the intention of clarifying the fundamental concepts and working out the basic formal tools through a case study focused on a target superspace whose geometric simplicity, as reflected in the (trivial) topology and (high) symmetry, gives hope for a relatively straightforward separation of that which is peculiar to such theories, and hence truly novel, from the standard intricacies and technical complexities of a higher-geometric and -algebraic analysis of a low-dimensional field theory with a topological charge. Thus, our endeavour is meant to be a prelude to more advanced studies in the direction that it sets, preparing the ground for subsequent developments in which the complexity of the geometries considered will no longer obscure the basic mechanisms at play in a supersymmetric $\si$-model.\smallskip

The (pre)history of supersymmetry starts with the works of Miyazawa \cite{Miyazawa:1966mfa}, largely overlooked at an early stage of development of the idea and its associated mathematical formalism, as laid out in the later works of Gervais, Golfand, Volkov and Akulov \cite{Gervais:1971ji,Golfand:1971iw,Volkov:1972jx,Volkov:1973ix,Akulov:1974xz}, and -- in particular -- Wess and Zumino \cite{Wess:1973kz,Wess:1974tw} in which the theoretical concept was rediscovered and boosted in the direction of the at that time much promising and exciting applications in the model-building of high-energy physics. The related theory of supergeometry, based on the notion of a supermanifold, was worked out a little later by Berezin, Le\"ites, Schwarz and Voronov \cite{Berezin:1975,Schwarz:1984,Voronov:1984}, its geometric content clarified by the structure theorem of Gaw\c{e}dzki and Batchelor \cite{Gawedzki:1977pb,Batchelor:1979a}. These new concepts were assimilated and adapted by the string-theoretic community very early on, and gave rise to a plethora of consistent models free of the pathologies of their purely bosonic counterparts, of which we name only the original breakthrough models of the superstring due to Green and Schwarz \cite{Green:1983wt,Green:1983sg}, their higher-dimensional analogues for super-$p$-branes \cite{Achucarro:1987nc}, the celebrated anti-de Sitter superstring models of Refs.\,\cite{Metsaev:1998it,Arutyunov:2008if,Gomis:2008jt,Fre:2008qc,DAuria:2008vov} and the M-brane models of \Rcite{Bandos:1997ui,deWit:1998yu,Claus:1998fh}, as well as the superstring \cite{Bergshoeff:1985su} and su\-per\-mem\-brane \cite{Bergshoeff:1987cm} theories in curved supergravity backgrounds. It borders on impossible to do justice to a vast area of research such as this one and to recapitulate its development over the decades in a concise form adequate for our purposes, and so instead of doing this, we refer the interested Reader to the excellent reviews and introductory materials on the subject, {\it e.g.}, Refs.\,\cite{Weinberg:1999,Martin:1997ns} for an introduction to the physical, even phenomenological, aspects of the idea of supersymmetry, and Refs.\,\cite{Deligne:1999sgn,Freed:1999,Varadarajan:2004} for the more mathematically oriented mind looking in the same direction, as well as Refs.\,\cite{DeWitt:1992,Rogers:2007} for a gentle introduction to supergeometry. That which renders such a solution all the more apposite is the current somewhat uncertain phenomenological status of supersymmetry as a feature of fundamental interactions, which seems to imply that more weight should be attached to the motivation for the study of field theories exhibiting supersymmetry in an unbroken form than to the standard historical retrospective. In our case, the general motivation is of three-fold nature. On the one hand, there is a purely mathematical argument: Supergeometry, and in particular the theory of Lie supergroups, is a field of a robust mathematical development and it seems only natural to transplant the ideas and methods of (bosonic) higher geometry unto it with view to furthering its progress, especially that this, in the theoretical context in hand, naturally leads to the emergence of a variety of mathematical structures interesting in their own right, such as, {\it e.g.}, the Lie-$n$-(super)algebras and $L_\infty$-(super)algebras of Baez and Huerta \cite{Baez:2010ye,Huerta:2011ic} that correspond to classes in higher groups of the Chevalley--Eilenberg cohomology encountered in the construction of super-$\si$-models. On the other hand, there is a physical argument: It is the supersymmetric string theory in the distinguished supergravitational backgrounds of the anti-de Sitter type that forms the basis of one of the very few and important direct applications of string theory in the {\it predictive} description of observable phenomena involving strongly interacting elementary coloured particles (outside the perturbative r\'egime) {\it via} the conjectural AdS/CFT `correspondence' -- the super-$\si$-models of loop dynamics of relevance in this context, originally advanced by Metsaev and Tseytlin in \Rcite{Metsaev:1998it}, are precisely of the distinguished type described above as the corresponding supermanifolds with the anti-de Sitter space as the body are homogeneous spaces of certain Lie supergroups, {\it e.g.},
\qq\nn
&{\rm s}\bigl({\rm AdS}_5\x\bS^5\bigr)\cong{\rm SU}(2,2\,\vert\,4)/\bigl({\rm SO}(1,4)\x{\rm SO}(5)\bigr)\,,&\cr\cr
&{\rm s}\bigl({\rm AdS}_4\x\bS^7\bigr)\cong{\rm OSp}(8\,\vert\,4)/\bigl({\rm SO}(1,3)\x{\rm SO}(7)\bigr)\,,\quad\qquad{\rm s}\bigl({\rm AdS}_7\x\bS^4\bigr)\cong{\rm OSp}(6,2\,\vert\,4)/\bigl({\rm SO}(1,6)\x{\rm SO}(4)\bigr)\,,&
\qqq
and so developing new geometric tools for these models might shed some light on the fundamental nature of the still incompletely understood correspondence of much physical relevance. Finally, there is a mixed mathematical-physical argument: Given the successes of the gerbe-theoretic paradigm established for the bosonic two-dimensional $\si$-model with the topological charge, it is tempting to test its universality by attempting to adapt it to an environment in which cohomological mechanisms altogether different from the previously encountered sheaf-theoretic and purely de Rham ones are at work and demand geometrisation, namely, the Cartan--Eilenberg {\it supersymmetry-invariant} cohomology of superdifferential forms on a Lie supergroup resp.\ its homogeneous space.\medskip

Our choice of target supermanifolds to be studied, that is to say homogeneous spaces of Lie supergroups, has far-reaching field-theoretic consequences. In the simplest case of the super-Minkowskian spacetime\footnote{The notation will be clarified in the main text.} $\,{\rm sMink}(d,1\,\vert\,D_{d,1})$,\ the corresponding super-$\si$-models are simply super-counterparts of the bosonic WZW $\si$-models mentioned earlier \cite{Henneaux:1984mh}, and more generally they can be thought of as super-variants of gauged WZW $\si$-models, in conformity with the findings of \Rcite{Gawedzki:1988hq}. In the bosonic setting, there are simple geometric mechanisms that effect a quantisation of the topological charge and fix the relative normalisation of the topological and `metric' terms in the action functional. In the former case, and for a {\it compact} Lie group, Dirac's argument is usually adduced, which explicitly expresses integrality of the periods of the curvature of the gerbe whose holonomy along the embedded worldsheet defines the topological Wess--Zumino term of the $\si$-model action functional (while secretly capturing the coherence condition imposed upon the groupoid structure of the gerbe). In the latter case, it is the requirement of existence of a (bi-)chiral (centrally extended) loop-group symmetry induced by left- and right-regular translations on the group manifold, and hence also of the (bi-)chiral Virasoro symmetry obtained from it {\it via} the standard Sugawara construction, that does the job. In the supergeometric setting at hand, we are confronted with the following obstacles that get in our way if we try to imitate the bosonic scheme: The non-compactness and topological triviality of the target supermanifold, and of the underlying (super)symmetry group, render Dirac's argument ineffective, hence no quantisation of the topological charge is observed and the de Rham cohomology behind the topological term is as trivial as that of the bosonic body of the supermanifold\footnote{In general, this follows from a theorem by Kostant \cite{Kostant:1975}. In the cases studied, it can be checked directly.}. Hence, apparently, the super-$\si$-models of interest seem to have no non-trivial gerbe-theoretic content. Furthermore, the local symmetry fixing the relative normalisation of the two terms in the action functional of the super-$\si$-model, known as $\k$-symmetry \cite{deAzcarraga:1982njd,Siegel:1983hh,Siegel:1983ke}, while readily shown to have a simple geometric origin in the linearised (and further constrained) right-regular action of the Lie supergroup on itself, has a rather cumbersome and peculiar field-theoretic realisation in that it necessarily mixes the metric and topological (that is, gerbe-theoretic) components of the standard (Nambu--Goto resp.\ Polyakov) action functional and -- on top of that -- requires for the closure of its (commutator) algebra not only an enhancement by worldsheet diffeomorphisms (which is understandable in view of its origin and relation to the chiral symmetries of the bosonic WZW $\si$-model -- this is simply a super-instantiation of the Sugawara mechanism) but also the imposition of field equations of the super-$\si$-model \cite{McArthur:1999dy}, which seems to preclude its geometrisation in the form of an equivariant structure on the object geometrising the de Rham super-cocycle that determines the topological term of the action functional. A moment's thought reveals that both obstacles can and therefore ought to be circumnavigated, and it is the purpose of the present paper to demonstrate how to do it and to study the ramifications.\smallskip

The triviality of the de Rham cohomology does not imply -- in consequence of the same non-compactness of the supersymmetry group that kills it, but with it also the implications of the Cartan--Eilenberg theorem for the relation between the standard de Rham cohomology and its invariant version -- triviality of the supersymmetric ({\it i.e.}, supersymmetry-invariant) de Rham cohomology, and -- indeed -- the Green--Schwarz super-$(p+2)$-cocycles on the super-Minkowskian spacetime defining the Wess--Zumino terms of the respective super-$p$-brane super-$\si$-models bear witness to that. A simple argument due to Rabin and Crane \cite{Rabin:1984rm,Rabin:1985tv} then shows that the invariant de Rham cohomology actually encodes information on the nontrivial {\it topology} of a supermanifold of the same type as $\,{\rm sMink}(d,1\,\vert\,D_{d,1})\,$ ({\it i.e.}, modelled on the same vector bundle in the sense of the Gaw\c{e}dzki--Batchelor Theorem), namely, an orbifold of the super-Minkowskian spacetime by the natural action of the discrete Kosteleck{\'y}--Rabin supersymmetry group constructed in \Rcite{Kostelecky:1983qu} in the context of the supersymmetric lattice field theory. This implies that the Green--Schwarz super-$\si$-model should be understood as a theory of embeddings of the super-$p$-brane worldvolume in the topologically nontrivial supertarget, and puts the topological term of that model on equal footing with the topological term of the bosonic WZW $\si$-model with a compact (and topologically nontrivial) Lie-group target. This means, in particular, that we should look for an appropriate geometrisation of the Green--Schwarz super-$(p+2)$-cocycles that define the topological term. Following this line of reasoning to its logical conclusion, we readily realise that in the present context `appropriate' is equivalent to `supersymmetry-(left-)invariant', which simply means that we may reproduce the geometrisation procedure of cohomological descent that associates a (bosonic) $p$-gerbe with a standard de Rham $(p+2)$-cocycle (to be detailed shortly) as long as we ensure that each supermanifold obtained in the procedure and -- as part of it -- surjectively submersed onto $\,{\rm sMink}(d,1\,\vert\,D_{d,1})\,$ is equipped with a Lie-supergroup structure that projects, along the surjective submersion, to the original Lie-supergroup structure on $\,{\rm sMink}(d,1\,\vert\,D_{d,1})$,\ and -- finally -- that the superdifferential forms defined on these supermanifolds and employed in the said procedure are left-invariant with respect to the natural (left) action of the respective Lie supergroups on their support (that is, on themselves). The success of a (super)geometrisation project thus outlined hinges on two classic cohomological results that carry over from the bosonic world to the supergeometric setting (as demonstrated in App.\,\ref{app:LieAlgCohom}), to wit, the equivalence between the Cartan--Eilenberg invariant cohomology of the Lie (super)group and the Chevalley--Eilenberg cohomology of its Lie (super)algebra with values in the trivial module $\,\bR\,$ in conjunction with the correspondence between classes in the second cohomology group of the latter cohomology and equivalence classes of supercentral extensions of the Lie (super)algebra by that module. These results translate the original geometric problem of finding a surjective submersion over the original supermanifold equipped with a Lie-supergroup structure and such that the pullback of the original Cartan--Eilenberg super-cocycle to it trivialises in the corresponding Cartan--Eilenberg cohomology into a purely algebraic one: In a systematic procedure laid out by de Azc\'arraga {\it et al.} in \Rcite{Chryssomalakos:2000xd}, we identify various Cartan--Eilenberg super-2-cocycles engendered by the Green--Schwarz super-$(p+2)$-cocycles and associate with them supercentral extensions of the underlying super-Minkowskian algebra, subsequently demonstrated to integrate to supercentral extensions of the Lie supergroup $\,\bR(d,1\,\vert\,D_{d,1})\equiv{\rm sMink}(d,1\,\vert\,D_{d,1})\,$ on which the pullbacks of the respective super-$(p+2)$-cocycles trivialise partially, whereupon the procedure can be repeated with respect to these partial (supersymmetric) trivialisations. This leads to a family of so-called extended superspacetimes of the type first considered in \Rcite{Chryssomalakos:2000xd} which we then take to be {\it the} surjective submersions of the gerbe-theoretic geometrisation scheme. This basic idea is then reapplied at higher levels of Murray's geometrisation ladder \cite{Murray:1994db}, ultimately leading to the emergence of a new (super)geometric species -- the {\bf Green--Schwarz super-$p$-gerbe}, the central result of the work reported herein (explicited for $\,p\in\{0,1,2\}$).\smallskip

At this stage, the structural affinity with the bosonic WZW $\si$-model becomes a rich source of intuitions concerning anticipated properties of the newly constructed (super)geometric objects -- their verification seems to provide the right measure of evidence in support of our claim of naturality of the construction postulated in the paper. The first of these properties is the amenability of a {\it distinguished} realisation of the {\it rigid} (or {\it global}) supersymmetry of the super-$\si$-model under consideration to gauging, as reflected in the existence of an appropriate supersymmetry-equivariant structure on the associated super-$p$-gerbe, in conformity with the findings of Refs.\,\cite{Gawedzki:2010rn,Gawedzki:2012fu,Suszek:2011,Suszek:2012ddg,Suszek:2013}. Here, as before, `appropriate' means `supersymmetry-(left-)-invariant' but the concept has to be adapted to the changed circumstances in which the spaces on which the supersymmetry group acts are components of the nerve of the action groupoid of the group subject to gauging. This line of research will be developed in an upcoming paper (Part II). We emphasise once more that it provides us with a nontrivial consistency check of our main proposal. \smallskip

Finally, we come to the second apparent obstacle indicated above: the obstruction to the geometrisation of the gauge supersymmetry of the super-$\si$-model in the form of a full-fledged standard equivariant structure on the super-$p$-gerbe. The relevance of this issue follows from the field-theoretic r\^ole played by the supersymmetry, which is that of a mechanism effectively removing the fermionic Goldstone modes and thereby restoring an actual balance between the bosonic and fermionic physical degrees of freedom in the vaccum of the theory, and of an algebraic structure extending the worldvolume-diffeomorphism algebra. The absence of its analysis in any rigorous discussion of the higher-geometric content of the Green--Schwarz super-$\si$-models renders such a discussion fundamentally incomplete. The first problem with $\k$-symmetry, which is the mixing of the metric and gerbe-theoretic components of the supergeometric background, can be remedied easily by passing to an equivalent formulation of the super-$\si$-model, originally due to Hughes and Polchinski \cite{Hughes:1986dn}, in which an enlargement of the covariant configuration bundle (or the `space of lagrangean fields') is accompanied by a replacement of the metric term in the original (Nambu--Goto resp.\ Polyakov) action functional with the pullback, along the lagrangean embedding field, of a distinguished super-$(p+1)$-form on the enlarged target supermanifold ({\it cp} Refs.\,\cite{McArthur:1999dy,Gomis:2006wu}), with the topological term left unchanged, {\it i.e.}, pulled back from the original target supermanifold. The very structure of this reformulation suggests a straightforward extension of the geometrisation scheme to the new setting, and the latter turns out to be manifestly compatible with the corresponding realisation of $\k$-symmetry, as will be reported in Part II. The second problem is the intricate algebra behind $\k$-symmetry and the constraints that it introduces in the original higher-geometric picture that render the geometrisation of the symmetry a nontrivial task. While far from being fully understood, this construction lends additional and highly nontrivial support to the main claim of the present work, which is that the (super)geometrisation of the Green--Schwarz super-$(p+2)$-cocycles postulated hereunder should be regarded as the proper counterpart of the well-established geometrisation scheme for de Rham $(p+2)$-cocycles, to be considered in the setting of the supersymmetric supergeometry of homogeneous spaces of Lie supergroups. We shall address this issue extensively in Part II.

{\it Addendum:} The notion of the supergerbe, understood as a geometrisation of the Green--Schwarz super-$(p+2)$-cocycle, was discussed from a formal point of view by Fiorenza, Sati and Schreiber in \Rcite{Fiorenza:2013nha}. The Author is grateful to Urs Schreiber for kindly drawing his attention to that article.
\medskip

The paper is organised as follows:
\bit
\item In Section \ref{sec:Bose}, we recapitulate those elements of gerbe theory that become essential in subsequent analyses, and review the resulting canonical description of the bosonic two-dimensional $\si$-model with a topological term that the gerbe-theoretic approach naturally provides, with special emphasis on the geometric (and cohomological) structures that describe symmetries of the $\si$-model induced by automorphisms of the target space, and in particular those amenable to gauging; we complement the introductory part with a definition of a bundle 2-gerbe for the sake of handy reference in a later supersymmetric generalisation.
\item In Section \ref{sec:stensor}, we introduce the broad class of supergeometries of direct interest to us in the present work and its planned continuation. These are supermanifolds endowed with the structure of a homogeneous space of a Lie supergroup and a distinguished representative of a class in the corresponding Cartan--Eilenberg cohomology (of that Lie supergroup) that determines the topological term in the action functional of the supersymmetric $\si$-model to be studied, that is the Green--Schwarz super-$\si$-model that describes the geometrodynamics of standard super-$p$-branes. The Nambu--Goto formulation of the super-$\si$-model and its building blocks are recalled and discussed at some length.
\item In Section \ref{sec:sMinktarget}, we zoom in on the specific backgrounds of super-$p$-brane propagation that constitute the main subject of the study reported herein, to wit, the super-Minkowskian backgrounds equipped with the Green--Schwarz super-$(p+2)$-cocycles that we list, alongside their non-supersymmetric primitives used in subsequent analysis.
\item Section \ref{sec:GSgerbe} contains the main proposals and results of the study reported herein: the geometrisation scheme for the previously introduced Green--Schwarz super-$(p+2)$-cocyles (for $\,p\in\{0,1,2\}$), resulting in the general definition (Defs.\,\ref{def:CaEs0g} and \ref{def:CaEs1g}) and explicit construction (Defs.\,\ref{def:s0gerbe}, \ref{def:s1gerbe} and \ref{def:s2gerbe} and Props.\,\ref{prop:s0gerbe} and \ref{prop:s1gerbe}) of the corresponding Green--Schwarz super-$p$-gerbes on the super--Minkowskian spacetime, understood as examples of a more general notion of a Cartan--Eilenberg super-$p$-gerbe from Defs.\,\ref{def:CaEs0g}, \ref{def:CaEs1g} and, finally, Def.\,\ref{def:CaEspg}. 
\item Section \ref{ref:CandO} summarises the main constructions and findings reported in the present paper and indicates directions of potential future research based upon them. 
\item Appendices \ref{app:conv}--\ref{app:homformsgroup} contain introductory material and certain ancillary results on Lie superalgebras and their (Chevalley--Eilenberg) cohomology, as well as some technical proofs of statements articulated in the main text.
\eit

\bigskip

\noindent{\bf Acknowledgements:}  This work is a humble tribute to a Friend and Teacher, Professor Krzysztof Gaw\c{e}dzki, on the occasion of His seventieth birthday. Any attempt at a concise verbalisation of a meaningful and yet -- of necessity -- sufficiently formal acknowledgement of His r\^ole in the scientific and extra-scientific formation of the author is bound to fail short of the sincere intention, and so shall be omitted.

This leaves the author with the pleasurable obligation of expressing a deep and true thankfulness to his Colleagues and Friends at and outside the Department of Mathematical Methods in Physics of the Faculty of Physics at the University of Warsaw for creating and maintaining an inspiring atmosphere of scientific work and human interaction in which the spirit of the late Professor Krzysztof Maurin finds its very fitting incarnation, as well as for their understanding of the author's other passions, including that for the defense of fundamental civil rights and liberties of his fellow citizens, in which understanding their human sensitivity and decency is congenially reflected -- a rare source of satisfaction and relief in these sad times.

Finally, the author cannot but acknowledge, without the least gratitude but with, instead, deepest civil despondency and a poignant awareness of a rapidly growing cultural alienation within a largely indifferent and populism-prone society of the post-truth era, the steadfast and disquietingly methodical efforts on the part of the current pro-authoritarian government of the Republic of Poland, of a truly bewildering intensity and scope and devastating sociological ramifications, to keep him, alongside many other active members of the Polish civil society, as occupied -- be it with acts of civil disobedience, stubborn street protests, confrontation with the party-controlled prosecution, the incessant (and sadly ineffective) write-up of petitions and letters of grievance or various activities aimed at raising social consciousness of the government's heinous wrongdoing and the complex context of the current civilisational devolution -- and consequently as withdrawn from research as a passion-driven individual with a non-trivial charge of civil sensitivity and a rich historical memory of the villainy of totalitarian r\'egimes can ever be made by a government with the intelectual deficiencies, cultural ignorance, moral depravity and documented propensity for increasingly frequent abysmal paroxysms of barbarism pure of this one.

\bigskip

\section{Recapitulation of the gerbe theory behind the bosonic $\si$-model}\label{sec:Bose}

In this opening section, we consider the monophase bosonic two-dimensional non-linear $\si$-model with a spacetime $\,\Si$,\ termed the \textbf{worldsheet} and given by a closed two-dimensional manifold $\,\Si\,$ with a metric induced along a section of the covariant configuration bundle $\,\Si\x M\too\Si\,$ whose fibre $\,M$,\ termed the \textbf{target space}, is a differentiable manifold of class\footnote{Strictly speaking, formulation of the $\si$-model requires the target space to be of class $\,C^2\,$ only, or even \emph{patchwise} $\,C^2$,\ but we shall assume higher degree of smoothness for the sake of simplicity.} $\,C^\infty\,$ endowed with a metric tensor $\,\txg\in\G(\sfT^*M\ox^{\rm sym}_{M,\bR}\sfT^*M)$.\ The model is defined by an action functional $\,S_\si\,$ with domain 
\qq\nn
C^\infty(\Si,M)\equiv[\Si,M] 
\qqq
whose stationary points are (generalised) harmonic maps $\,x\in[\Si,M]$.

A rigorous formulation of the monophase $\si$-model calls for an additional structure on $\,M$,\ to wit, an abelian bundle gerbe $\,\cG\,$ (with connection and curving) of curvature $\,\txH\equiv\curv(\cG)\in Z^3_{\rm{dR}}(M)\,$ with periods in $\,2\pi\bZ$.\ The two tensors $\,\txg\,$ and $\,\txH\,$ are related by the requirement of the vanishing of the Weyl anomaly\footnote{The anomaly is usually computed and presented as a perturbative series in the string tension $\,\a'$.} of the $\si$-model ($\nabla_{\rm{LC}}^\txg\,$ is the Levi-Civita connection of $\,\txg$)
\qq\nn
R_{\mu\nu}(\nabla_{\rm{LC}}^\txg)-\tfrac{1}{4}\,\left(\txg^{-1}\right)^{\a\g}\,\left(\txg^{-1}\right)^{\b\d}\,\txH_{\mu\a\b}\,\txH_{\nu\g\d}+O(\a')=0\,,
\qqq
a prerequisite of a non-anomalous realisation of the conformal (gauge) symmetry of the classical field theory in the quantum r\'egime. The metric on $\,M\,$ determines -- through induction of the first fundamental form $\,x^*\txg\,$ on $\,\Si\,$ along $\,x\,$ -- the so-called metric term in $\,S_\si$,\ which we choose -- with hindsight -- to write in the Nambu--Goto form\footnote{There exists an alternative, and classically essentially equivalent form of the metric term, termed the Polyakov form, which, however, will not be employed in the present work.} 
\qq\nn
S_{\rm{metr, NG}}[x]:=\int_\Si\,\Vol(\Si)\,\sqrt{\bigl\vert\det_{(2)}\,\bigl(x^*\txg\bigr)\bigr\vert}\,,
\qqq
whereas the gerbe defines the topological Wess--Zumino term that exponentiates to a Cheeger--Simons differential character $\,\Hol_\cG\,$ termed the (surface) holonomy of the gerbe $\,\cG\,$ (and computed along the map $\,x$), altogether giving rise to a well-defined Dirac--Feynman amplitude (written for $\,\hbar=1$)
\qq\label{eq:sibos}
\cA_{\rm{DF}}[x]:=\exp\left(\sfi\,S_\si[x]\right)\equiv\exp\left(\sfi\,S_{\rm{metr, NG}}[x]\right)\cdot\Hol_\cG(x)\,.
\qqq
The holonomy can most concisely be described as the image of the isoclass of the flat pullback gerbe $\,x^*\cG\,$ under the composite isomorphism\footnote{The isomorphism can readily be derived by examining a sheaf-theoretic description of the flat gerbe (note that every gerbe over $\,\Si\,$ is flat for dimensional reasons) and following the long exact sequence in the sheaf cohomology of $\,\Si\,$ induced by the standard exponential short exact sequence $\,0\too\unl\bZ\xrightarrow{\ 2\pi\cdot\ }\unl\bR\xrightarrow{\ \exp(\sfi\,\cdot)\ }\unl\uj\too 0\,$ of sheaves of locally constant maps on $\,\Si$.}
\qq\nn
\cW^3(\Si;0)\cong\vH^2\bigl(\Si,\unl\uj\bigr)\cong\uj
\qqq
between the group $\,\cW^3(\Si;0)\,$ of isoclasses of flat gerbes over $\,\Si\,$ (with the class of the tensor product of representatives as the binary operation) and $\,\uj\,$ (we assume $\,\Si\,$ to be connected). The intermediate group is the second \Cv ech-cohomology group of $\,\Si\,$ with values in the sheaf of constant maps to $\,\uj$.

It stands to reason that a structural (non-na\"ive) supersymmetrisation of the $\si$-model affects the various components $\,\txg,\cG\,$ of the geometric backgound of the loop propagation. While the candidate extension of the tensor $\,\txg\,$ under such supersymmetrisation is not -- as shall be elucidated shortly -- difficult to conceive and quantify, at least in geometrically simple circumstances, it is not at all clear even how to approach the supergeometric counterpart of $\,\cG$.\ Therefore, it seems apposite to first present a number of equivalent descriptions and fundamenal properties of the gerbe and its field-theoretic guises with view to establishing a vast scope of constructions from which to choose those that generalise to the supergeometric setting naturally and usefully. Below, we demonstrate the many faces of the gerbe $\,\cG\,$ with a fixed {\bf curvature} $\,\txH\in Z^3_{\rm{dR}}(M,2\pi\bZ)$,\ to be understood as a geometrisation of the de Rham 3-cocycle $\,\txH\,$ on the base $\,M$,\ much in the same manner as a line bundle (with connection) is to be understood as a geometrisation of the de Rham 2-cocycle $\,\txF\in Z^2_{\rm{dR}}(M,2\pi\bZ)\,$ of its curvature.

\subsection{Gerbe theory in a nutshell}

The point of departure of our recapitulation is the cohomological description of the gerbe. Thus, any local trivialisation of the (co)homology of $\,M\,$ yields a presentation of $\,\cG\,$ in terms of its sheaf-theoretic data: Given a good open cover\footnote{A good open cover is an open cover $\,\{\cO_i\}_{i\in\xcI}\,$ with all non-empty (finite) multiple intersections $\,\cO_{i_1}\cap\cO_{i_2}\cap\cdots\cap\cO_{i_N},\ i_1,i_2,\ldots,i_N\in\xcI,\ N\in\bN^\x\,$ contractible. In the light of Weil's proof of the Weil--de Rham Theorem, reported in \Rcite{Weil:goc1952}, such a cover always exists on a differentiable manifold of class $\,C^2$,\ a property implicitly assumed in constructing the $\si$-model.} $\,\cO_M:=\{\cO_i\}_{i\in\xcI}\,$ ($\xcI\,$ is an index set and we introduce, for any $\,N\in\bN^\x$,\ sets $\,\xcI_N:=\{\ (i_1,i_2,\ldots,i_N)\in\xcI^{\x N}\ |\ \exists\ \cO_{i_1 i_2\cdots i_N}\equiv\cO_{i_1}\cap\cO_{i_2}\cap\cdots\cap\cO_{i_N}\neq\emptyset\ \}$), the gerbe is identified with a class
\qq\nn
\left[(B_i,A_{jk},g_{lmn})_{i\in\xcI,\ (j,k)\in\xcI_2,\ (l,m,n)\in\xcI_3}\right]
\qqq
of a \Cv ech--de Rham 2-cocycle trivialising the de Rham 3-cocycle $\,\txH\,$ over $\,\cO_M$,\ with data $\,(B_i,A_{jk},g_{lmn})\in\Om^2(\cO_i)\x\Om^1(\cO_{jk})\x C^\infty(\cO_{lmn},\uj)$ defined by the relations\footnote{Here, and in what follows, in the local sheaf-theoretic description of a (super-)$p$-gerbe, the inverses on functions are implicitly point-wise inverses, and \emph{not} the functional inverses.}
\qq\nn
&\sfd B_i=\txH\rstr_{\cO_i}\,,\qquad\sfd A_{jk}=(B_k-B_j)\rstr_{\cO_{jk}}\,,\qquad\sfi\,\sfd\log g_{lmn}=(A_{mn}-A_{ln}+A_{lm})\rstr_{\cO_{lmn}}\,,&\cr\cr
&\bigl(g_{pqr}\cdot g_{oqr}^{-1}\cdot g_{opr}\cdot g_{opq}^{-1}\bigr)\rstr_{\cO_{opqr}}=1\,,\quad(o,p,q,r)\in\xcI_4&
\qqq
up to redefinitions, for arbitrary $\,(C_i,h_{jk})\in\Om^1(\cO_i)\x C^\infty(\cO_{jk},\uj)$,
\qq
(B_i,A_{jk},g_{lmn})\longmapsto\bigl(B_i+\sfd C_i,A_{jk}+(C_k-C_j)\rstr_{\cO_{jk}}-\sfi\,\sfd\log h_{jk},g_{lmn}\cdot \bigl(h_{mn}^{-1}\cdot h_{ln}\cdot h_{lm}^{-1}\bigr)\rstr_{\cO_{lmn}}\bigr)\,,\cr\label{eq:1iso}
\qqq
in the 2nd real Deligne--Beilinson hypercohomology group $\,\bH^2\left(M,\cD(2)^\bullet \right)$,\ {\it i.e.}, the cohomology of the total complex of the bicomplex formed by an extension of the bounded Deligne complex 
\qq\nn
\cD(n)^\bullet\quad\equiv\quad \unl\uj_M\xrightarrow{\ \tfrac{1}{\sfi}\,\sfd\log\ }\unl{\Om^1(M)}\xrightarrow{\ \sfd\ }\unl{\Om^2(M)}\xrightarrow{\ \sfd\ }\cdots\xrightarrow{\ \sfd\ }\unl{\Om^n(M)}
\qqq 
of sheaves of locally smooth maps and $p$-forms (for $\,p\in\ovl{1,n}\,$ with, in the present case, $\,n=2$) in the direction of the \Cv ech cohomology associated with $\,\cO_M$,\ {\it cp}\ \Rcite{Johnson:2003}. Of course, a given gerbe may -- just like a line bundle -- trivialise over an open cover $\,\cO_M\,$ that is not good in the sense specified above -- in any event, we call the latter a trivialising open cover in the present context.

The gerbe may also, and equivalently, be realised as a purely geometric object 
\qq\nn
\cG=(\sfY M,\pi_{\sfY M},\txB,L,\pi_L,\nabla_L,\mu_L)\,,
\qqq
known also as the bundle gerbe: Given an arbitrary surjective submersion 
\qq\nn
\pi_{\sfY M}\ :\ \sfY M\too M
\qqq 
on whose total space there exists a globally smooth primitive 
\qq\nn
\txB\in\Om^2(\sfY M)
\qqq 
of the pullback 
\qq\nn
\pi_{\sfY M}^*\txH=\sfd\txB 
\qqq
(termed the {\bf curving} of the gerbe), we erect, over the double fibred product $\,\sfY^{[2]}M\equiv\sfY M\x_M\sfY M\,$ described by the commutative diagram
\qq\nn
\alxydim{@C=.75cm@R=1cm}{ & \sfY^{[2]}M \ar[rd]^{\pr_2} \ar[ld]_{\pr_1} & \\ \sfY M \ar[rd]_{\pi_{\sfY M}} & & \sfY M \ar[ld]^{\pi_{\sfY M}} \\ & M & }\,,
\qqq
a principal $\bC^\x$-bundle 
\qq\nn
\alxydim{@C=.75cm@R=1cm}{\bC^\x \ar[r] & L \ar[d]^{\pi_L} \\ & \sfY^{[2]}M }
\qqq 
with a principal $\bC^\x$-connection (termed the {\bf connection} of the gerbe and represented by the covariant derivative) $\,\nabla_L\,$ of curvature 
\qq\nn
\curv(\nabla_L)=\bigl(\pr_2^*-\pr_1^*\bigr)\txB\,,
\qqq
endowed with a fibrewise groupoid structure, {\it i.e.}, a connection-preserving principal-bundle isomorphism\footnote{The tensor product $\,L_1\ox L_2\,$ of principal $\bC^\x$-bundles $\,L_\a,\ \a\in\{1,2\}\,$ over a common base $\,X\,$ is defined, after \Rcite{Brylinski:1993ab}, as the (principal) bundle $\,(L_1\x_X L_2)/\bC^\x\,$ associated with $\,L_1\,$ through the defining $\bC^\x$-action on $\,L_2$,\ to be denoted by $\,\ract$,\ {\it cp} Remark \ref{rem:tensprinc} for details.\label{foot:Cxprintens}} (termed the {\bf groupoid structure})  
\qq\nn
\mu_L\ :\ \pr_{1,2}^*L\ox\pr_{2,3}^*L\xrightarrow{\ \cong\ }\pr_{1,3}^*L
\qqq
over the triple fibred product $\,\sfY^{[3]}M\equiv\sfY M\x_M\sfY M\x_M\sfY M\,$ described by the commutative diagram
\qq\nn
\alxydim{@C=1.5cm@R=1cm}{ & \sfY^{[3]}M \ar[rd]^{\pr_3} \ar[d]_{\pr_2} \ar[ld]_{\pr_1} & \\ \sfY M \ar[rd]_{\pi_{\sfY M}} & \sfY M \ar[d]_{\pi_{\sfY M}} & \sfY M \ar[ld]^{\pi_{\sfY M}} \\ & M & }\,,
\qqq
(with its canonical projections $\,\pr_{i,j}\equiv(\pr_i,\pr_j)\ :\ \sfY^{[3]}M\too \sfY^{[2]}M,\ (i,j)\in\{(1,2),(2,3),(1,3)\}$),\ subject to the associativity constraint 
\qq\label{eq:mugrpd}
\pr_{1,2,4}^*\mu_L\circ(\id_{\pr_{1,2}^*L}\ox\pr_{2,3,4}^*\mu_L)=\pr_{1,3,4}^*\mu_L\circ(\pr_{1,2,3}^*\mu_L\ox \id_{\pr_{3,4}^*L})
\qqq
over the quadruple fibred product $\,\sfY^{[4]}M\equiv\sfY M\x_M\sfY M\x_M\sfY M\x_M\sfY M\,$ described by the commutative diagram
\qq\nn
\alxydim{@C=1.5cm@R=1cm}{ & & \sfY^{[4]}M \ar[rrd]^{\pr_4} \ar[rd]_{\pr_3} \ar[ld]^{\pr_2} \ar[lld]_{\pr_1} & & \\ \sfY M \ar[rrd]_{\pi_{\sfY M}} & \sfY M \ar[rd]^{\pi_{\sfY M}} & & \sfY M \ar[ld]_{\pi_{\sfY M}}  & \sfY M \ar[lld]^{\pi_{\sfY M}} \\ & & M & & }\,,
\qqq
(with its canonical projections $\,\pr_{i,j,k}\equiv(\pr_i,\pr_j,\pr_k)\ :\ \sfY^{[4]}M\too \sfY^{[3]}M\,$ and $\,\pr_{i,j}\equiv(\pr_i,\pr_j)\ :\ \sfY^{[4]}M\too \sfY^{[2]}M,\ i,j\in\{1,2,3,4\}$), {\it cp}\ Refs.\,\cite{Murray:1994db,Murray:1999ew,Stevenson:2000wj}.

Equivalence between the two pictures: the cohomological one and the geometric one is established, going in one direction, with the help of the construction of the nerve of the trivialising open cover $\,\cO_M$,\ whose components become the $M$-fibred powers of the surjective submersion $\,\bigsqcup_{i\in\xcI}\,\cO_i\too M\,$ over which the various collections of local data define smooth geometric objects (for the standard differentiable structure of a disjoint union of manifolds), and, going in the opposite direction, with the help of local sections of the various surjective submersions employed in the geometric description: $\,\pi_{\sfY M},\ \pi_L\,$ and those derived from them, providing us with local data of the geometric objects $\,\txB,\nabla_L\,$ and $\,\mu_L$.\ Thus, the geometric objects determined by the \Cv ech--Deligne 2-cocycle $\,(B_i,A_{ij},g_{ijk})_{i,j,k\in\xcI}\,$ are 
\qq\nn
&\pi_{\sfY M}\ :\ \bigsqcup_{i\in\xcI}\,\cO_i\too M\ :\ (x,i)\longmapsto x\,,\qquad\txB\rstr_{\cO_i}=B_i\,,&\cr\cr &\pi_L=\pr_1\ :\ L=\bigl(\bigsqcup_{(j,k)\in\xcI_2}\,\cO_{jk}\bigr)\x\bC^\x\too\bigsqcup_{(j,k)\in\xcI_2}\,\cO_{jk}\equiv\sfY^{[2]}M\,,\qquad\nabla_L\rstr_{\cO_{jk}}=\sfd+\tfrac{1}{\sfi}\,A_{jk}\,,&\cr\cr
&\mu_L\bigl((x,l,m,z_1)\ox(x,m,n,z_2)\bigr)=\bigl(x,l,n,g_{lmn}(x)\cdot z_1\cdot z_2\bigr)\,,\quad (x,l,m,n)\in\bigsqcup_{(i,j,k)\in\xcI_3}\,\cO_{ijk}\equiv\sfY^{[3]}M\,.&
\qqq
Conversely, given geometric data $\,(\sfY M,\pi_{\sfY M},\txB,L,\pi_L,\nabla_L,\mu_L)\,$ and a choice of an open cover $\,\{\cO_i\}_{i\in\xcI}\,$ of $\,M\,$ with local sections $\,\si_i\ :\ \cO_i\too\sfY M\,$ giving rise to local sections $\,\si_{i_1 i_2\ldots i_N}\equiv(\si_{i_1},\si_{i_2},\ldots,\si_{i_N})\ :\ \cO_{i_1 i_2\ldots i_N}\too\sfY^{[N]}M\,$ and sufficiently fine for the sets $\,\si_{ij}(\cO_{ij})\subset\sfY^{[2]}M\,$ to support flat (unital) local sections $\,s_{ij}=s^{-1}_{ji}\circ\t\ :\ \si_{ij}(\cO_{ij})\too L$,\ with $\,\t\ :\ \sfY^{[2]}M\too\sfY^{[2]}M\ :\ (y_1,y_2)\longmapsto(y_2,y_1)$,\ we define local data
\qq\nn
B_i=\si_i^*\txB\,,\qquad A_{ij}\ox s_{ij}\circ\si_{ij}=\sfi\,\si_{ij}^*(\nabla_L s_{ij})\,,\qquad\mu\bigl(s_{ij}\circ\si_{ij}\ox s_{jk}\circ\si_{jk}\bigr)=(s_{ik}\circ\si_{ik})\ract g_{ijk}\,.
\qqq
Under the correspondence, the cohomological equivalence relation behind the definition of the class $\,\left[(B_i,A_{jk},g_{lmn})_{i\in\xcI,\ (j,k)\in\xcI_2,\ (l,m,n)\in\xcI_3}\right]\,$ translates into the notion of an isomorphism between bundle gerbes: Given two such gerbes $\,\cG_\a=(\sfY_\a M,\pi_{\sfY_\a M},\txB_\a,L_\a,\pi_{L_\a},\nabla_{L_\a},\mu_{L_\a}),\ \a\in\{1,2\}$,\ we call them {\bf 1-isomorphic} if there exists a sextuple 
\qq\nn
\Phi=(\sfY\sfY_{1,2}M,\pi_{\sfY\sfY_{1,2}M},E,\pi_E,\nabla_E,\a_E)
\qqq
termed a {\bf 1-isomorphism} and composed of a surjective submersion 
\qq\nn
\pi_{\sfY\sfY_{1,2}M}\ :\ \sfY\sfY_{1,2}M\too\sfY_1 M\x_M\sfY_2 M\equiv\sfY_{1,2}M
\qqq
over the fibred product $\,\sfY_{1,2}M$,\ described by the commutative diagram
\qq\nn
\alxydim{@C=.75cm@R=1cm}{ & \sfY_{1,2}M \ar[rd]^{\pr_2} \ar[ld]_{\pr_1} & \\ \sfY_1 M \ar[rd]_{\pi_{\sfY_1 M}} & & \sfY_2 M \ar[ld]^{\pi_{\sfY_2 M}} \\ & M & }\,;
\qqq 
of a principal $\bC^\x$-bundle
\qq\nn
\alxydim{@C=.75cm@R=1cm}{\bC^\x \ar[r] & E \ar[d]^{\pi_E} \\ & \sfY\sfY_{1,2}M }
\qqq 
with a principal $\bC^\x$-connection $\,\nabla_E\,$ of curvature 
\qq\nn
\curv(\nabla_E)=\pi_{\sfY\sfY_{1,2}M}^*\bigl(\pr_2^*\txB_2-\pr_1^*\txB_1\bigr)\,,
\qqq
and of a connection-preserving principal-bundle isomorphism
\qq\nn
\a_E\ :\ \pi_{\sfY\sfY_{1,2}M}^{\x 2\,*}\pr_{1,3}^*L_1\ox\pr_2^*E\xrightarrow{\ \cong\ }\pr_1^*E\ox\pi_{\sfY\sfY_{1,2}M}^{\x 2\,*}\pr_{2,4}^*L_2
\qqq
over the fibred product $\,\sfY^{[2]}\sfY_{1,2}M=\sfY\sfY_{1,2}M\x_M\sfY\sfY_{1,2}M\,$ described by the commutative diagram
\qq\nn
\alxydim{@C=.75cm@R=1cm}{ & \sfY^{[2]}\sfY_{1,2}M \ar[rd]^{\pr_2} \ar[ld]_{\pr_1} & \\ \sfY\sfY_{1,2}M \ar[rd]_{\pi_{\sfY_1 M}\circ\pr_1\circ\pi_{\sfY\sfY_{1,2}M}\quad} & & \sfY\sfY_{1,2}M \ar[ld]^{\quad\pi_{\sfY_2 M}\circ\pr_2\circ\pi_{\sfY\sfY_{1,2}M}} \\ & M & }\,,
\qqq 
subject to the coherence constraint expressed by the commutative diagram of connection-pre\-serv\-ing principal-bundle isomorphisms
\qq
\alxydim{@C=.15cm@R=1.5cm}{ & \pi_{1,2}^*\pr_{1,3}^*L_1\ox\pi_{2,3}^*\pr_{1,3}^*L_1\ox\pr_3^*E \ar[rd]^{\qquad\pi_{1,2,3}^*\pr_{1,3,5}^*\mu_{L_1}\ox\id_{\pr_3^*E}} \ar[ld]_{\id_{\pi_{1,2}^*\pr_{1,3}^*L_1}\ox\pr_{2,3}^*\a_E\qquad} & \\ \pi_{1,2}^*\pr_{1,3}^*L_1\ox\pr_2^*E\ox\pi_{2,3}^*\pr_{2,4}^*L_2 \ar[d]_{\pr_{1,2}^*\a_E\ox\id_{\pi_{2,3}^*\pr_{2,4}^*L_2}} & & \pi_{1,3}^*\pr_{1,3}^*L_1\ox\pr_3^*E \ar[d]^{\pr_{1,3}^*\a_E} \\ \pr_1^*E\ox\pi_{1,2}^*\pr_{2,4}^*L_2\ox\pi_{2,3}^*\pr_{2,4}^*L_2 \ar[rr]_{\id_{\pr_1^*E}\ox\pi_{1,2,3}^*\pr_{2,4,6}^*\mu_{L_2}} & & \pr_1^*E\ox\pi_{1,3}^*\pr_{2,4}^*L_2 }\cr\label{diag:grb1isocoh}
\qqq
over the fibred product $\,\sfY^{[3]}\sfY_{1,2}M\equiv\sfY\sfY_{1,2}M\x_M\sfY\sfY_{1,2}M\x_M\sfY\sfY_{1,2}M\,$ described by the commutative diagram
\qq\nn
\alxydim{@C=4cm@R=1cm}{ & \sfY^{[3]}\sfY_{1,2}M \ar[rd]^{\pr_3} \ar[d]_{\pr_2} \ar[ld]_{\pr_1} & \\ \sfY\sfY_{1,2}M \ar[rd]_{\pi_{\sfY_1 M}\circ\pr_1\circ\pi_{\sfY\sfY_{1,2}M}\qquad} & \sfY\sfY_{1,2}M \ar[d]_{\pi_{\sfY_2 M}\circ\pr_2\circ\pi_{\sfY\sfY_{1,2}M}} & \sfY\sfY_{1,2}M \ar[ld]^{\qquad\pi_{\sfY_1 M}\circ\pr_1\circ\pi_{\sfY\sfY_{1,2}M}} \\ & M & }\,,
\qqq
with
\qq\nn
&\pi_{i,j}=\pi_{\sfY\sfY_{1,2}M}^{\x 2}\circ\pr_{i,j}\,,\quad(i,j)\in\{(1,2),(2,3),(1,3)\}\,,&\cr\cr &\pi_{1,2,3}=\pi_{\sfY\sfY_{1,2}M}\x\pi_{\sfY\sfY_{1,2}M}\x\pi_{\sfY\sfY_{1,2}M}\,.&
\qqq
In view of the results obtained in \Rcite{Waldorf:2007mm}, we may always assume the surjective submersion of the 1-isomorphism to be of the distinguished form $\,\pi_{\sfY\sfY_{1,2}M}=\id_{\sfY_{1,2}M}$,\ which we do in what follows unless expressly stated otherwise. The situation just described is concisely denoted as 
\qq\nn
\Phi\ :\ \cG_1\xrightarrow{\ \cong\ }\cG_2\,.
\qqq
In fact, the transformation \eqref{eq:1iso} is left unchanged by secondary redefinitions 
\qq\nn
(C_i,h_{jk})\longmapsto\bigl(C_i-\sfi\,\sfd\log f_i,h_{jk}\cdot\bigl(f_k^{-1}\cdot f_j\bigr)\rstr_{\cO_{jk}}\bigr)\,,
\qqq
which indicates the existence of isomorphisms between 1-isomorphisms, or {\bf 2-isomorphisms}, with local data $\,[(f_i)_{i\in\xcI}]\,$ (defined up to local constants). In the geometric language, and for a given pair of 1-isomorphisms $\,\Phi_\b=(\sfY^\b\sfY_{1,2}M,\pi_{\sfY^\b\sfY_{1,2}M},E_\b,\pi_{E_\b},\nabla_{E_\b},\a_{E_\b}),\ \b\in\{1,2\}\,$ between bundle gerbes $\,\cG_\a=(\sfY_\a M,\pi_{\sfY_\a M},\txB_\a,L_\a,\pi_{L_\a},\nabla_{L_\a},\mu_{L_\a}),\ \a\in\{1,2\}$,\ a 2-isomorphism is represented\footnote{Strictly speaking, we should consider classes of such triples with respect to the following equivalence relation: $\,(\sfY_1\sfY^{1,2}\sfY_{1,2}M,\pi_{\sfY_1\sfY^{1,2}\sfY_{1,2}M},\b_1)\sim(\sfY_2\sfY^{1,2}\sfY_{1,2}M,\pi_{\sfY_2\sfY^{1,2}\sfY_{1,2}M},\b_2)\,$ iff there exist surjective submersions $\,\pi_\a\ :\ Z\too\sfY_\a\sfY^{1,2}\sfY_{1,2}M,\ \a\in\{1,2\}\,$ with the property $\,\pi_{\sfY_1\sfY^{1,2}\sfY_{1,2}M}\circ\pi_1=\pi_{\sfY_2\sfY^{1,2}\sfY_{1,2}M}\circ\pi_2$,\ and such that $\,\pi_1^*\b_1=\pi_2^*\b_2$.} by a triple
\qq\nn
\varphi=(\sfY\sfY^{1,2}\sfY_{1,2}M,\pi_{\sfY\sfY^{1,2}\sfY_{1,2}M},\b)
\qqq
composed of a surjective submersion
\qq\nn
\pi_{\sfY\sfY^{1,2}\sfY_{1,2}M}\ :\ \sfY\sfY^{1,2}\sfY_{1,2}M\too\sfY^1\sfY_{1,2}M\x_{\sfY_{1,2}M}\sfY^2\sfY_{1,2}M\equiv\sfY^{1,2}\sfY_{1,2}M
\qqq
and a connection-preserving principal-bundle isomorphism
\qq\nn
\b\ :\ \pi_{\sfY\sfY^{1,2}\sfY_{1,2}M}^*\pr_1^*E_1\xrightarrow{\ \cong\ }\pi_{\sfY\sfY^{1,2}\sfY_{1,2}M}^*\pr_2^*E_2
\qqq
subject to the coherence constraint expressed by the commutative diagram of connection-preserving principal-bundle isomorphism
\qq\label{diag:betacohalpha}
\alxydim{@C=2.5cm@R=1.75cm}{ p_{1,1}^*L_1\ox\pi_{1,2}^*E_1 \ar[r]^{\pi_1^{\x 2\,*}\a_{E_1}} \ar[d]_{\id_{p_{1,1}^*L_1}\ox\pr_2^*\b} & \pi_{1,1}^*E_1\ox p_{2,1}^*L_2 \ar[d]^{\pr_1^*\b\ox\id_{p_{2,1}^*L_2}} \\ p_{1,1}^*L_1\ox\pi_{2,2}^*E_2\equiv p_{1,2}^*L_1\ox\pi_{2,2}^*E_2 \ar[r]_{\pi_2^{\x 2\,*}\a_{E_2}} & \pi_{2,1}^*E_2\ox p_{2,1}^*L_2\equiv\pi_{2,1}^*E_2\ox p_{2,2}^*L_2 }
\qqq 
over $\,\sfY\sfY^{1,2}\sfY_{1,2}M\x_M\sfY\sfY^{1,2}\sfY_{1,2}M$,\ with 
\qq\nn
&\pi_i=\pr_i\circ\pi_{\sfY\sfY^{1,2}\sfY_{1,2}M}\,,\qquad\pi_{j,k}=\pi_j\circ\pr_k\,,\quad i,j,k\in\{1,2\}\,,&\cr\cr
&p_{l,m}=\pr_l\circ\pi_{\sfY^m\sfY_{1,2}M}\circ\pi_m\x\pr_l\circ\pi_{\sfY^m\sfY_{1,2}M}\circ\pi_m\,,\quad l,m\in\{1,2\}\,.&
\qqq
We denote the 2-isomorphism as
\qq\nn
\varphi\ :\ \Phi_1\xLongrightarrow{\ \cong\ }\Phi_2\,.
\qqq
For details of the correspondence indicated above, consult, {\it e.g.}, Refs.\,\cite{Murray:1999ew,Gawedzki:2002se} and \cite{Waldorf:2007mm}.

Our subsequent discussion calls for several additional elementary objects and constructions of the theory of gerbes. The first among them is the trivial gerbe over $\,M\,$ which is none other than a de Rham 3-coboundary $\,\txH=\sfd\txB\,$ with a globally smooth primitive $\,\txB\in\Om^2(M)$,\ with an obvious cohomological representation (associated with an arbitrary open cover $\,\cO_M$)
\qq\nn
\cI_\txB=[(\txB\rstr_{\cO_i},0,1)_{i\in\xcI}]\,,
\qqq
and a simple geometrisation  
\qq\nn
\cI_\txB=(M,\id_M,\txB,M\x\bC^\x,\pr_1,\sfd,\mu)
\qqq
with the trivial groupoid structure
\qq\nn
\mu\ :\ (M\x\bC^\x)\ox(M\x\bC^\x)\too M\x\bC^\x\ :\ (x,z_1)\ox(x,z_2)\longmapsto(x,z_1\cdot z_2)\,.
\qqq
A trivial 1-isomorphism is defined analogously as a trivial principal $\bC^\x$-bundle with a global (base component of) principal $\bC^\x$-connection.

We shall also need an explicit description of the distinguished identity 1-isomorphism,
\qq\nn
\id_\cG\ :\ \cG\xrightarrow{\ \cong\ }\cG\,,
\qqq
of a given gerbe $\,\cG=(\sfY M,\pi_{\sfY M},\txB,L,\pi_L,\nabla_L,\mu_L)$.\ This is readily seen to admit the geometrisation
\qq\nn
\id_\cG=\bigl(\sfY^{[2]}M,\id_{\sfY^{[2]}M},L,\pi_L,\nabla_L,\bigl(\id_{\pr_{1,2}^*L}\ox\pr_{2,3,4}^*\mu_L\bigr)\circ\bigl(\pr_{1,2,3}^*\mu_L^{-1}\ox\id_{\pr_{3,4}^*L}\bigr)\bigr)\,.
\qqq

The next concept is that of the tensor product $\,\cG_1\ox\cG_2\,$ of (bundle) gerbes $\,\cG_\a,\ \a\in\{1,2\}\,$ over a common base $\,M$.\ This has a simple cohomological description over a common trivialising open cover, to wit, given the respective local data $\,\left[(B^\a,A^\a_{jk},g^\a_{lmn})_{i\in\xcI,\ (j,k)\in\xcI_2,\ (l,m,n)\in\xcI_3}\right]$,
\qq\nn
&&\left[(B^1_i,A^1_{jk},g^1_{lmn})_{i\in\xcI,\ (j,k)\in\xcI_2,\ (l,m,n)\in\xcI_3}\right]\ox\left[(B^2_i,A^2_{jk},g^2_{lmn})_{i\in\xcI,\ (j,k)\in\xcI_2,\ (l,m,n)\in\xcI_3}\right]\cr\cr
&=&\left[(B^1_i+B^2_i,A^1_{jk}+A^2_{jk},g^1_{lmn}\cdot g^2_{lmn})_{i\in\xcI,\ (j,k)\in\xcI_2,\ (l,m,n)\in\xcI_3}\right]\,.
\qqq
The geometric counterpart of this construction, for the choice of respective geometrisations $\,\cG_\a=(\sfY_\a M,\pi_{\sfY_\a M},\txB_\a,L_\a,\pi_{L_\a},\nabla_{L_\a},\mu_{L_\a}),\ \a\in\{1,2\}$,\ is the bundle gerbe
\qq\nn
\cG_1\ox\cG_2&=&\bigl(\sfY_{1,2}M,\pi_{\sfY_1 M}\circ\pr_1,\pr_1^*\txB_1+\pr_2^*\txB_2,\pr_{1,3}^*L_1\ox\pr_{2,4}^*L_2,[\pi_{\pr_{1,3}^*L_1}\circ\pr_1],\pr_{1,3}^*\nabla_{L_1}\ox\id_{\pr_{2,4}^*L_2}\cr\cr
&&+\id_{\pr_{1,3}^*L_1}\ox\pr_{2,4}^*\nabla_{L_2},\pr_{1,3,5}^*\mu_{L_1}\ox\pr_{2,4,6}^*\mu_{L_2}\bigr)
\qqq
(with $\,\pi_{\pr_{1,3}^*L_1}\,$ the projection to the base of the pullback bundle $\,\pr_{1,3}^*L_1$,\ and $\,[\pi_{\pr_{1,3}^*L_1}\circ\pr_1]\,$ the corresponding projection to the base of the first factor in the tensor product). The construction of the tensor product descends naturally to (stable) isomorphisms between (bundle) gerbes: Given gerbes $\,\cG_\a,\ \a\in\{1,2,3,4\}\,$ and isomorphisms $\,\Phi_\b\ :\ \cG_\b\xrightarrow{\ \cong\ }\cG_{\b+2},\ \b\in\{1,2\}$,\ we may define a tensor-product isomorphism $\,\Phi_1\ox\Phi_2\ :\ \cG_1\ox\cG_2\xrightarrow{\ \cong\ }\cG_3\ox\cG_4$.\ If the respective local data are $\,[(C^\b_i,h^\b_{jk})_{i\in\xcI,\ (j,k)\in\xcI_2}]$,\ we have
\qq\nn
[(C^1_i,h^1_{jk})_{i\in\xcI,\ (j,k)\in\xcI_2}]\ox[(C^2_i,h^2_{jk})_{i\in\xcI,\ (j,k)\in\xcI_2}]=[(C^1_i+C^2_i,h^1_{jk}\cdot h^2_{jk})_{i\in\xcI,\ (j,k)\in\xcI_2}]\,.
\qqq
When expressed in terms of the respective geometrisations $\,\Phi_\b=(\sfY\sfY_{\b,\b+2}M,\pi_{\sfY\sfY_{\b,\b+2}M},E_\b,\pi_{E_\b},\nabla_{E_\b},$ $\a_{E_\b})$,\ the tensor product takes the form
\qq\nn
\Phi_1\ox\Phi_2&=&\bigl(\sfY\sfY_{1,3}M\x_M\sfY\sfY_{2,4}M,(\id_{\sfY_1 M}\x\tau_{\sfY_3 M,\sfY_2 M}\x\id_{\sfY_4 M})\circ(\pi_{\sfY\sfY_{1,3}M}\x\pi_{\sfY\sfY_{2,4}M}),\pr_1^*E_1\ox\pr_2^*E_2,&\cr\cr
&&\quad[\pi_{\pr_1^*E_1}\circ\pr_1],\pr_1^*\nabla_{E_1}\ox\id_{\pr_2^*E_2}+\id_{\pr_1^*E_1}\ox\pr_2^*\nabla_{E_2},\pr_{1,3}^*\a_{E_1}\ox\pr_{2,4}^*\a_{E_2}\bigr)\,,
\qqq
with ($[\pi_{\pr_1^*E_1}\circ\pr_1]\,$ defined similarly as $\,[\pi_{\pr_{1,3}^*L_1}\circ\pr_1]\,$ and) the fibred product $\,\sfY\sfY_{1,3}M\x_M\sfY\sfY_{2,4}M\,$ described by the commutative diagram
\qq\nn
\alxydim{@C=.75cm@R=1cm}{ & \sfY\sfY_{1,3}M\x_M\sfY\sfY_{2,4}M \ar[rd]^{\pr_2} \ar[ld]_{\pr_1} & \\ \sfY\sfY_{1,3}M \ar[rd]_{\pi_{\sfY_1 M}\circ\pr_1\circ\pi_{\sfY\sfY_{1,3}M}\qquad} & & \sfY\sfY_{2,4}M \ar[ld]^{\qquad\pi_{\sfY_4 M}\circ\pr_2\circ\pi_{\sfY\sfY_{2,4}M}} \\ & M & }\,,
\qqq 
and with
\qq\nn
\tau_{\sfY_3 M,\sfY_2 M}\ :\ \sfY_3 M\x_M\sfY_2 M\too\sfY_2 M\x_M\sfY_3 M\ :\ (y_3,y_2)\longmapsto(y_2,y_3)\,.
\qqq
We may also conceive the tensor product of a pair of 2-isomorphisms $\,\varphi_\g\ :\ \Phi^1_\g\xLongrightarrow{\ \cong\ }\Phi^2_\g,\ \g\in\{1,2\}\,$ between isomorphisms $\,\Phi^\b_\g\ :\ \cG_\g\xrightarrow{\ \cong\ }\cG_{\g+2},\ \b\in\{1,2\}$.\ For the (respective) local data $\,[(f_i^\g)_{i\in\xcI}]$,\ we obtain
\qq\nn
[(f_i^1)_{i\in\xcI}]\ox[(f_i^2)_{i\in\xcI}]=[(f_i^1\cdot f_i^2)_{i\in\xcI}]\,,
\qqq
whereas in the language of the respective geometrisations $\,\varphi_\g=(\sfY\sfY^{1,2}\sfY_{\g,\g+2}M,\pi_{\sfY\sfY^{1,2}\sfY_{\g,\g+2}M},\b_\g)$,\ the tensor product is the 2-isomorphism
\qq\nn
\varphi_1\ox\varphi_2&=&\bigl(\sfY\sfY^{1,2}\sfY_{1,3}M\x_M\sfY\sfY^{1,2}\sfY_{2,4}M,\cr\cr
&&\ (\id_{\sfY^1\sfY_{1,3}M}\x\tau_{\sfY^2\sfY_{1,3}M,\sfY^1\sfY_{2,4}M}\x\id_{\sfY^2\sfY_{2,4}M})\circ(\pi_{\sfY\sfY^{1,2}\sfY_{1,3}M}\x\pi_{\sfY\sfY^{1,2}\sfY_{2,4}M}),\pr_1^*\b_1\ox\pr_2^*\b_2\bigr)\,,
\qqq
with the fibred product $\,\sfY\sfY^{1,2}\sfY_{1,3}M\x_M\sfY\sfY^{1,2}\sfY_{2,4}M\,$ described by the commutative diagram
\qq\nn
\alxydim{@C=.75cm@R=1cm}{ & \sfY\sfY^{1,2}\sfY_{1,3}M\x_M\sfY\sfY^{1,2}\sfY_{2,4}M \ar[rd]^{\pr_2} \ar[ld]_{\pr_1} & \\ \sfY\sfY^{1,2}\sfY_{1,3}M \ar[rd]_{\pi_{\sfY_1 M}\circ\pr_1\circ\pi_{\sfY^1\sfY_{1,3}M}\circ\pr_1\circ\pi_{\sfY\sfY^{1,2}\sfY_{1,3}M}\qquad\qquad\qquad} & & \sfY\sfY^{1,2}\sfY_{2,4}M \ar[ld]^{\qquad\qquad\qquad\pi_{\sfY_4 M}\circ\pr_2\circ\pi_{\sfY^2\sfY_{2,4}M}\circ\pr_2\circ\pi_{\sfY\sfY^{1,2}\sfY_{2,4}M}} \\ & M & }\,,
\qqq 
and with
\qq\nn
\tau_{\sfY^2\sfY_{1,3}M,\sfY^1\sfY_{2,4}M}\ :\ \sfY^2\sfY_{1,3}M\x_M\sfY^1\sfY_{2,4}M\too\sfY^1\sfY_{2,4}M\x_M\sfY^2\sfY_{1,3}M\ :\ (y^2_{1,3},y^1_{2,4})\longmapsto(y^1_{2,4},y^2_{1,3})\,.
\qqq

Stable isomorphisms and 2-isomorphisms can be not only tensored, but also composed. Given 1-isomorphisms $\,\Phi_\b\ :\ \cG_\b\xrightarrow{\ \cong\ }\cG_{\b+1},\ \b\in\{1,2\}\,$ between gerbes $\,\cG_\a,\ \a\in\{1,2,3\}$,\ we may define the composite 1-isomorphism $\,\Phi_2\circ\Phi_1\ :\ \cG_1\xrightarrow{\ \cong\ }\cG_3\,$ with local data 
\qq\nn
[(C^2_i,h^2_{jk})_{i\in\xcI,\ (j,k)\in\xcI_2}]\circ[(C^1_i,h^1_{jk})_{i\in\xcI,\ (j,k)\in\xcI_2}]=[(C^1_i+C^2_i,h^1_{jk}\cdot h^2_{jk})_{i\in\xcI,\ (j,k)\in\xcI_2}]\,,
\qqq
and with a geometrisation
\qq\nn
\Phi_2\circ\Phi_1&=&\bigl(\sfY\sfY_{1,2}M\x_{\sfY_2 M}\sfY\sfY_{2,3}M,\pr_{1,4}\circ(\pi_{\sfY\sfY_{1,2}M}\x\pi_{\sfY\sfY_{2,3}M}),\pr_1^*E_1\ox\pr_2^*E_2,[\pi_{\pr_1^*E_1}\circ\pr_1],\cr\cr
&&\quad\pr_1^*\nabla_{E_1}\ox\id_{\pr_2^*E_2}+\id_{\pr_1^*E_1}\ox\pr_2^*\nabla_{E_2},(\id_{\pr_1^*\pr_1^*E_1}\ox\pr_{2,4}^*\a_{E_2})\circ(\pr_{1,3}^*\a_{E_1}\ox\id_{\pr_2^*\pr_2^*E_2})\bigr)\,,
\qqq
where the fibred product $\,\sfY\sfY_{1,2}M\x_{\sfY_2 M}\sfY\sfY_{2,3}M\,$ is described by the commutative diagram
\qq\nn
\alxydim{@C=.75cm@R=1cm}{ & \sfY\sfY_{1,2}M\x_{\sfY_2 M}\sfY\sfY_{2,3}M \ar[rd]^{\pr_2} \ar[ld]_{\pr_1} & \\ \sfY\sfY_{1,2}M \ar[rd]_{\pr_2\circ\pi_{\sfY\sfY_{1,2}M}\qquad} & & \sfY\sfY_{2,3}M \ar[ld]^{\qquad\pr_1\circ\pi_{\sfY\sfY_{2,3}M}} \\ & \sfY_2 M & }\,.
\qqq 
In the case of 2-isomorphisms, we encounter two types of composition. Given two pairs of 1-isomorphisms $\,\Phi_\g^\b\ :\ \cG_\g\xrightarrow{\ \cong\ }\cG_{\g+1},\ \b,\g\in\{1,2\}\,$ between gerbes $\,\cG_\a,\ \a\in\{1,2,3\}\,$ and two 2-isomorphisms $\,\varphi_\g\ :\ \Phi_\g^1\xLongrightarrow{\ \cong\ }\Phi_\g^2\,$ between the former, we define the {\bf horizontal composition} $\,\varphi_2\circ\varphi_1\ :\ \Phi_2^1\circ\Phi_1^1\xLongrightarrow{\ \cong\ }\Phi_2^2\circ\Phi_1^2\,$ as the 2-isomorphism with local data 
\qq\nn
[(f_i^2)_{i\in\xcI}]\circ[(f_i^1)_{i\in\xcI}]=[(f_i^2\cdot f_i^1)_{i\in\xcI}]
\qqq
and -- for $\,\varphi_\g=(\sfY\sfY^{1,2}\sfY_{\g,\g+1}M,\pi_{\sfY\sfY^{1,2}\sfY_{\g,\g+1}M},\b_\g)\,$ -- a geometrisation 
\qq\nn
\varphi_2\circ\varphi_1&=&\bigl((\sfY^1\sfY_{1,2}M\x_{\sfY_2 M}\sfY^1\sfY_{2,3}M)\x_{\sfY_{1,3}M}(\sfY\sfY^{1,2}\sfY_{1,2}M\x_{\sfY_2 M}\sfY\sfY^{1,2}\sfY_{2,3}M)\x_{\sfY_{1,3}M}\cr\cr
&&\quad(\sfY^2\sfY_{1,2}M\x_{\sfY_2 M}\sfY^2\sfY_{2,3}M),\pr_{1,2,5,6},\pi^{2\,*}_{1,2,3}d_{\Phi_2^2\circ\Phi_1^2}\circ(\pr_3^*\b_1\ox\pr_4^*\b_2)\circ\pi^{1\,*}_{1,2,3}d_{\Phi_2^1\circ\Phi_1^1} \bigr)\,,
\qqq
written in terms of the surjective submersions
\qq\nn
\pi^1_{1,2,3}&=&\bigl(\id_{\sfY^1\sfY_{1,2}M\x_{\sfY_2 M}\sfY^1\sfY_{2,3}M}\x(\pr_1\circ\pi_{\sfY\sfY^{1,2}\sfY_{1,2}M})\x(\pr_1\circ\pi_{\sfY\sfY^{1,2}\sfY_{2,3}M})\bigr)\circ \pr_{1,2,3,4}\,,\cr\cr
\pi^2_{1,2,3}&=&\bigl((\pr_2\circ\pi_{\sfY\sfY^{1,2}\sfY_{1,2}M})\x(\pr_2\circ\pi_{\sfY\sfY^{1,2}\sfY_{2,3}M})\x\id_{\sfY^2\sfY_{1,2}M\x_{\sfY_2 M}\sfY^2\sfY_{2,3}M}\bigr)\circ \pr_{3,4,5,6}
\qqq
and of the canonical (connection-preserving principal-bundle) isomorphisms
\qq\nn
d_{\Phi_2^\b\circ\Phi_1^\b}\ :\ \pr_1^*(\pr_1^*E_1^\b\ox\pr_2^*E_2^\b)\xrightarrow{\ \cong\ }\pr_2^*(\pr_1^*E_1^\b\ox\pr_2^*E_2^\b)\,,\quad\b\in\{1,2\}
\qqq
over the respective fibred products $\,(\sfY^1\sfY_{1,2}M\x_{\sfY_2 M}\sfY^1\sfY_{2,3}M)\x_{\sfY_{1,3}M}(\sfY^1\sfY_{1,2}M\x_{\sfY_2 M}\sfY^1\sfY_{2,3}M)$,\ derived in \Rcite{Waldorf:2007mm}. For any pair $\,\varphi_\d\ :\ \Phi_\d\xLongrightarrow{\ \cong\ }\Phi_{\d+1},\ \d\in\{1,2\}\,$ of 2-isomorphisms between 1-isomorphisms $\,\Phi_\d,\Phi_{\d+1}\ :\ \cG_1\xrightarrow{\ \cong\ }\cG_2\,$ between given gerbes $\,\cG_1\,$ and $\,\cG_2$,\ on the other hand, we may define their {\bf vertical composition} $\,\varphi_2\bullet\varphi_1\ :\ \Phi_1\xLongrightarrow{\ \cong\ }\Phi_3\,$ as the 2-isomorphism with local data 
\qq\nn
[(f_i^2)_{i\in\xcI}]\bullet[(f_i^1)_{i\in\xcI}]=[(f_i^2\cdot f_i^1)_{i\in\xcI}]
\qqq
and -- for $\,\Phi_\d=(\sfY^\d\sfY_{1,2}M,\pi_{\sfY^\d\sfY_{1,2}M},E_\d,\pi_{E_\d},\nabla_{E_\d},\a_{E_\d})\,$ and $\,\varphi_\d=(\sfY\sfY^{\d,\d+1}\sfY_{1,2}M,\pi_{\sfY\sfY^{\d,\d+1}\sfY_{1,2}M},\b_\d)\,$ -- a geometrisation 
\qq\nn
\varphi_2\bullet\varphi_1=\bigl(\sfY\sfY^{1,2}\sfY_{1,2}M\x_{\sfY^2\sfY_{1,2}M}\sfY\sfY^{2,3}\sfY_{1,2}M,\pr_1\circ\pi_{\sfY\sfY^{1,2}\sfY_{1,2}M}\x\pr_2\circ\pi_{\sfY\sfY^{2,3}\sfY_{1,2}M},\pr_2^*\b_2\circ\pr_1^*\b_1\bigr)\,.
\qqq
The above structure can be organised into a (weak) 2-category with (bundle) gerbes as 0-cells (or objects), 1-isomorphisms as 1-cells and 2-isomorphisms as 2-cells, which puts us in the higher-categorial context of the loop (quantum) mechanics.

Finally, we should mention the pullback, along smooth maps between the bases, of the various structures introduced heretofore. It is completely straightforward to present it in the local cohomological description. Indeed, let $\,f\in C^\infty(M_1,M_2)\,$ and let 
\qq\nn
\bigl[\bigl(X^p_{i^1_1},X^{p-1}_{i^2_1 i^2_2},\ldots,X^0_{i^{p+1}_1 i^{p+1}_2\ldots i^{p+1}_{p+1}}\bigr)_{i^1_1\in\xcI^2,\ (i^2_1,i^2_2)\in\xcI^2_2,\ldots,(i^{p+1}_1,i^{p+1}_2,\ldots,i^{p+1}_{p+1})\in\xcI^2_{p+1}}\bigr]\,,\quad p\in\{0,1,2\}
\qqq
be local data of an object (a gerbe for $\,p=2$,\ a 1-isomorphism for $\,p=1$,\ and a 2-isomorphism for $\,p=0$) on $\,M_2\,$ associated with an open cover $\,\cO_{M_2}=\{\cO^2_i\}_{i\in\xcI^2}\,$ (of $\,M_2$). In order to define the pullback of that object to $\,M_1\,$ in terms of its local data, we need to fix an open cover $\,\{\cO^1_j\}_{j\in\xcI^1}\,$ of $\,M_1\,$ together with a map $\,\phi\ :\ \xcI^1\too\xcI^2\,$ subordinate to $\,f\,$ in the sense expressed by the condition
\qq\nn
\forall_{i^1\in\xcI^1}\ :\ f\bigl(\cO^1_{i^1}\bigr)\subset\cO^2_{\phi(i^1)}
\qqq
(which may require passing to a refinement of $\,\cO_{M_2}$), whereupon we define 
\qq\nn
&&f^*[\bigl(X^p_{i^1_1},X^{p-1}_{i^2_1 i^2_2},\ldots,X^0_{i^{p+1}_1 i^{p+1}_2\ldots i^{p+1}_{p+1}}\bigr)_{i^1_1\in\xcI^2,\ (i^2_1,i^2_2)\in\xcI^2_2,\ldots,(i^{p+1}_1,i^{p+1}_2,\ldots,i^{p+1}_{p+1})\in\xcI^2_{p+1}}\bigr]\cr\cr
&\equiv&[\bigl(Y^p_{j^1_1},Y^{p-1}_{j^2_1 j^2_2},\ldots Y^0_{j^{p+1}_1 j^{p+1}_2\ldots j^{p+1}_{p+1}}\bigr)_{j^1_1\in\xcI^1,\ (j^2_1,j^2_2)\in\xcI^1_2,\ldots,(j^{p+1}_1,j^{p+1}_2,\ldots,j^{p+1}_{p+1})\in\xcI^1_{p+1}}\bigr]
\qqq
by the formul\ae
\qq\nn
Y^{p-k}_{j^{k+1}_1 j^{k+1}_2\ldots j^{k+1}_{k+1}}:=f^*X^{p-k}_{\phi(j^{k+1}_1)\phi(j^{k+1}_2)\ldots\phi(j^{k+1}_{k+1})}\,,\quad k\in\ovl{0,p}\,.
\qqq
We complete our presentation by giving definitions of pullbacks of the geometrisations of the local data that we introduced earlier. Thus, given a gerbe $\,\cG=(\sfY M_2,\pi_{\sfY M_2},\txB,L,\pi_L,\nabla_L,\mu_L)\,$ over the codomain of $\,f$,\ we first erect an arbitrary surjective submersion $\,\pi_{\sfY M_1}\ :\ \sfY M_1\too M_1\,$ endowed with a smooth map $\,\widehat f\ :\ \sfY M_1\too\sfY M_2\,$ that covers $\,f\,$ in the sense specified by the commutative diagram
\qq\nn
\alxydim{@C=1.5cm@R=1.5cm}{\sfY M_1 \ar[r]^{\widehat f} \ar[d]_{\pi_{\sfY M_1}} & \sfY M_2 \ar[d]^{\pi_{\sfY M_2}} \\ M_1 \ar[r]_{f} & M_2 }
\qqq
(we may, {\it e.g.}, take $\,\sfY M_1=M_1\x_{M_2}\sfY M_2\,$ with $\,\pi_{\sfY M_1}=\pr_1\,$ and $\,\widehat f=\pr_2$), and subsequently define
\qq\nn
f^*\cG=\bigl(\sfY M_1,\pi_{\sfY M_1},\widehat f^*\txB,\bigl(\widehat f^{\x 2}\rstr_{\sfY^{[2]}M_1}\bigr)^*L,\pi_{(\widehat f^{\x 2}\rstr_{\sfY^{[2]}M_1})^*L},\bigl(\widehat f^{\x 2}\rstr_{\sfY^{[2]}M_1}\bigr)^*\nabla_L,\bigl(\widehat f^{\x 3}\rstr_{\sfY^{[3]}M_1}\bigr)^*\mu_L\bigr)\,,
\qqq
the base projection $\,\pi_{(\widehat f^{\x 2}\rstr_{\sfY^{[2]}M_1})^*L}\,$ of the pullback bundle $\,(\widehat f^{\x 2}\rstr_{\sfY^{[2]}M_1})^*L\,$ being the mapping that closes the commutative diagram 
\qq\nn
\alxydim{@C=1.5cm@R=1.5cm}{\bigl(\widehat f^{\x 2}\rstr_{\sfY^{[2]}M_1}\bigr)^*L \ar[r]^{\quad\qquad L\widehat f^{\x 2}} \ar[d]_{\pi_{(\widehat f^{\x 2}\rstr_{\sfY^{[2]}M_1})^*L}} & L \ar[d]^{\pi_L} \\ \sfY^{[2]}M_1 \ar[r]_{\widehat f^{\x 2}} & \sfY^{[2]}M_2 }
\qqq
in which $\,L\widehat f^{\x 2}\,$ is a map covering $\,\widehat f^{\x 2}$.\ Similarly, in order to pull back a 1-isomorphism $\,\Phi=(\sfY\sfY_{1,2}M_2,$ $\pi_{\sfY\sfY_{1,2}M_2},E,\pi_E,\nabla_E,\a_E)\,$ between gerbes $\,\cG_\a=(\sfY_\a M_2,\pi_{\sfY_\a M_2},\txB_\a,L_\a,\pi_{L_\a},\nabla_{L_\a},\mu_{L_\a}),\ \a\in\{1,2\}\,$ along $\,f$,\ we choose a surjective submersion $\,\pi_{\sfY\sfY_{1,2}M_1}\ :\ \sfY\sfY_{1,2}M_1\too\sfY_{1,2}M_1\equiv\sfY_1 M_1\x_{M_1}\sfY_2 M_1\,$ alongside a map $\,\check f_{1,2}\ :\ \sfY\sfY_{1,2}M_1\too\sfY\sfY_{1,2}M_2\,$ satisfying the condition described by the commutative diagram 
\qq\nn
\alxydim{@C=1.5cm@R=1.5cm}{\sfY\sfY_{1,2}M_1 \ar[r]^{\check f_{1,2}} \ar[d]_{\pi_{\sfY\sfY_{1,2}M_1}} & \sfY\sfY_{1,2}M_2 \ar[d]^{\pi_{\sfY\sfY_{1,2}M_2}} \\ \sfY_{1,2}M_1 \ar[r]_{\widehat f_1\x\widehat f_2} & \sfY_{1,2}M_2 }
\qqq
(for $\,\widehat f_\a\,$ the respective covers of $\,f$), whereupon we define
\qq\nn
f^*\Phi=\bigl(\sfY\sfY_{1,2}M_1,\pi_{\sfY\sfY_{1,2}M_1},\check f_{1,2}^*E,\pi_{\check f_{1,2}^*E},\check f_{1,2}^*\nabla_E,(\check f_{1,2}\x\check f_{1,2})\rstr_{\sfY^{[2]}\sfY_{1,2}M_1}^*\a_E\bigr)\ :\ f^*\cG_1\xrightarrow{\ \cong\ }f^*\cG_2\,.
\qqq 
We complete our construction of the pullback functor between the (weak) 2-categories of gerbes over the two manifolds related by the smooth map $\,f\,$ by taking, for any 2-isomorphism $\,\varphi=(\sfY\sfY^{1,2}\sfY_{1,2}M_2,$ $\pi_{\sfY\sfY^{1,2}\sfY_{1,2}M_2},\b)\,$ between 1-isomorphisms $\,\Phi_\b=(\sfY^\b\sfY_{1,2}M,\pi_{\sfY^\b\sfY_{1,2}M},E_\b,\pi_{E_\b},\nabla_{E_\b},\a_{E_\b}),\ \b\in\{1,2\}\,$ between gerbes $\,\cG_\a=(\sfY_\a M,\pi_{\sfY_\a M},\txB_\a,L_\a,\pi_{L_\a},\nabla_{L_\a},\mu_{L_\a}),\ \a\in\{1,2\}$,\ a surjective submersion $\,\pi_{\sfY\sfY^{1,2}\sfY_{1,2}M_1}\ :\ \sfY\sfY^{1,2}\sfY_{1,2}M_1\too\sfY^{1,2}\sfY_{1,2}M_1\,$ together with a map $\,\widetilde f^{1,2}_{1,2}\ :\ \sfY\sfY^{1,2}\sfY_{1,2}M_1\too\sfY\sfY^{1,2}\sfY_{1,2}M_2\,$ that renders the following diagram commutative,
\qq\nn
\alxydim{@C=1.5cm@R=1.5cm}{\sfY\sfY^{1,2}\sfY_{1,2}M_1 \ar[r]^{\widetilde f^{1,2}_{1,2}} \ar[d]_{\pi_{\sfY\sfY^{1,2}\sfY_{1,2}M_1}} & \sfY\sfY^{1,2}\sfY_{1,2}M_2 \ar[d]^{\pi_{\sfY\sfY^{1,2}\sfY_{1,2}M_2}} \\ \sfY^{1,2}\sfY_{1,2}M_1 \ar[r]_{\check f^1_{1,2}\x\check f^2_{1,2}} & \sfY^{1,2}\sfY_{1,2}M_2 }
\qqq
(for $\,\check f^\b_{1,2}\,$ the respective covers of $\,\widehat f_1\x\widehat f_2$), and then write
\qq\nn
f^*\varphi=\bigl(\sfY\sfY^{1,2}\sfY_{1,2}M_1,\pi_{\sfY\sfY^{1,2}\sfY_{1,2}M_1},\widetilde f^{1,2\,*}_{1,2}\b\bigr)\ :\ f^*\Phi^1\xLongrightarrow{\ \cong\ }f^*\Phi^2\,.
\qqq
This exhausts the list of rudimentary concepts and constructions of the standard gerbe theory that we shall have a need for in the main part of our subsequent discussion. 

Prior to passing to the field-theoretic applications of the formalism recapitulated above, we close this section by giving -- after \Rcite{Stevenson:2001grb2} -- one last definition which is going to serve as a reference for our supergeometric constructions. Thus, we consider an object one degree higher in the natural hierarchy of geometrisations of de Rham classes, to wit, a {\bf bundle 2-gerbe} over a manifold $\,M\,$ with connection of {\bf curvature} given by a de Rham 4-cocycle with periods in $\,2\pi\bZ$,\ 
\qq\nn
\txJ\in Z^4_{\rm dR}(M,2\pi\bZ)\,.
\qqq
This is to be understood as a quintuple 
\qq\nn
\cG^{(2)}=(\sfY M,\pi_{\sfY M},\txC,\cG,\cM_\cG,\mu_\cG)\,,
\qqq
composed of a surjective submersion 
\qq\nn
\pi_{\sfY M}\ :\ \sfY M\too M
\qqq 
supporting a global primitive 
\qq\nn
\txC\in\Om^3(\sfY M)
\qqq
of the pullback 
\qq\nn
\pi_{\sfY M}^*\txJ=\sfd\txC\,,
\qqq
alongside a bundle gerbe $\,\cG\,$ over the fibred square $\,\sfY^{[2]}M\,$ with connection of curvature 
\qq\nn
\txH=\bigl(\pr_2^*-\pr_1^*\bigr)\txC
\qqq
together with a 1-isomorphism (termed the {\bf product} of the 2-gerbe)
\qq\nn
\cM_\cG\ :\ \pr_{1,2}^*\cG\ox\pr_{2,3}^*\cG\xrightarrow{\ \cong \ }\pr_{1,3}^*\cG
\qqq
of bundle gerbes over the fibred cube $\,\sfY^{[3]}M\,$ and a 2-isomorphism (termed the {\bf associator} of the 2-gerbe)
\qq\nn
\alxydim{@C=4.cm@R=2cm}{\pr_{1,2}^*\cG\ox\pr_{2,3}^*\cG\ox\pr_{3,4}^*\cG \ar[r]^{\pr_{1,2,3}^*\cM_{\cG}\ox\id_{\pr_{3,4}^*\cG}} \ar[d]_{\id_{\pr_{1,2}^*\cG}\ox\pr_{2,3,4}^*\cM_{\cG}} & \pr_{1,3}^*\cG\ox\pr_{3,4}^*\cG \ar[d]^{\pr_{1,3,4}^*\cM_{\cG}} \ar@{=>}[dl]|{\ \mu_{\cG}\ } \\ \pr_{1,2}^*\cG\ox\pr_{2,4}^*\cG \ar[r]_{\pr_{1,2,4}^*\cM_{\cG}} & \pr_{1,4}^*\cG }
\qqq
between the 1-isomorphisms of bundle gerbes over the fourfold fibred product $\,\sfY^{[4]}M$,\ subject to the coherence constraints expressed by the commutative diagram of 2-isomorphisms (here, $\,X_{i_1 i_2\ldots i_k}\equiv\pr_{i_1 i_2\ldots i_k}^*X\,$ for any $\,i_1,i_2,\ldots,i_k\in\{1,2,3,4,5\},\ k\in\{2,3,4\}$)
{\tiny\qq\nn
\alxydim{@C=-3.3cm@R=1.5cm}{ & \cM_{\cG\,1,4,5}\circ(\cM_{\cG\,1,3,4}\ox\id_{\cG_{4,5}})\circ(\cM_{\cG\,1,2,3}\ox\id_{\cG_{3,4}}\ox\id_{\cG_{4,5}}) \ar@{=>}[ld]_{\id_{\cM_{\cG\,1,4,5}}\circ(\mu_{\cG\,1,2,3,4}\ox\id_{\id_{\cG_{4,5}}})\qquad\quad} \ar@{=>}[rd]^{\qquad\quad\mu_{\cG\,1,3,4,5}\circ\id_{\cM_{\cG\,1,2,3}\ox\id_{\cG_{3,4}}\ox\id_{\cG_{4,5}}}} & \\ \cM_{\cG\,1,4,5}\circ(\cM_{\cG\,1,2,4}\ox\id_{\cG_{4,5}})\circ(\id_{\cG_{1,2}}\ox\cM_{\cG\,2,3,4}\ox\id_{\cG_{4,5}}) \ar@{=>}[d]_{\mu_{\cG\,1,2,4,5}\circ\id_{\id_{\cG_{1,2}}\ox\cM_{\cG\,2,3,4}\ox\id_{\cG_{4,5}}}} & & \cM_{\cG\,1,3,5}\circ(\id_{\cG_{1,3}}\ox\cM_{\cG\,3,4,5})\circ(\cM_{\cG\,1,2,3}\ox\id_{\cG_{3,4}}\ox\id_{\cG_{4,5}}) \ar@{=}[d] \\ 
\cM_{\cG\,1,2,5}\circ(\id_{\cG_{1,2}}\ox\cM_{\cG\,2,4,5})\circ(\id_{\cG_{1,2}}\ox\cM_{2,3,4}\ox\id_{\cG_{4,5}})
\ar@{=>}[rd]_{\id_{\cM_{\cG\,1,2,5}}\circ(\id_{\id_{\cG_{1,2}}}\ox\mu_{\cG\,2,3,4,5})\qquad\qquad\ } & & \cM_{\cG\,1,3,5}\circ(\cM_{\cG\,1,2,3}\ox\id_{\cG_{3,5}})\circ(\id_{\cG_{1,2}}\ox\id_{\cG_{2,3}}\ox\cM_{\cG\,3,4,5}) \ar@{=>}[ld]^{\ \qquad\qquad\mu_{\cG\,1,2,3,5}\circ\id_{\id_{\cG_{1,2}}\ox\id_{\cG_{2,3}}\ox\cM_{\cG\,3,4,5}}} \\ & \cM_{\cG\,1,2,5}\circ(\id_{\cG_{1,2}}\ox\cM_{\cG\,2,3,5})\circ(\id_{\cG_{1,2}}\ox\id_{\cG_{2,3}}\ox\cM_{3,4,5}) & }
\qqq}
between the 1-isomorphisms of bundle gerbes over the fifthfold fibred product $\,\sfY^{[5]}M$.
\medskip

\subsection{A rigorous definition, canonical description \& geometric quantisation of the $\si$-model}\label{sub:defcanquant}

The most immediate application of the above formalism is an explicit formula for the holonomy, determined by local data of the gerbe. The formula calls for an extra technical ingredient, to wit, a choice of a tessellation $\,\triangle(\Si)\,$ of the worldsheet, consisting of plaquettes (whose set will be denoted as $\,\Pgt_{\triangle(\Si)}$), edges and vertices, subordinate, for a given map $\,x\in[\Si,M]$,\ to the open cover $\,\cO_M=\{\cO_i\}_{i\in\xcI}$,\ by which we mean that there exists a map $\,i_\cdot\ :\ \triangle(\Si)\too\xcI\,$ with the property
\qq\nn
\forall_{\xi\in\triangle(\Si)}\ :\ x(\xi)\subset\cO_{i_\xi}\,.
\qqq
With all the requisite data in place, we write 
\qq\nn
-\sfi\,\log\Hol_\cG(x)&=&\sum_{p\in\Pgt_{\triangle(\Si)}}
\left[\int_p\,(x\rstr_p)^*B_{i_p}+\sum_{e\subset\p p}\left(\int_e\,(x\rstr_e)^*
A_{i_p i_e}-\sfi\,\sum_{v\in\p e}\,\vep_{pev}\,\log g_{i_p i_e i_v}\bigl(x(v)\bigr)\right)\right]\,,
\qqq
where $\,\vep_{pev}=1\,$ if $\,v\,$ sits at the end of $\,e\,$ with respect to the orientation of the edge induced (as the orientation of the boundary) from that of $\,p$,\ and $\,\vep_{pev}=-1\,$ otherwise. The above formula is a natural point of departure for the analysis that establishes the nature of geometric objects that complement the monophase background $\,(M,\txg,\cG)\,$ in the presence of world-sheet defects, {\it cp}\ \Rcite{Runkel:2008gr}.
 
The presence of the gerbe on the target space $\,M\,$ is also reflected directly in the canonical description of the $\si$-model. The latter description is readily established in the first-order formalism of Tulczyjew, Gaw\c{e}dzki, Kijowski and Szczyrba ({\it cp}\ Refs.\,\cite{Gawedzki:1972ms,Kijowski:1973gi,Kijowski:1974mp,Kijowski:1976ze,Szczyrba:1976,Kijowski:1979dj}, and also \Rcite{Saunders:1989jet} for an elementary modern treatment) that provides us with a (pre)symplectic structure on the space of states of the monophase theory and a Poisson bracket on the set of smooth (hamiltonian) functions on it. The first step towards its derivation consists in associating with the lagrangean density $\,\xcL_\si\ :\ J^1(\Si\x M)\too|\bigwedge|^1\sfT^*\Si\,$ of the $\si$-model, defined on the total space of the first-jet bundle $\,J^1(\Si\x M)\,$ of its covariant configuration bundle, with standard (adapted) coordinates $\,(x^a,\xi^b_i)\,$ on the fibre $\,J^1_{(\si^1,\si^2)}(\Si\x M)\,$ over a point $\,(\si^1,\si^2)\in\Si\,$ in the worldsheet, the Poincar\'e--Cartan form on $\,J^1(\Si\x M)\,$ (written in the standard notation that employs the symbol $\,\d\,$ for vertical differentials on $\,J^1(\Si\x M)\,$ and has $\,\p_i\equiv\frac{\p\ }{\p\si^i}\,$ in the local coordinate system $\,\{\si^i\}^{i\in\{1,2\}}\,$ on $\,\Si$) 
\qq\nn
\Theta_\si(x^a,\xi^b_i)&=&\left(\xcL_\si(x^a,\xi^b_i)-\xi^a_i\,\tfrac{\p\xcL_\si}{\p\xi^a_i}(x^a,\xi^b_i)\right)\,\Vol(\Si)+\d x^a\,\tfrac{\p\xcL_\si}{\p\xi^a_i}(x^a,\xi^b_i)\wedge\left(\p_i\con\Vol(\Si)\right)\,.
\qqq
It is not difficult to see that the extremals of the functional 
\qq\nn
\ee^{\sfi\,S_{\Theta_\si}}\ :\ \G\bigl(J^1(\Si\x M)\bigr)\too\uj\ :\ \Psi\longmapsto\ee^{\sfi\,\int_\Si\,\Psi^*\Theta_\si}
\qqq
are first jets of extremals of $\,\cA_{\rm DF}$,\ and this observation justifies the definition of a {\bf presymplectic form} $\,\Om_\si\,$ on the space of states (of a single loop) $\,\sfP_\si=\sfT^*\sfL M\,$ of the $\si$-model, the space itself being coordinatised by Cauchy data $\,\Psi\rstr_\xcC\equiv(x^a,p_b)\,$ of extremals $\,\Psi\,$ of $\,\ee^{\sfi\,S_{\Theta_\si}}\,$ supported on a model Cauchy section (or equitemporal slice) $\,\xcC\equiv\bS^1\subset\Si\,$ of the worldsheet. The definition reads
\qq\nn
\Om_\si[\Psi\rstr_\xcC]=\d\int_\xcC\,\left(\Psi\rstr_\xcC\right)^*\Theta_\si\,,
\qqq
and it depends only on the homotopy class of $\,\xcC\,$ within $\,\Si$.\ Through direct computation, we arrive at the explicit form 
\qq\nn
(\sfP_\si,\Om_\si)=\left(\sfT^*\sfL M,\d\theta_{\sfT^*\sfL M}+\pi_{\sfT^*\sfL M}^*\int_\xcC\,\ev^*\txH \right)\,,
\qqq
expressed in terms of the bundle projection $\,\pi_{\sfT^*\sfL M}\ :\ \sfT^*\sfL M\too\sfL M$,\ the canonical (kinetic-action) 1-form $\,\theta_{\sfT^*\sfL M}\,$ on $\,\sfT^*\sfL M\,$ (with local presentation $\,\theta_{\sfT^*\sfL M}[x,p]=\int_\xcC\,\Vol(\xcC)\,p_a(\cdot)\,\d x^a(\cdot)$), and the standard evaluation map
\qq\nn
\ev\ :\ \xcC\x\sfL M\too M\ :\ (\varphi,\g)\longmapsto\g(\varphi)\,.
\qqq
The 2-form serves to define a Poisson bracket of hamiltonians on $\,\sfP_\si$,\ {\it i.e.}, of those smooth functionals $\,h\,$ on $\,\sfP_\si\,$ for which there exist smooth vector fields $\,\cV$,\ termed ({\bf globally}) {\bf hamiltonian}, satisfying the relation
\qq\label{eq:canHamcond}
\cV\con\Om_\si=-\d h\,.
\qqq
Indeed, for any two such functionals $\,h_A,\ A\in\{1,2\}$,\ and the corresponding vector fields $\,\cV_A$,\ we may define a skew bracket
\qq\nn
\{h_1,h_2\}_{\Om_\si}[\Psi\rstr_\xcC]:=\cV_2\con\cV_1\con\Om_\si[\Psi\rstr_\xcC]\,,
\qqq
and the Jacobi identity follows automatically from the closedness of $\,\Om_\si$.\ A detailed discussion of the thus defined canonical description of the $\si$-model and its adaptations to the multi-phase setting can be found in Refs.\,\cite{Suszek:2011hg,Suszek:2012ddg}.

The ultimate confirmation of the naturality and functionality of gerbe theory in the analysis of the bosonic $\si$-model comes with the derivation of a quantisation scheme from that theory. The latter scheme employs Gaw\c{e}dzki's transgression map (extended to the polyphase setting and subsequently used in the analysis of symmetries and dualities of the $\si$-model in Refs.\,\cite{Suszek:2011hg,Suszek:2012ddg})
\qq\nn
\tau\ :\ \bH^2\left(M,\cD(2)^\bullet\right)\too\bH^1\left(\sfL M,\cD(1)^\bullet\right) 
\qqq
that canonically associates with (the isomorphism class of) the gerbe $\,\cG\,$ (the isomorphism class of) a principal $\bC^\x$-bundle
\qq\label{eq:transbund}
\alxydim{@C=.5cm@R=1cm}{\bC^\x \ar[r] & \xcL_\cG \ar[d]^{\pi_{\xcL_\cG}} \\ & \sfL M\equiv[\bS^1,M] }
\qqq 
over the configuration space $\,\sfL M\,$ of the $\si$-model, with connection $\,\nabla_{\xcL_\cG}\,$ of curvature 
\qq\nn
\curv(\nabla_{\xcL_\cG})=\int_{\bS^1}\, \ev^*\txH\,,
\qqq
termed the {\bf transgression bundle}, and thus induces over the phase space $\,\sfT^*\sfL M\,$ of the monophase $\si$-model a {\bf pre-quantum bundle} of the $\si$-model 
\qq\nn
\alxydim{@C=1.cm@R=1cm}{\bC \ar[r] & \xcL_\si \ar[d]^{\pi_{\xcL_\si}} \\ & \sfT^*\sfL M }
\qqq 
with the corresponding\footnote{The vector bundle can be obtained from its frame bundle in the standard construction as the associated bundle $\,(\sfF\xcL_\si\x\bC)/\bC^\x\cong\xcL_\si$.} frame bundle
\qq\label{eq:preqbund}
\sfF\xcL_\si:=(\sfT^*\sfL M\x\bC^\x)\ox\pi_{\sfT^*\sfL M}^*\xcL_\cG\,,
\qqq
where the first (trivial) tensor factor is taken to carry the global (base) connection 1-form $\,\theta_{\sfT^*\sfL M}$.\ It ought to be emphasised that the transgression bundle $\,\xcL_\cG\,$ can be reconstructed explicitly, on the basis of The Clutching Theorem, by sewing together its local trivialisations given in terms of the local data $\,(B_i,A_{jk},g_{lmn})_{i\in\xcI,\ (j,k)\in\xcI_2,\ (l,m,n)\in\xcI_3}\,$ of $\,\cG\,$ over the pullback along $\,\pi_{\sfT^*\sfL M}\,$ of an overcomplete basis 
\qq\nn
\cO_\igt\equiv\cO_{\triangle(\bS^1),i_\cdot}=\{\ x\in\sfL M \quad\vert\quad \forall_{(e,v)\in\Egt_{\triangle(\bS^1)}\x\Vgt_{\triangle(
\bS^1)}}\ :\ x(e)\subset\cO^M_{i_e}\quad\land\quad x(v)\in
\cO^M_{i_v} \ \}\,,
\qqq
of the compact-open topology of the Fr\'echet manifold $\,\sfL M\,$ indexed by pairs $\,\igt\equiv(\triangle(\bS^1)
,i_\cdot)\,$ composed of a tessellation $\,\triangle(\bS^1)\,$ of the unit circle, with its set of edges $\,\Egt_{\triangle(\bS^1)}\,$ and its set of vertices $\,\Vgt_{\triangle(\bS^1)}$,\ and a choice $\,i_\cdot\ :\ \triangle(\bS^1)\too\xcI\ :\ \xi
\longmapsto i_\xi\,$ of assignment of indices of $\,\cO_M\,$ to elements of $\,\triangle(\bS^1)$.\ By varying these two choices arbitrarily, whereby an index set $\,\xcI_{\sfL M}\ni\igt\,$ is formed, we cover all of $\,\sfL M$,\ thus forming an open cover $\,\cO_{\sfL M}=\{\cO_\igt \}_{\igt\in\xcI_{\sfL M}}\,$ of the free-loop space $\,\sfL M$.\ It is straightforward to describe intersections of elements of the open cover $\,\cO_{\sfL M}$,\ {\it cp}\ \Rcite{Gawedzki:1987ak}. Given a pair $\,\cO_{\igt^\a},\ \a\in\{1,2\}\,$ with the respective triangulations $\,\triangle_\a(\bS^1)\,$ (consisting of edges $\,e_\a\in\Egt_{\triangle_\a(\bS^1)}\,$ and vertices $\,v_\a\in\Egt_{\triangle_\a(\bS^1)}$) and index assignments $\,i^\a_\cdot\ :\ \triangle_\a(\bS^1)\too\xcI\ :\ \xi_\a\longmapsto i^\a_{\xi_\a}$,\ we consider the tessellation $\,\ovl\triangle(\bS^1)\,$ obtained by intersecting $\,\triangle_1(\bS^1)\,$ with $\,\triangle_2(\bS^1)$,\ by which we mean that the edges $\,\ovl e\,$ of $\,\ovl\triangle(\bS^1)\,$ are the non-empty intersections of the edges of the $\,\triangle_\a(\bS^1)$,\ and its vertices $\,\ovl v\,$ are taken from $\,\Vgt_{\triangle_1(\bS^1)}\cup\Vgt_{\triangle_2(\bS^1)}$.\ A non-empty double intersection $\,\cO_{\igt^1}\cap\cO_{\igt^2}=:\cO_{\igt^1
\igt^2}\,$ is then labelled by the tessellation $\,\ovl\triangle(\bS^1)$,\ taken together with the indexing convention such that $\,i^\a_{\ovl e}\,$ is the \v Cech index assigned -- via $\,\igt^\a\,$ -- to the edge of $\,\triangle_\a(\bS^1)\,$ containing $\,\ovl e\in\ovl\triangle(\bS^1)$,\ and $\,i^\a_{\ovl v}\,$ is the \v Cech index assigned -- via $\,\igt^\a\,$ -- to $\,\ovl v\,$ if $\,\ovl v\in\triangle_\a(\bS^1)$,\ or the \v Cech index assigned -- also via $\,\igt^\a\,$ -- to the edge of $\,\triangle_\a(\bS^1)\,$ containing $\,\ovl v\,$ otherwise. With the foregoing description in hand, we may finally write out explicit formul\ae ~for local data of the transgression bundle: we begin with local connection 1-forms (written for $\,x\in\cO_\igt\,$ and $\,\igt=(\triangle(\bS^1),i_\cdot)$)
\qq\nn
E_\igt[x]=-\sum_{e\in\Egt_{\triangle(\bS^1)}}\,\int_e\,(x\rstr_e)^*B_{i_e}-
\sum_{v\in\Vgt_{\triangle(\bS^1)}}\,x^*A_{i_{e_+(v)}i_{e_-(v)}}(v)\,,
\qqq
where $\,e_+(v)\,$ and $\,e_-(v)\,$ denote the incoming and the outgoing edge meeting at $\,v$,\ respectively; this (leads to and) is augmented with the definition of $\uj$-valued transition maps (written for $\,y\in\cO_{\igt\jgt}\,$ with $\,(\igt,\jgt)\in\xcI_{\sfL M\,2}$)
\qq\nn
G_{\igt\jgt}[y]=\prod_{\ovl e\in\Egt_{\ovl\triangle(\bS^1)}}\,\ee^{-\sfi\,
\int_{\ovl e}\,(y\rstr_{\ovl e})^*A_{i_{\ovl e}j_{\ovl e}}}\,\prod_{\ovl v
\in\Vgt_{\ovl\triangle(\bS^1)}}\,g_{i_{\ovl e_+(\ovl v)}i_{\ovl
e_-(\ovl v)}j_{\ovl e_+(\ovl v)}}\bigl(y(\ovl v)\bigr)\cdot g_{j_{\ovl e_+(\ovl v)}
j_{\ovl e_-(\ovl v)}i_{\ovl e_-(\ovl v)}}\bigl(y(\ovl v)\bigr)^{-1}\,,
\qqq
in which the $\,\ovl e\,$ are edges and the $\,\ovl v\,$ are vertices of the tessellation $\,\ovl\triangle(\bS^1)\,$ described above. As previously, the incoming (resp.\ outgoing) edge of $\,\ovl\triangle(\bS^1)\,$ at the vertex $\,\ovl v\,$ is denoted by $\,\ovl e_+(\ovl v)\,$ (resp.\ $\,\ovl e_-(\ovl v)$). The data satisfy the standard cohomological identities (written for $\,(\igt,\jgt,\kgt)\in\xcI_{\sfL M\,3}$)
\qq\nn
(E_\jgt-E_\igt)\rstr_{\cO_{\igt\jgt}}=\sfi\,\d\log G_{\igt\jgt}\,,\qquad\qquad\bigl(G_{\jgt\kgt}
\cdot G_{\igt\kgt}^{-1}\cdot G_{\igt\jgt}\bigr)\rstr_{\cO_{\igt\jgt\kgt}}=1\,.
\qqq
Under a gauge transformation of the gerbe $\,\cG\,$ with local data $\,(C_i,h_{jk})_{i\in\xcI\,,\ (j,k)\in\xcI_2}$ of \Reqref{eq:1iso}, the local connection 1-forms undergo the induced gauge transformation
\qq\nn
E_\igt\longmapsto E_\igt-\sfi\,\d\log H_\igt\,,
\qqq
where
\qq\nn
H_\igt[x]=\prod_{e\in\Egt_{\triangle(\bS^1)}}\,\ee^{\sfi\,\int_e\,(x\rstr_e)^*C_{i_e}}\,\prod_{v\in\Vgt_{\triangle(\bS^1)}}\,h_{i_{e_+(v)}i_{e_-(v)}}\bigl(x(v)\bigr)^{-1}\,.
\qqq
With the help of these data, we define those of the pre-quantum bundle over elements $\,\cO^*_\igt\equiv\pi_{\sfT^*\sfL M}^{-1}(\cO_\igt)\,$ of the pullback cover, to wit, the local symplectic potentials
\qq\nn
\vartheta_{\si\,\igt}=\theta_{\sfT^*\sfL M}\rstr_{\cO^*_\igt}+\pi_{\sfT^*\sfL M}^*E_\igt
\qqq
and the corresponding gluing maps
\qq\nn
\g_{\si\,\igt\jgt}=\pi_{\sfT^*\sfL M}^*G_{\igt\jgt}\,.
\qqq

The construction of the transgression bundle is a key step towards Dirac's geometric quantisation of the model in what can be regarded as an explicit realisation of Segal's abstract categorial quantisation paradigm. In it, the Hilbert space 
\qq\nn
\ceH_\si:=\G_{\rm{pol}}(\xcL_\si)
\qqq
assigned to a loop is the space of suitably polarised sections of the pre-quantum bundle on which hamiltonians are realised as certain sections of the sheaf of first-order differential operators. The Dirac--Feynman amplitudes for surfaces $\,\Sigma\,$ with boundaries are now readily seen to play the r\^ole of (linear) transport operators between Hilbert spaces assigned to the cobordant loops of $\,\Si\,$ -- {\it cp}\ \Rcite{Gawedzki:1987ak}, but also \Rcite{Suszek:2011hg} for more details. In this picture, a wave functional $\,\Psi\in\G_{\rm{pol}}(\xcL_\si)\,$ in the position polarisation admits -- at least formally\footnote{In the case of target manifolds given by homogeneous spaces of Lie groups, the formal construction can be concretised with the help of an invariant Haar measure, {\it cp}, {\it e.g.}, Refs.\,\cite{Felder:1988sd,Gawedzki:1999bq}, and one may anticipate that an analogous construction works for homogeneous spaces of Lie supergroups, {\it cp}\ \Rcite{Williams:1984}.} -- a path-integral presentation
\qq\nn
\Psi[\phi]=\int_{x\rstr_{\p\Si_{\rm in}}=\phi}\,\xcD x\,\ee^{\sfi\,S_\si[x]}
\qqq
written for a worldsheet $\,\Si_{\rm in}=\bD^2$,\ parameterising the trajectory of an `incoming' state, {\it cp}\ \Rcite{Gawedzki:1987ak} (such formal expressions are also considered in the framework of perturbative quantisation of a lagrangean field theory, cp \Rcite{Cattaneo:2012}).

We may also analyse, in the above framework, global (or rigid) symmetries of the field theory under study, induced by automorphisms of the target space. Let $\,\txG_\si\,$ be the subgroup of $\,\Diff(M)\,$ composed of all those automorphisms that preserve the Dirac--Feynman amplitude. Thus, upon denoting the natural action (evaluation) of $\,\Diff(M)\,$ on $\,M\,$ as
\qq\nn
\ell_\cdot\ :\ \Diff(M)\x M\too M\ :\ (g,m)\longmapsto g(m)\equiv\ell_g(m)\,,
\qqq
we have
\qq\nn
\forall_{g\in\txG_\si}\ :\ \cA_{\rm DF}[\ell_g\circ x]=\cA_{\rm DF}[x]\,.
\qqq
Under the assumption of the existence of a measure $\,\xcD x\,$ on $\,[\Si,M]\ni x\,$ \emph{invariant} under the induced action $\,(g,x)\longmapsto\ell_g\circ x$,\ the above presentation enables us to discuss quantum lifts of (global) symmetries of the classical theory in an explicit manner. Indeed, let the induced action preserve the integrand of the action functional of the $\si$-model up to a total derivative (which is necessary for the action functional for the closed worldsheet to remain invariant under symmetry transformations),
\qq\nn
\xcL_\si\left(\ell_g\circ x,\p(\ell_g\circ x)\right)\,\Vol(\Si)-\xcL_\si(x,\p x)\,\Vol(\Si)=\sfd J_g(x,\p x)\,,
\qqq
for some $\,J_g(x,\p x)\in\Om^1(\Si)$.\ We then obtain the induced realisation of $\,\txG_\si\,$ on the quantum space of states in the form
\qq\nn
\left(R(g)\Psi\right)[\phi]:=\int_{\ell_g\circ x\rstr_{\p\Om_{\rm in}}=\phi}\,\xcD(\ell_g\circ x)\,\ee^{\sfi\,S_\si[\ell_g\circ x]}=
\int_{x\rstr_{\p\Om_{\rm in}}=\ell_{g^{-1}}\circ\phi}\,\xcD x\,\ee^{\sfi\,S_\si[x]}\cdot\ee^{\sfi\,\int_{\p\Si_{\rm in}}\,J_g(x,\p x)}\,.
\qqq
If, furthermore, 
\qq\label{eq:jpullJ}
J_g=x^*\jmath_g
\qqq
for some {\bf target symmetry current} $\,\jmath_g\in\Om^1(M)$,\ then we may rewrite the above definition as
\qq\nn
\left(R(g)\Psi\right)[\phi]=c_g[\phi]\cdot\Psi\bigl[\ell_{g^{-1}}\circ\phi\bigr]\,,\qquad\qquad c_g[\phi]:=\ee^{\sfi\,\int_{\p\Om_{\rm in}}\,\bigl(\ell_{g^{-1}}\circ\phi\bigr)^*\jmath_g}\,.
\qqq
Thus, to a realisation of the classical (global-)symmetry group on the quantum space of states, there is associated an {\bf action 1-cochain} on $\,\txG_\si\,$ with values in $\uj$-valued functionals on the classical space of states. The space of such functionals carries the structure of a $\txG_\si$-module with a (left) $\txG_\si$-action
\qq\nn
(g_2\lact c_{g_1})[\phi]:=c_{g_1}\bigl[\ell_{g_2^{-1}}\circ\phi\bigr]\,.
\qqq
In order to have an actual representation of the symmetry group on quantum states, we must demand that the 1-cochain be a 1-cocycle. Indeed, we have
\qq\nn
\bigl(R(g_1)\circ R(g_2)\Psi\bigr)[\phi]&=&c_{g_1}[\phi]\cdot\bigl(R(g_2)\Psi\bigr)\bigl[\ell_{g_1^{-1}}\circ\phi\bigr]=c_{g_1}[\phi] \cdot c_{g_2}\bigl[\ell_{g_1^{-1}}\circ\phi\bigr]\cdot\Psi\bigl[\ell_{g_2^{-1}}\circ\ell_{g_1^{-1}}\circ\phi\bigr]\cr\cr
&=&c_{g_1}[\phi] \cdot c_{g_2}\bigl[\ell_{g_1^{-1}}\circ\phi\bigr]\cdot\Psi\bigl[\ell_{(g_1\cdot g_2)^{-1}}\circ\phi\bigr]\cr\cr
&=&(\d_\txG c)_{g_1,g_2}[\phi]\cdot\bigl(R(g_1\cdot g_2)\Psi\bigr)[\phi]
\qqq
with the {\bf homomorphicity 2-cocycle}
\qq\nn
(\d_\txG c)_{g_1,g_2}[\phi]&=&c_{g_2}\bigl[\ell_{g_1^{-1}}\circ\phi\bigr]\cdot c_{g_1\cdot g_2}[\phi]^{-1}\cdot c_{g_1}[\phi]\cr\cr
&=&\ee^{\sfi\,\int_{\p\Si_{\rm in}}\,[(\ell_{g_2^{-1}}\circ(\ell_{g_1^{-1}}\circ\phi))^*\jmath_{g_2}-(\ell_{(g_1\cdot g_2)^{-1}}\circ\phi)^*\jmath_{g_1\cdot g_2}+(\ell_{g_1^{-1}}\circ\phi)^*\jmath_{g_1}]}\cr\cr
&=&\ee^{\sfi\,\int_{\p\Si_{\rm in}}\,(\ell_{(g_1\cdot g_2)^{-1}}\circ\phi)^*(\jmath_{g_2}-\jmath_{g_1\cdot g_2}+\ell_{g_2}^*\jmath_{g_1})}\,,
\qqq
the latter being determined by the {\bf current 2-cocycle}
\qq\nn
(\d_\txG\jmath)_{g_1,g_2}:=\jmath_{g_1}\ract g_2-\jmath_{g_1\cdot g_2}+\jmath_{g_2}\,,\qquad\qquad\jmath_{g_1}\ract g_2:=\ell_{g_2}^*\jmath_{g_1}
\qqq
whose triviality in the de Rham cohomology of $\,M\,$ is a necessary and sufficient condition for the coclosedness of $\,c_g$.\ The existence of a \emph{projective} representation (and so also of a standard linear representation of a (super)central extension of $\,\txG_\si$), on the other hand, requires only that the group-coboundary of the above 1-cochain be a 2-cocycle on $\,\txG_\si\,$ with values in the trivial $\txG_\si$-module $\,\uj$,
\qq\nn
d_{g_1,g_2}:=(\d_{\txG}c)_{g_1,g_2}\in Z^2\bigl(\txG_\si,\uj\bigr)\,,
\qqq
which is to say that it satisfies the identity
\qq\nn
(\d_\txG d)_{g_1,g_2,g_3}:=d_{g_1,g_2}\cdot d_{g_1,g_2\cdot g_3}^{-1}\cdot d_{g_1\cdot g_2,g_3}\cdot d_{g_2,g_3}^{-1}=1
\qqq
for arbitrary $\,g_1,g_2,g_3\in\txG_\si$.\ Indeed, given such a 2-cocycle, we may define a standard action of the (super)central extension
\qq\nn
\bd1\too\uj\too\widehat{\txG}_\si:=\txG_\si\lx\uj\too\txG_\si\too\bd1
\qqq
of the symmetry group $\,\txG_\si$,\ with the group operation determined by the 2-cocycle as
\qq\label{eq:probasext}
\widehat\txG_\si\x\widehat\txG_\si\too\widehat\txG_\si\ :\ \bigl((g_1,u_1),(g_2,u_2)\bigr)\longmapsto(g_1\cdot g_2,d_{g_1,g_2}\cdot u_1\cdot u_2)\,.
\qqq
The action is given by the formula
\qq\nn
\bigl(R(g,u)\Psi\bigr)[\phi]:=u\cdot\bigl(R(g)\Psi\bigr)[\phi]\,.
\qqq
These considerations will play an important r\^ole in the fundamental construction developed in the present article, that is in the (super)geometrisation scheme for {\it supergroup-invariant} de Rham cohomology of super-$\si$-model targets.\medskip

\section{Tensorial super-$\si$-model backgrounds -- generalities}\label{sec:stensor}

We shall be concerned with the by now well-established Green--Schwarz-type models of dynamics of extended supersymmetric objects, also known as \textbf{super-$p$-branes}, whose classical configurations are generalised superharmonic embeddings $\,\xi\in[\Om,\xcM]\,$ of the {\bf worldvolume} $\,\Om$,\ a standard manifold of dimension $p+1\in\ovl{1,11}\,$ parametrising the history of a charged point-like particle, loop, membrane {\it etc.}, in a {\bf target supermanifold} $\,\xcM$,\ to be termed the {\bf supertarget} in what follows. A general (real) \textbf{supermanifold} of \textbf{superdimension} $\,m\,|\,n\,$ is a ringed space $\,\xcM=(M,\cO_M)\,$ composed of a (second countable Hausdorff) topological space $\,M\,$ (termed the {\bf body} of $\,\xcM$) and a sheaf $\,\cO_M\,$ of (real) associative unital supercommutative algebras on $\,M\,$ (termed the {\bf structure sheaf} and to be thought of as a generalisation of the sheaf of real functions on a manifold), locally modelled on $\,(\bR^{\x m},C^\infty(\cdot,\bR)\ox\bigwedge^\bullet\bR^{\x n})\,$ -- here, the pair $\,(m,n)$,\ is assumed to be constant over the entire $M$.\ Accordingly, a morphism $\,\xcF\ :\ (M_1,\cO_{M_1})\too(M_2,\cO_{M_2})\,$ between supermanifolds $\,(M_A,\cO_{M_A}),\ A\in\{1,2\}\,$ is a pair $\,\xcF\equiv(f,\phi)\,$ composed of a continuous map $\,f\ :\ M_1\too M_2\,$ and a morphism of structure sheaves $\,\phi\ :\ \cO_{M_2}\too f_*\cO_{M_1}$,\ \emph{i.e.}, a family of algebra homomorphisms $\,\phi_\cU\ :\ \cO_{M_2}(\cU)\too\cO_{M_1}(f^{-1}(\cU))\,$ indexed by the topology $\,\xcT(M_2)\ni\cU\,$ of $\,M_2$.\ The global geometry of such a structure is identified in the fundamental Gaw\c{e}dzki--Batchelor Theorem of Refs.\,\cite{Gawedzki:1977pb,Batchelor:1979a} which states that $\,\xcM\,$ is (globally, but non-canonically) isomorphic with the ringed space $\,(M,\G(\bigwedge^\bullet\bV))\,$ for $\,(\bV,M,\pi_\bV,\bR^{\x n})\,$ a real vector bundle of rank $n$ over the body. Supermanifolds admit (local) coordinate descriptions, and in this work we shall deal exclusively with supermanifolds with {\it global} coordinate systems, so that there will be no need for the abstract theory of supermanifolds beyond the above definition. The presence of global coordinate systems helps to simplify our treatment of the differential calculus on the supermanifolds of interest, which will be seen to play an instrumental r\^ole in the field-theoretic constructions. Thus, we shall use the fact that the \textbf{tangent sheaf} $\,\cT\xcM\equiv{\rm sDer}(\cO_M)\,$ of superderivations of the structure sheaf (whose sections are to be thought of as (super)vector fields on $\,\xcM$), as well as the dual \textbf{cotangent sheaf} $\,\cT^*\xcM\equiv\Hom_{\Mod_{\cO_M}}(\cT\xcM,\cO_M)\,$ (whose sections are to be thought of as super-1-forms on $\,\xcM$) are in general locally, and in our case also globally free, with generators given by -- respectively -- coordinate superderivations and coordinate superdifferentials. All this will enable us to develop our discussion in a far-reaching structural analogy with the standard ({\it i.e.}, Gra\ss mann even-)geometric approach to $\si$-models, with the graded nature of the geometry under consideration reflected solely -- on the computational level -- in the elementary sign conventions tabulated in Conv.\,\ref{conv:SignManifesto}.

Passing to the supergeometries of interest, we shall further assume the supermanifold to be endowed with a (left) transitive action of a Lie supergroup $\,\txG\,$ ({\it i.e.}, a group object in the \textbf{category of supermanifolds} $\,{\rm sMan}$,\ the latter having $\,\bR^{0\,|\,0}\,$ as the terminal object). The supergroup will play the r\^ole of the global-(super)symmetry group of the field theory in question. As such the supertarget will be presentable as (or equivariantly superdiffeomorphic with) a supercoset $\,\txG/\txH\cong\xcM\,$ of that supergroup relative to a Lie group $\,\txH\,$ embedded in the body of $\,\txG$.\ Such a presentation of the target supermanifold puts us in the framework of Cartan geometry\footnote{An extensive discussion of the supergeometric counterparts of the standard constructions from the theory of Lie groups and manifolds with smooth Lie-group actions, and in particular -- their homogeneous spaces, can be found in \Rcite{Kostant:1975}, {\it cp} also \Rcite{Koszul:1982}, and \Rcite{Carmeli:2011} for a modern perspective. }, which, in turn, affords a neat description of the additional tensorial {\bf superbackground} of the super-$p$-brane propagation in $\,\xcM$,\ composed of a $\txG$-invariant metric tensor $\,\txg\,$ on $\,\xcM\,$ (typically degenerate in the Gra\ss mann-odd directions) and a left-$\txG$-invariant de Rham super-$(p+2)$-cocycle $\,\underset{\tx{\ciut{(p+2)}}}{\chi}\in Z^{p+2}_{\rm dR}(\xcM)^\txG$.\ Thus, we construct the action functionals of the models of interest in terms of components of the left-invariant Maurer--Cartan super-1-form $\,\theta_{\rm L}\,$ on $\txG\,$ with values in the Lie superalgebra ({\it cp} App.\,\ref{app:LieAlgCohom}) $\,\ggt\,$ of that supergroup as well as of invariant superdifferential forms on the latter, and the lagrangean fields of the theory are identified -- {\it via} flows of appropriate left-invariant supervector fields -- with normal coordinates on the image, within $\,\txG$,\ of a section 
\qq\nn
\g\in\G(\txG)
\qqq 
of the principal $\txH$-bundle $\,\txG\too\txG/\txH\,$ modelling the supertarget\footnote{In what follows, we presuppose that either the entire homogeneous space is mapped into the Lie supergroup $\,\txG\,$ by a single (global) section $\,\g\,$ (as is the case, {\it e.g.}, for $\,{\rm sMink}(d,1\,\vert\,D_{d,1})$,\ the main protagonist in the concrete analysis to be presented) or that the field theory under study has been restricted to field configurations contained in the superdomain $\,\cO\subset\txG/\txH\,$ of a single section, in which case further restrictions have to be imposed upon the (super)symmetry transformations $\,g\in\txG\,$ considered, to wit, these are taken from a vicinity of the group unit. These assumptions can be abandoned, at the expense of the simplicity of the treatment which (the expense!) we want to avoid in the present paper, and do not invalidate the ensuing conclusions -- a fact that will be demonstrated in a future work.}. 

Taking into account the structure of the super-Poincar\'e algebra that serves as the local model for the geometries under consideration, as well as that of the Lie superalgebras associated with the distinguished anti-de Sitter superbackgrounds to be explored in subsequent studies, we shall restrict our attention to the so-called {\bf reductive} homogeneous spaces $\,\txG/\txH$,\ {\it i.e.}, those for which the direct-sum (supervector-space) complement $\,\tgt\,$ of the tangent Lie algebra $\,\hgt\,$ of $\,\txH\,$ within the tangent Lie superalgebra 
\qq\label{eq:gdecomp}
\ggt=\tgt\oplus\hgt
\qqq
of $\,\txG\,$ has the $\hgt$-module property 
\qq\label{eq:hgtasmod}
[\hgt,\tgt]\subset\tgt\,.
\qqq
The prototype of the said structure is an extension of the super-point algebra of anti-commuting {\bf supercharges} $\,Q_{\a I},\ (\a,I)\in\ovl{1,D}\x\ovl{1,N}\,$ (with $\,N\,$ denoting the number of supersymmetries, {\it i.e.}, of distinct Majorana spinors entering the definition of the relevant GS model, and $\,D\,$ the dimension of the Majorana-spinor representation of the underlying Clifford algebra, the two numbers being constrained severly by the requirement of existence of the corresponding GS model) by the algebra of Gra\ss mann-even translations $\,P_a,\ a\in\ovl{0,d}\,$ ($d+1\,$ is the spacetime dimension of the body of the supertarget), further enhanced -- as a spinor/vector-module algebra -- by the Lorentz algebra\footnote{Further enhancements are possible and, indeed, physically relevant, {\it e.g.}, by generators of dilations and special conformal transformations in the supersymmetric anti-de Sitter setting.} $\,\hgt\,$ with generators $\,J_{ab}=J_{[ab]},\ a,b\in\ovl{0,d}\,$ to form the Lie superalgebra $\,\ggt\,$ with the defining supercommutation relations
\qq\nn
&\{Q_{\a I},Q_{\b J}\}=f_{\a I,\b J}^{\hspace{24pt}a}\,P_a+f_{\a I,\b J}^{\hspace{24pt}ab}\,J_{ab}\,,\qquad[Q_{\a I},P_a]=f_{\a I,a}^{\hspace{15pt}\b J}\,Q_{\b J}\,,\qquad[P_a,P_b]=f_{a,b}^{\hspace{10pt}cd}\,J_{cd}\,,&\cr\cr
&[J_{ab},J_{cd}]=\eta_{ad}\,J_{bc}-\eta_{ac}\,J_{bd}+\eta_{bc}\,J_{ad}-\eta_{bd}\,J_{ac}\,,&\cr\cr
&[J_{ab},P_c]=\eta_{bc}\,P_a-\eta_{ac}\,P_b\,,\qquad[J_{ab},Q_{\a I}]=\tfrac{1}{2}\,(\G_{[ab]})^\b_{\ \a}\,Q_{\b I}\,,&
\qqq
written in terms of the Minkowskian metric $\,\eta=\diag(-1,\underbrace{1,1,\ldots,1}_{d\ {\rm times}})\,$ on the body of the supertarget, and in terms of the generators $\,\G_a,\ a\in\ovl{0,d}\,$ of the corresponding Clifford algebra, {\it cp}\ App.\,\ref{app:conv}. For instance, for $\,N=1\,$ and $\,d=3$,\ the choice of the structure constants
\qq\nn
f_{\a,\b}^{\hspace{12pt}a}=\ovl\G{}^a_{\a\b}\,,\qquad f_{\a,\b}^{\hspace{12pt}ab}=\tfrac{\la_1}{R}\,\bigl(\ovl\G{}^{[ab]}\bigr)_{\a\b}\,,\qquad f_{\a,a}^{\hspace{11pt}\b}=\tfrac{\la_2}{R}\,(\G_a)^\b_{\ \a}\,,\qquad f_{a,b}^{\hspace{10pt}cd}=\tfrac{\la_3}{R^2}\,\d^c_{\ [a}\,\d^d_{\ b]}
\qqq
yields, for certain (normalisation-dependent) values of the numerical constants $\,\la_1,\la_2\in\bR\,$ and $\,\la_3\in\bR_{>0}$,\ the $\,d+1=4\,$ {\bf super-anti-de Sitter algebra} at radius $\,R\in\bR_{>0}\,$ of \Rcite{Freedman:2012zz}, and reduces, {\it via} the \.In\"on\"u--Wigner contraction $\,R\to\infty$,\ to the standard ($N=1$) {\bf super-Minkowski algebra} 
\qq\nn
f_{\a,\b}^{\hspace{12pt}a}=\ovl\G{}^a_{\a\b}\,,\qquad\qquad f_{\a,\b}^{\hspace{12pt}ab}\,,\ f_{\a,a}^{\hspace{11pt}\b}\,,\ f_{a,b}^{\hspace{10pt}cd}=0\,.
\qqq
In this latter setting, $\,\hgt\,$ is the Lie algebra of the isotropy Lie group $\,\txH={\rm SO}(3,1)\,$ that defines the homogeneous space $\,\txG/\txH$,\ {\it i.e.}, the (asymptotic) super-Minkowski space ($D_{3,1}\,$ is the dimension of the Majorana-spinor module of the Clifford algebra of $\,\bR^{3,1}$)
\qq\nn
{\rm sMink}(3,1\,\vert\,D_{3,1})={\rm sISO}(3,1\,\vert\,D_{3,1})/{\rm SO}(3,1)
\qqq 
resp.\ the super-anti-de Sitter space
\qq\nn
{\rm sAdS}_4={\rm SO}(3,2)/{\rm SO}(3,1)\,.
\qqq

We shall denote the homogeneous basis vectors (generators) of $\,\tgt\,$ as $\,\{t_{\unl A}\}_{\unl A\in\ovl{1,\dim\,\tgt}}$,\ and among them those of $\,\tgt^{(0)}\,$ (the Gra\ss mann-even subspace) as $\,\{P_a\}_{a\in\ovl{1,\dim\,\tgt^{(0)}}}$,\ and those of $\,\tgt^{(1)}\,$ (the Gra\ss mann-odd subspace) as $\,\{Q_{\widehat\a}\}_{\widehat\a\in\ovl{1,\dim\,\tgt^{(1)}}}$.\ The generators of $\,\hgt\,$ will be written as $\,\{J_\k\}_{\k\in\ovl{1,\dim\,\hgt}}$.\ In the specific examples listed above, $\,\tgt\,$ is the linear span of supertranslations, and so -- in particular -- it is promoted to the rank of a Lie sub-superalgebra in the Minkowskian setting. In this notation, we may write the Maurer--Cartan super-1-form as
\qq\label{eq:sMC}
\theta_{\rm L}=\theta_{\rm L}^A\ox t_A=\theta_{\rm L}^{\unl A}\ox t_{\unl A}+\theta_{\rm L}^\k\ox J_\k
\qqq
and subsequently formulate the physical theory of interest in terms of the $\,\theta_{\rm L}^{\unl A}$.

With all the ingredients in place, we may finally write down the Dirac--Feynman amplitude for the mappings\footnote{{\it Cp} Rem.\,\ref{rem:Spoints} for an elucidation.} 
\qq\nn
\widetilde\xi\equiv\g\circ\xi\in[\Om,\txG]
\qqq 
of the field theory of interest, embedding the worldvolume $\,\Om\,$ within a section of the principal bundle $\,\txG\too\txG/\txH\,$ introduced earlier and precomposed with a mapping 
\qq\nn
\xi\in[\Om,\txG/\txH]\,.
\qqq
Generically, it takes the familiar form
\qq\label{eq:GSmodel}
\cA_{{\rm DF,GS},p}[\xi]=\exp\bigl(\sfi\,S_{{\rm metr,GS},p}[\widetilde\xi]\bigr)\cdot\exp\bigg(\sfi\,\tint_\Om\,\widetilde\xi^*\bigl(\sfd^{-1}\underset{\tx{\ciut{(p+2)}}}{\chi}\bigr)\bigg)
\qqq
in which the first factor $\,S_{{\rm metr,GS},p}\,$ computes the ($(p+1)$-)volume of the embedded hypersurface $\,\widetilde\xi(\Om)\,$ measured in the metric induced from $\,\txg\,$ along $\,\widetilde\xi\,$ on the worldvolume $\,\Om\,$ and can assume various forms, depending on the choice of the supertarget $\,\xcM\,$ (or, more to the point, on the choice of the embedding of the physical spacetime in $\,\xcM\,$ -- {\it cp}\ below) and that of the attendant choice of formulation of the field theory, and in which the second factor, tentatively respresenting the geometric coupling of the external field $\,\underset{\tx{\ciut{(p+2)}}}{\chi}\,$ to the charge current defined by the trajectory of the super-$p$-brane in $\,\xcM$,\ is {\it locally} (over the body of $\,\xcM$) expressed as the integral 
\qq\nn
S_{{\rm top,GS},p}[\xi]\equiv\int_\Om\,\widetilde\xi^*\bigl(\sfd^{-1}\underset{\tx{\ciut{(p+2)}}}{\chi}\bigr)\xrightarrow[\ {\rm loc.}\ ]{}\int_\Om\,\widetilde\xi^*\underset{\tx{\ciut{(p+1)}}}{\b}
\qqq
of a (local) de Rham primitive $\,\underset{\tx{\ciut{(p+1)}}}{\b}\,$ of $\,\underset{\tx{\ciut{(p+2)}}}{\chi}$,
\qq\nn 
\underset{\tx{\ciut{(p+2)}}}{\chi}=\sfd\underset{\tx{\ciut{(p+1)}}}{\b}\,.
\qqq
In the most studied examples, {\it i.e.}, on supercosets of the super-Minkowskian type (body) $\,\bR^{d,1}\,$ ({\it cp}\ Refs.\,\cite{Brink:1981nb,Green:1983wt,Bergshoeff:1985su,Hughes:1986fa,Achucarro:1987nc}) and of the super-anti-de Sitter type $\,{\rm AdS}_{p+2}\x\bS^{d-p-2},\ p\in\ovl{0,d-2},\ d\in\{10,11\}\,$ ({\it cp}\ Refs.\,\cite{Metsaev:1998it,Arutyunov:2008if,deWit:1998yu,Claus:1998fh}), the primitives of various physically relevant GS super-$(p+2)$-cocycles exist (although in the latter class, they are often given in an implicit integral form), and yet they are not supersymmetric (or, in our language, not left-$\txG$-invariant) in general, {\it cp}\ Refs.\,\cite{Chryssomalakos:2000xd,Sakaguchi:1999fm} and Refs.\,\cite{Hatsuda:2001pp,Hatsuda:2002hz} for interesting analyses in the super-Minkowskian and super-anti-de Sitter settings, respectively. 

Hence, in an attempt to grasp the geometric meaning and thereupon properly define the WZ term, we arrive at a crossroads -- we are confronted with the choice between the standard de Rham cohomology $\,H^\bullet_{\rm dR}(\xcM)\,$ of the supertarget $\,\xcM\,$ and the $\txG$-invariant (or at the very least supertranslationally-invariant) de Rham cohomology $\,H^\bullet_{\rm dR}(\xcM)^\txG\,$ of the same space. In many examples of $\si$-models with standard ({\it i.e.}, Gra\ss mann-even) manifolds as targets, there is either no distinguished symmetry group identified in the canonical analysis, or that group is compact, as is the case for the WZW $\si$-model on a compact Lie group -- the mother of all rational conformal field theories in two dimensions. In the fomer situation, the question of choice does not even come up, whereas in the latter one, it is answered by the Chevalley--Eilenberg Theorem of \Rcite{Chevalley:1948} that states the equivalence of the two cohomologies (that is, the existence of an isomorphism of the corresponding cohomology groups). The supergeometric setting of interest does not fall into either category as the symmetry (super)group is built into the definition of the supertarget and the latter group is (assumed) non-compact, which precludes the application of the Chevalley--Eilenberg Theorem. Thus, we do have to choose the cohomology with which we work, and the physical relevance of supersymmetry enforces the choice of the supersymmetric refinement of the de Rham cohomology upon us. However, while $\,H^\bullet_{\rm dR}(\xcM)\,$ encodes an obvious topological information on the (super)manifold $\,\xcM\,$ {\it via} its relation to the homology thereof ({\it i.e.}, of its body), there is, on the face of it, no geometry behind the refinement defined by $\,H^\bullet_{\rm dR}(\xcM)^\txG$.\ There seems to be, accordingly, no rationale for a geometrisation of classes in $\,H^\bullet_{\rm dR}(\xcM)^\txG\,$ understood as a resolution of the underlying topology. Intuition drawn from the study of gauged field theories then suggests a potential way out of the conundrum: We ought to replace $\,\xcM\,$ with its quotient by a subgroup of the supersymmetry group $\,\txG\,$ to which the invariant superdifferential forms descend and such that the nontrivial descended cocycles dualise nontrivial cycles in the quotient. But then another obstruction arises as the na\"ive choice of the subgroup, to wit, the subgroup generated by arbitrary translations in the soul directions, takes us out of the original geometric category of supermanifolds, as illustrated in \Rcite{Rabin:1984rm} -- this is analogous to the descent to a non-smooth space of orbits of the action of a finite group on a smooth manifold. Objections against geometrisation of the supersymmetry-invariant cohomology encountered along the way would be lifted, though, if we could find a subgroup $\,\sfT\subset\txG\,$ with the following properties:
\bit
\item[(i)] $\sfT$-invariance of a differential form on $\,\xcM\,$ implies its $\txG$-invariance;
\item[(ii)] the orbit space $\,\xcM/\sfT\,$ is a supermanifold locally modelled on the Gra\ss mann bundle of the same vector bundle over the body manifold of $\,\xcM\,$ as that of the supermanifold $\,\xcM\,$ itself.
\eit
The above properties would legitimise thinking of the original super-$\si$-model with the supertarget $\,\xcM\,$ as one with the supertarget $\,\xcM/\sfT\,$ on which the GS super-cocycle defines -- by construction -- a non-trivial de Rham class, and this would, in turn, mean that the topology of $\,\xcM/\sfT\,$ encodes the non-trivial supersymmetric cohomology of $\,\xcM\,$ and justify a geometrisation of the GS super-cocycle in a manner completely analogous with that employed in the Gra\ss mann-even setting. 

The existence of the relevant subgroup $\,\sfT\,$ in the case of the super-Minkowski space was demonstrated explicitly by Rabin and Crane in Refs.\,\cite{Rabin:1984rm,Rabin:1985tv}, and is therefore anticipated (but has to be proven on a case-by-case basis) on a generic supermanifold (of the type under consideration) in the light of the Gaw\c{e}dzki--Batchelor Theorem of Refs.\,\cite{Gawedzki:1977pb,Batchelor:1979a}. In the former setting, condition (ii) rules out the obvious candidate for $\,\sfT\,$ given by the full supersymmetry group -- indeed, the resulting quotient is not of the same type as the original supermanifold. It is then readily checked that the Kosteleck\'y--Rabin discrete supersymmetry subgroup of \Rcite{Kostelecky:1983qu}, to be thought of as a lattice variant of the continuous supergroup of supertranslations, is a suitable choice -- it yields a quotient supermanifold with the fundamental group generated by unital (in the lattice spacing) translations in the Gra\ss mann-odd (or {\bf soul}) directions, and the only nontrivial supercommutator in the underlying Lie superalgebra (the anticommutator of the supercharges) gives rise to a torsion component in the ensuing homology, {\it cp}\ \Rcite{Rabin:1984rm}. It is also worth noting that Witten's trick\footnote{Defining the Wess--Zumino term for the $\si$-model on $\,\xcM/\sfT\,$ as an integral of the GS 3-cocycle over a filling 3-manifold (a solid handlebody) of the worldsheet $\,\Si$.} does not work in this setting, {\it cp} \Rcite{Rabin:1985tv}, which is another reason to look for a geometrisation of the GS cocycle. We shall construct such a geometrisation explicitly in what follows. Since, moreover, we want to study, at a later stage, its equivariance under actions of subgroups of $\,\txG\,$ and subalgebras of $\,\ggt\,$ induced from the underlying left- and right-regular actions of $\,\txG\,$ on itself upon restriction to the section of the principal bundle $\,\txG\too\txG/\txH\,$ referred to previously, it will be important to gain a better understanding of the induction scheme first.

Our introductory remarks concerning the general structure of the supertargets of interest essentially determine the nature of the action of the symmetry supergroup $\,\txG\,$ to be considered, and so also -- in particular -- the implementation of supersymmetries. These will be realised nonlinearly in the scheme originally conceived by Schwinger and Weinberg in Refs.\,\cite{Schwinger:1967tc,Weinberg:1968de} in the context of effective field theories with chiral symmetries, and subsequently elaborated in Refs.\,\cite{Coleman:1969sm,Callan:1969sn,Salam:1969rq}, to be adapted to the study of spacetime symmetries by Salam, Strathdee and Isham in Refs.\,\cite{Salam:1970qk,Isham:1971dv}. The scheme was successfully employed in the setting of the supersymmetric field theory by Akulov and Volkov {\it et al.} in Refs.\,\cite{Volkov:1972jx,Volkov:1973ix,Ivanov:1978mx,Lindstrom:1979kq,Uematsu:1981rj,Ivanov:1982bpa,Samuel:1982uh,Ferrara:1983fi,Bagger:1983mv}, and this is the variant that we encounter below. Adapting the general theory to the field-theoretic context of interest, and taking into account the reasoning presented in Refs.\,\cite{West:2000hr,Gomis:2006wu} and the results derived therefrom, we introduce (local) coordinates $\,\xi^{\unl A},\ \unl A\in\ovl{1,\dim\,\tgt}\,$ on the homogeneous space\footnote{Again, we assume that either there exists a global coordinate system of the type indicated on $\,\txG/\txH\,$ (as is the case for $\,{\rm sMink}(d,1\,\vert\,D_{d,1})$) or restrict to field configurations that map the spacetime of the field theory to the domain of a single coordinate chart in a superdomain in the vicinity of the unital coset in the body. Also this simplification can (and shall) be abandoned easily.} and subsequently parametrise the aforementioned section $\,\g\in\G(\txG)\,$ as
\qq\label{eq:expcoord}
\g(\xi)=\ee^{\xi^{\unl A}\,t_{\unl A}}\,.
\qqq
Accordingly, the lagrangean field of the super-$\si$-model is of the form 
\qq\nn
\xi\equiv\bigl(\xi^{\unl A}\bigr)(\cdot)\in[\Om,\txG/\txH]\,,
\qqq
and we consider composite mappings
\qq\nn
\widetilde\xi\equiv\g\circ\xi\in[\Om,\txG]\,,\qquad\qquad\widetilde\xi(\si)=\ee^{\xi^{\unl A}(\si)\,t_{\unl A}}\,.
\qqq
Here, the Gra\ss mann-even components $\,\{x^a\equiv\xi^a\}_{a\in\ovl{1,\dim\,\tgt^{(0)}}}\,$ are to be thought of as coordinates on a physical spacetime $\,M=|\xcM|\,$ in which the super-$p$-brane propagates in a manner dictated by the Green--Schwarz super-$\si$-model. The Gra\ss mann-odd components $\,\{\xi^{\widehat\a}\}_{{\widehat\a}\in\ovl{1,\dim\,\tgt^{(1)}}}$,\ on the other hand, map the spinorial (super-charge) directions. 

The first type of an induced action of $\,\txG\ni g\,$ on the supertarget $\,\txG/\txH\,$ can be read off from the (local, in general) multiplication rule
\qq\label{eq:cosetlreg}
g\cdot\g(\xi)=:\g\bigl(\xi_l(\xi,g)\bigr)\cdot h_l(\xi,g)^{-1}
\qqq
in which $\,\xi_l\ :\ \txG/\txH\x\txG\too\txG/\txH\,$ is a certain (non-linear, in general) mapping, and the last element $\,h_l(\xi,g)\in\txH\,$ translates the product $\,g\cdot\g(\xi)\,$ back into the section $\,\g$,\ defining therewith an effective (non-linear) transformation 
\qq\nn
\unl\ell_\cdot\ :\ \txG\x\xcM\too\xcM\ :\ (g,\xi)\longmapsto\xi_l(\xi,g)
\qqq
on the base $\,\xcM\equiv\txG/\txH\,$ of the principal $\txH$-bundle. By construction, this action captures the rigid symmetry of the super-$\si$-model. Besides it, the theory has {\it infinitesimal} gauge symmetries that can be modelled -- after Refs.\,\cite{McArthur:1999dy,Gomis:2006wu} -- on infinitesimal right-regular translations of the lagrangean section in the directions of the subspace $\,\tgt\subset\ggt\,$ subject to certain constraints, to be established through a direct calculation in Part II.\medskip

Finally, passing to the class of field theories of interest, we recall the most common formulation of the super-$\si$-model in which the worldvolume of the super-$p$-brane is embedded entirely in a super-extension of the physical metric spacetime $\,(M,\unl\txg)$,\ with (local) coordinates $\,\{\xi^{\unl A}\equiv\xi^{\unl A}\}^{\unl A\in\ovl{1,\dim\,\tgt}}$.\ The body metric $\,\unl\txg\,$ defines an $\hgt$-horizontal metric on $\,\ggt\,$ which, in turn, lifts to a degenerate metric tensor  
\qq\nn
\txg=\txg_{AB}\,\theta^A_{\rm L}\ox\theta^B_{\rm L}\equiv\txg_{ab}\,\theta^a_{\rm L}\ox\theta^b_{\rm L}
\qqq
The influence of the background gravitational field on the dynamics of the super-$p$-brane is encoded in the action functional in the form of the induced-metric volume of the embedded worldvolume $\,\Om$,\ that is (the $\,\p_i,\ i\in\ovl{0,p}\,$ are the coordinate derivations in a (local) coordinate system $\,\{\si^i\}^{i\in\ovl{0,p}}\,$ on $\,\Om$)
\qq\label{eq:SmetrNG}
S^{({\rm NG})}_{{\rm metr,GS},p}[\xi]&=&\int_\Om\,\Vol(\Om)\,\sqrt{\det_{(p+1)}\,\bigl(\txg_{ab}\bigl(\g\circ\xi\bigr)\,\bigl(\p_i\con(\g\circ\xi)^*\theta^a_{\rm L}\bigr)\,\bigl(\p_j\con(\g\circ\xi)^*\theta^b_{\rm L}\bigr)\bigr)}\cr\cr
&\equiv&\int_\Om\,\Vol(\Om)\,\xcL^{({\rm NG})}_{{\rm metr,GS},p}(\xi,\p\xi)
\qqq
Thus, the metric term alone favours minimal hypersurfaces. The condition of minimality receives a correction from the topological term
\qq\label{eq:StopNG}
S^{({\rm NG})}_{{\rm top,GS},p}[\xi]:=\int_\Om\,(\g\circ\xi)^*\bigl(\sfd^{-1}\underset{\tx{\ciut{(p+2)}}}{\chi}\bigr)
\qqq
Together, the two terms yield the Green--Schwarz action functional in the Nambu--Goto form
\qq\label{eq:NGGS}
S^{({\rm NG})}_{{\rm GS},p}[\xi]=S^{({\rm NG})}_{{\rm metr,GS},p}[\xi]+S^{({\rm NG})}_{{\rm top,GS},p}[\xi]\,,
\qqq
with the reference section $\,\g\,$ considered fixed once and for all.

\brem\label{rem:Spoints}
At this stage, a word is well due regarding the mathematical status of the above intuitively acceptable field-theoretic formul\ae ~in which we cavalierly defined functionals on the `space of mappings' $\,[\Om,\xcM]\,$ in terms of integrals with integrands composed, in particular, of Gra\ss mann-\emph{odd} (local) objects $\,\xi^{\widehat\a}(\cdot),\ \widehat\a\in\ovl{1,\dim\,\tgt^{(1)}}$,\ and wrote various functional expressions as though we were dealing with standard ({\it i.e.}, purely Gra\ss mann-even) geometries. Assigning a rigorous meaning to such expressions calls for the introduction of the \textbf{functor of points} and the \textbf{internal $\,{\rm Hom}\,$} in our discussion of supermanifolds -- a potent tool that enables us to efficiently and conveniently transcribe the standard concepts and methods of functional analysis into the supergeometric domain in a manner that justifies subsequent formal manipulations on supergeometric objects, carried out in full analogy with those familar from the standard (purely Gra\ss mann-even) geometry. Below, we recapitulate the relevant categorial construction whose details can be found, {\it e.g.}, in \Rcite{Varadarajan:2004}. 

The point of departure is the Yoneda Lemma which gives us a fully faithful functor (the so-called Yoneda embedding) 
\qq\nn
{\rm Yon}\ :\ {\rm sMan}\too{\rm Fun}\left({\rm sMan}^{\rm opp},{\rm Set}\right)
\qqq
that realises the previously introduced category of supermanifolds $\,{\rm sMan}\,$ as a full subcategory in the category $\,{\rm Fun}({\rm sMan}^{\rm opp},{\rm Set})\,$ of presheaves on that category ({\it i.e.}, in the category of contravariant functors from $\,{\rm sMan}\,$ to the category $\,{\rm Set}\,$ of sets). The embedding assigns to a supermanifold $\,\xcM\,$ the corresponding (representable) functor
\qq\nn
{\rm Yon}(\xcM)\equiv\Hom_{\rm sMan}\bigl(-,\xcM\bigr)
\qqq
and gives us a handy model of (the set of) morphisms between supermanifolds, to wit,
\qq\nn
\Hom_{{\rm Fun}({\rm sMan}^{\rm opp},{\rm Set})}\bigl({\rm Yon}(\xcM_1),{\rm Yon}(\xcM_2)\bigr)\cong\Hom_{\rm sMan}(\xcM_1,\xcM_2)\,.
\qqq
Thus, instead of the original morphisms, we may, equivalently, consider natural transformations between the corresponding functors $\,{\rm Yon}(\xcM_1)\,$ and $\,{\rm Yon}(\xcM_2)$.\ Upon fixing a reference supermanifold $\,\xcS$,\ we may, next, speak -- in analogy with the standard topology -- of the \textbf{$\xcS$-points} of a given supermanifold $\,\xcM$,\ composing the set (of $\xcS$-points in $\,\xcM$)
\qq\nn
{\rm Yon}(\xcM)(\xcS)\equiv\Hom_{\rm sMan}(\xcS,\xcM)\,.
\qqq
As the reference object $\,\xcS\,$ varies, we collect complete information on $\,{\rm Yon}(\xcM)$,\ and so also on $\,\xcM$.\ In particular, the $\bR^{0\,\vert\,0}$-points are (as a set) 
\qq\nn
{\rm Yon}(\xcM)\bigl(\bR^{0\,\vert\,0}\bigr)=|\xcM|\,.
\qqq
Taking into account the existence of local models $\,\xcW_A\equiv(W_A,C^\infty(\cdot,\bR)\ox\wedge^\bullet\bR^{q_A}),\ W_A\subset\bR^{\x p_A},\ A\in\{1,2\}\,$ (with the $\,W_A\,$ open) for the supermanifolds $\,\xcM_A\equiv(M_A,\cO_{M_A})$,\ as encoded by local coordinate charts
\qq\nn
\k_A\ :\ \xcU_A\equiv\bigl(U_A,\cO_{M_A}\rstr_{U_A}\bigr)\xrightarrow{\ \cong\ }\bigl(W_A,C^\infty(\cdot,\bR)\ox\wedge^\bullet\bR^{q_A}\bigr)
\qqq
over superdomains $\,\xcU_A\subset\xcM_A$,\ we thus arrive at a local-coordinate description 
\qq\nn
{\rm Yon}\bigl(\k_2\circ\phi\circ\circ\k_1^{-1}\bigr)(\xcS)\ :\ {\rm Yon}(\xcW_1)(\xcS)\too{\rm Yon}(\xcW_1)(\xcS)
\qqq
of supermanifold morphisms $\,\phi\in\morf_{\rm sMan}(\xcM_1,\xcM_2)\,$ between them. It is in this picture that formul\ae ~such as, {\it e.g.}, \eqref{eq:expcoord} and \eqref{eq:cosetlreg}, ought to be viewed, with the understanding that there is a proper sheaf-theoretic mapping behind each of them. In particular, an explicit sheaf-theoretic discussion of the intuitive exponential parametrisation \eqref{eq:expcoord} shall be provided in a future work.

Note, as an aside, that the above description paves the way to a natural definition of the (cartesian) product of supermanifolds $\,\xcM_A,\ A\in\{1,2\}\,$ -- this is just the functor 
\qq\nn
\bigl({\rm Yon}(\xcM_1),{\rm Yon}(\xcM_2)\bigr)
\qqq
with
\qq\nn
\bigl({\rm Yon}(\xcM_1),{\rm Yon}(\xcM_2)\bigr)(\xcS)={\rm Yon}(\xcM_1)(\xcS)\x{\rm Yon}(\xcM_2)(\xcS)\,,
\qqq
or -- equivalently --
\qq\nn
\xcM_1\x\xcM_2\equiv{\rm Yon}^{-1}\bigl({\rm Yon}(\xcM_1),{\rm Yon}(\xcM_2)\bigr)\,.
\qqq 
Parenthetically, we add that arbitrary presheaves on $\,{\rm sMan}\,$ are sometimes referred to as \textbf{generalised supermanifolds}. We distinguish those isomorphic with objects from the image of the Yoneda embedding by the term \textbf{representable supermanifolds}.

We may now define the (generalised) mapping supermanifold 
\qq\nn
[\xcM_1,\xcM_2]:=\unl\Hom_{\rm sMan}(\xcM_1,\xcM_2)\in{\rm Obj}\,{\rm Fun}\left({\rm sMan}^{\rm opp},{\rm Set}\right)
\qqq
as the internal $\,\Hom$
\qq\nn
\unl\Hom_{\rm sMan}(\xcM_1,\xcM_2)\equiv\Hom_{\rm sMan}(-\x\xcM_1,\xcM_2)
\qqq
with $\xcS$-points
\qq\nn
\unl\Hom_{\rm sMan}(\xcM_1,\xcM_2)(\xcS)=\Hom_{\rm sMan}(\xcS\x\xcM_1,\xcM_2)\,.
\qqq
Whenever the mapping supermanifold $\,[\xcM_1,\xcM_2]\,$ is representable, its basic property
\qq\nn
[\xcM_1,\xcM_2](\xcS)\cong\Hom_{\rm sMan}\bigl(\xcS,[\xcM_1,\xcM_2]\bigr)
\qqq
mimics the analogous property of function sets, and so we take it as the appropriate definition of the `space of mappings' alluded to at the beginning of this remark. 

Following \Rcite{Freed:1999} ({\it cp} also Refs.\,\cite{Rabin:1984rm,Rabin:1985tv}), we now think of the super-$\si$-model as a \emph{family} of field theories indexed by the distinguished reference supermanifolds: the odd hyperplanes $\,\bR^{0\,\vert\,N},N\in\bN^\x$.\ More specifically, we give a concrete functional form to the Dirac--Feynman amplitude of the super-$\si$-model only upon fixing $\,N\in\bN^\x\,$ and decomposing the lagrangean `fields' $\,\xi^{\unl A},\ \unl A\in\ovl{1,\dim\,\tgt}\,$ in the basis $\,\eta^i,\ i\in\ovl{1,N}\,$ of the space of global sections $\,\bR[\eta^1,\eta^2,\ldots,\eta^N]\,$ of the structure sheaf of $\,\bR^{0\,\vert\,N}\,$ as
\qq\nn
\xi^a&=&\xi^a_0+\xi^a_{i_1 i_2}\,\eta^{i_1}\,\eta^{i_2}+\cdots+\xi^a_{i_1 i_2\ldots i_{2[\frac{N}{2}]}}\,\eta^{i_1}\,\eta^{i_2}\,\ldots\,\eta^{i_{2[\frac{N}{2}]}}\cr\cr
\xi^{\widehat\a}&=&\xi^{\widehat\a}_{i_1}\,\eta^{i_1}+\xi^{\widehat\a}_{i_1 i_2 i_3}\,\eta^{i_1}\,\eta^{i_2}\,\eta^{i_3}+\cdots+\xi^{\widehat\a}_{i_1 i_2\ldots i_{2[\frac{N-1}{2}]+1}}\,\eta^{i_1}\,\eta^{i_2}\,\ldots\,\eta^{i_{2[\frac{N-1}{2}]+1}}\,.
\qqq
Here, the Gra\ss mann-\emph{even} (functional) coefficients of the decompositions acquire the interpretation of the (component) physical fields of the lagrangean field theory. We shall work in the $\xcS$-point picture throughout the present paper.
\erem

\section{The super-Minkowskian background}\label{sec:sMinktarget}

In the present section, we restrict our considerations to one of the simplest supertargets, to wit, the {\bf super-Minkowski spacetime with $N$ supersymmetries}, and specify its tensorial data necessary for the definition of the relevant super-$\si$-model.

\subsection{The Cartan supergeometry of the super-Minkowskian target}\label{sec:CartMink}

As a supermanifold, the super-Minkowski spacetime with $N$ supersymmetries is the previously introduced model ringed space 
\qq\nn
\bigl(\bR^{\x d+1},C^\infty(\cdot,\bR)\ox\bigwedge\bR^{\x ND_{d,1}}\bigr)\equiv{\rm sMink}^{d,1\,\vert\,ND_{d,1}}\,,\qquad D_{d,1}=\dim\,S_{d,1}\,,
\qqq
where $\,S_{d,1}\,$ denotes the Majorana-spinor representation of the spin group $\,{\rm Spin}(d,1)\,$ of the Clifford algebra $\,\Cliff(\bR^{d,1})\,$ of the standard Minkowski (quadratic) space $\,\bR^{d,1}\equiv(\bR^{\x d+1},\eta),\ \eta=\diag(-,\underbrace{+,+,\ldots,+}_{d\ {\rm times}})$.\ The supertarget will be conveniently described as a homogeneous space of the natural action of the $N$-extended super-Poincar\'e Lie supergroup, the latter being defined as the semidirect product of the $N$-extended {\bf supertranslation group}\footnote{We adopt mathematicians' notation in which the supertranslation group is denoted as $\,\bR^{d,1\,\vert\,ND_{d,1}}$,\ while physicists would have it in the form $\,\bR(d,1\,\vert\,D_{d,1})$.} $\,\bR^{d,1\,\vert\,ND_{d,1}}\equiv{\rm sMink}^{d,1\,\vert\,ND_{d,1}}\,$ with the spin group $\,{\rm Spin}(d,1)$,
\qq\label{eq:sPoinc}
{\rm sISO}(d,1\,\vert\,N D_{d,1})=\bR^{d,1\,\vert\,ND_{d,1}}\rx{\rm Spin}(d,1)\,,
\qqq
with respect to the standard vector-spinor representation of $\,{\rm Spin}(d,1)\,$ on $\,\bR^{d,1\,\vert\,ND_{d,1}}$.\ 
The supergroup, and so also the supertarget, admits homogeneous coordinates: the Gra\ss mann-even ones $\,\{x^a\}^{a\in\ovl{0,d}}\,$ associated with the left-invariant vector fields $\,\{P_a\}_{a\in\ovl{0,d}}\,$ generating translations and $\,\{\phi^{bc}=-\phi^{cb}\}_{b,c\in\ovl{0,d}}\,$ (local) associated with the left-invariant vector fields $\,\{J_{bc}=-J_{cb}\}^{b,c\in\ovl{0,d}}\,$ generating Lorentz transformations, as well as the Gra\ss mann-odd ones $\,\{\theta^{\a I}\}^{(\a,I)\in\ovl{1,D_{d,1}}\x\ovl{1,N}}\,$ associated with left-invariant (super)vector fields $\,\{Q_{\a I}\}_{(\a,I)\in\ovl{1,D_{d,1}}\x\ovl{1,N}}\,$ generating spinorial translations. The Lie-supergroup structure on the above supermanifold is determined by the binary operation
\qq\nn
\txm\ &:&\ {\rm sISO}(d,1\,\vert\,N D_{d,1})\x{\rm sISO}(d,1\,\vert\,N D_{d,1})\too{\rm sISO}(d,1\,\vert\,N D_{d,1})\cr\cr
\ &:&\ \bigl(\bigl(\theta^{\a I}_1,x_1^a,\phi_1^{bc}\bigr),\bigl(\theta^{\b J}_2,x_2^d,\phi_2^{ef}\bigr)\bigr)\longmapsto\bigl(\theta_1^{\a I}+S_I(\phi_1)_{\ \b}^\a\,\theta_2^{\b I},x_1^a+ L(\phi_1)^a_{\ b}\,x_2^b\cr\cr
&&\hspace{5.75cm}-\tfrac{1}{2}\,\theta_1^{\a I}\,\d_{IJ}\,\bigl(C\,\G^a\,S_J(\phi_1)\bigr)_{\a\b}\,\theta_2^{\b J},\widetilde\phi(\phi_1,\phi_2)^{cd}\bigr)\,,
\qqq
written in terms of the vector representation $\,L\ :\ {\rm Spin}(d,1)\too\End\,(\bR^{\x d+1})\,$ and of the $I$-th Majorana-spinor representation $\,S_I\ :\ {\rm Spin}(d,1)\too\End\,(S_{d,1})\,$ of $\,{\rm Spin}(d,1)$,\ in which we also take the relevant charge-conjugation matrix and the generators of the Clifford algebra (with contributions from the representations resummed over the range $\,I,J\in\ovl{1,N}\,$ in the vectorial Gra\ss mann-even component), and in terms of the standard non-linear group law $\,\widetilde\phi\,$ for elements of group $\,{\rm Spin}(d,1)$.\ Upon one-sided restriction to $\,{\rm sMink}^{d,1\,\vert\,ND_{d,1}}\equiv\bR^{d,1\,\vert\,ND_{d,1}}\subset{\rm sISO}(d,1\,\vert\,N D_{d,1})\,$ in the above group law, we recover the natural (left) action of $\,{\rm sISO}(d,1\,\vert\,N D_{d,1})\,$ on the super-Minkowski space (coset),
\qq\nn
\ell_\cdot\ &:&\ {\rm sISO}(d,1\,\vert\,N D_{d,1})\x{\rm sMink}^{d,1\,\vert\,ND_{d,1}}\too{\rm sMink}^{d,1\,\vert\,ND_{d,1}}\cr\cr
\ &:&\ \bigl(\bigl(\vep^{\a I},y^a,\psi^{bc}\bigr),\bigl(\theta^{\b J},x^d\bigr)\bigr)\longmapsto\bigl( S_I(\psi)_{\ \b}^\a\,\theta^{\b I}+\vep^{\a I},L(\psi)_{\ b}^a\,x^b+y^a-\tfrac{1}{2}\,\vep^{\a I}\,\d_{IJ}\,\bigl(C\,\G^a\,S_J(\psi)\bigr)_{\a\b}\,\theta^{\b J}\bigr)\,.
\qqq
The right action of the supertranslation group on the super-Minkowski spacetime is defined analogously,
\qq\label{eq:sMinksMink}
\wp\ &:&\ {\rm sMink}^{d,1\,\vert\,ND_{d,1}}\x\bR^{d,1\,\vert\,ND_{d,1}}\too{\rm sMink}^{d,1\,\vert\,ND_{d,1}}\cr\cr
\ &:&\ \bigl(\bigl(\theta^{\a I},x^a\bigr),\bigl(\vep^{\b J},y^b\bigr)\bigr)\longmapsto\bigl(\theta^{\a I}+\vep^{\a I},x^a+y^a-\tfrac{1}{2}\,\theta^{\a I}\,\d_{IJ}\,\bigl(C\,\G^a\bigr)_{\a\b}\,\vep^{\b J}\bigr)\,.
\qqq
It is to be noted at this stage that the generators of the Clifford algebra are equivariant with respect to the two representations of $\,{\rm Spin}(d,1)\,$ introduced above, as expressed by the identities
\qq\label{eq:Gammasinter}
S_I(\phi)\cdot\G^a\cdot S_I(-\phi)= L(-\phi)^a_{\ b}\,\G^b\,.
\qqq
Here, $\,S_I(-\phi)\equiv S_I(\phi)^{-1}\,$ and, similarly, $\, L(-\phi)\equiv L(\phi)^{-1}$,\ and we have the defining identity
\qq\label{eq:LorintMink}
 L(\phi)^c_{\ b}\,\eta_{ca}= L(-\phi)^c_{\ a}\,\eta_{cb}\,.
\qqq
In consequence of the symmetry properties of the said generators listed in Conv.\,\ref{conv:Cliff}, we also obtain the useful equality (writing $\,\phi_{ab}\equiv\phi^{cd}\,\eta_{ca}\,\eta_{db}\,$ where necessary)
\qq
C\cdot S_I(\phi)\cdot C^{-1}&\equiv&C\cdot\exp\bigl(\tfrac{1}{2}\,\phi_{ab}\,\G^a\cdot\G^b\bigr)\cdot C^{-1}=\exp\bigl(\tfrac{1}{2}\,\phi_{ab}\,C\cdot\G^a\cdot C^{-1}\cdot C\cdot\G^b\cdot C^{-1}\bigr)\cr\cr
&=&\exp\bigl(\tfrac{1}{2}\,\phi_{ab}\,\bigl(-\G^{a\,{\rm T}}\bigr)\cdot\bigl(-\G^{b\,{\rm T}}\bigr)\bigr)=\exp\bigl(\tfrac{1}{2}\,\phi_{ab}\,\bigl(\G^b\cdot\G^a\bigr)^{\rm T}\bigr)\cr\cr
&=&\exp\bigl(\tfrac{1}{2}\,\phi_{ab}\,\G^b\cdot\G^a\bigr)^{\rm T}\equiv\exp\bigl(\tfrac{1}{2}\,\phi_{ab}\,\bigl(\{\G^b,\G^a\}-\G^a\cdot\G^b\bigr)\bigr)^{\rm T}\cr\cr
&=&\exp\bigl(\tfrac{1}{2}\,\phi_{ab}\,\bigl(2\eta^{ba}\,\bd1_{D_{d,1}}-\G^a\cdot\G^b\bigr)\bigr)^{\rm T}=\exp\bigl(-\tfrac{1}{2}\,\phi_{ab}\,\G^a\cdot\G^b\bigr)^{\rm T}\cr\cr
&\equiv&S_I(-\phi)^{\rm T}\,.\label{eq:LorintC}
\qqq
We may finally write out the left-invariant supervector fields on $\,{\rm sISO}(d,1\,\vert\,N D_{d,1})$:
\qq\nn
&P_a(\theta,x,\phi)= L(\phi)^b_{\ a}\,\tfrac{\p\ }{\p x^b}\,,\qquad\qquad Q_{\a I}(\theta,x,\phi)=S_I(\phi)^\b_{\ \a}\,\bigl(\tfrac{\overrightarrow\p\ }{\p\theta^{\b I}}+\tfrac{1}{2}\,\theta^{\g I}\,\bigl(C\,\G^a\bigr)_{\g\b}\,\tfrac{\p\ }{\p x^a}\bigr)\,,&\cr\cr
&J_{ab}(\theta,x,\phi)=\tfrac{\sfd\ }{\sfd t}\rstr_{t=0}\,\widetilde\phi(\phi,t\,\phi_{ab})\,,\quad(\phi_{ab})^{cd}=\d^{\ c}_a\,\d^{\ d}_b-\d^{\ c}_b\,\d^{\ d}_a\,.&
\qqq
These satisfy the familiar super-Poincar\'e (super)algebra
\qq
&[P_a,P_b]=0\,,\qquad\qquad\{Q_{\a I},Q_{\b J}\}=\d_{IJ}\,\bigl(C\,\G^a\bigr)_{\a\b}\,P_a\,,\qquad\qquad[P_a,Q_{\a I}]=0\,,&\cr\cr
&[J_{ab},J_{cd}]=\eta_{ad}\,J_{bc}-\eta_{ac}\,J_{bd}+\eta_{bc}\,J_{ad}-\eta_{bd}\,J_{ac}\,,&\label{eq:sPoincalg}\\ \cr
&[J_{ab},P_c]=\eta_{bc}\,P_a-\eta_{ac}\,P_b\,,\qquad\qquad[J_{ab},Q_{\a I}]=\tfrac{1}{2}\,\bigl(\G_{ab}\bigr)^\b_{\ \a}\,Q_{\b I}\,.&\nonumber
\qqq
We shall also need the right-invariant supervector fields on $\,{\rm sISO}(d,1\,\vert\,N D_{d,1})$:
\qq\nn
&\xcP_a(\theta,x,\phi)=\tfrac{\p\ }{\p x^a}\,,\qquad\qquad\xcQ_{\a I}(\theta,x,\phi)=\tfrac{\overrightarrow\p\ }{\p\theta^{\a I}}-\tfrac{1}{2}\,\theta^{\b I}\,\bigl(C\,\G^a\bigr)_{\b\a}\,\tfrac{\p\ }{\p x^a}\,,&\cr\cr
&\xcJ_{ab}(\theta,x,\phi)=x^c\,\bigl(\eta_{cb}\,\tfrac{\p\ }{\p x^a}-\eta_{ca}\,\tfrac{\p\ }{\p x^b}\bigr)+\tfrac{1}{2}\,\bigl(\G_{ab}\bigr)^\a_{\ \b}\,\theta^{\b I}\,\tfrac{\p\ }{\p\theta^{\a I}}+\tfrac{\sfd\ }{\sfd t}\rstr_{t=0}\,\widetilde\phi(t\,\phi_{ab},\phi)\,,&
\qqq 
with the corresponding super-Poincar\'e (super)algebra
\qq
&[\xcP_a,\xcP_b]=0\,,\qquad\qquad\{\xcQ_{\a I},\xcQ_{\b J}\}=-\d_{IJ}\,\bigl(C\,\G^a\bigr)_{\a\b}\,\xcP_a\,,\qquad\qquad[\xcP_a,\xcQ_{\a I}]=0\,,&\cr\cr
&[\xcJ_{ab},\xcJ_{cd}]=-\eta_{ad}\,\xcJ_{bc}+\eta_{ac}\,\xcJ_{bd}-\eta_{bc}\,\xcJ_{ad}+\eta_{bd}\,\xcJ_{ac}\,,&\label{eq:sPoincalgR}\\ \cr
&[\xcJ_{ab},\xcP_c]=-\eta_{bc}\,\xcP_a+\eta_{ac}\,\xcP_b\,,\qquad\qquad[\xcJ_{ab},\xcQ_{\a I}]=-\tfrac{1}{2}\,\bigl(\G_{ab}\bigr)^\b_{\ \a}\,\xcQ_{\b I}\,.&\nonumber
\qqq
In their derivation, we employ the explicit vector and spinor representations
\qq\label{eq:Lorexpl}
(J_{cd})^a_{\ b}=\d^a_{\ c}\,\eta_{db}-\d^a_{\ d}\,\eta_{cb}\,,\qquad\qquad(J_{cd})^\a_{\ \b}=\tfrac{1}{2}\,\bigl(\G_{cd}\bigr)^\a_{\ \b}
\qqq
of the Lorentz generators.

The above data enable us to describe and manipulate, in a particularly convenient manner, the dual left-invariant superdifferential forms given by components of the Maurer--Cartan super-1-form. These are instrumental in defining the super-$\si$-models. Thus, we parametrise the group (locally) with the help of the exponential mapping ($t_A\equiv t_A(0,0,0)$) as
\qq\nn
g(\theta,x,\phi)=\ee^{x^a\,P_a}\cdot\ee^{\theta^{\a I}\,Q_{\a I}}\cdot\ee^{\frac{1}{2}\,\phi^{bc}\,J_{bc}}\in{\rm sISO}(d,1\,\vert\,N D_{d,1})
\qqq
and obtain the desired decomposition
\qq\nn
g^*\theta_{\rm L}(\theta,x,\phi)&=&\ee^{-\frac{1}{2}\,\phi^{bc}\,J_{bc}}\cdot\ee^{-\theta^{\a I}\,Q_{\a I}}\cdot\ee^{-x^a\,P_a}\,\sfd\bigl(\ee^{x^a\,P_a}\cdot\ee^{\theta^{\a I}\,Q_{\a I}}\cdot\ee^{\frac{1}{2}\,\phi^{bc}\,J_{bc}}\bigr)\cr\cr
&=&\sfd x^a\ox\sfT_e\Ad_{\ee^{-\frac{1}{2}\,\phi^{bc}\,J_{bc}}}(P_a)+\bigl(\id_{\Om^1({\rm sISO}(d,1;ND_{d,1}))}\ox\sfT_e\Ad_{\ee^{-\frac{1}{2}\,\phi^{bc}\,J_{bc}}}\bigr)\bigl(\ee^{-\theta^{\a I}\,Q_{\a I}}\,\sfd\ee^{\theta^{\a I}\,Q_{\a I}}\bigr)\cr\cr
&&+\ee^{-\frac{1}{2}\,\phi^{bc}\,J_{bc}}\,\sfd\ee^{\frac{1}{2}\,\phi^{bc}\,J_{bc}}\cr\cr
&=& L(-\phi)^a_{\ b}\,\sfd x^b\ox P_a+\tfrac{1}{2}\,\theta^{\a I}\,\bigl(C\,\G^a\bigr)_{\a\b}\,\sfd\theta^{\b I}\ox\sfT_e\Ad_{\ee^{-\frac{1}{2}\,\phi^{bc}\,J_{bc}}}(P_a)\cr\cr
&&+\sfd\theta^{\a I}\ox\sfT_e\Ad_{\ee^{-\frac{1}{2}\,\phi^{bc}\,J_{bc}}}(Q_{\a I})+\frac{1}{2}\, L(-\phi)^b_{\ d}\,\sfd L(\phi)^c_{\ e}\,\eta^{de}\ox J_{bc}\cr\cr
&=& L(-\phi)^a_{\ b}\,\bigl(\sfd x^b+\tfrac{1}{2}\,\theta^{\a I}\,\bigl(C\,\G^b\bigr)_{\a\b}\,\sfd\theta^{\b I}\bigr)\ox P_a+S_I(-\phi)^\a_{\ \b}\,\sfd\theta^{\b I}\ox Q_{\a I}\cr\cr
&&+\frac{1}{2}\, L(-\phi)^b_{\ d}\,\sfd L(\phi)^d_{\ e}\,\eta^{ec}\ox J_{bc}
\qqq
of the Maurer--Cartan super-1-form. In its derivation, we use the following identity (in which we have fixed $\,n\in\bN^\x\,$ and suppressed the representation label $\,I\,$ for the sake of transparency):
\qq\nn
\sfd(\theta^{\a_1}\,Q_{\a_1}\,\theta^{\a_2}\,Q_{\a_2}\,\cdots\,\theta^{\a_n}\,Q_{\a_n})&=&\bigl(n(n-1)\,\theta^{\a_1}\,Q_{\a_1}\,\theta^{\a_2}\,Q_{\a_2}\,\cdots\,\theta^{\a_{n-2}}\,Q_{\a_{n-2}}\bigr)\,\tfrac{1}{2}\,\theta^\a\,\bigl(C\,\G^a\bigr)_{\a\b}\,\sfd\theta^\b\,P_a\cr\cr
&&+n\,\theta^{\a_1}\,Q_{\a_1}\,\theta^{\a_2}\,Q_{\a_2}\,\cdots\,\theta^{\a_{n-1}}\,Q_{\a_{n-1}}\,\sfd\theta^\a\,Q_\a\,.
\qqq
In this manner, we identify the sought-after component left-invariant super-1-forms in the decomposition
\qq\nn
g^*\theta_{\rm L}(\theta,x,\phi)=\theta_{\rm L}^a(\theta,x,\phi)\ox P_a+\theta_{\rm L}^{\a I}(\theta,x,\phi)\ox Q_{\a I}+\theta_{\rm L}^{bc}(\theta,x,\phi)\ox J_{bc}
\qqq
as
\qq
\theta_{\rm L}^a(\theta,x,\phi)&=& L(-\phi)^a_{\ b}\,\bigl(\sfd x^b+\tfrac{1}{2}\,\theta^{\a I}\,\d_{IJ}\,\bigl(C\,\G^b\bigr)_{\a\b}\,\sfd\theta^{\b J}\bigr)\,,\cr\cr
\theta_{\rm L}^{\a I}(\theta,x,\phi)&=&S_I(-\phi)^\a_{\ \b}\,\sfd\theta^{\b I}\,,\label{eq:thetaLdef}\\\cr
\theta_{\rm L}^{bc}(\theta,x,\phi)&=& L(-\phi)^b_{\ d}\,\sfd L(\phi)^d_{\ e}\,\eta^{ec}\,.\nonumber
\qqq
Their invariance with respect to left translations on $\,{\rm sISO}(d,1\,\vert\,N D_{d,1})\,$ can be checked directly. They satisfy the Maurer--Cartan equations 
\qq\nn
\sfd\theta_{\rm L}^a&=&-\eta_{bc}\,\theta_{\rm L}^{ab}\wedge\theta_{\rm L}^c+\tfrac{1}{2}\, L(-\phi)^a_{\ b}\,S_I(\phi)^\a_{\ \d}\,S_J(\phi)^\b_{\ \g}\,\theta_{\rm L}^{\d I}\wedge\d_{IJ}\,\bigl(C\,\G^b\bigr)_{\a\b}\,\theta_{\rm L}^{\g J}\cr\cr
&=&-\eta_{bc}\,\theta_{\rm L}^{ab}\wedge\theta_{\rm L}^c+\tfrac{1}{2}\, L(-\phi)^a_{\ b}\,S_J(-\phi)^\b_{\ \d}\,S_J(\phi)^\g_{\ \ep}\,\theta_{\rm L}^{\a I}\wedge\d_{ab}\,C_{\a\b}\,\bigl(\G^a\bigr)^\d_{\ \g}\,\theta_{\rm L}^{\ep J}\cr\cr
&=&-\eta_{bc}\,\theta_{\rm L}^{ab}\wedge\theta_{\rm L}^c+\tfrac{1}{2}\, L(-\phi)^a_{\ b}\, L(\phi)^b_{\ c}\,\theta_{\rm L}^{\a I}\wedge\d_{IJ}\,\bigl(C\,\G^c\bigr)_{\a\b}\,\theta_{\rm L}^{\b J}\cr\cr
&=&-\eta_{bc}\,\theta_{\rm L}^{ab}\wedge\theta_{\rm L}^c+\tfrac{1}{2}\,\theta_{\rm L}^{\a I}\wedge\d_{IJ}\,\bigl(C\,\G^a\bigr)_{\a\b}\,\theta_{\rm L}^{\b J}\,,\cr\cr\cr
\sfd\theta_{\rm L}^{\a I}&=&\sfd S_I(-\phi)^\a_{\ \b}\wedge\sfd\theta^{\b I}\equiv\sfd\bigl(\ee^{-\frac{1}{2}\,\phi^{bc}\,J_{bc}}\bigr)^\a_{\b}\wedge\sfd\theta^{\b I}\cr\cr
&=&-\bigl(\ee^{-\frac{1}{2}\,\phi^{bc}\,J_{bc}}\,\sfd\ee^{\frac{1}{2}\,\phi^{bc}\,J_{bc}}\,\ee^{-\frac{1}{2}\,\phi^{bc}\,J_{bc}}\bigr)^\a_{\b}\wedge\sfd\theta^{\b I}=-\bigl(\ee^{-\frac{1}{2}\,\phi^{bc}\,J_{bc}}\,\sfd\ee^{\frac{1}{2}\,\phi^{bc}\,J_{bc}}\bigr)^\a_{\b}\wedge\theta_{\rm L}^{\b I}\cr\cr
&=&-\tfrac{1}{2}\,\theta_{\rm L}^{bc}\,\bigl(J_{bc}\bigr)^\a_{\b}\wedge\theta_{\rm L}^{\b I}=-\tfrac{1}{4}\,\theta_{\rm L}^{bc}\wedge\bigl(\G_{bc}\bigr)^\a_{\ \b}\,\theta_{\rm L}^{\b I}\,,\cr\cr\cr
\sfd\theta_{\rm L}^{bc}&=&-\eta_{de}\,\theta_{\rm L}^{bd}\wedge\theta_{\rm L}^{ce}\,,
\qqq
dictated by the algebra \eqref{eq:sPoincalg} in what is the most fundamental manifestation of the standard (Free Differential-Algebraic) relation between the Chevalley--Eilenberg model of the Lie-(super)algebra cohomology and the Cartan--Eilenberg model of the Lie (super)group-invariant de Rham cohomology. Above, we used \Reqref{eq:Lorexpl} to compute the differential 
\qq\nn
\sfd L(-\phi)^a_{\ b}&=&-\bigl(\ee^{-\frac{1}{2}\,\phi^{cd}\,J_{cd}}\,\sfd\ee^{\frac{1}{2}\,\phi^{cd}\,J_{cd}}\,\ee^{-\frac{1}{2}\,\phi^{cd}\,J_{cd}}\bigr)^a_{\ b}=-\tfrac{1}{2}\,\theta_{\rm L}^{cd}\,\bigl(J_{cd}\bigr)^a_{\ e}\, L(-\phi)^e_{\ b}\cr\cr
&=&-\theta_{\rm L}^{ac}\,\eta_{cd}\, L(-\phi)^d_{\ b}
\qqq
and its spinorial variant.

Finally, we are ready to specify the (super)group-theoretic form of the lagrangean fields of the Nambu--Goto formulation of the Green--Schwarz super-$\si$-model, regarded as a close cousin of the Wess--Zumino--Witten model of \Rcite{Witten:1983ar}. Thus, we take the ($\g$-shifted) lagrangean field in the form
\qq\nn
\widetilde\xi\equiv\g\circ\xi\in[\Om,{\rm sISO}(d,1\,\vert\,N D_{d,1})]\,,\qquad\qquad\widetilde\xi(\si)=\ee^{x^a(\si)\,P_a}\cdot\ee^{\theta^{\a I}(\si)\,Q_{\a I}}\,.
\qqq
From now onwards, we shall, in our analysis, use the shorthand notation (and make the assumptions) of Conv.\,\ref{conv:Cliff} and consider non-extended supersymmetry, 
\qq\nn
N=1\,, 
\qqq
for the sake of clarity, so that the generation index $\,I\,$ can be suppressed. In particular, we denote the corresponding left-invariant super-1-forms as
\qq\nn
S(-\phi)^\a_{\ \b}\,\si^\b(\theta)=\Si^\a_{\rm L}(\theta,x,\phi)\equiv\theta^\a_{\rm L}(\theta,x,\phi)
\qqq
in order to distinguish them clearly from their spacetime-indexed counterparts 
\qq\label{eq:eIdef}
\theta_{\rm L}^a(\theta,x,\phi)\equiv L(-\phi)^a_{\ b}\,e^b(\theta,x) 
\qqq
in index-free expressions with contracted spinorial indices, such as, {\it e.g.}, the following one
\qq\nn
\theta^\a_{\rm L}\wedge\ovl\G{}^a_{\a\b}\,\theta^\b_{\rm L}\equiv\ovl\Si_{\rm L}\wedge\G^a\,\Si_{\rm L}\,.
\qqq 

\brem
We are now ready to concretise the gometric contents of the Rabin--Crane argument invoked in the opening paragraphs of Sec.\,\ref{sec:stensor} and thus give some flesh to the idea underlying the geometrisation of the Green--Schwarz super-$(p+2)$-cocycles to be considered in what follows. To this end, we should specialise the discussion presented in Remark \ref{rem:Spoints} to the present supergeometric setting. Upon fixing a reference supermanifold $\,\bR^{0\,\vert\,N},\ N\in\bN^\x$,\ we obtain lagrangean fields 
\qq\nn
x^a&=&x^a_0+x^a_{i_1 i_2}\,\eta^{i_1}\,\eta^{i_2}+\cdots+x^a_{i_1 i_2\ldots i_{2[\frac{N}{2}]}}\,\eta^{i_1}\,\eta^{i_2}\,\ldots\,\eta^{i_{2[\frac{N}{2}]}}\cr\cr
\theta^\a&=&\theta^\a_{i_1}\,\eta^{i_1}+\theta^\a_{i_1 i_2 i_3}\,\eta^{i_1}\,\eta^{i_2}\,\eta^{i_3}+\cdots+\theta^\a_{i_1 i_2\ldots i_{2[\frac{N-1}{2}]+1}}\,\eta^{i_1}\,\eta^{i_2}\,\ldots\,\eta^{i_{2[\frac{N-1}{2}]+1}}\,.
\qqq
Taking into account the corresponding decomposition of parameters of (rigid) supersymmetry transformations,
 \qq\nn
y^a&=&y^a_0+y^a_{i_1 i_2}\,\eta^{i_1}\,\eta^{i_2}+\cdots+y^a_{i_1 i_2\ldots i_{2[\frac{N}{2}]}}\,\eta^{i_1}\,\eta^{i_2}\,\ldots\,\eta^{i_{2[\frac{N}{2}]}}\cr\cr
\vep^\a&=&\vep^\a_{i_1}\,\eta^{i_1}+\vep^\a_{i_1 i_2 i_3}\,\eta^{i_1}\,\eta^{i_2}\,\eta^{i_3}+\cdots+\vep^\a_{i_1 i_2\ldots i_{2[\frac{N-1}{2}]+1}}\,\eta^{i_1}\,\eta^{i_2}\,\ldots\,\eta^{i_{2[\frac{N-1}{2}]+1}}\,,
\qqq
we readily identify the Kosteleck\'y--Rabin discrete supersymmetry supergroup $\,\G_{\rm KR}\,$ as the subgroup of $\,\bR(d,1\,\vert\,D_{d,1})\,$ generated, in the $\bR^{0\,\vert\,N}$-parametrisation adopted, by supertranslations $\,(y^a,\vep^\a)\,$ with integer coefficients,
\qq\nn
y^a_0,y^a_{i_1 i_2\ldots i_k}\in\bZ\,,\qquad k\in\ovl{1,2[\tfrac{N}{2}]}\,,\qquad\qquad\vep^\a_{i_1 i_2\ldots i_l}\in\bZ\,,\qquad l\in\ovl{1,2[\tfrac{N-1}{2}]+1}\,.
\qqq
An example of a quotient $\,\bR(d,1\,\vert\,D_{d,1})/\G_{\rm KR}$,\ with a nontrivial topology in the Gra\ss mann-odd directions, was worked out completely explicitly by Rabin and Crane in \Rcite{Rabin:1984rm}.
\erem

\subsection{The $\,N=1\,$ GS super-$(p+2)$-cocycles and the ensuing ``old branescan''}\label{ref:GScocyc}

The super-$p$-branes whose dynamics we intend to geometrise carry topological charge, and so their propagation defines a charge current to which a gauge field couples in the usual geometric manner, that is, through pullback (of the gauge potential) to the worldvolume of the charged object. The coupling gives rise to corrections to the condition of minimality of the classical embedding that follows from minimising the metric term of the (super-)$\si$-model action functional, and the corrections are determined by the field strength of the said gauge field. As was announced at the beginning of Sec.\,\ref{sec:stensor}, these field strengths are certain distinguished ${\rm sISO}(d,1\,|\,D_{d,1})$-invariant de Rham super-$(p+2)$-cocycles that -- owing to the topological triviality of the super-Minkowski space -- admit global primitives, none of which, however, is ${\rm sISO}(d,1\,|\,D_{d,1})$-invariant. The super-$(p+2)$-cocycles that we want to consider take the general form
\qq\label{eq:GScurv}\qquad
\underset{\tx{\ciut{(p+2)}}}{\chi}=\ovl\Si_{\rm L}\wedge\G_{a_1 a_2\ldots a_p}\,\Si_{\rm L}\wedge\theta_{\rm L}^{a_1 a_2\ldots a_p}\,,\qquad\qquad\theta_{\rm L}^{a_1 a_2\ldots a_p}=\theta_{\rm L}^{a_1}\wedge\theta_{\rm L}^{a_2}\wedge\cdots\wedge\theta_{\rm L}^{a_p}\,,\qquad p>0\,,
\qqq
with the sole exception\footnote{There are also other possibilities in the distinguished case $\,p=0$.}
\qq\label{eq:GScurv0}
\underset{\tx{\ciut{(2)}}}{\chi}=\ovl\Si_{\rm L}\wedge\G_{11}\,\Si_{\rm L}
\qqq
occuring for $\,p=0\,$ and defined in terms of the volume element $\,\G_{11}\,$ of the Clifford algebra $\,\Cliff(\bR^{9,1})$,\ described in App.\,\ref{conv:Cliff}. These are readily seen to descend to $\,{\rm sMink}(d,1\,\vert\,D_{d,1})\,$ as
\qq\nn
\underset{\tx{\ciut{(2)}}}{\chi}(\theta,x,\phi)&\equiv&\si^{\rm T}(\theta)\,\wedge S(-\phi)^{\rm T}\cdot C\cdot\G_{11}\cdot S(-\phi)\,\si(\theta)=\ovl\si(\theta)\wedge S(\phi)\cdot\G_{11}\cdot S(-\phi)\,\si(\theta)\cr\cr
&=&\det_{(10)}\,\bigl(L(\phi)\bigr)\cdot\bigl(\ovl\si\wedge\G_{11}\,\si\bigr)(\theta)=\bigl(\ovl\si\wedge\G_{11}\,\si\bigr)(\theta)\equiv\underset{\tx{\ciut{(2)}}}{\chi}(\theta,x,0)=:\underset{\tx{\ciut{(2)}}}{\txH}(\theta,x)
\qqq
and -- for $\,p>0\,$ --
\qq\nn
\underset{\tx{\ciut{(p+2)}}}{\chi}(\theta,x,\phi)&\equiv&\si^{\rm T}(\theta)\wedge S(-\phi)^{\rm T}\cdot C\cdot\G_{a_1 a_2\ldots a_p}\cdot S(-\phi)\,\si(\theta)\wedge\theta_{\rm L}^{a_1 a_2\ldots a_p}(\theta,x,\phi)\cr\cr
&=&\si^{\rm T}(\theta)\wedge C\cdot S(\phi)\cdot\G_{a_1 a_2\ldots a_p}\cdot S(-\phi)\,\si(\theta)\wedge\theta_{\rm L}^{a_1 a_2\ldots a_p}(\theta,x,\phi)\cr\cr
&=&\si^{\rm T}(\theta)\wedge\ovl\G{}^{c_1 c_2\ldots c_p}\,\si(\theta)\cr\cr
&&\wedge\eta_{a_1 b_1}\,\eta_{a_2 b_2}\,\cdots\,\eta_{a_p b_p}\, L(-\phi)^{b_1}_{\ c_1}\, L(-\phi)^{b_2}_{\ c_2}\,\cdots\, L(-\phi)^{b_p}_{\ c_p}\,\theta_{\rm L}^{a_1 a_2\ldots a_p}(\theta,x,\phi)\cr\cr
&=&\si^{\rm T}(\theta)\wedge\ovl\G{}^{c_1 c_2\ldots c_p}\,\si(\theta)\wedge\prod_{k=1}^p\,\bigl(\eta_{a_k b_k}\, L(-\phi)^{b_k}_{\ c_k}\, L(-\phi)^{a_k}_{\ d_k}\bigr)\,e^{d_1 d_2\ldots d_p}(\theta,x)\cr\cr
&=&\ovl\si(\theta)\wedge\G_{a_1 a_2\ldots a_p}\,\si(\theta)\wedge e^{a_1 a_2\ldots a_p}(\theta,x)\equiv\underset{\tx{\ciut{(p+2)}}}{\chi}(\theta,x,0)=:\underset{\tx{\ciut{(p+2)}}}{\txH}(\theta,x)\,,
\qqq
with
\qq\nn
e^{a_1 a_2\ldots a_p}\equiv e^{a_1}\wedge e^{a_2}\wedge\cdots\wedge e^{a_p}\,,
\qqq
and, for all $\,a,b\in\ovl{0,d}$,
\qq\nn
J_{ab}\con\underset{\tx{\ciut{(p+2)}}}{\chi}(\theta,x,\phi)=0\,,
\qqq
which means that the super-$(p+2)$-forms are not only Lorentz-invariant but also (Lorentz-)horizontal, and so, altogether, basic. The descended super-$(p+2)$-forms $\,\underset{\tx{\ciut{(p+2)}}}{\txH}\,$ will be jointly referred to as the {\bf Green--Schwarz} ({\bf GS}) {\bf super-$(p+2)$-cocycles} in what follows. 

For $\,p>0$,\ their closedness, 
\qq\nn
0\must\sfd\underset{\tx{\ciut{(p+2)}}}{\chi}&=&\tfrac{p}{2}\,\Si_{\rm L}^\a\wedge\bigl(\ovl\G_{a_1 a_2 a_3\ldots a_p}\bigr)_{\a\b}\,\Si_{\rm L}^\b\wedge\Si_{\rm L}^\g\wedge\bigl(\ovl\G{}^{a_1}\bigr)_{\g\d}\,\Si_{\rm L}^\d\wedge\theta_{\rm L}^{a_2 a_3\ldots a_p}\cr\cr
&\equiv&\tfrac{p}{2}\,\bigl(\ovl\G_{a_1 a_2 a_3\ldots a_p}\bigr)_{\a\b}\,\bigl(\ovl\G{}^{a_1}\bigr)_{\g\d}\,\Si_{\rm L}^\a\wedge\Si_{\rm L}^\b\wedge\Si_{\rm L}^\g\wedge\Si_{\rm L}^\d\wedge\theta_{\rm L}^{a_2 a_3\ldots a_p}\,,
\qqq
is ensured by a suitable choice of the relevant representation of the Clifford algebra such that the symmetry constraints
\qq\label{eq:ClifFierz}
\ovl\G{}^{a_1}_{(\a\b}\,(\ovl\G_{a_1 a_2\ldots a_p})_{\g\d)}=0\,,
\qqq 
implied by the previous condition, are obeyed. Note that for $\,p=1\,$ the latter reduces to the (more) familiar identity
\qq\label{eq:ClifFierz1}
\ovl\G{}^a_{\a(\b}\,(\ovl\G_a)_{\g\d)}=0
\qqq
due to the assumed symmetry of the $\,\ovl\G{}^a$,\ {\it cp}\ Conv.\,\ref{conv:Cliff}. The admissible pairs $\,(d,p)\,$ for which the above constraints can be solved and a super-$\si$-model with the appropriate local supersymmetry ({\it cp}\ \cite[Sec.\,3]{Suszek:2018sM}) can be written down were found in \Rcite{Achucarro:1987nc} and constitute the so-called ``old branescan''. Closedness of the GS super-$(p+2)$-cocycles implies -- in consequence of the (de Rham-)cohomological triviality of their support (which follows directly from The Kostant Theorem of \Rcite{Kostant:1975}) -- the existence of smooth primitives. These were derived in \Rcite{Hughes:1986fa}, albeit in a different convention, and so we rederive them in App.\,\ref{app:GSprim} through an adaptation of the original method to the current algebraic setting.
\berop\label{prop:GSprim}
For any $\,p>0$,\ the GS super-$(p+2)$-cocycle $\,\underset{\tx{\ciut{(p+2)}}}{\chi}\,$ of \Reqref{eq:GScurv} admits a manifestly ${\rm ISO}(d,1)$-invariant primitive
\qq
\underset{\tx{\ciut{(p+1)}}}{\b}(\theta,x)=\tfrac{1}{p+1}\,\sum_{k=0}^p\,\ovl\theta\,\G_{a_1 a_2\ldots a_k a_{k+1} a_{k+2}\ldots a_p}\,\si(\theta)\wedge\sfd x^{a_1}\wedge\sfd x^{a_2}\wedge\cdots\wedge\sfd x^{a_k}\wedge e^{a_{k+1} a_{k+2}\ldots a_p}(\theta,x)\,.\cr\label{eq:GSprim}
\qqq
A primitive of the super-2-form $\,\underset{\tx{\ciut{(2)}}}{\chi}\,$ of \Reqref{eq:GScurv0} can be chosen in the ${\rm ISO}(9,1)$-invariant form
\qq\nn
\underset{\tx{\ciut{(1)}}}{\b}(\theta,x)=\ovl\theta\,\G_{11}\,\si(\theta)\,.
\qqq
\eerop
\beroof
A proof is given in App.\,\ref{app:GSprim}.
\eroof
\noindent The above primitives are manifestly non-supersymmetric. In fact, this cannot be repaired -- it was demonstrated in \Rcite{deAzcarraga:1989vh} that the GS super-$(p+2)$-cocycles do {\it not} admit ${\rm sISO}(d,1\,|\,D_{d,1})$-invariant primitives. This result puts us naturally in the framework of the ($\bR$-valued) Chevalley--Eilenberg cohomology of the super-Poincar\'e algebra ({\it cp}\ \Rcite{Chevalley:1948}), which we shall exploit in the present treatment, whence a recapitulation thereof in App.\,\ref{app:LieAlgCohom} in the superalgebraic context of interest. 

On the other hand, the manifest left-invariance of the GS super-$(p+2)$-cocycle $\,\underset{\tx{\ciut{(p+2)}}}{\chi}\,$ itself, in conjunction with the triviality of the standard de Rham cohomology of $\,{\rm sMink}(d,1\,\vert\,D_{d,1})$,\ ensures that condition \Reqref{eq:jpullJ} is satisfied as the supersymmetry variation of the global primitive $\,\underset{\tx{\ciut{(p+1)}}}{\b}\,$ is exact. Indeed, we have the identity
\qq\nn
\sfd\underset{\tx{\ciut{(p+1)}}}{\b}\equiv\underset{\tx{\ciut{(p+2)}}}{\chi}=\ell_{(\vep,y)}^*\underset{\tx{\ciut{(p+2)}}}{\chi}\equiv\sfd\bigl(\ell_{(\vep,y)}^*\underset{\tx{\ciut{(p+1)}}}{\b}\bigr)
\qqq
which implies the existence of a super-$p$-form $\,\underset{\tx{\ciut{(p)}}}{\jmath}{}_{(\vep,y)}\in\bigwedge^p\cT^*{\rm sMink}(d,1\,\vert\,D_{d,1})\,$ satisfying the condition
\qq\nn
\bigl(\d\underset{\tx{\ciut{(p+1)}}}{\b}\bigr)_{(\vep,y)}\equiv\bigl(\d_{\bR(d,1\,\vert\,D_{d,1})}\underset{\tx{\ciut{(p+1)}}}{\b}\bigr)_{(\vep,y)}=\ell_{(\vep,y)}^*\underset{\tx{\ciut{(p+1)}}}{\b}-\underset{\tx{\ciut{(p+1)}}}{\b}=\sfd\underset{\tx{\ciut{(p)}}}{\jmath}{}_{(\vep,y)}\,.
\qqq
We shall call $\,\underset{\tx{\ciut{(p)}}}{\jmath}{}_{(\vep,y)}\,$ the {\bf target supercurrent}. The action 1-cochain now takes the explicit form
\qq\nn
c_{(\vep,y)}[\xi]=\ee^{\sfi\,\int_{\p\Om}\,\left(\ell_{(-\vep,-y)}\circ\g\circ\xi\rstr_{\p\Om}\right)^*\underset{\tx{\ciut{(p)}}}{\jmath_{(\vep,y)}}}\,.
\qqq\medskip

By way of a closing remark, we note that the GS super-$(p+2)$-cocycles give rise -- as revealed by inspection -- to a host of supersymmetric super-2-cocycles that play a fundamental r\^ole in our geometrisation of the ${\rm sISO}(d,1\,\vert\,D_{d,1})$-invariant cohomology classes of the $\,\underset{\tx{\ciut{(p+2)}}}{\chi}$.\ These will be obtained through contraction of (certain) $p$-tuples of fundamental (left-invariant) vector fields $\,\cK_{A_i}\in\{Q_\a,P_a\}_{(\a,a)\in\ovl{1,D_{d,1}}\x\ovl{0,d}},\ i\in\ovl{1,p}\,$ (the $\,A_i\,$ are indices of the supersymmetry algebra) with the super-$(p+2)$-cocycle $\,\underset{\tx{\ciut{(p+2)}}}{\chi}\,$ of \Reqref{eq:GScurv},
\qq\label{eq:LI2bas}
\underset{\tx{\ciut{(2)}}}{h}{}_{\la^\cdot}:=\la^{A_1 A_2\ldots A_p}\,\cK_{A_1}\con\cK_{A_2}\con\cdots\cK_{A_p}\con\underset{\tx{\ciut{( p+2)}}}{\chi}\,,\qquad\la^{A_1 A_2\ldots A_p}\in\bR\,.
\qqq
The condition of closedness of the above left-invariant super-2-form is tantamount to certain linear constraints on the coefficients $\,\la^{A_1 A_2\ldots A_p}\,$ which involve (also linearly) the structure constants of the Lie superalgebra under consideration, and so it is far from obvious that such super-2-cocycles exist. Specific examples will be examined closely in Sec.\,\ref{sec:GSgerbe}, where we also adapt this generating mechanism to the super-cocycles on extended spacetimes.

\section{Supergerbes  for the Nambu--Goto super-$p$-branes from extensions of $\,{\rm sMink}(d,1\,\vert\,D_{d,1})$}\label{sec:GSgerbe}

The aim of this section, and the main objective of the present paper motivated amply in Sec.\,\ref{sec:stensor}, is to work out for the GS super-$(p+2)$-cocycles $\,\underset{\tx{\ciut{(p+2)}}}{\txH}\,$ of Sec.\,\ref{ref:GScocyc} a supergeometric analogon of the standard scheme of geometrisation of de Rham cocycles known from the theory of fibre bundles with connection and the theory of bundle ($n$-)gerbes with connection and recalled concisely in Sec.\,\ref{sec:Bose} in the context of the 2d bosonic $\si$-model with the Wess--Zumino term. The conceptual basis of our construction is the relation between algebra and geometry of the Lie (super)group established -- in the manner delineated in Thm.\,\ref{thm:sCEmodelIdR} -- by the Chevalley--Eilenberg model of Lie-(super)algebra cohomology (with values in the trivial module $\,\bR$) in conjunction with the interpretation -- expressed in Props.\,\ref{prop:ExtoCE} and \ref{prop:CEtoExt} -- of the second cohomology group in that model in terms of equivalence classes of (super)central extensions of the underlying Lie-(super)algebra. More specifically, the said cohomological results enable us to associate with the super-$(p+2)$-cocycles\footnote{Strictly speaking, we present an explicit analysis for the cases $\,p\in\{0,1,2\}$.\ However, the structural nature of our construction turns it into a tenable proposal for a completely general geometrisation scheme.} a tower of Lie-supergroup extensions of the Lie supergroup $\,\bR(d,1\,\vert\,D_{d,1})\equiv{\rm sMink}(d,1\,\vert\,D_{d,1})\,$ (of the kind originally discovered in \Rcite{Chryssomalakos:2000xd}) that are readily verified to play the r\^ole of the various surjective submersions encountered in the geometric definition of the (0-, 1- and 2-)gerbe and thus give us a natural definition of a supergerbe with curvature $\,\underset{\tx{\ciut{(p+2)}}}{\chi}$,\ conceived along the lines of the fundamental Principle of Categorial Descent of \Rcite{Stevenson:2000wj}. 

Rudimentary aspects of the Lie-superalgebra (to be abbreviated as \textbf{LSA} in what follows) cohomology and its Chevalley--Eilenberg (to be abbreviated as \textbf{CE}) model, as well as the link with the Cartan--Eilenberg (to be abbreviated as \textbf{CaE}) cohomology of supersymmetric superdifferential forms on the Lie supergroup that are of relevance to the subsequent discussion have been recalled in App.\,\ref{app:LieAlgCohom}.\medskip

By way of preparation for the systematic (super)geometric resolution of the super-$(p+2)$-cocycles of interest, we should first review -- after Refs.\,\cite{Aldaya:1984gt,Chryssomalakos:2000xd}) -- the construction of the super-Minkowski spacetime $\,{\rm sMink}(d,1\,\vert\,D_{d,1})\,$ as a supercentral extension of the purely Gra\ss mann-odd superspace $\,\bR^{0\,\vert\,D_{d,1}}\,$ (the so-called superpoint, also known as the odd hyperplane), determined by a canonical super-2-cocycle on the supercommutative Lie superalgebra $\,\bR^{0\,\vert\,D_{d,1}}\,$ with values in its trivial module $\,\bR^{d,1}$.\ A natural point of departure for our general discussion is the manifestly closed left-invariant (to be abbreviated as \textbf{LI} in what follows) super-2-form
\qq\nn
\underset{\tx{\ciut{(2)}}}{\chi}^a:=\tfrac{1}{2}\,\ovl\si\wedge\G^a\,\si\,,\qquad I\in\ovl{0,d}
\qqq
on the supermanifold $\,\cM^{(0)}\equiv\bR^{0\,\vert\,D_{d,1}}$,\ with global Gra\ss mann-odd coordinates $\,\{\theta^\a\}^{\a\in\ovl{1,D_{d,1}}}\,$ and the associated LI vector fields (the arrow over the symbol of the partial derivative indicates that the latter acts on sections of the structure sheaf from the \emph{left})
\qq\nn
 Q_\a^{(0)}(\theta)=\tfrac{\overrightarrow\p\ }{\p\theta^\a}
\qqq 
furnishing a realisation of the supercommutative LSA 
\qq\nn
\{ Q_\a^{(0)}, Q_\b^{(0)}\}=0\,.
\qqq
The de Rham super-2-cocycle $\,\underset{\tx{\ciut{(2)}}}{\chi}^a\,$ does not admit a primitive on $\,\cM^{(0)}\,$ invariant with respect to the (left) regular action of $\,\bR^{0\,\vert\,D_{d,1}}\,$ on (itself) $\,\cM^{(0)}$,
\qq\nn
\ell^{(0)}_\cdot\ :\ \bR^{0\,\vert\,D_{d,1}}\x\cM^{(0)}\too\cM^{(0)}\ :\ (\vep^\a,\theta^\a)\longmapsto\theta^\a+\vep^\a\,,
\qqq
and so -- arguing along the lines of App.\,\ref{app:LieAlgCohom} -- we are led to consider a supercentral extension $\,\cM^{(1)}:={\rm sMink}(d,1\,\vert\,D_{d,1})\,$ of the Lie supergroup $\,\cM^{(0)}$,\ the former being canonically surjectively submersed onto the latter as a rank-$(d+1)$ (real) vector bundle\footnote{For a detailed account of the fibre-bundle structure on the extended superspacetime(s), consult Refs.\,\cite{Aldaya:1984gt,Chryssomalakos:2000xd}.}
\qq\nn
\pi_0\equiv\pr_1\ :\ \cM^{(1)}\too\cM^{(0)}\ :\ (\theta^\a,x^a)\longmapsto\theta^\a
\qqq
with fibre coordinates $\,x^a,\ a\in\ovl{0,d}$.\ The pullback of the GS super-2-cocycle $\,\underset{\tx{\ciut{(2)}}}{\chi}^a\,$ to $\,\cM^{(1)}\,$ trivialises in the associated CaE cohomology as ({\it cp}\ Remark \ref{rem:LSApulltriv})
\qq\nn
\pi_0^*\underset{\tx{\ciut{(2)}}}{\chi}^a=\sfd e^a\,,
\qqq 
for the $\,e^a\,$ as defined by Eqs.\,\eqref{eq:thetaLdef} and \eqref{eq:eIdef}. The corresponding centrally extended LSA of the equivariant lifts 
\qq\label{eq:Qal1}
 Q^{(1)}_\a(\theta,x):= Q^{(0)}_\a(\theta)+\tfrac{1}{2}\,\ovl\G{}^a_{\a\b}\,\theta^\b\,\tfrac{\p\ }{\p x^a}
\qqq
of the $\, Q^{(0)}_\a\,$ and of the coordinate vector fields 
\qq\label{eq:Pa1}
 P^{(1)}_a(\theta,x):=\tfrac{\p\ }{\p x^a}\,,
\qqq
the two families making up a basis of the tangent sheaf dual to that of the cotangent sheaf formed by the LI super-1-forms $\,\si^\a,\ \a\in\ovl{1,D_{d,1}}\,$ and $\,e^a,\ a\in\ovl{0,d}$,\ reads
\qq
\{ Q^{(1)}_\a, Q^{(1)}_\b\}=\ovl\G{}^a_{\a\b}\, P^{(1)}_a\,,\qquad\qquad[ P^{(1)}_a, P^{(1)}_b]=0\,,\qquad\qquad[ P^{(1)}_a, Q^{(1)}_\a]=0\,.\cr\label{eq:sMincdSA}
\qqq
The action of the original supergroup $\,\bR^{0\,\vert\,D_{d,1}}\,$ on $\,\cM^{(1)}\,$ follows from the demand that it project to $\,\ell^{(0)}_\cdot\,$ and that the super-1-forms $\,\si^\a\,$ and $\,e^a,\ a\in\ovl{0,d}\,$ be invariant with respect to it, and we may extend it to a full-blown structure of a Lie supergroup on $\,\cM^{(1)}\,$ by requiring that it yield the above supervector fields $\, Q^{(1)}_\a,\ \a\in\ovl{1,D_{d,1}}\,$ and $\, P^{(1)}_a,\ a\in\ovl{0,d}\,$ as the fundamental left-invariant supervector fields and that it leave the super-1-forms intact when treated as an action of $\,\bR(d,1\,\vert\,D_{d,1})\,$ on (itself) $\,\cM^{(1)}\,$ -- this determines the said action in the familiar form 
\qq
\ell^{(1)}_\cdot\equiv\txm^{(1)}\ &:&\ \bR(d,1\,\vert\,D_{d,1})\x\cM^{(1)}\too\cM^{(1)}\cr\cr 
&:&\ \left((\vep^\a,y^a),(\theta^\b,x^b)\right)\longmapsto\left(\theta^\a+\vep^\a,x^a+y^a-\tfrac{1}{2}\,\ovl\vep\,\G^a\,\theta\right)\,,\label{eq:sact-sMink}
\qqq
equivalent to the one given in \Reqref{eq:sMinksMink} (for $\,N=1$). Clearly, we could have equivalently derived it from the LSA \eqref{eq:sMincdSA} by exponentiating the generators and computing, with the help of the standard Baker--Campbell--Hausdorff formula,
\qq\nn
\ee^{\vep^\a\, Q^{(1)}_\a+y^a\, P^{(1)}_a}\cdot\ee^{\theta^\b\, Q^{(1)}_\b+x^b\, P^{(1)}_b}&=&\ee^{(\vep^\a+\theta^\a)\, Q^{(1)}_\a+(y^a+x^a)\, P^{(1)}_a+\frac{1}{2}\,[\vep^\a\, Q^{(1)}_\a,\theta^\b\, Q^{(1)}_\b]}\cr\cr
&=&\ee^{(\vep^\a+\theta^\a)\, Q^{(1)}_\a+(y^a+x^a)\, P^{(1)}_a-\frac{1}{2}\,\vep^\a\,\theta^\b\,\{ Q^{(1)}_\a, Q^{(1)}_\b\}}\cr\cr
&=&\ee^{(\vep^\a+\theta^\a)\, Q^{(1)}_\a+(y^a+x^a-\frac{1}{2}\,\ovl\vep\,\G^a\,\theta)\, P^{(1)}_a}\,.
\qqq

Thus, if we take $\,\cM^{(0)}\,$ as the basis of our supergeometry for the sake of illustrating the extension principle, the CaE/CE-cohomological trivialisation leads us quite naturally to a surjective submersion over it, with the Gra\ss mann-even typical fibre $\,\bR^{d,1}$.\ In the remainder of this section, we assume, instead, the super-Minkowski spacetime $\,\cM^{(1)}\,$ to be the actual basis of subsequent extensions necessitated by the trivialisation of the GS super-$(p+2)$-cocycles. However, in order to indicate the relation between $\,\cM^{(0)}\,$ and $\,\cM^{(1)}$,\ we pedantically pull back the LI super-1-forms $\,\si\,$ to $\,\cM^{(1)}\,$ along $\,\pi_0$.\medskip

\subsection{The super-0-gerbe of the super-$0$-brane} 

The GS super-2-cocycle on the ten-dimensional super-Minkowski spacetime $\,\cM^{(1)}\,$ reconstructed above that codetermines the dynamics of the super-$0$-brane has the simple form
\qq\label{eq:GS0form}
\underset{\tx{\ciut{(2)}}}{\txH}=\pi_0^*\bigl(\si\wedge\ovl\G_{11}\,\si\bigr)\,,
\qqq
implicitly written in a spin representation of the Clifford algebra in which the product of the charge-conjugation matrix and the volume element $\,\G_{11}\,$ is symmetric,
\qq\label{eq:Csymm}
\bigl(C\cdot\G_{11}\bigr)^{\rm T}=C\cdot\G_{11}\,.
\qqq 
The super-2-form is manifestly LI but does not possess a primitive with this property. Indeed, such a global primitive $\,\underset{\tx{\ciut{(1)}}}{\b}{}_{\rm L}\,$ of $\,\underset{\tx{\ciut{(2)}}}{\txH}\,$ would be a linear combination of the basis LI super-1-forms 
\qq\nn
\underset{\tx{\ciut{(1)}}}{\b}{}_{\rm L}=\la_\a\,\pi_0^*\si^\a+2\mu_a\,e^a\,,
\qqq
with (constant) ${\rm Spin}(d,1)$-invariant tensors $\,\la_\a\,$ and $\,\mu_a\,$ as coefficients. Setting aside the issue of their (existence and) identification among available algebraic objects, we may directly impose the defining identity
\qq\nn
\si\wedge\ovl\G_{11}\,\si(\theta)\equiv\underset{\tx{\ciut{(2)}}}{\txH}(\theta,x)\must\sfd\underset{\tx{\ciut{(1)}}}{\b}{}_{\rm L}(\theta,x)=2\mu_a\,\sfd e^a(\theta,x)=\mu_a\,\si\wedge\ovl\G{}^a\,\si(\theta)
\qqq
to obtain the constraints
\qq\nn
\ovl\G_{11}-\mu_a\,\ovl\G{}^a\equiv\bigl(\ovl\G_{11}-\mu_a\,\ovl\G{}^a\bigr)^{\rm T}\must-\bigl(\ovl\G_{11}-\mu_a\,\ovl\G{}^a\bigr)\,,
\qqq
or, equivalently, the contradiction
\qq\nn
\G_{11}=\mu_a\,\G{}^a\,.
\qqq
Prior to proceeding with the extension, we pause to take a closer look at the (super)symmetry properties of a primitive of $\,\underset{\tx{\ciut{(2)}}}{\txH}\,$ in its most general form consistent with the (de Rham-)cohomological triviality of the super-Minkowski space,
\qq\label{eq:spartprimgen}
\underset{\tx{\ciut{(1)}}}{\b}(\theta,x)=\theta\,\ovl\G_{11}\,\sfd\theta+\sfd\sfa(\theta,x)\,.
\qqq
with view to understanding the nature of the extension to be constructed. Reasoning along the lines of the (pre-)quantum-symmetry analysis presented at the end of Sec.\,\ref{sub:defcanquant}, we are led to consider the expression
\qq\nn
(\d_{\bR(d,1\,\vert\,D_{d,1})}\underset{\tx{\ciut{(1)}}}{\b})_{(\vep,y)}(\theta,x)\equiv(\d\underset{\tx{\ciut{(1)}}}{\b})_{(\vep,y)}(\theta,x)=\sfd\bigl(\vep\,\ovl\G_{11}\,\theta+(\d_{\bR(d,1\,\vert\,D_{d,1})}\sfa)_{(\vep,y)}(\theta,x)\bigr)\,,
\qqq
which gives us the {\bf target supercurrent}
\qq\label{eq:tgt-curr-0}
\underset{\tx{\ciut{(0)}}}{\jmath}{}_{(\vep,y)}(\theta,x)=\vep\,\ovl\G_{11}\,\theta+(\d_{\bR(d,1\,\vert\,D_{d,1})}\sfa)_{(\vep,y)}(\theta,x)+\txc_{(\vep,y)}\,,\qquad\txc_{(\vep,y)}\in\bR\,.
\qqq
Here, the Gra\ss mann-even constants $\,\txc_{(\vep,y)}\,$ quantify the residual freedom of redefinition of the current. With the latter, we associate, in the manner structurally identical with that discussed in Sec.\,\ref{sub:defcanquant},
the {\bf current super-2-cocycle},
\qq
(\d_{\bR(d,1\,\vert\,D_{d,1})}\underset{\tx{\ciut{(0)}}}{\jmath})_{(\vep_1,y_1),(\vep_2,y_2)}(\theta,x)&=&\vep_1\,\ovl\G_{11}\,(\theta+\vep_2)-(\vep_1+\vep_2)\,\ovl\G_{11}\,\theta+\vep_2\,\ovl\G_{11}\,\theta\cr\cr
&&+(\d_{\bR(d,1\,\vert\,D_{d,1})}^2\sfa)_{(\vep_1,y_1),(\vep_2,y_2)}(\theta,x)+\txc_{(\vep_1,y_1)}-\txc_{(\vep_1,y_1)\cdot(\vep_2,y_2)}+\txc_{(\vep_2,y_2)}\cr\cr
&=&\vep_1\,\ovl\G_{11}\,\vep_2+(\d_{\bR(d,1\,\vert\,D_{d,1})}\txc)_{(\vep_1,y_1),(\vep_2,y_2)}\,, \label{eq:delphi}
\qqq
whose \emph{non}triviality is readily verified. We begin the proof by rephrasing the question about the triviality of $\,\d\underset{\tx{\ciut{(0)}}}{\jmath}\,$ -- this is equivalent to the existence of a 1-cochain $\,\txc_\cdot\,$ with the property that 
\qq\label{eq:spartcurrtriv}
(\d_{\bR(d,1\,\vert\,D_{d,1})}\txc)_{(\vep_1,0),(\vep_2,0)}=-\vep_1\,\ovl\G_{11}\,\vep_2\,.
\qqq
Given the nilpotence of the Gra\ss mann-odd coordinates, it makes sense to  write the Maclaurin expansion of the parameterised constants,
\qq\nn
\txc_{(\vep,y)}=2\txC_0(y)+\txC_1(y)\,\vep+\tfrac{1}{2}\,\vep\,\txC_2(y)\,\vep+\D_3(\vep,y)
\qqq
where $\,\D_3(\vep,y)\,$ is a rest trilinear in $\,\vep$,\ and where, for all $\,y\in\bR^{d,1}$,
\qq\nn
\txC_2(y)^{\rm T}=-\txC_2(y)\,.
\qqq
We now obtain
\qq\nn
(\d_{\bR(d,1\,\vert\,D_{d,1})}\txc)_{(\vep_1,0),(\vep_2,0)}&\equiv&\txc_{(\vep_1,0)}-\txc_{(\vep_1+\vep_2,-\frac{1}{2}\,\vep_1\,\ovl\G{}^\cdot\,\vep_2)}+\txc_{(\vep_2,0)}\cr\cr
&=&4\txC_0(0)+\txC_1(0)(\vep_1+\vep_2)+\tfrac{1}{2}\,\vep_1\,\txC_2(0)\,\vep_1+\tfrac{1}{2}\,\vep_2\,\txC_2(0)\,\vep_2-2\txC_0\bigl(-\tfrac{1}{2}\,\vep_1\,\ovl\G{}^\cdot\,\vep_2\bigr)\cr\cr
&&-\txC_1\bigl(-\tfrac{1}{2}\,\vep_1\,\ovl\G{}^\cdot\,\vep_2\bigr)(\vep_1+\vep_2)-\tfrac{1}{2}\,(\vep_1+\vep_2)\,\txC_2\bigl(-\tfrac{1}{2}\,\vep_1\,\ovl\G{}^\cdot\,\vep_2\bigr)\,(\vep_1+\vep_2)\cr\cr
&&+\D_3(\vep_1,0)+\D_3(\vep_2,0)-\D_3\bigl(\vep_1+\vep_2,-\tfrac{1}{2}\,\vep_1\,\ovl\G{}^\cdot\,\vep_2\bigr)\cr\cr
&=&2\txC_0(0)+\vep_1\,\bigl(\p_a\txC_0(0)\,\ovl\G{}^a-\txC_2(0)\bigr)\,\vep_2+\widetilde\D(\vep_1,\vep_2)
\qqq
in which the last term depends at least cubically on the $\,\vep_a\,$ and hence cannot cancel $\,\ovl\vep_1\,\G_{11}\,\vep_2$.\ The relevant equality \eqref{eq:spartcurrtriv} implies
\qq\nn
\ovl\G_{11}=-\p_a\txC_0(0)\,\ovl\G{}^a+\txC_2(0)\,.
\qqq
In view of the assumed symmetricity of $\,\ovl\G_{11}\,$ and of the $\,\ovl\G{}^a\,$ ({\it cp}\ \Reqref{eq:CGamSym}), the above yields the equality 
\qq\nn
\txC_2(0)=0
\qqq
and further reduces to
\qq\nn
\G_{11}=-\p_a\txC_0(0)\,\G^a\,,
\qqq
which admits no solutions. Thus convinced of the nontriviality of the supersymmetry current 2-cocycle, we conclude that the ensuing {\bf homomorphicity super-2-cocycle}
\qq\nn
d^{(0)}\bigl((\vep_1,y_1),(\vep_2,y_2)\bigr)=\ee^{\sfi\,\vep_1\,\ovl\G_{11}\,\vep_2}
\qqq
is also nontrivial, and therefore predicts a projective nature of the realisation of supersymmetry on the Hilbert space of the super-$0$-brane. This suggests that it is, in fact, the supercentral extension of $\,\bR(d,1\,\vert\,D_{d,1})\,$ determined by the above super-2-cocycle, and not the supersymmetry group $\,\bR(d,1\,\vert\,D_{d,1})\,$ itself, that will lift to the extension of $\,\cM^{(1)}\,$ that we are about to derive. We may now return to our geometric construction and seek corroboration of our expectations.

Consider the trivial principal $\bC^\x$-bundle 
\qq\label{eq:spartbndl}
\pi_{\xcL^{(0)}}\equiv\pr_1\ :\ \xcL^{(0)}:=\cM^{(1)}\x\bC^\x\too\cM^{(1)}\ :\ \bigl(\theta^\a,x^a,z\bigr)\longmapsto\bigl(\theta^\a,x^a\bigr)
\qqq
with the principal $\bC^\x$-connection 
\qq\label{eq:conn0grb}
\nabla_{\xcL^{(0)}}=\sfd+\tfrac{1}{\sfi}\,\underset{\tx{\ciut{(1)}}}{\b}\,,
\qqq
or -- equivalently -- the principal $\bC^\x$-connection 1-form
\qq\nn
\underset{\tx{\ciut{(1)}}}{\b}^{(2)}(\theta,x,z)=\tfrac{\sfi\,\sfd z}{z}+\underset{\tx{\ciut{(1)}}}{\b}(\theta,x)\,,
\qqq
where we fix the primitive of $\,\underset{\tx{\ciut{(2)}}}{\txH}\,$ to be
\qq\nn
\underset{\tx{\ciut{(1)}}}{\b}(\theta,x)=\theta\,\ovl\G_{11}\,\si(\theta)\,,
\qqq
and demand that a lift of the geometric action $\,\ell^{(1)}_\cdot\,$ of \Reqref{eq:sact-sMink} to the total space $\,\xcL^{(0)}\,$ be a connection-preserving automorphism. In the light of our analysis of the supersymmetry properties of the primitive $\,\underset{\tx{\ciut{(1)}}}{\b}\,$ of $\,\underset{\tx{\ciut{(2)}}}{\txH}$,\ it is justified to leave open the possibility of inducing the said lift (through restriction) from the Lie supergroup structure on $\,\xcL^{(0)}\,$ determined by the binary operation
\qq
\txm^{(2)}_0\ :\ \xcL^{(0)}\x\xcL^{(0)}\too\xcL^{(0)}\ :\ \bigl(\bigl(m_1^1,z_1\bigr),\bigl(m_1^2,z_2\bigr)\bigr)\longmapsto\bigl(\txm^{(1)}\bigl(m_1^1,m_1^2\bigr),\ee^{\sfi\,\la(m_1^1,m_1^2)}\cdot z_1\cdot z_2\bigr)\,,\cr\label{eq:spartLonL}
\qqq
in whose definition $\,m_1^A=(\theta_A^\a,x_A^a),\ A\in\{1,2\}\,$ and $\,\la(\cdot,\cdot)\,$ is required to give rise to a homomorphicity 2-cocycle on $\,\bR(d,1\,\vert\,D_{d,1})\,$ with values in $\,\bR/2\pi\bZ\,$ so that the induced action by the bundle automorphisms
\qq
\xcL^{(0)}\ell_\cdot^{(1)}\ &:&\ \bR(d,1\,\vert\,D_{d,1})\x\xcL^{(0)}\too\xcL^{(0)}\ :\ \bigl(n_1,(m_1,z)\bigr)\longmapsto\txm^{(2)}_0\bigl((n_1,1),(m_1,z)\bigr)\label{eq:spartindact}
\qqq
lifts to a standard action of a supercentral extension of the supertranslations group. The requirement that it preserve the connection \eqref{eq:conn0grb} is equivalent to the imposition of the constraints
\qq\nn
\sfd\la(n_1,m_1)=\sfd(\vep\,\ovl\G_{11}\,\theta)\,,\qquad(n_1,m_1)\equiv\bigl((\vep,y),(\theta,x)\bigr)\,,
\qqq
to which the solution reads, in conformity with our expectations,
\qq\nn
\la(n_1,m_1)=\vep\,\ovl\G_{11}\,\theta+\D(\vep,y)\,,\qquad\qquad\D(\vep,y)\in\bR/2\pi\bZ\,.
\qqq
The ensuing 2-cochain 
\qq\nn
d^{(0)}\bigl((\vep_1,y_1),(\vep_2,y_2)\bigr)&\equiv&\ee^{\sfi\,(\la((\vep_1,y_1),(\vep_1,y_1)\cdot(\theta,x))-\la((\vep_1,y_1)\cdot(\vep_2,y_2),(\theta,x))+\la((\vep_2,y_2),(\theta,x)))}\cr\cr
&=&\ee^{\sfi\,(\vep_1\,\ovl\G_{11}\,\vep_2+\D(\vep_1,y_1)-\D((\vep_1,y_1)\cdot(\vep_2,y_2))+\D(\vep_2,y_2))}\cr\cr
&\equiv&\ee^{\sfi\,\vep_1\,\ovl\G_{11}\,\vep_2}\cdot\bigl(\d_{\bR(d,1\,\vert\,D_{d,1})}\ee^{\sfi\,\D(\cdot,\cdot)}\bigr)\bigl((\vep_1,y_1),(\vep_2,y_2)\bigr)
\qqq
is a 2-cocycle for any $\,\D\,$,\ the latter contributing trivially (a 2-coboundary). Consequently, we may set 
\qq\nn
\la(n_1,m_1)=\ovl\vep\,\G_{11}\,\theta\,,
\qqq
and so also 
\qq\label{eq:homscocyc}
d^{(0)}\bigl((\vep_1,y_1),(\vep_2,y_2)\bigr)=\ee^{\sfi\,\vep_1\,\ovl\G_{11}\,\vep_2}\,.
\qqq
With the phases thus fixed, we arrive at
\berop\label{prop:L0group}
The principal $\bC^\x$-bundle $\,\xcL^{(0)}\,$ of \Reqref{eq:spartbndl} equipped with the binary operation
\qq
\txm^{(2)}_0\ :\ \xcL^{(0)}\x\xcL^{(0)}\too\xcL^{(0)}\ :\ \bigl(\bigl(m_1^1,z_1\bigr),\bigl(m_1^2,z_2\bigr)\bigr)\longmapsto\bigl(\txm^{(1)}\bigl(m_1^1,m_1^2\bigr),\ee^{\sfi\,\la(m_1^1,m_1^2)}\cdot z_1\cdot z_2\bigr)\,,\cr\label{eq:spartLonLfix}
\qqq
with the inverse
\qq\nn
\Inv_0^{(2)}\ :\ \xcL^{(0)}\too\xcL^{(0)}\ :\ \bigl(\theta^\a,x^a,z\bigr)\longmapsto\bigl(-\theta^\a,-x^a,z^{-1}\bigr)
\qqq
and the neutral element
\qq\nn
e^{(2)}_0=(0,0,1)
\qqq
is a Lie supergroup. It is a supercentral extension 
\qq\nn
\bd1\too\bC^\x\too\bR(d,1\,\vert\,D_{d,1})\lx\bC^\x\equiv\widehat\bR(d,1\,\vert\,D_{d,1})\xrightarrow{\ \pi_{\xcL^{(0)}}\ }\bR(d,1\,\vert\,D_{d,1})\too\bd1
\qqq
of the super-Minkowski group $\,\bR(d,1\,\vert\,D_{d,1})\,$ determined by the homomorphicity super-2-cocycle $\,d^{(0)}_{\cdot,\cdot}\,$ of \Reqref{eq:homscocyc}.
\eerop
\beroof
Obvious, through inspection\footnote{The existence of the structure of a Lie supergroup on this and many other supercentral extensions of Lie supergroups derived from the underlying super-Minkowski group $\,\bR(d,1\,\vert\,D_{d,1})\,$ through consecutive extensions determined by CaE super-2-cocycles was noted and discussed at great length in \Rcite{Chryssomalakos:2000xd}, and follows from the general theory of central group extensions, {\it cp}\ also \Rcite{deAzcarraga:1995}. Our results, augmented with detailed derivations, should therefore be compared with those obtained in the paper.}.
\eroof
Using the above binary operation $\,\txm^{(2)}_0\,$ in \Reqref{eq:spartindact}, we obtain the composition law of the induced action:
\qq\nn
\xcL^{(0)}\ell^{(1)}_{(\vep_2,y_2)}\circ\xcL^{(0)}\ell^{(1)}_{(\vep_1,y_1)}=\bigl(\id_{\xcM^{(1)}}\x\sfm\bigl(\ee^{\sfi\,\vep_1\,\ovl\G_{11}\,\vep_2},\cdot\bigr)\bigr)\circ\xcL^{(0)}\ell^{(1)}_{(\vep_2,y_2)\cdot(\vep_1,y_1)}\,,
\qqq
with the supergroup structure on $\,\xcL^{(0)}\,$ defining a projective realisation of supersymmetry on $\,\xcL^{(0)}$.\ It is only upon defining the action of the full supercentral extension $\,\widehat\bR(d,1\,\vert\,D_{d,1})\,$ that the realisation becomes a regular action, as anticipated. The definition is deduced from \Reqref{eq:spartLonLfix} and reads
\qq\nn
\widehat\ell^{(0)}_\cdot\equiv\txm^{(2)}_0\ &:&\ \widehat\bR(d,1\,\vert\,D_{d,1})\x\xcL^{(0)}\too\xcL^{(0)}\,.
\qqq

Our discussion leads us naturally to
\bedef\label{def:s0gerbe}
The \textbf{Green--Schwarz super-0-gerbe} over $\,\cM^{(1)}\equiv{\rm sMink}(d,1\,\vert\,D_{d,1})\,$ of curvature $\,\underset{\tx{\ciut{(2)}}}{\txH}\,$ is the triple
\qq\nn
\sG^{(0)}_{\rm GS}:=\bigl(\xcL^{(0)},\pi_{\xcL^{(0)}},\underset{\tx{\ciut{(1)}}}{\b}^{(2)}\bigr)
\qqq
constructed in the preceding paragraphs.
\exdef
\noindent We may now restate the results of our analysis in the form of
\berop\label{prop:s0gerbe}
The Green--Schwarz super-0-gerbe of Definition \ref{def:s0gerbe} is a principal $\bC^\x$-bundle with connection over the super-Minkowski space $\,{\rm sMink}(d,1\,\vert\,D_{d,1})$.\ The bundle admits the natural projective action $\,\xcL^{(0)}\ell_\cdot^{(1)}\,$ of \Reqref{eq:spartindact} of the supersymmetry group $\,\bR(d,1\,\vert\,D_{d,1})\,$ by connection-preserving principal-bundle automorphisms, induced, through restriction, by the group structure $\,\txm^{(2)}_0\,$ of \Reqref{eq:spartLonL} on the total space of the bundle. The said group structure defines an action of the supercentral extension $\,\widehat\bR(d,1\,\vert\,D_{d,1})\,$ of $\,\bR(d,1\,\vert\,D_{d,1})\,$ determined by the homomorphicity super-2-cocycle $\,d^{(0)}_{\cdot,\cdot}\,$ on $\,\bR(d,1\,\vert\,D_{d,1})\,$ specified in \Reqref{eq:homscocyc}.
\eerop

With view to subsequent discussion, and to potential future applications, we abstract from the above
\bedef\label{def:CaEs0g}
Let $\,\txG\,$ be a Lie supergroup. Denote the binary operation on $\,\txG\,$ as  
\qq\nn
\txm\ :\ \txG\x\txG\too\txG
\qqq
and the corresponding left regular action of $\,\txG\,$ on itself as 
\qq\nn
\ell_\cdot\equiv\txm\ :\ \txG\x\txG\too\txG\ :\ (h,g)\longmapsto\txm_\txG(h,g)\equiv\ell_h(g)\,.
\qqq
Let $\,\underset{\tx{\ciut{(2)}}}{\txh}\,$ be a super-2-cocycle on $\,\txG\,$ representing a class in its (left) CaE cohomology. A {\bf Cartan--Eilenberg super-0-gerbe} over $\,\txG\,$ with curvature $\,\underset{\tx{\ciut{(2)}}}{\txh}\,$ is a triple  
\qq\nn
\sG^{(0)}_{\rm CaE}:=\bigl(\sfY\txG,\pi_{\sfY\txG},\underset{\tx{\ciut{(1)}}}{\txa}\bigr)
\qqq
composed of 
\bit
\item a principal $\bC^\x$-bundle 
\qq\nn
\alxydim{@C=1cm@R=1cm}{\bC^\x \ar[r] & \sfY\txG \ar[d]^{\pi_{\sfY\txG}} \\ & \txG}\,;
\qqq 
\item a principal connection 1-form $\,\underset{\tx{\ciut{(1)}}}{\txa}\,$ on it, trivialising the pullback of the curvature super-2-form $\,\,$ along the projection to the base,
\qq\nn
\pi_{\sfY\txG}^*\underset{\tx{\ciut{(2)}}}{\txh}=\sfd\underset{\tx{\ciut{(1)}}}{\txa}\,,
\qqq
\eit
in which the total space $\,\sfY\txG\,$ of the bundle carries the structure of a Lie supergroup (with the binary operation $\,\sfY\txm$) that extends that on $\,\txG$,\ as captured by the short exact sequence of Lie supergroups
\qq\nn
\bd1\too\bC^\x\too\sfY\txG\xrightarrow{\ \pi_{\sfY\txG}\ }\txG\too\bd1
\qqq
that integrates the short exact sequence of Lie superalgebras determined by $\,\underset{\tx{\ciut{(2)}}}{\txh}\,$ along the lines of Prop.\,\ref{prop:CEtoExt}, and such that the left regular action of the extended Lie supergroup $\,\sfY\txG\,$ upon itself,
\qq\nn
\sfY\ell_\cdot\ :\ \sfY\txG\x\sfY\txG\too\sfY\txG\ :\ (h,g)\longmapsto\sfY\txm(h,g)\equiv\sfY\ell_h(g)\,,
\qqq
preserves the connection 1-form,
\qq\nn
\sfY\ell_g^*\underset{\tx{\ciut{(1)}}}{\txa}=\underset{\tx{\ciut{(1)}}}{\txa}\,,\qquad g\in\sfY\txG\,.
\qqq

Given CaE super-0-gerbes $\,\sG^{(0)\,A}_{\rm CaE}=\bigl(\sfY_A\txG,\pi_{\sfY_A\txG},\underset{\tx{\ciut{(1)}}}{\txa_A}\bigr),\ A\in\{1,2\}\,$ over a common base $\,\txG$,\ an {\bf isomorphism} between them is an isomorphism of the principal $\bC^\x$-bundles
\qq\nn
\varphi\ :\ \sfY_1\txG\xrightarrow{\ \cong\ }\sfY_2\txG
\qqq
that preserves the connection,
\qq\nn
\varphi^*\underset{\tx{\ciut{(1)}}}{\txa_2}=\underset{\tx{\ciut{(1)}}}{\txa_1}\,,
\qqq
and is, at the same time, an isomorphism of the respective Lie supergroups that defines an equivalence between the two extensions, as described by the commutative diagram
\qq\nn
\alxydim{@C=1cm@R=1cm}{ & & \sfY_1\txG \ar[dd]^{\varphi}_{\cong} \ar[dr]^{\pi_{\sfY_1\txG}} & & \\ \bd1 \ar[r] & \bC^\x \ar[ur] \ar[dr] & & \txG \ar[r] & \bd1 \\ & & \sfY_2\txG \ar[ur]_{\pi_{\sfY_2\txG}} & & }
\qqq
in the category of Lie supergroups.
\exdef
\brem\label{rem:tensprinc}
Note that the tensor product of principal $\bC^\x$-bundles described in the footnote on p.\,\pageref{foot:Cxprintens} gives rise to a tensor product of Cartan--Eilenberg super-0-gerbes. Indeed, given super-0-gerbes $\,\sG^{(0)\,A}_{\rm CaE},\ A\in\{1,2\}\,$ as above, there exists a natural associative binary operation on the total space of the principal $\bC^\x$-bundle
\qq\nn
\bigl(\sfY_1\txG\x_\txG\sfY_2\txG\bigr)/\bC^\x\equiv\sfY_1\txG\ox\sfY_2\txG
\qqq
associated to the principal $\bC^\x$-bundle $\,\sfY_1\txG\,$ through the defining action (obtained through partial restriction of the multiplication in $\,\sfY_2\txG$) 
\qq\nn
\sfY_2 r_\cdot\ :\ \sfY_2\txG\x\bC^\x\too\sfY_2\txG\,,
\qqq
so that, for any $\,(y_1,y_2)\in\sfY_1\txG\x\sfY_2\txG\,$ such that $\,\pi_{\sfY_1\txG}(y_1)=\pi_{\sfY_2\txG}(y_2)$,
\qq\nn
\sfY_1\txG\ox\sfY_2\txG\ni[(y_1,y_2)]=\bigl[\bigl(\sfY_1 r_{z^{-1}}(y_1),\sfY_2 r_z(y_2)\bigr)\bigr]\equiv\bigl[\bigl(\sfY_1\txm\bigl(y_1,z^{-1}\bigr),\sfY_2\txm(y_2,z)\bigr)\bigr]\,,
\qqq
to wit
\qq\nn
[\sfY_1\txm\x\sfY_2\txm]\ &:&\ \bigl(\sfY_1\txG\ox\sfY_2\txG\bigr)^{\x 2}\too\sfY_1\txG\ox\sfY_2\txG\cr\cr &:&\ \bigl([(y_1,y_2)],[(y_1',y_2')]\bigr)\longmapsto\bigl[\bigl(\sfY_1\txm(y_1,y_1'),\sfY_2\txm(y_2,y_2')\bigr)\bigr]\,.
\qqq
That the latter is well-defined follows straightforwardly from the centrality of both extensions.
\erem
\brem
Cartan--Eilenberg super-0-gerbes admit pullback along Lie-supergroup homomorphisms. Indeed, given a super-0-gerbe $\,\sG^{(0)}_{\rm CaE}\,$ as above and a Lie-supergroup homomorphism
\qq\nn
\psi\ :\ \widetilde\txG\too\txG
\qqq
from a Lie supergroup $\,\widetilde\txG\,$ with the binary operation $\,\widetilde\txm\,$ to its base $\,\txG$,\ the pullback
\qq\nn
\alxydim{@C=2cm@R=1cm}{ \psi^*\sfY\txG\equiv\widetilde\txG\hspace{2pt}{}_\psi\hspace{-2pt}\x\sfY\txG \ar[r]^{\qquad\pr_2} \ar[d]_{\pr_1} & \sfY\txG \ar[d]^{\pi_{\sfY\txG}} \\ \widetilde\txG \ar[r]_{\psi} & \txG}
\qqq
inherits, through restriction, a natural Lie-supergroup structure from the product Lie supergroup $\,\widetilde\txG\x\sfY\txG\supset\widetilde\txG\hspace{2pt}{}_\psi\hspace{-2pt}\x\sfY\txG$.\ This is to say that the binary operation 
\qq\nn
\widetilde\txm\hspace{2pt}{}_\psi\hspace{-2pt}\x\sfY\txm\ :\ \bigl(\widetilde\txG\hspace{2pt}{}_\psi\hspace{-2pt}\x\sfY\txG\bigr)^{\x 2}\too\widetilde\txG\x\sfY\txG\ :\ \bigl((\widetilde g_1,y_1),(\widetilde g_2,y_2)\bigr)\longmapsto\bigl(\widetilde\txm(\widetilde g_1,\widetilde g_2),\sfY\txm(y_1,y_2)\bigr)
\qqq
has the subspace $\,\widetilde\txG\hspace{2pt}{}_\psi\hspace{-2pt}\x\sfY\txG\,$ as its codomain since
\qq\nn
\psi\circ\widetilde\txm(\widetilde g_1,\widetilde g_2)=\txm\bigl(\psi(\widetilde g_1),\psi(\widetilde g_2)\bigr)=\txm\bigl(\pi_{\sfY\txG}(y_1),\pi_{\sfY\txG}(y_2)\bigr)=\pi_{\sfY\txG}\circ\sfY\txm(y_1,y_2)\,.
\qqq
\erem
\medskip

\subsection{The super-1-gerbe of the Green--Schwarz superstring} 

At the next level in cohomology, which is where the super-$\si$-model for the superstring is constructed, we find the GS super-3-cocycle 
\qq\label{eq:GS3form}
\underset{\tx{\ciut{(3)}}}{\txH}=e^a\wedge\pi_0^*\bigl(\si\wedge\ovl\G_a\,\si\bigr)\,.
\qqq 
This is a closed and manifestly LI  super-3-form on $\,\cM^{(1)}$,\ with no smooth LI primitive on the latter space. Indeed, any such primitive would, of necessity, take the form of a linear combination
\qq\nn
\underset{\tx{\ciut{(2)}}}{\b}{}_{\rm L}=\la_{\a\b}\,\pi_0^*(\si^\a\wedge\si^\b)+2\mu_{\a a}\,\pi_0^*\si^\a\wedge e^a+\nu_{ab}\,e^a\wedge e^b\,,
\qqq
with certain \emph{constant} tensors $\,\la_{\a\b}\equiv\la_{(\a\b)},\mu_{\a a}\,$ and $\,\nu_{ab}\equiv\nu_{[ab]}\,$ as coefficients, leading to a condition
\qq\nn
e^a\wedge\pi_0^*\bigl(\si\wedge\ovl\G_a\,\si\bigr)\must 2\sfd e^a\wedge\bigl(\nu_{ab}\,e^b-\mu_{\a a}\,\pi_0^*\si^\a\bigr)=\pi_0^*\bigl(\si\wedge\ovl\G{}^a\,\si\bigr)\wedge\bigl(\nu_{ab}\,e^b-\mu_{\a a}\,\pi_0^*\si^\a\bigr)\,.
\qqq
The latter implies, in particular, the equality
\qq\nn
\nu_{ab}\equiv\eta_{ab}\,,
\qqq
manifestly inconsistent with the assumption $\,\nu_{ba}=-\nu_{ab}$.

The stepwise procedure of trivialisation of $\,\underset{\tx{\ciut{(3)}}}{\txH}\,$ in the CaE cohomology through pullback to consecutive supercentral extensions of the underlying Lie supergroup $\,\bR(d,1\,\vert\,D_{d,1})\,$ begins over the latter space, which is also where we look for LI de Rham super-2-cocycles of the type \eqref{eq:LI2bas}. The distinguished members of the family that we shall examine with view to solving the trivialisation problem are 
\qq\label{eq:CaEscocyc1}
\underset{\tx{\ciut{(2)}}}{h}{}^{(1)}_\a=-\tfrac{1}{2} Q_\a\con\underset{\tx{\ciut{(3)}}}{\txH}=-\ovl\G_{a\,\a\b}\,\pi_0^*\si^\b\wedge e^a\,,\qquad\a\in\ovl{1,D_{d,1}}\,.
\qqq
Their closedness (in the CaE cohomology) follows -- just as that of the super-3-cocycle \eqref{eq:GS3form} -- directly from the (de Rham) closedness and left-invariance of the super-3-cocycle, and so -- technically -- from the assumed Fierz identity \eqref{eq:ClifFierz1}. 

In order to construct a suitable common extension of their support $\,\cM^{(1)}\,$ on which their pullbacks trivialise, we first derive their non-LI primitives on $\,\cM^{(1)}$.\ To this end, we compute
\qq\nn
\underset{\tx{\ciut{(2)}}}{h}{}^{(1)}_\a(\theta,x)=-\sfd\bigl(\ovl\G_{a\,\a\b}\,\theta^\b\,\sfd x^a\bigr)-\tfrac{1}{2}\,\ovl\G_{a\,\a\b}\,\ovl\G{}^a_{\g\d}\,\sfd\theta^\b\wedge\,\theta^\g\,\sfd\theta^\d\,.
\qqq
Using identity \eqref{eq:ClifFierz1}, we readily find
\qq\nn
\ovl\G_{a\,\a\b}\,\ovl\G{}^a_{\g\d}\,\sfd\theta^\b\wedge\,\theta^\g\,\sfd\theta^\d&=&\sfd\left(\ovl\G_{a\,\a\b}\,\ovl\G{}^a_{\g\d}\,\theta^\b\,\theta^\g\,\sfd\theta^\d\right)-\ovl\G_{a\,\a\b}\,\ovl\G{}^a_{\g\d}\,\theta^\b\,\sfd\theta^\g\wedge\sfd\theta^\d\cr\cr
&=&\sfd\left(\ovl\G_{a\,\a\b}\,\ovl\G{}^a_{\g\d}\,\theta^\b\,\theta^\g\,\sfd\theta^\d\right)+2\ovl\G_{a\,\a\g}\,\ovl\G{}^a_{\b\d}\,\theta^\b\,\sfd\theta^\g\wedge\sfd\theta^\d\,,
\qqq
so that
\qq\nn
\underset{\tx{\ciut{(2)}}}{h}{}^{(1)}_\a(\theta,x)=\sfd\bigl(-\ovl\G_{a\,\a\b}\,\theta^\b\,\sfd x^a-\tfrac{1}{6}\,\ovl\G_{a\,\a\b}\,\ovl\G{}^a_{\g\d}\,\theta^\b\,\theta^\g\,\sfd\theta^\d\bigr)\,.
\qqq
Drawing on our hitherto experience, we may now conceive a trivial vector bundle 
\qq\nn
\pi^{(2)}_1\equiv\pr_1\ :\ \cM^{(2)}_1=\cM^{(1)}\x\bR^{0\,\vert\,D_{d,1}}\too\cM^{(1)} 
\qqq
with the purely Gra\ss mann-odd fibre $\,\bR^{0\,\vert\,D_{d,1}}\,$ and a Lie supergroup structure that projects to the previously considered supersymmetry-group structure on the base and so lifts the action \eqref{eq:sact-sMink} of that supersymmetry group in a manner that we fix by demanding invariance under this lift of the primitives $\,e^{(2)}_\a\in\Om^1(\cM^{(2)}_1)\,$ of the distinguished super-2-cocycles,
\qq\nn
\pi^{(2)\,*}_1\underset{\tx{\ciut{(2)}}}{h}{}^{(1)}_\a=:\sfd e^{(2)}_\a\,.
\qqq
Let us denote the (global) coordinates on the fibre of $\,\cM^{(2)}_1\,$ as $\,\xi_\a,\ \a\in\ovl{1,D_{d,1}}$.\ We then take
\qq\nn
e^{(2)}_\a(\theta,x,\xi)=\sfd\xi_\a-\ovl\G_{a\,\a\b}\,\theta^\b\,\bigl(\sfd x^a+\tfrac{1}{6}\,\theta\,\ovl\G{}^a\,\si(\theta)\bigr)\,.
\qqq
The supersymmetry variation of the non-LI primitive $\,\unl e^{(2)}_\a:=e^{(2)}_\a-\sfd\xi_\a\,$ of $\,\underset{\tx{\ciut{(2)}}}{h}{}^{(1)}_\a\,$ is exact and may be cast in the form
\qq\nn
\unl e^{(2)}_\a\bigl(\theta+\vep,x+y-\tfrac{1}{2}\,\vep\,\ovl\G{}^\cdot\,\theta\bigr)-\unl e^{(2)}_\a(\theta,x)&=&\tfrac{1}{3}\,\ovl\G{}^a_{\a\b}\,\theta^\b\,\bigl(\vep\,\ovl\G_a\,\si(\theta)\bigr)-\ovl\G_{a\,\a\b}\,\vep^\b\,\bigl(\sfd x^a+\tfrac{1}{6}\,\theta\,\ovl\G{}^a\,\si(\theta)-\tfrac{1}{3}\,\vep\,\ovl\G{}^a\,\si(\theta)\bigr)\cr\cr
&=&\sfd\bigl(-x^a\,\ovl\G_{a\,\a\b}\,\vep^\b+\tfrac{1}{6}\,\ovl\G{}^a_{\a\b}\,\bigl(2\vep^\b+\theta^\b\bigr)\,\vep\,\ovl\G_a\,\theta\bigr)
\qqq
with the help of the Fierz identity \eqref{eq:ClifFierz1}. In this manner, we arrive at
\berop\label{prop:M2group}
The above-described vector bundle $\,\cM^{(2)}_1\,$ equipped with the binary operation
\qq\nn
\txm_1^{(2)}\ &:&\ \cM_1^{(2)}\x\cM_1^{(2)}\too\cM_1^{(2)}\cr\cr
&:&\ \bigl(\bigl(m_1^1,\xi_{1\,\a}\bigr),\bigl(m_1^2,\xi_{2\,\b}\bigr)\bigr)\longmapsto\bigl(\txm^{(1)}\bigl(m_1^1,m_1^2\bigr),\xi_{1\,\a}+\xi_{2\,\a}+\ovl\G_{a\,\a\b}\,\theta_1^\b\,x_2^a-\tfrac{1}{6}\,\bigl(\theta_1\,\ovl\G_a\,\theta_2\bigr)\,\ovl\G{}^a_{\a\b}\,\bigl(2\theta_1^\b+\theta_2^\b\bigr)\bigr)\,,
\qqq
with the inverse
\qq\nn
\Inv_1^{(2)}\ :\ \cM_1^{(2)}\too\cM_1^{(2)}\ :\ (\theta^\a,x^a,\xi_\b)\longmapsto\bigl(-\theta^\a,-x^a,-\xi_\b+x^b\,\ovl\G_{b\,\b\g}\,\theta^\g\bigr)
\qqq
and the neutral element
\qq\nn
e_1^{(2)}=(0,0,0)
\qqq
is a Lie supergroup. It is a supercentral extension 
\qq\nn
\bd1\too\bR^{0\,\vert\,D_{d,1}}\too\bR(d,1\,\vert\,D_{d,1})\lx\bR^{0\,\vert\,D_{d,1}}\equiv\cM_1^{(2)}\xrightarrow{\ \pi^{(2)}_1\ }\bR(d,1\,\vert\,D_{d,1})\too\bd1
\qqq
of the super-Minkowski group $\,\bR(d,1\,\vert\,D_{d,1})\,$ determined by the family of CE super-2-cocycles corresponding to the CaE super-2-cocycles $\,\{\underset{\tx{\ciut{(2)}}}{h}{}^{(1)}_\a\}_{\a\in\ovl{1,D_{d,1}}}\,$ of \Reqref{eq:CaEscocyc1}.
\eerop
\beroof
Through inspection. In particular, the associativity of $\,\txm_1^{(2)}\,$ hinges upon identity \eqref{eq:ClifFierz1}.
\eroof

Upon pullback to $\,\cM_1^{(2)}$,\ we obtain the sought-after trivialisation 
\qq\nn
\pi^{(2)\,*}_1\underset{\tx{\ciut{(3)}}}{\txH}=\sfd\underset{\tx{\ciut{(2)}}}{\b}^{(2)}\,,\qquad\qquad\underset{\tx{\ciut{(2)}}}{\b}^{(2)}:=\pi_{01}^{(2)\,*}\si^\a\wedge e^{(2)}_\a\,,
\qqq
written in the shorthand notation
\qq\nn
\pi^{(2)}_{01}:=\pi_0\circ\pi^{(2)}_1
\qqq
that we adapt in our subsequent considerations. The relation of the above trivialisation to the previously found (in Prop.\,\ref{prop:GSprim}) non-LI one on $\,\cM^{(1)}\,$ reads
\qq\label{eq:betobet3}
\underset{\tx{\ciut{(2)}}}{\b}{}^{(2)}=\pi^{(2)\,*}_1\underset{\tx{\ciut{(2)}}}{\b}+\sfd\txB\,,\qquad\qquad
\txB(\theta,x,\xi):=\theta^\a\,\sfd\xi_\a\,.
\qqq
~\medskip

Structurally, the construction of the supercentral extension 
\qq\nn
\pi_{\sfY_1\cM^{(1)}}:=\pi^{(2)}_1\ :\ \sfY_1\cM^{(1)}:=\cM^{(2)}_1\too\cM^{(1)}\ :\ (m_1,\xi_\b)\longmapsto m_1
\qqq
in the present geometric context plays a r\^ole fully analogous to that of the surjective submersion $\,\pi_{\sfY M}:\sfY M\too M\,$ from Section \ref{sec:Bose}, to wit, it yields an epimorphism, in the geometric category of interest, onto the support of a non-trivial 3-cocycle on which, upon pullback along that epimorphism, the 3-cocycle trivialises in the same cohomology (in which it would not, at least in general, trivialise on the base/codomain of the epimorphism). The last observation motivates our subsequent attempt at establishing a (super)geometric realisation of the super-3-cocycle $\,\underset{\tx{\ciut{(3)}}}{\txH}\,$ in the CaE cohomology through a procedure closely imitating the one that defines its analogon $\,(\sfY M,\pi_{\sfY M},\txB,L,\pi_L,\nabla_L,\mu_L)\,$ in the standard (purely Gra\ss mann-even) setting.

To this end, let us first consider the fibred square, represented by the commutative diagram
\qq\nn
\alxydim{@C=.75cm@R=1cm}{& \sfY_1^{[2]}\cM^{(1)} \ar[rd]^{\pr_2} \ar[ld]_{\pr_1} & \\ \sfY_1\cM^{(1)} \ar[rd]_{\pi_{\sfY_1\cM^{(1)}}} & &  \sfY_1\cM^{(1)} \ar[ld]^{\pi_{\sfY_1\cM^{(1)}}} \\ & \cM^{(1)} & }\,.
\qqq
The difference of the pullbacks of the primitive $\,\underset{\tx{\ciut{(2)}}}{\b^{(2)}}\,$ along the two canonical projections reads
\qq\nn
(\pr_2^*-\pr_1^*)\underset{\tx{\ciut{(2)}}}{\b}{}^{(2)}=(\pi_0\circ\pi_{\sfY_1\cM^{(1)}}\circ\pr_1)^*\si^\a\wedge(\pr_2^*-\pr_1^*)e^{(2)}_\a\,.
\qqq
This is, by construction, an LI super-2-cocycle, and so we may seek to trivialise it, or -- if necessary -- its pullback to a suitable supercentral extension of $\,\sfY_1^{[2]}\cM^{(1)}$,\ in the CaE cohomology. Inspection of the coordinate expression 
\qq\nn
(\pr_2^*-\pr_1^*)\underset{\tx{\ciut{(2)}}}{\b}{}^{(2)}\bigl(m_2^1,m_2^2\bigr)=\sfd\left(\theta^\a\,\sfd\xi^{21}_\a\right)\,,\qquad m_2^A\equiv(\theta^\a,x^a,\xi^A_\b)\,,
\qqq
written in terms of the variables $\,\xi^{21}_\a:=\xi^2_\a-\xi^1_\a,\ \a\in\ovl{1,D_{d,1}}$,\ convinces us that there is no LI primitive of the above super-2-cocycle on $\,\sfY_1^{[2]}\cM^{(1)}$,\ and so, invoking our results from the analysis of the GS super-2-cocycle, we are led to associate with it a trivial principal $\bC^\x$-bundle
\qq
\pi_{\xcL^{(1)}}\equiv\pr_1\ :\ \xcL^{(1)}:=\sfY_1^{[2]}\cM^{(1)}\x\bC^\x\too\sfY_1^{[2]}\cM^{(1)}\ :\ \bigl(\bigl(m_2^1,m_2^2\bigr),z\bigr)\longmapsto\bigl(m_2^1,m_2^2\bigr) \label{eq:sstringbndl}
\qqq
with a principal $\bC^\x$-connection 
\qq\nn
\nabla_{\xcL^{(1)}}=\sfd+\tfrac{1}{\sfi}\,\txA\,,
\qqq
or -- equivalently -- a principal $\bC^\x$-connection super-1-form
\qq\nn
\cA\bigl(\bigl(m_2^1,m_2^2\bigr),z\bigr)=\tfrac{\sfi\,\sfd z}{z}+\txA\bigl(m_2^1,m_2^2\bigr)\,,
\qqq 
with the base component
\qq\label{eq:ABB}
\txA\bigl(m_2^1,m_2^2\bigr)\equiv(\pr_2^*-\pr_1^*)\txB\bigl(m_2^1,m_2^2\bigr)=\theta^\a\,\sfd\xi^{21}_\a\,,
\qqq

The fibred product $\,\sfY_1^{[2]}\cM^{(1)}\,$ of Lie supergroups inherits from $\,\sfY_1\cM^{(1)}\equiv\cM^{(2)}_1\,$ a Lie-supergroup structure determined by the binary operation
\qq
\txm_1^{(2)\,[2]}\ :\ \sfY_1^{[2]}\cM^{(1)}\x\sfY_1^{[2]}\cM^{(1)}\too\sfY_1^{[2]}\cM^{(1)}\ :\ \bigl(\bigl(m_2^1,m_2^2\bigr),\bigl(n_2^1,n_2^2\bigr)\bigr)\longmapsto\bigl(\txm_1^{(2)}\bigl(m_2^1,n_2^1\bigr),\txm_1^{(2)}\bigl(m_2^2,n_2^2\bigr)\bigr)\,.\cr\label{eq:sstringY2Lie}
\qqq
In analogy with the case of the superparticle, we endow $\,\xcL^{(1)}\,$ with the structure of a Lie supergroup determined by the requirement of left-invariance of the principal $\bC^\x$-connection super-1-form $\,\cA$.
\berop\label{prop:L1}
The principal $\bC^\x$-bundle $\,\xcL^{(1)}\,$ of \Reqref{eq:sstringbndl} equipped with the binary operation
\qq
\txm_1^{(3)}\ &:&\ \xcL^{(1)}\x\xcL^{(1)}\too\xcL^{(1)}\ :\ \bigl(\bigl(\bigl(m_2^1,m_2^2\bigr),z_1\bigr),\bigl(\bigl(n_2^1,n_2^2\bigr),z_2\bigr)\bigr)\longmapsto\cr\cr
&&\longmapsto\bigl(\txm_1^{(2)\,[2]}\bigl(\bigl(m_2^1,m_2^2\bigr),\bigl(n_2^1,n_2^2\bigr)\bigr),d^{(1)}\bigl(\bigl(m_2^1,m_2^2\bigr),\bigl(n_2^1,n_2^2\bigr)\bigr)\cdot z_1\cdot z_2\bigr)\,,\label{eq:sstringLonLfix}
\qqq
the latter being defined in terms of the super-2-cocycle
\qq\nn
d^{(1)}\bigl(\bigl(m_2^1,m_2^2\bigr),\bigl(n_2^1,n_2^2\bigr)\bigr)=\ee^{\sfi\,\theta_1^\a\,\xi_{2\,\a}^{21}}\,,
\qqq
written for $\,m_2^A=(\theta_1^\a,x_1^a,\xi_{1\,\b}^A)\,$ and $\,n_2^A=(\theta_2^\a,x_2^a,\xi_{2\,\b}^A),\ A\in\{1,2\}$,\ 
with the inverse
\qq\nn
\Inv_1^{(3)}\ :\ \xcL^{(1)}\too\xcL^{(1)}\ :\ \bigl(\bigl(m_2^1,m_2^2\bigr),z\bigr)\longmapsto\bigl(\Inv_1^{(3)}\bigl(m_2^1\bigr),\Inv_1^{(3)}\bigl(m_2^2\bigr),\ee^{\sfi\,\theta^\a\,\xi^{21}_\a}\cdot z^{-1}\bigr)
\qqq
and the neutral element
\qq\nn
\si^{(3)}_1=(0,0,1)
\qqq
is a Lie supergroup. It is a supercentral extension 
\qq\nn
\bd1\too\bC^\x\too\sfY_1^{[2]}\cM^{(1)}\lx\bC^\x\equiv\widehat{\sfY_1^{[2]}\cM^{(1)}}\xrightarrow{\ \pi_{\xcL^{(1)}}\ }\sfY_1^{[2]}\cM^{(1)}\too\bd1
\qqq
of the Lie supergroup $\,\sfY_1^{[2]}\cM^{(1)}\,$ of \Reqref{eq:sstringY2Lie} determined by $\,d^{(1)}_{\cdot,\cdot}$.
\eerop
\beroof
Through inspection.
\eroof
The above structure can be employed to lift the original action of the supersymmetry group $\,\bR(d,1\,\vert\,D_{d,1})\,$ all the way up to the total space of $\,\xcL^{(1)}\,$ as per
\qq\nn
\widehat\ell^{(1)}_\cdot\equiv\txm_1^{(3)}\ :\ \widehat{\sfY_1^{[2]}\cM^{(1)}}\x\xcL^{(1)}\too\xcL^{(1)}\,.
\qqq
By construction, the lift preserves the connection on $\,\xcL^{(1)}$.

In the last step, we consider the cartesian cube of the surjective submersion $\,\sfY_1\cM^{(1)}\,$ fibred over $\,\cM^{(1)}$,\ with its canonical projections $\,\pr_{i,j}\equiv(\pr_i,\pr_j),\ (i,j)\in\{(1,2),(2,3),(1,3)\}\,$ to $\,\sfY_1^{[2]}\cM^{(1)}\,$ that render the diagram 
\qq\nn
\alxydim{@C=1cm@R=1cm}{& & \sfY_1^{[3]}\cM^{(1)} \ar[rd]^{\pr_{1,3}} \ar[ld]_{\pr_{1,2}} \ar[d]_{\pr_{2,3}} & & \\ & \sfY_1^{[2]}\cM^{(1)} \ar[ld]_{\pr_1} \ar[d]_{\pr_2} & \sfY_1^{[2]}\cM^{(1)} \ar[ld]_{\pr_1} \ar[rd]^{\pr_2} & \sfY_1^{[2]}\cM^{(1)} \ar[d]^{\pr_2} \ar[rd]^{\pr_1} & \\ \sfY_1\cM^{(1)} \ar@/_5.0pc/@{=}[rrrr] \ar[rrd]_{\pi_{\sfY_1\cM^{(1)}}} & \sfY_1\cM^{(1)} \ar[rd]^{\pi_{\sfY_1\cM^{(1)}}} &  & \sfY_1\cM^{(1)} \ar[ld]_{\pi_{\sfY_1\cM^{(1)}}} & \sfY_1\cM^{(1)} \ar[lld]^{\pi_{\sfY_1\cM^{(1)}}} \\ & & \cM^{(1)} & & }\cr\cr
\qqq
commutative, and, over it, look for a connection-preserving isomorphism
\qq\nn
\mu_{\xcL^{(1)}}\ :\ \pr_{1,2}^*\xcL^{(1)}\ox\pr_{2,3}^*\xcL^{(1)}\xrightarrow{\ \cong\ }\pr_{1,3}^*\xcL^{(1)}\,.
\qqq
Comparison of the pullbacks of the (global) connection 1-forms 
\qq\nn
(\pr_{1,2}^*+\pr_{2,3}^*-\pr_{1,3}^*)\txA\bigl(m_2^1,m_2^2,m_2^3\bigr)=0\,,
\qqq
in conjunction with inspection of the invariance of the relevant combination $\,z_{1,2}\cdot z_{2,3}\cdot z_{1,3}^{-1}\,$ of the fibre coordinates lead us to set 
\qq\nn
&&\mu_{\xcL^{(1)}}\bigl(\bigl(\bigl(m_2^1,m_2^2,m_2^3\bigr),\bigl(m_2^1,m_2^2,z_{1,2}\bigr)\bigr)\ox\bigl(\bigl(m_2^1,m_2^2,m_2^3\bigr),\bigl(m_2^2,m_2^3,z_{2,3}\bigr)\bigr)\bigr)\bigr)\cr\cr
&:=&\bigl(\bigl(m_2^1,m_2^2,m_2^3\bigr),\bigl(m_2^1,m_2^3,z_{1,2}\cdot z_{2,3}\bigr)\bigr)\,. 
\qqq
A fibre-bundle map thus defined trivially satisfies the groupoid identity \eqref{eq:mugrpd} over $\,\sfY_1^{[4]}\cM^{(1)}$,\ and conforms with the previous definition of a super-0-gerbe isomorphism. Indeed, the map transports the natural Lie-supergroup structure on
\qq\nn
\pr_{1,2}^*\xcL^{(1)}\ox\pr_{2,3}^*\xcL^{(1)}\equiv\bigl(\sfY_1^{[3]}\cM^{(1)}\x^{(1,2)}_{\sfY_1^{[2]}\cM^{(1)}}\xcL^{(1)}\bigr)\x_{\id_{\sfY_1^{[3]}\cM^{(1)}}\x\txm_1^{(3)}}\bigl(\sfY_1^{[3]}\cM^{(1)}\x^{(2,3)}_{\sfY_1^{[2]}\cM^{(1)}}\xcL^{(1)}\bigr)\,,
\qqq
with 
\qq\nn
\alxydim{@C=1.5cm@R=1.5cm}{\sfY_1^{[3]}\cM^{(1)}\x^{(i,j)}_{\sfY_1^{[2]}\cM^{(1)}}\xcL^{(1)} \ar[r]^{\hspace{.75cm}\pr_2} \ar[d]_{\pr_1} & \xcL^{(1)} \ar[d]^{\pi_{\xcL^{(1)}}} \\ \sfY_1^{[3]}\cM^{(1)} \ar[r]_{\pr_{i,j}} &  \sfY_1^{[2]}\cM^{(1)} }\,,
\qqq
determined by the binary operation
\qq\nn
\bigl[\txm_1^{(2)\,[3]}\circ\pr_{1,5},\txm_1^{(3)}\circ\pr_{2,6},\txm_1^{(2)\,[3]}\circ\pr_{3,7},\txm_1^{(3)}\circ\pr_{4,8}\bigr]\ :\ \bigl(\pr_{1,2}^*\xcL^{(1)}\ox\pr_{2,3}^*\xcL^{(1)}\bigr)^{\x 2}\cr\cr
\too\pr_{1,2}^*\xcL^{(1)}\ox\pr_{2,3}^*\xcL^{(1)}\,,
\qqq
into that on 
\qq\nn
\pr_{1,3}^*\xcL^{(1)}\equiv\sfY_1^{[3]}\cM^{(1)}\hspace{2pt}{}_{\pr_{1,3}}\hspace{-2pt}\x\xcL^{(1)}\,,
\qqq
determined by
\qq\nn
\bigl(\txm_1^{(2)\,[3]}\circ\pr_{1,3},\txm_1^{(3)}\circ\pr_{2,4}\bigr)\ :\ \bigl(\pr_{1,3}^*\xcL^{(1)}\bigr)^{\x 2}\too\pr_{1,3}^*\xcL^{(1)}\,.
\qqq
This follows straightforwardly from the identity
\qq\nn
d^{(1)}\bigl(\bigl(m_2^1,m_2^2\bigr),\bigl(n_2^1,n_2^2\bigr)\bigr)\cdot d^{(1)}\bigl(\bigl(m_2^2,m_2^3\bigr),\bigl(n_2^2,n_2^3\bigr)\bigr)=d^{(1)}\bigl(\bigl(m_2^1,m_2^3\bigr),\bigl(n_2^1,n_2^3\bigr)\bigr)\,.
\qqq

We conclude our analysis with 
\bedef\label{def:s1gerbe}
The \textbf{Green--Schwarz super-1-gerbe} over $\,\cM^{(1)}\equiv{\rm sMink}(d,1\,\vert\,D_{d,1})\,$ of curvature $\,\underset{\tx{\ciut{(3)}}}{\txH}\,$ is the septuple
\qq\nn
\sG^{(1)}_{\rm GS}:=\bigl(\sfY_1\cM^{(1)},\pi_{\sfY_1\cM^{(1)}},\underset{\tx{\ciut{(2)}}}{\b}^{(2)},\xcL^{(1)},\pi_{\xcL^{(1)}},\nabla_{\xcL^{(1)}},\mu_{\xcL^{(1)}}\bigr)
\qqq
constructed in the preceding paragraphs.
\exdef
\noindent Our discussion is now concisely summarised in
\berop\label{prop:s1gerbe}
The Green--Schwarz super-1-gerbe $\,\sG^{(1)}_{\rm GS}\,$ of Definition \ref{def:s1gerbe} is an abelian bundle gerbe with connection over the super-Minkowski space $\,{\rm sMink}(d,1\,\vert\,D_{d,1})$.\ The action \eqref{eq:sact-sMink} of the supersymmetry group $\,\bR(d,1\,\vert\,D_{d,1})\,$ on the base of the gerbe lifts to an action, by connection-preserving principal-bundle automorphisms, of the supercentral extension $\,\widehat{\sfY_1^{[2]}\cM^{(1)}}$,\ detailed in Prop.\,\ref{prop:L1}, of the Lie supergroup $\,\sfY_1^{[2]}\cM^{(1)}\,$ defined through \Reqref{eq:sstringY2Lie} and itself being an extension, described in Prop.\,\ref{prop:M2group}, of $\,\bR(d,1\,\vert\,D_{d,1})$.
\eerop

Following the lower-dimensional example, we formulate
\bedef\label{def:CaEs1g}
Adopt the notation of Def.\,\ref{def:CaEs0g}. Let $\,\underset{\tx{\ciut{(3)}}}{\txh}\,$ be a super-3-cocycle on $\,\txG\,$ representing a class in its (left) CaE cohomology. A {\bf Cartan--Eilenberg super-1-gerbe} over $\,\txG\,$ with curvature $\,\underset{\tx{\ciut{(3)}}}{\txh}\,$ is a septuple  
\qq\nn
\sG^{(1)}_{\rm CaE}:=\bigl(\sfY\txG,\pi_{\sfY\txG},\underset{\tx{\ciut{(2)}}}{\txb},\Lx,\pi_\Lx,\underset{\tx{\ciut{(1)}}}{\txa_{\rm L}},\mu_\Lx\bigr)
\qqq
composed of 
\bit
\item a surjective submersion
\qq\nn
\pi_{\sfY\txG}\ :\ \sfY\txG\too\txG
\qqq
with a structure of a Lie supergroup on its total space mapped onto that on $\,\txG\,$ by the Lie-supergroup epimorphism $\,\pi_{\sfY\txG}$;
\item a global primitive $\,\underset{\tx{\ciut{(2)}}}{\txb}\,$ of the pullback of $\,\underset{\tx{\ciut{(3)}}}{\txh}\,$ to it,
\qq\nn
\pi_{\sfY\txG}^*\underset{\tx{\ciut{(3)}}}{\txh}=\sfd\underset{\tx{\ciut{(2)}}}{\txb}\,,
\qqq
which is LI with respect to the induced left-regular action of $\,\sfY\txG\,$ on itself
\qq\nn
\sfY\ell_\cdot\ :\ \sfY\txG\x\sfY\txG\too\sfY\txG\,,
\qqq
lifting $\,\ell_\cdot\,$ along $\,\pi_{\sfY\txG}$,
\qq\nn
\sfY\ell_y^*\underset{\tx{\ciut{(2)}}}{\txb}=\underset{\tx{\ciut{(2)}}}{\txb}\,,\qquad y\in\sfY\txG\,;
\qqq
\item a CaE super-0-gerbe 
\qq\nn
\bigl(\Lx,\pi_\Lx,\underset{\tx{\ciut{(1)}}}{\txa_{\rm L}}\bigr)
\qqq
over the fibred square $\,\sfY^{[2]}\txG\equiv\sfY\txG\x_\txG\sfY\txG\,$ (endowed with the natural Lie-supergroup structure induced from the product structure on $\,\sfY\txG^{\x 2}\,$ through restriction), with a principal connection 1-form $\,\underset{\tx{\ciut{(1)}}}{\txa_{\rm L}}\,$ of curvature $\,\underset{\tx{\ciut{(2)}}}{\txh_\Lx}$, 
\qq\nn
\pi_\Lx^*\underset{\tx{\ciut{(2)}}}{\txh}{}_\Lx=\sfd\underset{\tx{\ciut{(1)}}}{\txa_{\rm L}}\,,
\qqq
that satisfies the identity
\qq\nn
\underset{\tx{\ciut{(2)}}}{\txh}{}_\Lx=\bigl(\pr_2^*-\pr_1^*\bigr)\underset{\tx{\ciut{(2)}}}{\txb}\,;
\qqq
\item an isomorphism of CaE super-0-gerbes\footnote{Note that pullback along a canonical projection is consistent with the definition of a super-0-gerbe due to its equivariance.}
\qq\nn
\mu_\Lx\ :\ \pr_{1,2}^*\Lx\ox\pr_{2,3}^*\Lx\xrightarrow{\ \cong\ }\pr_{1,3}^*\Lx
\qqq
over the fibred cube $\,\sfY^{[3]}\txG\equiv\sfY\txG\x_\txG\sfY\txG\x_\txG\sfY\txG\,$ that satisfies the coherence (associativity) condition
\qq\nn
\pr_{1,2,4}^*\mu_\Lx\circ(\id_{\pr_{1,2}^*\Lx}\ox\pr_{2,3,4}^*\mu_\Lx)=\pr_{1,3,4}^*\mu_\Lx\circ(\pr_{1,2,3}^*\mu_\Lx\ox \id_{\pr_{3,4}^*\Lx})
\qqq
over the quadruple fibred product $\,\sfY^{[4]}\txG\equiv\sfY\txG\x_\txG\sfY\txG\x_\txG\sfY\txG\x_\txG\sfY\txG$.
\eit

Given CaE super-1-gerbes $\,\sG^{(1)\,A}_{\rm CaE}=\bigl(\sfY_A\txG,\pi_{\sfY_A\txG},\underset{\tx{\ciut{(2)}}}{\txb_A},\Lx_A,\pi_{\Lx_A},\underset{\tx{\ciut{(1)}}}{\txa_{\Lx_A}},\mu_{\Lx_A}\bigr),\ A\in\{1,2\}\,$ over a common base $\,\txG$,\ a 1-{\bf isomorphism} between them is a sextuple
\qq\nn
\Phi^{(1)}_{\rm CaE}:=\bigl(\sfY\sfY_{1,2}\txG,\pi_{\sfY\sfY_{1,2}\txG},\txE,\pi_\txE,\underset{\tx{\ciut{(1)}}}{\txa_\txE},\a_\txE\bigr)\ :\ \sG^{(1)\,1}_{\rm CaE}\xrightarrow{\ \cong\ }\sG^{(1)\,2}_{\rm CaE}
\qqq
composed of 
\bit
\item a surjective submersion 
\qq\nn
\pi_{\sfY\sfY_{1,2}\txG}\ :\ \sfY\sfY_{1,2}\txG\too\sfY_1\txG\x_\txG\sfY_2\txG\equiv\sfY_{1,2}\txG
\qqq
with a structure of a Lie supergroup on its total space that lifts the product Lie-supergroup structure on the fibred product $\,\sfY_{1,2}\txG\,$ along the Lie-supergroup epimorphism $\,\pi_{\sfY\sfY_{1,2}\txG}$;
\item a CaE super-0-gerbe
\qq\nn
\bigl(\txE,\pi_\txE,\underset{\tx{\ciut{(1)}}}{\txa_\txE}\bigr)
\qqq
over the total space $\,\sfY\sfY_{1,2}\txG$,\ with a principal $\bC^\x$-connection super-1-form $\,\underset{\tx{\ciut{(1)}}}{\txa_\txE}\,$ of curvature $\,\underset{\tx{\ciut{(2)}}}{\txh_\txE}$,
\qq\nn
\pi_\txE^*\underset{\tx{\ciut{(2)}}}{\txh_\txE}=\sfd\underset{\tx{\ciut{(1)}}}{\txa_\txE}\,,
\qqq
that satisfies the identity
\qq\nn
\underset{\tx{\ciut{(2)}}}{\txh_\txE}=\pi_{\sfY\sfY_{1,2}\txG}^*\bigl(\pr_2^*\underset{\tx{\ciut{(2)}}}{\txb_2}-\pr_1^*\underset{\tx{\ciut{(2)}}}{\txb_1}\bigr)\,;
\qqq
\item an isomorphism of CaE super-0-gerbes
\qq\nn
\a_\txE\ :\ \pi_{\sfY\sfY_{1,2}\txG}^{\x 2\,*}\pr_{1,3}^*\Lx_1\ox\pr_2^*\txE\xrightarrow{\ \cong\ }\pr_1^*\txE\ox\pi_{\sfY\sfY_{1,2}\txG}^{\x 2\,*}\pr_{2,4}^*\Lx_2
\qqq
over the fibred product $\,\sfY^{[2]}\sfY_{1,2}\txG=\sfY\sfY_{1,2}\txG\x_\txG\sfY\sfY_{1,2}\txG$,\ subject to the coherence constraint expressed by the commutative diagram \eqref{diag:grb1isocoh} of isomorphisms of CaE super-0-gerbes over the fibred product $\,\sfY^{[3]}\sfY_{1,2}\txG\equiv\sfY\sfY_{1,2}\txG\x_\txG\sfY\sfY_{1,2}\txG\x_\txG\sfY\sfY_{1,2}\txG$.
\eit

Given a pair of 1-isomorphisms $\,\Phi^{(1)\,B}_{\rm CaE}=(\sfY^B\sfY_{1,2}\txG,\pi_{\sfY^B\sfY_{1,2}\txG},\txE_B,\pi_{\txE_B},\underset{\tx{\ciut{(1)}}}{\txa_{\txE_B}},\a_{\txE_B}),\ B\in\{1,2\}\,$ between CaE super-1-gerbes $\,\cG^{(1)\,A}_{\rm CaE}=(\sfY_A \txG,\pi_{\sfY_A \txG},\underset{\tx{\ciut{(2)}}}{\txb_A},\Lx_A,\pi_{\Lx_A},\underset{\tx{\ciut{(1)}}}{\txa_{\Lx_A}},\mu_{\Lx_A}),\ A\in\{1,2\}$,\ a 2-isomorphism between the latter is represented by a triple
\qq\nn
\varphi^{(1)}_{\rm CaE}=(\sfY\sfY^{1,2}\sfY_{1,2}\txG,\pi_{\sfY\sfY^{1,2}\sfY_{1,2}\txG},\b)\ :\ \Phi^{(1)\,1}_{\rm CaE}\xLongrightarrow{\ \cong\ }\Phi^{(1)\,2}_{\rm CaE}
\qqq
composed of 
\bit
\item a surjective submersion
\qq\nn
\pi_{\sfY\sfY^{1,2}\sfY_{1,2}\txG}\ :\ \sfY\sfY^{1,2}\sfY_{1,2}\txG\too\sfY^1\sfY_{1,2}\txG\x_{\sfY_{1,2}\txG}\sfY^2\sfY_{1,2}\txG\equiv\sfY^{1,2}\sfY_{1,2}\txG
\qqq
with a structure of a Lie supergroup on its total space that lifts the product Lie-supergroup structure on the fibred product $\,\sfY^{1,2}\sfY_{1,2}\txG\,$ along the Lie-supergroup epimorphism $\,\pi_{\sfY\sfY^{1,2}\sfY_{1,2}\txG}$;
\item an isomorphism of CaE super-0-gerbes
\qq\nn
\b\ :\ (\pr_1\circ\pi_{\sfY\sfY^{1,2}\sfY_{1,2}\txG})^*\txE_1\xrightarrow{\ \cong\ }(\pr_2\circ\pi_{\sfY\sfY^{1,2}\sfY_{1,2}\txG})^*\txE_2
\qqq
subject to the coherence constraint expressed by the commutative diagram \eqref{diag:betacohalpha} of isomorphisms of CaE super-0-gerbes over $\,\sfY^{[2]}\sfY^{1,2}\sfY_{1,2}\txG$.
\eit
\exdef
\medskip

\subsection{The super-2-gerbe of the M-theory supermembrane}

In order to corroborate our claim as to the structural nature of the proposed geometrisation scheme in the setting of the $\si$-model super-$p$-brane dynamics, we discuss yet another example of a super-$p$-gerbe, to wit, the super-2-gerbe for the super-4-form field that couples to the uniformly charged supermembrane in the super-Minkowski space $\,{\rm sMink}(10,1\,\vert\,32)\,$ with the 11-dimensional body. In so doing, we encounter a slightly more involved extension mechanism than those dealt with heretofore. Thus, we are going to adapt the logic employed -- after Refs.\,\cite{Aldaya:1984gt,Chryssomalakos:2000xd} -- in the previous paragraph to the problem of trivialisation of the Green--Schwarz super-4-form 
\qq\nn
\underset{\tx{\ciut{(4)}}}{\txH}=\pi_0^*\bigl(\si\wedge\ovl\G_{ab}\,\si\bigr)\wedge e^a\wedge e^b
\qqq
whose closedness is implied by the particular variant 
\qq\label{eq:ClifFierz2}
\ovl\G{}^a_{(\a\b}\,\bigl(\ovl\G_{ab}\bigr)_{\g\d)}=0
\qqq
of the Fierz identity \eqref{eq:ClifFierz}, which can be conveniently rewritten as
\qq\nn
\ovl\G{}^a_{(\g\d}\,\bigl(\ovl\G_{ab}\bigr)_{\b)\a}=-\ovl\G{}^a_{\a(\b}\,\bigl(\ovl\G_{ab}\bigr)_{\g\d)}\,.
\qqq
On the tentative list \eqref{eq:LI2bas} of LI de Rham super-2-cocycles, we now find
\qq\label{eq:CaEscocyc2}
\underset{\tx{\ciut{(2)}}}{h}{}^{(1)}_{ab}=-\tfrac{1}{4}\, P_a\con P_b\con\underset{\tx{\ciut{(4)}}}{\txH}=\tfrac{1}{2}\,\pi_0^*\bigl(\si\wedge\ovl\G_{ab}\,\si\bigr)\,.
\qqq
Reasoning along the lines of our previous analyses, we erect over $\,\cM^{(1)}={\rm sMink}(10,1\,\vert\,32)\,$ a trivial rank-55 (real) vector bundle 
\qq\nn
\pi^{(2)}_2\equiv\pr_1\ :\ \cM^{(2)}_2:=\cM^{(1)}\x\bR^{\x 55}\too\cM^{(1)}\ :\ \bigl(\theta^\a,x^a,\z_{bc}=-\z_{cb}\bigr)\longmapsto\bigl(\theta^\a,x^a\bigr)
\qqq
endowed with the structure of a Lie supergroup that lifts the same structure on $\,\bR(10,1\,\vert\,32)\,$ in a manner that ensures left-invariance of the super-1-forms
\qq\nn
e^{(2)}_{ab}(\theta,x,\z)=\sfd\z_{ab}+\tfrac{1}{2}\,\theta\,\ovl\G_{ab}\,\si(\theta)\,,
\qqq
the latter satisfying the identities
\qq\nn
\pi^{(2)\,*}_2\underset{\tx{\ciut{(2)}}}{h}{}^{(1)}_{ab}=\sfd e^{(2)}_{ab}\,.
\qqq
The relevant structure is given in
\berop\label{prop:M22sgroup}
The above-described vector bundle $\,\cM^{(2)}_2\,$ equipped with the binary operation
\qq\nn
\txm^{(2)}_2\ :\ \cM^{(2)}_2\x\cM^{(2)}_2\too\cM^{(2)}_2\ :\ \bigl(\bigl(m_1^1,\z_{1\,ab}\bigr),\bigl(m_1^2,\z_{2\,cd}\bigr)\bigr)\longmapsto\bigl(m^{(1)}\bigl(m_1^1,m_1^2\bigr),\z_{1\,ab}+\z_{2\,ab}-\tfrac{1}{2}\,\theta_1\,\ovl\G_{ab}\,\theta_2\bigr)\,,
\qqq
written for $\,m_1^A=(\theta^\a_A,x^a_A),\ A\in\{1,2\}$,\ with the inverse
\qq\nn
\Inv^{(2)}_2\ :\ \cM^{(2)}_2\too\cM^{(2)}_2\ :\ \bigl(\theta^\a,x^a,\z_{bc}\bigr)\longmapsto\bigl(-\theta^\a,-x^a,-\z_{bc}\bigr)
\qqq
and the neutral element
\qq\nn
e^{(2)}_2=(0,0,0)
\qqq
is a Lie supergroup. It is a supercentral extension 
\qq\nn
\bd1\too\bR^{\x 55}\too\bR(10,1\,\vert\,32)\lx\bR^{\x 55}\equiv\cM_2^{(2)}\xrightarrow{\ \pi_2^{(2)}\ }\bR(10,1\,\vert\,32)\too\bd1 
\qqq
of the super-Minkowski group $\,\bR(10,1\,\vert\,32)\,$ determined by the family of CE super-2-cocycles corresponding to the CaE super-2-cocycles $\,\{\underset{\tx{\ciut{(2)}}}{h}{}^{(1)}_{ab}\}_{a,b\in\ovl{0,10}}\,$ of \Reqref{eq:CaEscocyc2}.
\eerop
\beroof
Through inspection. 
\eroof

In the next step, let us split the super-4-cocycle as
\qq\nn
\underset{\tx{\ciut{(4)}}}{\txH}=\la_1\,\pi_0^*\bigl(\si\wedge\ovl\G_{ab}\,\si\bigr)\wedge e^a\wedge e^b+\la_2\,\pi_0^*\bigl(\si\wedge\ovl\G_{ab}\,\si\bigr)\wedge e^a\wedge e^b\,,\qquad\la_1+\la_2=1
\qqq
and pull it back to $\,\cM_2^{(2)}$,\ whereupon we judiciously\footnote{Trivialising the factor $\,\pi_0^*\bigl(\si\wedge\ovl\G_{ab}\,\si\bigr)\,$ within $\,\pi_0^*\bigl(\si\wedge\ovl\G_{ab}\,\si\bigr)\wedge e^a\wedge e^b\,$ as a whole does not solve our problem.} rewrite it, using the shorthand notation 
\qq\nn
\pi^{(2)}_{02}\equiv\pi_0\circ\pi^{(2)}_2
\qqq 
along the way, as 
\qq\nn
\pi^{(2)\,*}_2\underset{\tx{\ciut{(4)}}}{\txH}&=&\sfd\bigl(2\la_1\,e_{ab}^{(2)}\wedge\pi^{(2)\,*}_2(e^a\wedge e^b)\bigr)+2\la_1\,e_{ab}^{(2)}\wedge\pi^{(2)\,*}_{02}\bigl(\si\wedge\ovl\G^a\,\si\bigr)\wedge\pi^{(2)\,*}_2 e^b\cr\cr
&&+\la_2\,\pi^{(2)\,*}_{02}\bigl(\si\wedge\ovl\G_{ab}\,\si\bigr)\wedge\pi^{(2)\,*}_2\bigl(e^a\wedge e^b\bigr)\,.
\qqq
We may now apply the generating technique that gave us \eqref{eq:LI2bas} to the partially corrected super-4-cocycle 
\qq\nn
\widetilde{\underset{\tx{\ciut{(4)}}}{\txH}}=\pi^{(2)\,*}_2\underset{\tx{\ciut{(4)}}}{\txH}-\sfd\bigl(2\la_1\,e_{ab}^{(2)}\wedge\pi^{(2)\,*}_2(e^a\wedge e^b)\bigr) 
\qqq
over the supermanifold $\,\cM^{(2)}_2$.\ For that, we need supervector fields dual to the LI super-1-forms from the set $\,\{\pi^{(2)\,*}_{02}\si^\a,\pi_0^*e^a,e^{(2)}_{bc}\}_{\a\in\ovl{1,32},\ a,b,c\in\ovl{0,10}}$.\ These are readily found to be given by
\qq\nn
& Q^{(2)}_\a(\theta,x,\z)= Q^{(1)}_\a(\theta,x)+\tfrac{1}{2}\,\ovl\G_{ab\,\a\b}\,\theta^\b\,\tfrac{\p\ }{\p\z_{ab}}\,,&\cr\cr
& P^{(2)}_a(\theta,x,\z)= P^{(1)}_a(\theta,x)\,,\qquad\qquad Z^{(2)\,ab}(\theta,x,\z)=\tfrac{\p\ }{\p\z_{ab}}&
\qqq
and furnish the LSA
\qq\nn
&\{ Q^{(2)}_\a, Q^{(2)}_\b\}=\ovl\G{}^a_{\a\b}\, P^{(2)}_a+\ovl\G_{ab\,\a\b}\, Z^{(2)\,ab}\,,\quad\quad[ P^{(2)}_a, P^{(2)}_b]=0\,,\quad\quad[ Z^{(2)\,ab}, Z^{(2)\,cd}]=0\,,&\cr\cr
&[ Q^{(2)}_\a, P^{(2)}_a]=0\,,\qquad\qquad[ Q^{(2)}_\a, Z^{(2)\,ab}]=0\,,\qquad\qquad[ P^{(2)}_a, Z^{(2)\,bc}]=0\,.&
\qqq
We may, next, contract the super-4-cocycle $\,\widetilde{\underset{\tx{\ciut{(4)}}}{\txH}}\,$ with the vector fields $\, P^{(2)}_a\,$ and $\, Q^{(2)}_\a$,\ whereby, for suitably adjusted normalisation constants ($4\la_1=1=4\la_2$), we obtain the combination 
\qq\label{eq:CaEscocyc3}\hspace{1cm}
\widetilde{\underset{\tx{\ciut{(2)}}}{h}}{}_{a\a}:= Q^{(2)}_\a\con P^{(2)}_a\con\widetilde{\underset{\tx{\ciut{(4)}}}{\txH}}=\ovl\G{}^b_{\a\b}\,e_{ba}^{(2)}\wedge\pi_{02}^{(2)\,*}\si^\b+\ovl\G_{ba\,\a\b}\,\pi^{(2)\,*}_2e^b\wedge\pi_{02}^{(2)\,*}\si^\b\,.
\qqq
The latter is closed in virtue of identity \eqref{eq:ClifFierz2}. Upon rewriting the super-2-cocycle as
\qq\nn
\left(\ovl\G{}^b_{\a\b}\,e_{ba}^{(2)}(\theta,x,\z)+\ovl\G_{ba\,\a\b}\,e^b(\theta,x)\right)\wedge\si^\b(\theta)&=&\sfd\bigl[\bigl(\ovl\G{}^b_{\a\b}\,e_{ab}^{(2)}(\theta,x,\z)+\ovl\G_{ab\,\a\b}\,e^b(\theta,x)\bigr)\,\theta^\b\bigr]\cr\cr
&&-\tfrac{1}{2}\,(\ovl\G{}^b_{\a\b}\,\ovl\G_{ab\,\g\d}+\ovl\G{}^b_{\g\d}\,\ovl\G_{ab\,\a\b})\,\theta^\b\,\bigl(\si^\g\wedge\si^\d\bigr)(\theta)\,,
\qqq
and taking into account the identity (also following from \Reqref{eq:ClifFierz2})
\qq\nn
-(\ovl\G{}^b_{\a\b}\,\ovl\G_{ab\,\g\d}+\ovl\G{}^b_{\g\d}\,\ovl\G_{ab\,\a\b})\,\theta^\b\,\bigl(\si^\g\wedge\si^\d\bigr)(\theta)=2\bigl(\ovl\G{}^b_{\a\g}\,\ovl\G_{ab\,\b\d}+\ovl\G{}^b_{\b\d}\,\ovl\G_{ab\,\a\g}\bigr)\,\bigl[\sfd\bigl(\theta^\b\,\theta^\g\,\si^\d(\theta)\bigr)+\theta^\g\,\bigl(\si^\b\wedge\si^\d\bigr)(\theta)\bigr]\,,
\qqq
we may write
\qq\nn
\widetilde{\underset{\tx{\ciut{(2)}}}{h}}{}_{a\a}(\theta,x,\z)=\sfd\bigl[\bigl(\ovl\G{}^b_{\a\b}\,e_{ab}^{(2)}(\theta,x,\z)+\ovl\G_{ab\,\a\b}\,e^b(\theta,x)\bigr)\,\theta^\b+\tfrac{1}{3}\,\bigl(\ovl\G{}^b_{\a\g}\,\ovl\G_{ab\,\b\d}+\ovl\G{}^b_{\b\d}\,\ovl\G_{ab\,\a\g}\bigr)\,\theta^\b\,\theta^\g\,\si^\d(\theta)\bigr]\,,
\qqq
and so we see that the super-2-cocycle admits a manifestly non-LI de Rham primitive. Its trivialisation in the CaE cohomology necessitates the construction of a trivial vector bundle 
\qq\nn
\pi_2^{(3)}\equiv\pr_1\ :\ \cM_2^{(3)}:=\cM_2^{(2)}\x\bR^{0\,\vert\,(d+1)\,D_{d,1}}\too\cM_2^{(2)}\ :\ \bigl(\theta^\a,x^a,\z_{bc},\psi_{d\b}\bigr)\longmapsto\bigl(\theta^\a,x^a,\z_{bc}\bigr)
\qqq  
with the purely Gra\ss mann-odd fibre $\,\bR^{0\,\vert\,352}\,$ and a Lie-supergroup structure that extends the previously established structure on its base $\,\cM_2^{(2)}\,$ in such a way that the super-1-forms 
\qq\nn
\si^{(3)}_{a\a}(\theta,x,\z,\psi)=\sfd\psi_{a\a}+\bigl[\ovl\G{}^b_{\a\b}\,\bigl(e_{ab}^{(2)}(\theta,x,\z)-\tfrac{1}{3}\,\ovl\theta\,\G_{ab}\,\si(\theta)\bigr)+\ovl\G_{ab\,\a\b}\,\bigl(e^b(\theta,x)-\tfrac{1}{3}\,\ovl\theta\,\G^b\,\si(\theta)\bigr)\bigr]\,\theta^\b
\qqq
are LI with respect to this extension. We thus obtain
\berop\label{prop:M23sgroup}
The above-described vector bundle $\,\cM^{(3)}_2\,$ equipped with the binary operation
\qq\nn
\txm^{(3)}_2\ &:&\ \cM^{(3)}_2\x\cM^{(3)}_2\too\cM^{(3)}_2\ :\ \bigl(\bigl(m_2^1,\psi_{1\,a\a}\bigr),\bigl(m_2^2,\psi_{2\,b\b}\bigr)\bigr)\longmapsto\bigl(m_2^{(2)}\bigl(m_2^1,m_2^2\bigr),\cr\cr
&&\psi_{1\,a\a}+\psi_{2\,a\a}+\bigl(\ovl\G{}^b_{\a\b}\,\z_{2\,ba}+x_2^b\,\ovl\G_{ba\,\a\b}\bigr)\,\theta_1^\b-\tfrac{1}{6}\,\bigl(\ovl\G{}^b_{\a\b}\,\ovl\G_{ba\,\g\d}+\ovl\G{}^b_{\g\d}\,\ovl\G_{ba\,\a\b}\bigr)\bigl(2\theta_1^\b+\theta_2^\b\bigr)\,\theta_1^\g\,\theta_2^\d\bigr)\,,
\qqq
written for $\,m_2^A=(\theta^\a_A,x^a_A,\z_{A\,bc}),\ A\in\{1,2\}$,\ 
with the inverse
\qq\nn
\Inv^{(3)}_2\ :\ \cM^{(3)}_2\too\cM^{(3)}_2\ :\ \bigl(\theta^\a,x^a,\z_{bc},\psi_{d\b}\bigr)\longmapsto\bigl(-\theta^\a,-x^a,-\z_{bc},-\psi_{d\b}+\bigl(\ovl\G{}^e_{\b\g}\,\z_{ed}+x^e\,\ovl\G_{ed\,\b\g}\bigr)\,\theta^\g\bigr)
\qqq
and the neutral element
\qq\nn
e^{(2)}_2=(0,0,0,0)
\qqq
is a Lie supergroup. It is a supercentral extension 
\qq\nn
\bd1\too\bR^{0\,\vert\,352}\too\cM_2^{(2)}\lx\bR^{0\,\vert\,352}\xrightarrow{\ \pi_2^{(3)}\ }\cM_2^{(2)}\too\bd1
\qqq
of the Lie supergroup $\,\cM_2^{(2)}\,$ of Prop.\,\ref{prop:M22sgroup} determined by the family of CE super-2-cocycles corresponding to the CaE super-2-cocycles $\,\{\widetilde{\underset{\tx{\ciut{(2)}}}{h}}{}_{a\a}\}_{(a,\a)\in\ovl{0,10}\x\ovl{1,32}}\,$ of \Reqref{eq:CaEscocyc3}.
\eerop
\beroof
Through inspection. As previously, the associativity of $\,\txm_2^{(3)}\,$ hinges upon identity \eqref{eq:ClifFierz2}. The only slightly less obvious element of the proof is the derivation of the very last term in the formula for the product of two points in $\,\cM^{(3)}_2$.\ Indeed, one has to take into account the relevant Fierz identities in order to identify the primitive and this, while in no way dramatic in the present case, may take up more time than necessary. Therefore, we pause to indicate a simple calculation method that will come handy in subsequent computations. 

When considering the effect of a (lifted) supersymmetry transformation on the super-1-form \qq\nn
\widetilde\si^{(3)}_{a\a}(\theta,x,\z):=\si^{(3)}_{a\a}(\theta,x,\z,\psi)-\sfd\psi_{a\a}\,,
\qqq
we immediately arrive at the expression
\qq\nn
&&\widetilde\si^{(3)}_{a\a}\bigl(\txm^{(2)}_2\bigl((\vep,y,\xi),(\theta,x,\z)\bigr)\bigr)-\widetilde\si^{(3)}_{a\a}(\theta,x,\z)\cr\cr
&=&\sfd\bigl(\bigl(\ovl\G{}^b_{\a\b}\,\z_{ab}+x^b\,\ovl\G_{ab\,\a\b}-\tfrac{1}{3}\,\bigl(\ovl\G{}^b_{\a\b}\,\vep\,\ovl\G_{ab}\,\theta+\ovl\G_{ab\,\a\b}\,\vep\,\ovl\G{}^b\,\theta\bigr)\bigr)\,\vep^\g\bigr)\cr\cr
&&+\tfrac{1}{6}\,\bigl(\ovl\G{}^b_{\a\b}\,\ovl\G_{ab\,\g\d}+\ovl\G{}^b_{\g\d}\,\ovl\G_{ab\,\a\b}+2\ovl\G{}^b_{\a\g}\,\ovl\G_{ab\,\b\d}+2\,\ovl\G{}^b_{\b\d}\,\ovl\G_{ab\,\a\g}\bigr)\,\vep^\b\,\theta^\g\,\sfd\theta^\d
\qqq
in which the last term is closed by construction. Hence, we are led to consider the de Rham super-1-cocycle
\qq\nn
\underset{\tx{\ciut{(1)}}}{\eta}{}_{a\a}(\theta)=\tfrac{1}{6}\,\bigl(\ovl\G{}^b_{\a\b}\,\ovl\G_{ab\,\g\d}+\ovl\G{}^b_{\g\d}\,\ovl\G_{ab\,\a\b}+2\ovl\G{}^b_{\a\g}\,\ovl\G_{ab\,\b\d}+2\,\ovl\G{}^b_{\b\d}\,\ovl\G_{ab\,\a\g}\bigr)\,\vep^\b\,\theta^\g\,\sfd\theta^\d\,.
\qqq
In consequence of the cohomological triviality of the supermanifold under consideration (in fact, the super-1-cocycle descends to the odd hyperplane $\,\bR^{0\,\vert\,32}$,\ and so it is the triviality of the latter that matters here), the super-1-form has a global primitive given by a global section $\,F\,$ of the structure sheaf of $\,\bR^{0\,\vert\,32}$, 
\qq\nn
\underset{\tx{\ciut{(1)}}}{\eta}{}_{a\a}=\sfd F_{a\a}\,,
\qqq
which we derive with the help of the standard homotopy argument (the so-called `homotopy formula'). Thus, we consider a homotopy
\qq\nn
H\ :\ [0,1]\x\bR^{0\,\vert\,32}\too\bR^{0\,\vert\,32}\ :\ \bigl(t,\theta^\a\bigr)\longmapsto t\theta^\a
\qqq
that linearly retracts the odd hyperplane to its distinguished point $\,0$,\ and write the primitive in the form of the integral over the homotopy fibre
\qq\label{eq:shomform}
F_{a\a}(\theta)=\int_0^1\,\sfd t\,\p_t\con H^*\underset{\tx{\ciut{(1)}}}{\eta}{}_{a\a}(t,\theta)\,,
\qqq
to the effect
\qq\nn
F_{a\a}(\theta)&=&\tfrac{1}{6}\,\int_0^1\,\sfd t\,t\,\p_t\con\bigl(\ovl\G{}^b_{\a\b}\,\ovl\G_{ab\,\g\d}+\ovl\G{}^b_{\g\d}\,\ovl\G_{ab\,\a\b}+2\ovl\G{}^b_{\a\g}\,\ovl\G_{ab\,\b\d}+2\,\ovl\G{}^b_{\b\d}\,\ovl\G_{ab\,\a\g}\bigr)\,\vep^\b\,\theta^\g\,\bigl(t\,\sfd\theta^\d+\theta^\d\,\sfd t\bigr)\cr\cr
&=&\tfrac{1}{12}\,\bigl(\ovl\G{}^b_{\a\b}\,\ovl\G_{ab\,\g\d}+\ovl\G{}^b_{\g\d}\,\ovl\G_{ab\,\a\b}+2\ovl\G{}^b_{\a\g}\,\ovl\G_{ab\,\b\d}+2\,\ovl\G{}^b_{\b\d}\,\ovl\G_{ab\,\a\g}\bigr)\,\vep^\b\,\theta^\g\,\theta^\d\cr\cr
&=&\tfrac{1}{6}\,\bigl(\ovl\G{}^b_{\a\g}\,\ovl\G_{ab\,\b\d}+\ovl\G{}^b_{\b\d}\,\ovl\G_{ab\,\a\g}\bigr)\,\vep^\b\,\theta^\g\,\theta^\d\,.
\qqq
\eroof
Let us subsequently -- with hindsight, once again -- decompose the (partially trivialised) super-4-cocycle further by taking $\,\la_{11}+\la_{12}=\la_1\,$ and $\,\la_2=\la_{21}+\la_{22}\,$ with $\,\la_{21}=2\la_{11}$,\ whereupon the above manipulations advance the trivialisation of the  GS super-4-cocycle as
\qq\nn
\pi^{(2,3)\,*}_2\underset{\tx{\ciut{(4)}}}{\txH}&=&\sfd\bigl(2\la_1\,\pi^{(3)\,*}_2 e_{ab}^{(2)}\wedge\pi^{(2,3)\,*}_2(e^a\wedge e^b)\bigr)-2\la_{11}\,\sfd\si_{a\a}^{(3)}\wedge\pi^{(2,3)\,*}_2e^a\wedge\pi^{(2,3)\,*}_{02}\si^\a\cr\cr
&&+2\la_{12}\,\pi^{(3)\,*}_2e_{ab}^{(2)}\wedge\pi^{(2,3)\,*}_{02}\bigl(\si\wedge\ovl\G{}^a\,\si\bigr)\wedge\pi^{(2,3)\,*}_2e^b+\la_{22}\,\pi^{(2,3)\,*}_{02}\bigl(\si\wedge\ovl\G_{ab}\,\si\bigr)\wedge\pi^{(2,3)\,*}_2\bigl(e^a\wedge e^b\bigr)\cr\cr
&=&\sfd\bigl(2\la_1\,\pi^{(3)\,*}_2 e_{ab}^{(2)}\wedge\pi^{(2,3)\,*}_2(e^a\wedge e^b)-2\la_{11}\, \si_{a\a}^{(3)}\wedge\pi^{(2,3)\,*}_2e^a\wedge\pi^{(2,3)\,*}_{02}\si^\a\bigr)\cr\cr
&&-\la_{11}\,\si_{a\a}^{(3)}\wedge\pi^{(2,3)\,*}_{02}(\si\wedge\ovl\G{}^a\,\si\wedge \si^\a)+2\la_{12}\,\pi^{(3)\,*}_2 e_{ab}^{(2)}\wedge\pi^{(2,3)\,*}_{02}\bigl(\si\wedge\ovl\G{}^a\,\si\bigr)\wedge\pi^{(2,3)\,*}_2e^b\cr\cr
&&+\la_{22}\,\pi^{(2,3)\,*}_{02}\bigl(\si\wedge\ovl\G_{ab}\,\si\bigr)\wedge\pi^{(2,3)\,*}_2\bigl(e^a\wedge e^b\bigr)\cr\cr
&\equiv&\sfd\bigl(2\la_1\,\pi^{(3)\,*}_2 e_{ab}^{(2)}\wedge\pi^{(2,3)\,*}_2(e^a\wedge e^b)-2\la_{11}\, \si_{a\a}^{(3)}\wedge\pi^{(2,3)\,*}_2e^a\wedge\pi^{(2,3)\,*}_{02}\si^\a\bigr)\cr\cr
&&+\underset{\tx{\ciut{(2)}}}{\D}{}_{\a\b}\wedge\pi^{(2,3)\,*}_{02}(\si^\a\wedge\si^\b)\,,
\qqq
where we used the shorthand notation
\qq\nn
\pi^{(2,3)}_2\equiv\pi^{(2)}_2\circ\pi^{(3)}_2\,,\qquad\qquad\pi^{(2,3)}_{02}=\pi_0\circ\pi^{(2)}_2\circ\pi^{(3)}_2\,.
\qqq 
Upon setting $\,\la_{11}=\la_{111}+\la_{112}$,\ we may cast the factor $\,\underset{\tx{\ciut{(2)}}}{\D}{}_{\a\b}\,$ in the form
\qq
\underset{\tx{\ciut{(2)}}}{\D}{}_{\a\b}&=&-\bigl(\la_{111}\,\ovl\G{}^a_{\a\b}\,\si^{(3)}_{a\g}+\tfrac{1}{2}\,\la_{112}\,\ovl\G{}^a_{\a\g}\,\si^{(3)}_{a\b}+\tfrac{1}{2}\,\la_{112}\,\ovl\G{}^a_{\b\g}\,\si^{(3)}_{a\a}\bigr)\wedge\pi^{(2,3)\,*}_{02}\si^\g\cr\cr
&&+2\la_{12}\,\ovl\G{}^a_{\a\b}\,\pi^{(3)\,*}_2 e_{ab}^{(2)}\wedge\pi^{(2,3)\,*}_2e^b+\la_{22}\,\ovl\G_{ab\,\a\b}\,\pi^{(2,3)\,*}_2\bigl(e^a\wedge e^b\bigr)\,,\label{eq:scocDab}
\qqq
and enquire as to the existence of a choice of the parameters for which the latter is closed. Taking into account the definitions of the super-1-forms $\,e^a,e_{bc}^{(2)}\,$ and $\,\si^{(3)}_{a\a}$,\ we find the exterior derivative of $\,\underset{\tx{\ciut{(2)}}}{\D}{}_{\a\b}\,$ in the form
\qq\nn
\sfd\underset{\tx{\ciut{(2)}}}{\D}{}_{\a\b}&=&(\la_{111}-\la_{12})\,\ovl\G{}^a_{\a\b}\,\ovl\G{}^b_{\g\d}\,\widetilde\pi_2^*\bigl(e_{ab}^{(2)}\wedge\widetilde\pi_{01}^*(\si^\g\wedge \si^\d)\bigr)+\bigl[(\la_{111}+\la_{12})\,\ovl\G{}^a_{\a\b}\,\ovl\G_{ab\,\g\d}+\la_{22}\,\ovl\G{}^a_{\g\d}\,\ovl\G_{ab\,\a\b}\cr\cr
&&+\tfrac{1}{4}\,\la_{112}\,(\ovl\G{}^a_{\a\g}\,\ovl\G_{ab\,\b\d}+\ovl\G{}^a_{\b\g}\,\ovl\G_{ab\,\a\d}+\ovl\G{}^a_{\a\d}\,\ovl\G_{ab\,\b\g}+\ovl\G{}^a_{\b\d}\,\ovl\G_{ab\,\a\g})\bigr]\,\pi^{(2,3)\,*}_2\bigl(e^b\wedge\widetilde\pi_0^*(\si^\g\wedge \si^\d)\bigr)
\qqq
Thus, in order for the derivative to vanish identically in the given representation of the Clifford algebra (that is, with the Fierz identity \eqref{eq:ClifFierz2} in force), we have to impose the constraints 
\qq\nn
\la_{111}-\la_{12}\must 0\,,\qquad\qquad 4(\la_{111}+\la_{12})\must\la_{112}\must 4\la_{22}
\qqq
with the solution
\qq\nn
(\la_{112},\la_{12},\la_{22})=\la_{111}\cdot(8,1,2)\,.
\qqq
We fix the free coefficient by demanding consistency of the result derived above with the linear relations between the various coefficients, and in particular -- with $\,\la_1+\la_2=1$,\ whereupon we obtain
\qq\label{eq:scocDabnorm}
(\la_{111},\la_{112},\la_{12},\la_{21},\la_{22})=\tfrac{1}{30}\cdot(1,8,1,18,2)\,.
\qqq
Given the CaE super-2-cocycle $\,\underset{\tx{\ciut{(2)}}}{\D}{}_{\a\b}\,$ on $\,\cM^{(3)}_2$,\ we may -- following the same logic as usual -- construct a trivial vector bundle 
\qq\nn
\pi_2^{(4)}\equiv\pr_1\ &:&\ \cM_2^{(4)}:=\cM_2^{(3)}\x\bR^{\x 528}\too\cM_2^{(3)}\cr\cr 
&:&\ \bigl(\theta^\a,x^a,\z_{bc},\psi_{d\b},\upsilon_{\g\d}=\upsilon_{\d\g}\bigr)\longmapsto\bigl(\theta^\a,x^a,\z_{bc},\psi_{d\b}\bigr)
\qqq  
with the purely Gra\ss mann-even fibre $\,\bR^{\x 528}\,$ and a Lie-supergroup structure fixed by the requirement that the super-1-forms\footnote{The normalisation of the super-1-forms involved is arbitrary. We fix it by demanding that the result of the ensuing trivialisation of the GS super-4-cocycle reproduce the one obtained in \Rxcite{Eq.\,(73)}{Chryssomalakos:2000xd}.}
\qq\nn
\si^{(4)}_{\a\b}(\theta,x,\z,\psi,\upsilon)=\sfd\upsilon_{\a\b}-\tfrac{15}{2}\,\sfd^{-1}\underset{\tx{\ciut{(2)}}}{\D}{}_{\a\b}(\theta,x,\z,\psi)\,,
\qqq
defined in terms of some specific (non-LI) primitives $\,\sfd^{-1}\underset{\tx{\ciut{(2)}}}{\D}{}_{\a\b}\,$ of the respective super-2-forms $\,\underset{\tx{\ciut{(2)}}}{\D}{}_{\a\b}$,\ be LI with respect to this extension. We readily establish
\berop\label{prop:homformsgroup}
The super-2-cocycles $\,\underset{\tx{\ciut{(2)}}}{\D}{}_{\a\b}=\underset{\tx{\ciut{(2)}}}{\D}{}_{\b\a},\ \a,\b\in\ovl{1,32}\,$  of \Reqref{eq:scocDab} on $\,\cM_2^{(3)}\,$ corresponding to the choice of coefficients given in \Reqref{eq:scocDabnorm} admit primitives
\qq\nn
-30\,\sfd^{-1}\underset{\tx{\ciut{(2)}}}{\D}{}_{\a\b}(\theta,x,\z,\psi)&=&-\bigl(\ovl\G{}^a_{\a\b}\,\si^{(3)}_{a\g}+4\ovl\G{}^a_{\a\g}\,\si^{(3)}_{a\b}+4\ovl\G{}^a_{\b\g}\,\si^{(3)}_{a\a}\bigr)(\theta,x,\z,\psi)\,\theta^\g+2\ovl\G{}^a_{\a\b}\,e_{ab}^{(2)}(\theta,x,\z)\,x^b\cr\cr
&&-2\bigl(2\ovl\G{}^a_{\a\d}\,\ovl\G{}^b_{\b\g}\,e_{ab}^{(2)}(\theta,x,\z)+\bigl(\ovl\G{}^a_{\a\d}\,\ovl\G_{ab\,\b\g}+\ovl\G{}^a_{\b\d}\,\ovl\G_{ab\,\a\g}\bigr)\,e^b(\theta,x)\bigr)\,\theta^\d\,\theta^\g\cr\cr
&&-2\ovl\G_{ab\,\a\b}\,x^a\,e^b(\theta,x)-\bigl(\ovl\G{}^a_{\a\b}\,\ovl\G_{ab\,\g\d}+\ovl\G{}^a_{\g\d}\,\ovl\G_{ab\,\a\b}\bigr)\,x^b\,\theta^\g\,\si^\d(\theta)\cr\cr
&&-\D_{\a\b;\g\d\vep\eta}\,\theta^\g\,\theta^\d\,\theta^\ep\,\si^\eta(\theta)\,,
\qqq
written in terms of the expressions
\qq\nn
\D_{\a\b;\g\d\vep\eta}=\ovl\G{}^a_{\a\d}\,\ovl\G{}^b_{\b\g}\,\ovl\G_{ab\,\ep\eta}+\tfrac{1}{2}\,\bigl(\ovl\G{}^a_{\a\d}\,\ovl\G_{ab\,\b\g}+\ovl\G{}^a_{\b\d}\,\ovl\G_{ab\,\a\g}\bigr)\,\ovl\G{}^b_{\ep\eta}\,.
\qqq
These determine, in the manner detailed above, the structure of a Lie supergroup on the vector bundle $\,\cM_2^{(4)}\,$ with the binary operation 
\qq\nn
\txm_2^{(4)}\ &:&\ \cM_2^{(4)}\x\cM_2^{(4)}\too\cM_2^{(4)}\ :\ \bigl(\bigl(m_3^1,\upsilon_{1\,\a\b}\bigr),\bigl(m_3^2,\upsilon_{2\,\g\d}\bigr)\bigr)\longmapsto\bigl(\txm_2^{(3)}\bigl(m_3^1,m_3^2\bigr),\cr\cr
&&\upsilon_{1\,\a\b}+\upsilon_{2\,\a\b}+\bigl(\tfrac{1}{4}\,\ovl\G{}^a_{\a\b}\,\psi_{2\,a\g}+\ovl\G{}^a_{\a\g}\,\psi_{2\,a\b}+\ovl\G{}^a_{\b\g}\,\psi_{2\,a\a}\bigr)\,\theta_1^\g+\tfrac{1}{4}\,x_1^a\,\bigl(\ovl\G_{ab\,\a\b}\,\bigl(2x_2^b-\theta_1\,\ovl\G{}^b\,\theta_2\bigr)\cr\cr
&&+\ovl\G{}^b_{\a\b}\,\bigl(2\z_{2\,ab}-\theta_1\,\ovl\G_{ab}\,\theta_2\bigr)\bigr)+\tfrac{1}{4}\,x_2^b\,\bigl(\ovl\G{}^a_{\a\b}\,\ovl\G_{ab\,\g\d}+\ovl\G{}^a_{\g\d}\,\ovl\G_{ab\,\a\b}\bigr)\,\theta_1^\g\,\theta_2^\d+\bigl(\z_{2\,ab}\,\ovl\G{}^a_{\a\g}\,\ovl\G{}^b_{\b\d}\cr\cr
&&+\tfrac{1}{2}\,x_2^b\,\bigl(\ovl\G{}^a_{\a\g}\,\ovl\G_{ab\,\b\d}+\ovl\G{}^a_{\b\g}\,\ovl\G_{ab\,\a\d}\bigr)\bigr)\,\theta_1^\g\,\theta_1^\d-\tfrac{1}{24}\,\theta_2^\g\,\bigl(\theta_2^\d\,\bigl(2\D_{\a\b;\g\d\ep\eta}\,\theta_2^\ep+3\bigl(\D_{\a\b;\ep\g\d\eta}-\D_{\a\b;\g\ep\d\eta}\bigr)\,\theta_1^\ep\bigr)\cr\cr
&&+6\D_{\a\b;\ep\d\g\eta}\bigr)\,\theta_1^\eta\bigr)\,,
\qqq
written for $\,m_3^A=(\theta^\a_A,x^a_A,\z_{A\,bc},\psi_{A\,d\b}),\ A\in\{1,2\}$,\ with the inverse
\qq\nn
\Inv_2^{(4)}\ &:&\ \cM_2^{(4)}\too\cM_2^{(4)}\ :\ \bigl(\theta^\a,x^a,\z_{bc},\psi_{d\,\b},\upsilon_{\g\d}\bigr)\longmapsto\bigl(\Inv_2^{(3)}\bigl(\theta^\a,x^a,\z_{bc},\psi_{d\,\b}\bigr),-\upsilon_{\g\d}-\ovl\G{}^f_{\g\eta}\,\ovl\G{}^g_{\d\k}\,\z_{fg}\,\theta^\eta\,\theta^\k\cr\cr
&&+\tfrac{1}{2}\,x^f\,\bigl(\ovl\G{}^g_{\g\d}\,\z_{fg}+\bigl(\ovl\G{}^g_{\g\eta}\,\ovl\G_{fg\,\d\k}+\ovl\G{}^g_{\d\eta}\,\ovl\G_{fg\,\g\k}\bigr)\,\theta^\eta\,\theta^\k\bigr)+\bigl(\tfrac{1}{4}\,\ovl\G{}^f_{\g\d}\,\psi_{f\eta}+\ovl\G{}^f_{\g\eta}\,\psi_{f\d}+\ovl\G{}^f_{\d\eta}\,\psi_{f\g}\bigr)\,\theta^\eta\bigr)
\qqq
and the neutral element
\qq\nn
e_2^{(4)}=(0,0,0,0,0)\,.
\qqq
It is a supercentral extension 
\qq\nn
\bd1\too\bR^{\x 528}\too\cM_2^{(3)}\lx\bR^{\x 528}\xrightarrow{\ \pi_2^{(4)}\ }\cM_2^{(3)}\too\bd1
\qqq
of the Lie supergroup $\,\cM_2^{(3)}\,$ of Prop.\,\ref{prop:M23sgroup} determined by the family of CE super-2-cocycles corresponding to the CaE super-2-cocycles $\,\{\underset{\tx{\ciut{(2)}}}{\D}{}_{\a\b}\}_{\a,\b\in\ovl{1,32}}\,$ of \Reqref{eq:scocDab}.
\eerop
\beroof
Proofs of both statements made in the proposition are rather tedious but otherwise fairly straightforward. The former one requires some ingenuity, therefore, we detail it in App.\,\ref{app:homformsgroup}.
\eroof
\noindent The above analysis gives us an explicit formula for the new LI super-1-form
\qq\nn
\si^{(4)}_{\a\b}(\theta,x,\z,\psi,\upsilon)&=&\sfd\upsilon_{\a\b}-\bigl(\tfrac{1}{4}\,\ovl\G{}^a_{\a\b}\,\si^{(3)}_{a\g}+\ovl\G{}^a_{\a\g}\,\si^{(3)}_{a\b}+\ovl\G{}^a_{\b\g}\,\si^{(3)}_{a\a}\bigr)(\theta,x,\z,\psi)\,\theta^\g+\tfrac{1}{2}\,\ovl\G{}^a_{\a\b}\,e_{ab}^{(2)}(\theta,x,\z)\,x^b\cr\cr
&&-\bigl(\ovl\G{}^a_{\a\d}\,\ovl\G{}^b_{\b\g}\,e_{ab}^{(2)}(\theta,x,\z)+\tfrac{1}{2}\,\bigl(\ovl\G{}^a_{\a\d}\,\ovl\G_{ab\,\b\g}+\ovl\G{}^a_{\b\d}\,\ovl\G_{ab\,\a\g}\bigr)\,e^b(\theta,x)\bigr)\,\theta^\d\,\theta^\g\cr\cr
&&-\tfrac{1}{2}\,\ovl\G_{ab\,\a\b}\,x^a\,e^b(\theta,x)-\tfrac{1}{4}\,\bigl(\ovl\G{}^a_{\a\b}\,\ovl\G_{ab\,\g\d}+\ovl\G{}^a_{\g\d}\,\ovl\G_{ab\,\a\b}\bigr)\,x^b\,\theta^\g\,\si^\d(\theta)\cr\cr
&&-\tfrac{1}{4}\,\D_{\a\b;\g\d\vep\eta}\,\theta^\g\,\theta^\d\,\theta^\ep\,\si^\eta(\theta)\,.
\qqq

Altogether, we extract from our hitherto considerations a primitive for (the pullback of) the GS super-4-cocycle: 
\qq\nn
\pi^{(2,3,4)\,*}_{02}\underset{\tx{\ciut{(4)}}}{\txH}=\sfd\underset{\tx{\ciut{(3)}}}{\b}^{(4)}
\qqq
given by
\qq\nn
\underset{\tx{\ciut{(3)}}}{\b}^{(4)}&=&\tfrac{2}{3}\,\pi^{(3,4)\,*}_2\bigl(e^{(2)}_{ab}\wedge\pi^{(2)\,*}_2\bigl(e^a\wedge e^b\bigr)\bigr)-\tfrac{3}{5}\,\pi^{(4)\,*}_2\bigl(\si^{(3)}_{a\a}\wedge\pi^{(2,3)\,*}_2 e^a\wedge\pi^{(2,3)\,*}_{02}\si^\a\bigr)\cr\cr
&&-\tfrac{2}{15}\,\si_{\a\b}^{(4)}\wedge\pi^{(2,3,4)\,*}_{02}(\si^\a\wedge\si^\b)\,,
\qqq
where we used the self-explanatory shorthand notation
\qq\nn
\pi^{(3,4)}_2=\pi^{(3)}_2\circ\pi^{(4)}_2\,,\qquad\qquad
\pi^{(2,3,4)}_{02}=\pi_0\circ\pi^{(2,3,4)}_2\,,\qquad\qquad\pi^{(2,3,4)}_2=\pi^{(2)}_2\circ\pi^{(3)}_2\circ\pi^{(4)}_2\,.
\qqq
The primitive is {\it left-invariant} with respect to the lift $\,\ell^{(4)}_\cdot\,$ of the supersymmetry $\,\ell^{(1)}_\cdot\,$ induced from $\,\txm_2^{(4)}\,$ as {\it per}
\qq\nn
\ell^{(4)}_\cdot=\txm_2^{(4)}\,,
\qqq
in which the first component of the domain is to be regarded as the extended supersymmetry group. Guided by the intuition developed previously in our analysis of the GS super-2-cocycle, we take the complete extension 
\qq\nn
\pi_{\sfY_2\cM^{(1)}}:=\pi^{(2)}_2\circ\pi^{(3)}_2\circ\pi^{(4)}_2\ :\ \sfY_2\cM^{(1)}:=\cM^{(4)}_2\too\cM^{(1)}\ :\ \bigl(m_1,\z_{ab},\psi_{c\a},\upsilon_{\b\g}\bigr)\longmapsto m_1
\qqq
to be the surjective submersion of a super-geometrisation of the GS super-4-cocycle $\,\underset{\tx{\ciut{(4)}}}{\txH}\,$ that we now work out in detail. As a first step, we compare pullbacks of $\,\underset{\tx{\ciut{(3)}}}{\b}^{(4)}\,$ to the $\cM^{(1)}$-fibred square 
\qq\nn
\alxydim{@C=.75cm@R=1cm}{& \sfY_2^{[2]}\cM^{(1)} \ar[rd]^{\pr_2} \ar[ld]_{\pr_1} & \\ \sfY_2\cM^{(1)} \ar[rd]_{\pi_{\sfY_2\cM^{(1)}}} & &  \sfY_2\cM^{(1)} \ar[ld]^{\pi_{\sfY_2\cM^{(1)}}} \\ & \cM^{(1)} & }
\qqq
along the two canonical projections to $\,\sfY_2\cM^{(1)}$,\ whereby we obtain -- for $\,m^A_4:=(\theta^\a,x^a,\z^A_{bc},\psi^A_{d\b},\upsilon^A_{\g\d})\,,\ A\in\{1,2\}\,$ and $\,\z^{21}_{ab}:=\z^2_{ab}-\z^1_{ab}\,,\ \psi^{21}_{c\a}:=\psi^2_{c\a}-\psi^1_{c\a}\,$ and $\,\upsilon^{21}_{\b\g}:=\upsilon^2_{\b\g}-\upsilon^1_{\b\g}\,$ -- the expression
\qq\nn
(\pr_2^*-\pr_1^*)\underset{\tx{\ciut{(3)}}}{\b^{(4)}}\bigl(m_4^1,m_4^2\bigr)&=&\tfrac{2}{3}\,\sfd\z^{21}_{ab}\wedge(e^a\wedge e^b)(\theta,x)-\tfrac{3}{5}\,\bigl(\sfd\psi^{21}_{a\a}+\ovl\G{}^b_{\a\b}\,\theta^\b\,\sfd\z^{21}_{ab}\bigr)\wedge e^a(\theta,x)\wedge \si^\a(\theta)\cr\cr
&&-\tfrac{2}{15}\,\bigl[\sfd\upsilon_{\a\b}^{21}-\bigl(\tfrac{1}{4}\,\ovl\G{}^a_{\a\b}\,\bigl(\sfd\psi^{21}_{a\g}+\ovl\G{}^b_{\g\d}\,\theta^\d\,\sfd\z^{21}_{ab}\bigr)+\ovl\G{}^a_{\a\g}\,\bigl(\sfd\psi^{21}_{a\b}+\ovl\G{}^b_{\b\d}\,\theta^\d\,\sfd\z^{21}_{ab}\bigr)\cr\cr
&&+\ovl\G{}^a_{\b\g}\,\bigl(\sfd\psi^{21}_{a\a}+\ovl\G{}^b_{\a\d}\,\theta^\d\,\sfd\z^{21}_{ab}\bigr)\bigr)\,\theta^\g+\tfrac{1}{2}\,\ovl\G{}^a_{\a\b}\,x^b\,\sfd\z^{21}_{ab}\cr\cr
&&-\ovl\G{}^a_{\a\d}\,\ovl\G{}^b_{\b\g}\,\theta^\d\,\theta^\g\,\sfd\z^{21}_{ab}\bigr]\wedge(\si^\a\wedge\si^\b)(\theta)\,,
\qqq
in which the super-1-forms 
\qq\nn
\xcX_{ab}\bigl(m_4^1,m_4^2\bigr):=\sfd\z^{21}_{ab}\,,\qquad\qquad\xcY_{a\a}\bigl(m_4^1,m_4^2\bigr):=\sfd\psi^{21}_{a\a}+\ovl\G{}^b_{\a\b}\,\theta^\b\,\xcX_{ab}
\qqq
and
\qq\nn
\xcZ_{\a\b}\bigl(m_4^1,m_4^2\bigr)&:=&\sfd\upsilon^{21}_{\a\b}-\bigl(\tfrac{1}{4}\,\ovl\G{}^a_{\a\b}\,\xcY_{a\g}+\ovl\G{}^a_{\a\g}\,\xcY_{a\b}+\ovl\G{}^a_{\b\g}\,\xcY_{a\a}\bigr)\bigl(m_4^1,m_4^2\bigr)\,\theta^\g+\tfrac{1}{2}\,\ovl\G{}^a_{\a\b}\,x^b\,\xcX_{ab}\bigl(m_4^1,m_4^2\bigr)\cr\cr
&&-\ovl\G{}^a_{\a\g}\,\ovl\G{}^b_{\b\d}\,\theta^\g\,\theta^\d\,\xcX_{ab}\bigl(m_4^1,m_4^2\bigr)
\qqq
are -- by construction (as differences of pullbacks of LI super-1-forms along the canonical projections to the cartesian factors) -- invariant under the left-regular action of the (fibred-)product Lie supergroup $\,\sfY_2^{[2]}\cM^{(1)}\,$ upon itself. Following the standard procedure, we seek to trivialise the 3-cocycle in a LI manner by pulling it back to the total space of a suitable surjective submersion over (or, in other words, to a supercentral extension of) $\,\sfY_2^{[2]}\cM^{(1)}$.\ To this end, we first consider the collection 
\qq\label{eq:CaEscocyc5}
\widehat{\underset{\tx{\ciut{(2)}}}{h}}{}^{\a\b}:=\pr_1^*\pi^{(2,3,4)\,*}_{02}(\si^\a\wedge\si^\b) 
\qqq
of manifestly LI super-2-cocycles and associate with them a trivial vector bundle
\qq\nn
\widehat\pi_2{}^{(5)}\equiv\pr_1\ &:&\ \widehat{\sfY^{[2]}_2\cM^{(1)}}{}{}^{(5)}:=\sfY^{[2]}_2\cM^{(1)}\x\bR^{\x 528}\too\sfY^{[2]}_2\cM^{(1)}\cr\cr 
&:&\ \widehat m_5:=\bigl(m^1_4,m^2_4,X^{\a\b}=X^{\b\a}\bigr)\longmapsto\bigl(m^1_4,m^2_4\bigr)
\qqq
with the purely Gra\ss mann-even fibre $\,\bR^{\x 528}\,$ and a Lie-supergroup structure fixed -- as formerly -- by the requirement that the super-1-forms
\qq\nn
\widehat e^{(5)\,\a\b}(\widehat m_5)=\sfd X^{\a\b}+\tfrac{1}{2}\,\bigl(\theta^\a\,\sfd\theta^\b+\theta^\b\,\sfd\theta^\a\bigr)\,,
\qqq
be LI with respect to this extension. We have the obvious
\berop\label{prop:M25sgroup}
The above-described vector bundle $\,\widehat{\sfY^{[2]}_2\cM^{(1)}}{}{}^{(5)}\,$ equipped with the binary operation
\qq\nn
\widehat\txm{}^{(5)}_2\ &:&\ \widehat{\sfY^{[2]}_2\cM^{(1)}}{}^{(5)}\x\widehat{\sfY^{[2]}_2\cM^{(1)}}{}^{(5)}\too\widehat{\sfY^{[2]}_2\cM^{(1)}}{}^{(5)}\ :\ \bigl(\bigl(m_4^1,m_4^2,X_1^{\a\b}\bigr),\bigl(n_4^1,n_4^2,X_2^{\g\d}\bigr)\bigr)\cr\cr 
&&\longmapsto\bigl(\txm_2^{(4)}\bigl(m_4^1,n_4^1\bigr),\txm_2^{(4)}\bigl(m_4^2,n_4^2\bigr),X_1^{\a\b}+X_2^{\a\b}-\tfrac{1}{2}\,\bigl(\theta_1^\a\,\theta_2^\b+\theta_1^\b\,\theta_2^\a\bigr)\bigr)\,,
\qqq
written for $\,m^A_4:=(\theta_1^\a,x_1^a,\z^A_{1\,bc},\psi^A_{1\,d\b},\upsilon^A_{1\,\g\d})\,$ and $\,n^A_4:=(\theta_2^\a,x_2^a,\z^A_{2\,bc},\psi^A_{2\,d\b},\upsilon^A_{2\,\g\d})\,,\ A\in\{1,2\}$,\ with the inverse
\qq\nn
\widehat\Inv{}^{(5)}_2\ :\ \widehat{\sfY^{[2]}_2\cM^{(1)}}{}^{(5)}\too\widehat{\sfY^{[2]}_2\cM^{(1)}}{}^{(5)}\ :\ \bigl(m_4^1,m_4^2,X^{\a\b}\bigr)\longmapsto\bigl(\Inv^{(4)}_2\bigl(m_4^1\bigr),\Inv^{(4)}_2\bigl(m_4^2\bigr),-X^{\a\b}\bigr)
\qqq
and the neutral element
\qq\nn
\widehat e{}^{(5)}_2=(0,0,0)
\qqq
is a Lie supergroup. It is a supercentral extension 
\qq\nn
\bd1\too\bR^{\x 528}\too\sfY^{[2]}_2\cM^{(1)}\lx\bR^{\x 528}\xrightarrow{\ \widehat\pi_2{}^{(5)}\ }\sfY^{[2]}_2\cM^{(1)}\too\bd1
\qqq
of the (product) Lie supergroup $\,\sfY^{[2]}_2\cM^{(1)}$,\ the latter being formed from the Lie supergroup $\,\sfY_2\cM^{(1)}\,$ of Prop.\,\ref{prop:homformsgroup}. The supercentral extension is determined by the family of CE super-2-cocycles corresponding to the CaE super-2-cocycles $\,\{\widehat{\underset{\tx{\ciut{(2)}}}{h}}{}^{\a\b}\}^{\a,\b\in\ovl{1,32}}\,$ of \Reqref{eq:CaEscocyc5}.
\eerop
\beroof
Trivial.
\eroof
\noindent The LI 1-forms $\,\widehat e^{(5)\,\a\b}=\widehat e^{(5)\,\b\a}\,$ on the new Lie supergroup, satisfying the identity
\qq\nn
\sfd\widehat e^{(5)\,\a\b}=\widehat\pi^{(2,3,4,5)\,*}_{02}(\si^\a\wedge\si^\b)\,,
\qqq
written in the shorthand notation
\qq\nn
\widehat\pi^{(2,3,4,5)}_{02}=\pi^{(2,3,4)}_{02}\circ\pr_1\circ\widehat\pi_2{}^{(5)}
\qqq
(to be adapted to subsequent extensions in an obvious manner), enable us to partially trivialise the super-3-form $\,(\pr_2^*-\pr_1^*)\underset{\tx{\ciut{(3)}}}{\b^{(4)}}\,$ upon pullback to $\,\widehat{\sfY^{[2]}_2\cM^{(1)}}{}{}^{(5)}\ni\widehat m_5\equiv(m_4^1,m_4^2,X^{\a\b})\,$ as
\qq\nn
&&\widehat\pi_2^{(5)\,*}(\pr_2^*-\pr_1^*)\underset{\tx{\ciut{(3)}}}{\b^{(4)}}(\widehat m_5)\cr\cr
&=&\sfd\bigl(\tfrac{2}{15}\,\xcZ_{\a\b}\bigl(m_4^1,m_4^2\bigr)\wedge\widehat e^{(5)\,\a\b}(\widehat m_5)\bigr)-\tfrac{2}{15}\,\bigl[\bigl(\tfrac{1}{4}\,\ovl\G{}^a_{\a\b}\,\xcY_{a\g}+2\ovl\G{}^a_{\b\g}\,\xcY_{a\a}\bigr)\bigl(m_4^1,m_4^2\bigr)\wedge \si^\g(\theta)\cr\cr
&&-\tfrac{1}{2}\,\ovl\G{}^a_{\a\b}\,\xcX_{ab}\bigl(m_4^1,m_4^2\bigr)\wedge e^b(\theta,x)\bigr]\wedge\widehat e^{(5)\,\a\b}(\widehat m_5)+\tfrac{2}{3}\,\xcX_{ab}\bigl(m_4^1,m_4^2\bigr)\wedge(e^a\wedge e^b)(\theta,x)\cr\cr
&&-\tfrac{3}{5}\,\xcY_{a\a}\bigl(m_4^1,m_4^2\bigr)\wedge e^a(\theta,x)\wedge \si^\a(\theta)\cr\cr
&=&\sfd\bigl(\tfrac{2}{15}\,\xcZ_{\a\b}\bigl(m_4^1,m_4^2\bigr)\wedge\widehat e^{(5)\,\a\b}(\widehat m_5)\bigr)+\tfrac{1}{15}\,\xcX_{ab}\bigl(m_4^1,m_4^2\bigr)\wedge\bigl(10e^a(\theta,x)-\ovl\G{}^a_{\a\b}\,\widehat e^{(5)\,\a\b}(\widehat m_5)\bigr)\wedge e^b(\theta,x)\cr\cr
&&-\tfrac{1}{30}\,\xcY_{a\a}\bigl(m_4^1,m_4^2\bigr)\wedge\bigl(18\,e^a(\theta,x)\wedge \si^\a(\theta)-\ovl\G{}^a_{\b\g}\,\bigl(\widehat e^{(5)\,\b\g}(\widehat m_5)\wedge \si^\a(\theta)+8\,\widehat e^{(5)\,\a\b}(\widehat m_5)\wedge \si^\g(\theta)\bigr)\bigr)\,.
\qqq
In the next step, we readily verify that the manifestly LI super-2-form 
\qq\nn
\widehat{\underset{\tx{\ciut{(2)}}}{h}}{}^{a\a}:=18\,\widehat\pi^{(2,3,4,5)\,*}_2\bigl(e^a\wedge\pi^*_0\si^\a\bigr)-\ovl\G{}^a_{\b\g}\,\widehat\pi_2{}^{(5)\,*}\bigl(\widehat e^{(5)\,\b\g}\wedge\pr_1^*\pi^{(2,3,4)\,*}_{02}\si^\a+8\,\widehat e^{(5)\,\a\b}\wedge\pr_1^*\pi^{(2,3,4)\,*}_{02}\si^\g\bigr)\,,
\qqq
with
\qq\nn
\widehat\pi^{(2,3,4,5)}_2=\pi^{(2,3,4)}_2\circ\pr_1\circ\widehat\pi_2{}^{(5)}\,,
\qqq
is closed, and hence gives rise to yet another supercentral extension. This time, we take the trivial vector bundle
\qq\nn
\widehat\pi_2^{(6)}\equiv\pr_1\ :\ \widehat{\sfY^{[2]}_2\cM^{(1)}}{}^{(6)}:=\widehat{\sfY^{[2]}_2\cM^{(1)}}{}{}^{(5)}\x\bR^{0\,\vert\,352}\too\widehat{\sfY^{[2]}_2\cM^{(1)}}{}{}^{(5)}\ :\ \widehat m_6:=\bigl(\widehat m_5,Y^{a\a}\bigr)\longmapsto\widehat m_5
\qqq
with the purely Gra\ss mann-odd fibre $\,\bR^{0\,\vert\,352}\,$ and a Lie-supergroup structure that extends the previously established structure on its base $\,\widehat{\sfY^{[2]}_2\cM^{(1)}}{}{}^{(5)}\,$ so that the super-1-forms 
\qq\nn
\widehat\si^{(6)\,a\a}\bigl(\widehat m_6\bigr)=\sfd Y^{a\a}-18\,\theta^\a\,e^a(\theta,x)+\ovl\G{}^a_{\b\g}\,\bigl(\theta^\a\,\widehat e^{(5)\,\b\g}(\widehat m_5)+8\,\theta^\g\,\widehat e^{(5)\,\a\b}(\widehat m_5)+8\,\theta^\a\,\theta^\b\,\si^\g(\theta)\bigr)\,,
\qqq
satisfying the identities
\qq\nn
\sfd\widehat\si^{(6)\,a\a}&=&\widehat\pi_2^{(6)\,*}\widehat{\underset{\tx{\ciut{(2)}}}{h}}{}^{a\a}\,,
\qqq
are LI with respect to the action of this extension on itself. We have
\berop\label{prop:M26sgroup}
The above-described vector bundle $\,\widehat{\sfY^{[2]}_2\cM^{(1)}}{}{}^{(6)}\,$ equipped with the binary operation
\qq\nn
\widehat\txm{}^{(6)}_2\ &:&\ \widehat{\sfY^{[2]}_2\cM^{(1)}}{}^{(6)}\x\widehat{\sfY^{[2]}_2\cM^{(1)}}{}^{(6)}\too\widehat{\sfY^{[2]}_2\cM^{(1)}}{}^{(6)}\ :\ \bigl(\bigl(\widehat m_5,Y_1^{a\,\a}\bigr),\bigl(\widehat n_5,Y_2^{b\,\b}\bigr)\bigr)\longmapsto\cr\cr 
&&\longmapsto\bigl(\widehat\txm{}^{(5)}_2\bigl(\widehat m_5,\widehat n_5\bigr),Y^{a\g}_1+Y^{a\g}_2+18\,\theta_1^\g\,x_2^a-4(2\theta_1^\g+\theta_2^\g)\,\bigl(\theta_1\,\ovl\G{}^a\,\theta_2\bigr)-\ovl\G{}^a_{\d\ep}\,\bigl(\theta_1^\g\,X_2^{\d\ep}+8\,\theta_1^\d\,X_2^{\g\ep}\bigr)\bigr)\,,
\qqq
written for $\,\widehat m_5:=(\theta_1^\a,x_1^a,\z^A_{1\,bc},\psi^A_{1\,d\b},\upsilon^A_{1\,\g\d},X_1^{\ep\eta})\,$ and $\,\widehat n_5:=(\theta_2^\a,x_2^a,\z^A_{2\,bc},\psi^A_{2\,d\b},\upsilon^A_{2\,\g\d},X_2^{\ep\eta})\,,\ A\in\{1,2\}$,\ with the inverse
\qq\nn
\widehat\Inv{}^{(6)}_2\ &:&\ \widehat{\sfY^{[2]}_2\cM^{(1)}}{}^{(6)}\too\widehat{\sfY^{[2]}_2\cM^{(1)}}{}^{(6)}\cr\cr 
&:&\ \bigl(m_4^1,m_4^2,X^{\a\b},Y^{a\g}\bigr)\longmapsto\bigl(\widehat\Inv_2^{(5)}\bigl(m_4^1,m_4^2,X^{\a\b}\bigr),-Y^{a\g}+18\,x^a\,\theta^\g-\ovl\G{}^a_{\d\ep}\,\bigl(\theta^\g\,X^{\d\ep}+8\,\theta^\d\,X^{\g\ep}\bigr)\bigr)
\qqq
and the neutral element
\qq\nn
\widehat e{}^{(6)}_2=(0,0,0,0)
\qqq
is a Lie supergroup. It is a supercentral extension 
\qq\nn
\bd1\too\bR^{0\,\vert\,352}\too\widehat{\sfY^{[2]}_2\cM^{(1)}}{}{}^{(5)}\lx\bR^{0\,\vert\,352}\xrightarrow{\ \widehat\pi_2{}^{(6)}\ }\widehat{\sfY^{[2]}_2\cM^{(1)}}{}{}^{(5)}\too\bd1
\qqq
of the Lie supergroup $\,\widehat{\sfY^{[2]}_2\cM^{(1)}}{}{}^{(5)}\,$ of Prop.\,\ref{prop:M25sgroup}. The supercentral extension is determined by the family of CE super-2-cocycles corresponding to the CaE super-2-cocycles $\,\{\widehat{\underset{\tx{\ciut{(2)}}}{h}}{}^{a\a}\}_{(a,\a)\in\ovl{0,10}\x\ovl{1,32}}\,$ of \Reqref{eq:CaEscocyc5}.
\eerop
\beroof
Through inspection.
\eroof
\noindent Thus, upon pullback to $\,\widehat{\sfY^{[2]}_2\cM^{(1)}}{}^{(6)}\ni\widehat m_6\,$ (in the hitherto notation) along
\qq\nn
\widehat\pi_2^{(5,6)}=\widehat\pi_2^{(5)}\circ\widehat\pi_2^{(6)}\,,
\qqq
we obtain
\qq\nn
&&\widehat\pi_2^{(5,6)\,*}(\pr_2^*-\pr_1^*)\underset{\tx{\ciut{(3)}}}{\b^{(4)}}(\widehat m_6)\cr\cr
&=&\sfd\bigl(\tfrac{2}{15}\,\xcZ_{\a\b}\bigl(m_4^1,m_4^2\bigr)\wedge\widehat e^{(5)\,\a\b}(\widehat m_5)+\tfrac{1}{30}\,\xcY_{a\a}\bigl(m_4^1,m_4^2\bigr)\wedge\widehat\si^{(6)\,a\a}(\widehat m_6)\bigr)\cr\cr
&&+\tfrac{1}{30}\,\xcX_{ab}\bigl(m_4^1,m_4^2\bigr)\wedge\bigl(2\bigl(10\,e^a(\theta,x)-\ovl\G{}^a_{\a\b}\,\widehat e^{(5)\,\a\b}(\widehat m_5)\bigr)\wedge e^b(\theta,x)+\ovl\G{}^b_{\a\b}\,\si^\a(\theta)\wedge\widehat\si^{(6)\,a\b}(\widehat m_6)\bigr)\,,
\qqq
and it is easy to check (or deduce from the construction) that the LI super-2-form 
\qq
\widehat{\underset{\tx{\ciut{(2)}}}{h}}{}^{ab}&:=&20\,\widehat\pi^{(2,3,4,5,6)\,*}_2\bigl(e^a\wedge e^b\bigr)+\widehat\pi_2^{(6)\,*}\bigl(\widehat\pi^{(2,3,4,5)\,*}_2\bigl(\ovl\G{}^a_{\a\b}\,e^b-\ovl\G{}^b_{\a\b}\,e^a\bigr)\wedge\widehat e^{(5)\,\a\b}\bigr)\cr\cr
&&-\tfrac{1}{2}\,\widehat\pi^{(2,3,4,5,6)\,*}_{02}\si^\a\wedge\bigl(\ovl\G{}^a_{\a\b}\,\widehat\si^{(6)\,b\b}-\ovl\G{}^b_{\a\b}\,\widehat\si^{(6)\,a\b}\bigr)\,,\label{eq:CaEscocyc6}
\qqq
written in terms of the maps
\qq\nn
\widehat\pi^{(2,3,4,5,6)}_{02}=\widehat\pi^{(2,3,4,5)}_{02}\circ\widehat\pi_2{}^{(6)}\,,\qquad\qquad\widehat\pi^{(2,3,4,5,6)}_2=\widehat\pi^{(2,3,4,5)}_2\circ\widehat\pi_2{}^{(6)}\,,
\qqq
is closed, so that we may finally trivialise the difference of pullbacks in the CaE cohomology by constructing one last supercentral extension. Thus, take the trivial vector bundle
\qq\nn
\widehat\pi_2^{(7)}\equiv\pr_1\ &:&\ \widehat{\sfY^{[2]}_2\cM^{(1)}}{}^{(7)}:=\widehat{\sfY^{[2]}_2\cM^{(1)}}{}^{(6)}\x\bR^{\x 55}\too\widehat{\sfY^{[2]}_2\cM^{(1)}}{}^{(6)}\cr\cr
&:&\ \widehat m_7:=\bigl(\widehat m_6,Z^{bc}=-Z^{cb}\bigr)\longmapsto\widehat m_6
\qqq
with the purely Gra\ss mann-even fibre $\,\bR^{\x 55}\,$ and endow it with the structure of a Lie supergroup that lifts the previously established structure of the same type from its base in such a manner that the super-1-forms (written in the hitherto notation)
\qq\nn
\widehat e^{(7)\,ab}(\widehat m_7)&=&\sfd Z^{ab}+10\bigl(x^a\,\sfd x^b-x^b\,\sfd x^a\bigr)+4\ovl\G{}^a_{\a\b}\,\ovl\G{}^b_{\g\d}\,\theta^\a\,\theta^\g\,\widehat e^{(5)\,\b\d}(\widehat m_5)+\bigl(\ovl\G{}^a_{\a\b}\,x^b-\ovl\G{}^b_{\a\b}\,x^a\bigr)\,\sfd X^{\a\b}\cr\cr
&&+\tfrac{1}{2}\,\bigl(\ovl\G{}^a_{\a\b}\,\widehat\si^{(6)\,b\a}-\ovl\G{}^b_{\a\b}\,\widehat\si^{(6)\,a\a}\bigr)(\widehat m_6)\,\theta^\b\,,
\qqq
satisfying the identities
\qq\nn
\sfd\widehat e^{(7)\,ab}=\widehat\pi_2^{(7)\,*}\widehat{\underset{\tx{\ciut{(2)}}}{h}}{}^{ab}\,,
\qqq
are LI with respect to this new supergroup structure. Yet again, we obtain
\berop\label{prop:M27sgroup}
The above-described vector bundle $\,\widehat{\sfY^{[2]}_2\cM^{(1)}}{}{}^{(7)}\,$ equipped with the binary operation
\qq\nn
\widehat\txm{}^{(7)}_2\ &:&\ \widehat{\sfY^{[2]}_2\cM^{(1)}}{}^{(7)}\x\widehat{\sfY^{[2]}_2\cM^{(1)}}{}^{(7)}\too\widehat{\sfY^{[2]}_2\cM^{(1)}}{}^{(7)}\ :\ \bigl(\bigl(\widehat m_6,Z_1^{bc}\bigr),\bigl(\widehat n_6,Z_2^{ef}\bigr)\bigr)\longmapsto\bigl(\widehat\txm{}^{(7)}_2\bigl(\widehat m_6,\widehat n_6\bigr),\cr\cr
&&Z_1^{bc}+Z_2^{bc}-10\,(x_1^b\,x_2^c-x_2^b\,x_1^c)+4\bigl(\bigl(x_1^b+x_2^b\bigr)\,\bigl(\ovl\theta_1\,\G^c\,\theta_2\bigr)-\bigl(x_1^c+x_2^c\bigr)\,\bigl(\ovl\theta_1\,\G^b\,\theta_2\bigr)\bigr)\cr\cr
&&+\tfrac{1}{2}\,\theta_1^\eta\,\bigl(\ovl\G{}^b_{\eta\k}\,Y_2^{c\k}-\ovl\G{}^c_{\eta\k}\,Y_2^{b\k}\bigr)-\bigl(4\ovl\G{}^b_{\eta\k}\,\ovl\G{}^c_{\la\mu}\,\theta_1^\eta\,\theta_1^\la-x_1^b\,\ovl\G{}^c_{\k\mu}+x_1^c\,\ovl\G{}^b_{\k\mu}\bigr)\,X_2^{\k\mu}\bigr)
\qqq
written for $\,\widehat m_6:=(\theta_1^\a,x_1^a,\z^A_{1\,bc},\psi^A_{1\,d\b},\upsilon^A_{1\,\g\d},X_1^{\ep\eta},Y_1^{e\la})\,$ and $\,\widehat n_5:=(\theta_2^\a,x_2^a,\z^A_{2\,bc},\psi^A_{2\,d\b},\upsilon^A_{2\,\g\d},X_2^{\ep\eta},Y_2^{e\la})\,,\ A\in\{1,2\}$,\ with the inverse
\qq\nn
\widehat\Inv{}^{(7)}_2\ &:&\ \widehat{\sfY^{[2]}_2\cM^{(1)}}{}^{(7)}\too\widehat{\sfY^{[2]}_2\cM^{(1)}}{}^{(7)}\ :\ \bigl(m_4^1,m_4^2,X^{\a\b},Y^{a\g},Z^{bc}\bigr)\longmapsto\bigl(\widehat\Inv^{(6)}_2\bigl(m_4^1,m_4^2,X^{\a\b},Y^{a\g}\bigr),\cr\cr 
&&\hspace{1cm}-Z^{bc}+\tfrac{1}{2}\,\theta^\eta\,\bigl(\ovl\G{}^b_{\eta\k}\,Y^{c\k}-\ovl\G{}^c_{\eta\k}\,Y^{b\k}\bigr)+\bigl(x^b\,\ovl\G{}^c_{\k\mu}-x^c\,\ovl\G{}^b_{\k\mu}+4\ovl\G{}^b_{\eta\k}\,\ovl\G{}^c_{\la\mu}\,\theta^\eta\,\theta^\la\bigr)\,X^{\k\mu}\bigr)
\qqq
and the neutral element
\qq\nn
\widehat e{}^{(7)}_2=(0,0,0,0,0)
\qqq
is a Lie supergroup. It is a supercentral extension 
\qq\nn
\bd1\too\bR^{\x 55}\too\widehat{\sfY^{[2]}_2\cM^{(1)}}{}{}^{(6)}\lx\bR^{\x 55}\xrightarrow{\ \widehat\pi_2{}^{(7)}\ }\widehat{\sfY^{[2]}_2\cM^{(1)}}{}{}^{(6)}\too\bd1
\qqq
of the Lie supergroup $\,\widehat{\sfY^{[2]}_2\cM^{(1)}}{}{}^{(6)}\,$ of Prop.\,\ref{prop:M26sgroup}. The supercentral extension is determined by the family of CE super-2-cocycles corresponding to the CaE super-2-cocycles $\,\{\widehat{\underset{\tx{\ciut{(2)}}}{h}}{}^{ab}\}_{a,b\in\ovl{0,10}}\,$ of \Reqref{eq:CaEscocyc6}.
\eerop
\beroof
Straightforward, through inspection.
\eroof 
By the end of the long day, we are left with the desired result
\qq\nn
\widehat\pi_2^{(5,6,7)\,*}(\pr_2^*-\pr_1^*)\underset{\tx{\ciut{(3)}}}{\b^{(4)}}(\widehat m_7)&=&\sfd\bigl[\tfrac{2}{15}\,\xcZ_{\a\b}\bigl(m_4^1,m_4^2\bigr)\wedge\widehat e^{(5)\,\a\b}(\widehat m_5)+\tfrac{1}{30}\,\xcY_{a\a}\bigl(m_4^1,m_4^2\bigr)\wedge\widehat\si^{(6)\,a\a}(\widehat m_6)\cr\cr
&&-\tfrac{1}{30}\,\xcX_{ab}\bigl(m_4^1,m_4^2\bigr)\wedge\widehat e^{(7)\,ab}(\widehat m_7)\bigr]\,,
\qqq
where
\qq\nn
\widehat\pi_2^{(5,6,7)}=\widehat\pi_2^{(5)}\circ\widehat\pi_2^{(6)}\circ\widehat\pi_2^{(7)}\,.
\qqq
The above formula suggests that we should take the supercentral extension
\qq\nn
\pi_{\widehat\sfY\sfY_2^{[2]}\cM^{(1)}}:=\widehat\pi_2^{(5)}\circ\widehat\pi_2^{(6)}\circ\widehat\pi_2^{(7)}\ &:&\ \widehat\sfY\sfY_2^{[2]}\cM^{(1)}:=\widehat{\sfY^{[2]}_2\cM^{(1)}}{}^{(7)}\too\sfY_2^{[2]}\cM^{(1)}\cr\cr
&:&\ \bigl(m_4^1,m_4^2,X^{\a\b},Y^{a\g},Z^{bc}\bigr)\longmapsto\bigl(m_4^1,m_4^2\bigr)
\qqq
as the surjective submersion of the super-1-gerbe over $\,\sfY_2^{[2]}\cM^{(1)}\,$ with a connection of the LI curvature 
\qq\nn
\widehat{\underset{\tx{\ciut{(3)}}}{\txH}}=\tfrac{2}{3}\,\xcX_{ab}\wedge\pr_1^*\pi^{(2,3,4)\,*}_2(e^a\wedge e^b)-\tfrac{3}{5}\,\xcY_{a\a}\wedge\pr_1^*\pi^{(2,3,4)\,*}_2\bigl(e^a\wedge\pi_0^*\si^\a\bigr)-\tfrac{2}{15}\,\xcZ_{\a\b}\wedge\pi^{(2,3,4)\,*}_{02}(\si^\a\wedge\si^\b)
\qqq
and the LI curving given by the formula
\qq\nn
\widehat{\underset{\tx{\ciut{(2)}}}{\b}}=\tfrac{1}{30}\,\bigl(4\widehat\pi_2^{(6,7)\,*}\bigl(\widehat\pi_2^{(5)\,*}\xcZ_{\a\b}\wedge\widehat e^{(5)\,\a\b}\bigr)+\widehat\pi_2^{(7)\,*}\bigl(\widehat\pi_2^{(5,6)\,*}\xcY_{a\a}\wedge\widehat \si^{(6)\,a\a}\bigr)-\widehat\pi_2^{(5,6,7)\,*}\xcX_{ab}\wedge\widehat e^{(7)\,ab}\bigr)\,.
\qqq
In the next step, we compare pullbacks of that curving along the canonical projections to the $\sfY_2^{[2]}\cM^{(1)}$-fibred square 
\qq\nn
\widehat\sfY^{[2]}\sfY_2^{[2]}\cM^{(1)}\equiv\widehat{\sfY^{[2]}_2\cM^{(1)}}{}^{(7)}\x_{\sfY_2^{[2]}\cM^{(1)}}\widehat{\sfY^{[2]}_2\cM^{(1)}}{}^{(7)}\,,
\qqq
whereby we find -- for $\,\widehat m_7^A:=(m_4^1,m_4^2,X^A,Y^A,Z^A),\ A\in\{1,2\}\,$ and $\,X^{21}:=X^2-X^1\,,\ Y^{21}:=Y^2-Y^1\,$ and $\,Z^{21}:=Z^2-Z^1\,$ --
\qq\nn
&&(\pr_2^*-\pr_1^*)\widehat{\underset{\tx{\ciut{(2)}}}{\b}}\bigl(\widehat m_7^1,\widehat m_7^2\bigr)\cr\cr
&=&\tfrac{2}{15}\,\xcZ_{\a\b}\bigl(m_4^1,m_4^2\bigr)\wedge\sfd X^{21\,\a\b}+\tfrac{1}{30}\,\xcY_{a\a}\bigl(m_4^1,m_4^2\bigr)\wedge\bigl(\sfd Y^{21\,a\a}+\ovl\G{}^a_{\b\g}\,\bigl(\theta^\a\,\sfd X^{21\,\b\g}+8\theta^\g\,\sfd X^{21\,\a\b}\bigr)\bigr)\cr\cr
&&-\tfrac{1}{30}\,\xcX_{ab}\bigl(m_4^1,m_4^2\bigr)\wedge\bigl(\sfd Z^{21\,ab}+2\ovl\G{}^a_{\a\b}\,\bigl(x^b\,\sfd X^{21\,\a\b}+2\ovl\G{}^b_{\g\d}\,\theta^\b\,\theta^\g\,\sfd X^{21\,\a\d}\bigr)\cr\cr
&&+\ovl\G{}^a_{\a\b}\,\bigl(\sfd Y^{21\,b\a}+\ovl\G{}^b_{\g\d}\,\bigl(\theta^\a\,\sfd X^{21\,\g\d}+8\theta^\d\,\sfd X^{21\,\a\g}\bigr)\bigr)\,\theta^\b\bigr)\cr\cr
&\equiv&\tfrac{1}{30}\,\bigl[\bigl(4\xcZ_{\a\b}\bigl(m_4^1,m_4^2\bigr)+\bigl(\xcY_{a\g}\,\ovl\G{}^a_{\a\b}+8\xcY_{a\a}\,\ovl\G{}^a_{\b\g}\bigr)\bigl(m_4^1,m_4^2\bigr)\,\theta^\g+2\xcX_{ab}\bigl(m_4^1,m_4^2\bigr)\,\bigl(2\ovl\G{}^a_{\a\g}\,\ovl\G{}^b_{\b\d}\,\theta^\g\,\theta^\d\cr\cr
&&-\ovl\G{}^a_{\a\b}\,x^b\bigr)\bigr)\wedge\sfd X^{21\,\a\b}+\bigl(\xcY_{a\a}\bigl(m_4^1,m_4^2\bigr)-\xcX_{ab}\bigl(m_4^1,m_4^2\bigr)\,\ovl\G{}^b_{\a\b}\,\theta^\b\bigr)\wedge\sfd Y^{21\,a\a}-\xcX_{ab}\bigl(m_4^1,m_4^2\bigr)\wedge\sfd Z^{21\,ab}\bigr]\cr\cr
&=&\tfrac{1}{30}\,\bigl(4\sfd\upsilon^{21}_{\a\b}\wedge\sfd X^{21\,\a\b}+\sfd\psi^{21}_{a\a}\wedge\sfd Y^{21\,a\a}-\sfd\z^{21}_{ab}\wedge\sfd Z^{21\,ab}\bigr)\,.
\qqq
Thus, just as in the case of the GS super-1-gerbe, we obtain a trivial principal $\bC^\x$-bundle
\qq\label{eq:smembndl}\hspace{2cm}
\pi_{\widehat\xcL}\equiv\pr_1\ :\ \widehat\xcL:=\widehat\sfY^{[2]}\sfY_2^{[2]}\cM^{(1)}\x\bC^\x\too\widehat\sfY^{[2]}\sfY_2^{[2]}\cM^{(1)}\ :\ \bigl(\widehat m_7^1,\widehat m_7^2,\widehat z\bigr)\longmapsto\bigl(\widehat m_7^1,\widehat m_7^2\bigr)
\qqq
with a principal connection 
\qq\nn
\nabla_{\widehat\xcL}=\sfd+\tfrac{1}{\sfi}\,\widehat\txA\,,
\qqq
or -- equivalently -- a principal connection 1-form 
\qq\nn
\widehat\cA\bigl(\widehat m_7^1,\widehat m_7^2,\widehat z\bigr)=\tfrac{\sfi\,\sfd\widehat z}{\widehat z}+\widehat\txA\bigl(\widehat m_7^1,\widehat m_7^2\bigr)
\qqq
with the base component
\qq\nn
\widehat\txA\bigl(\widehat m_7^1,\widehat m_7^2\bigr)&=&\tfrac{1}{30}\,\bigl(Z^{21\,ab}\,\sfd\z^{21}_{ab}+Y^{21\,a\a}\,\sfd \psi^{21}_{a\a}-4X^{21\,\a\b}\,\sfd\upsilon^{21}_{\a\b}\bigr)\,.
\qqq 
With view to facilitation of subsequent calculations, we rewrite the above as
\qq\nn
\widehat\txA\bigl(\widehat m_7^1,\widehat m_7^2\bigr)=\widehat{\unl\txA}\bigl(\z^{21},\psi^{21},\upsilon^{21},X^2,Y^2,Z^2\bigr)-\widehat{\unl\txA}\bigl(\z^{21},\psi^{21},\upsilon^{21},X^1,Y^1,Z^1\bigr)
\qqq
in terms of the super-1-form
\qq\label{eq:hatunlA}
\widehat{\unl\txA}\bigl(\z^{21},\psi^{21},\upsilon^{21},X,Y,Z\bigr):=\tfrac{1}{30}\,\bigl(Z^{ab}\,\sfd\z^{21}_{ab}+Y^{a\a}\,\sfd \psi^{21}_{a\a}-4X^{\a\b}\,\sfd\upsilon^{21}_{\a\b}\bigr)\,.
\qqq
Following the by now well-established procedure, we determine the lift of the Lie-supergroup structure from the base of the bundle to its total space by imposing the requirement that the principal connection 1-form be LI with respect to the rigid lifted supersymmetry induced from the ensuing group law. In order to study its consequences, we first work out in detail how the various coordinate differences entering the definition of $\,\widehat\txA\,$ change under a rigid supersymmetry transformation with parameters $\,\widehat\d_7^A\equiv((\vep^\a,y^a,\xi^1_{bc},\phi^1_{d\b},\varpi^1_{\g\d}),(\vep^\a,y^a,\xi^2_{ef},\phi^2_{g\ep},\varpi^2_{\eta\k}),U^{A\,\la\mu},V^{A\,h\nu},W^{A\,ij})\in\widehat{\sfY^{[2]}_2\cM^{(1)}}{}^{(7)},\ A\in\{1,2\}\,$ induced, in the same manner, from $\,\widehat\txm{}^{(7)}_2$,\ in which we have taken into account the various fibrings involved in the construction (a point in the $A$-th factor of $\,\widehat\sfY^{[2]}\sfY_2^{[2]}\cM^{(1)}\,$ is transformed by the corresponding $\,\widehat\d_7^A$). The relevant supersymmetry transformations are obtained by restricting the standard product-(super)group structure on $\,\widehat{\sfY^{[2]}_2\cM^{(1)}}{}^{(7)}\x\widehat{\sfY^{[2]}_2\cM^{(1)}}{}^{(7)}\,$ to the subspace $\,\widehat\sfY^{[2]}\sfY_2^{[2]}\cM^{(1)}$.\ We readily find the following transformation laws
\qq\nn
\z^{21}_{ab}&\longmapsto&\z^{21}_{ab}+\xi^{21}_{ab}\,,\cr\cr
\psi^{21}_{a\a}&\longmapsto&\psi^{21}_{a\a}+\phi^{21}_{a\a}-\vep^\b\,\ovl\G{}^b_{\a\b}\,\z^{21}_{ab}\,,\cr\cr
\upsilon^{21}_{\a\b}&\longmapsto&\upsilon^{21}_{\a\b}+\varpi^{21}_{\a\b}-\vep^\g\,\bigl(\tfrac{1}{4}\,\ovl\G{}^a_{\a\b}\,\psi^{21}_{a\g}+\ovl\G{}^a_{\a\g}\,\psi^{21}_{a\b}+\ovl\G{}^a_{\b\g}\,\psi^{21}_{a\a}\bigr)+\bigl(\tfrac{1}{2}\,y^a\,\ovl\G{}^b_{\a\b}+\vep^\g\,\vep^\d\,\ovl\G{}^a_{\a\g}\,\ovl\G{}^b_{\b\d}\bigr)\,\z^{21}_{ab}\,,\cr\cr
X^{21\,\a\b}&\longmapsto&X^{21\,\a\b}+U^{21\,\a\b}\,,\cr\cr
Y^{21\,a\a}&\longmapsto&Y^{21\,a\a}+V^{21\,a\a}-\ovl\G{}^a_{\b\g}\,\bigl(\vep^\a\,X^{21\,\b\g}+8\,\vep^\b\,X^{21\,\a\g}\bigr)\,,\cr\cr
Z^{21\,ab}&\longmapsto&Z^{21\,ab}+W^{21\,ab}+\tfrac{1}{2}\,\vep^\a\,\bigl(\ovl\G{}^a_{\a\b}\,Y^{21\,b\b}-\ovl\G{}^b_{\a\b}\,Y^{21\,a\b}\bigr)-\bigl(4\ovl\G{}^a_{\a\b}\,\ovl\G{}^b_{\g\d}\,\vep^\a\,\vep^\g-y^a\,\ovl\G{}^b_{\b\d}+y^b\,\ovl\G{}^a_{\b\d}\bigr)\,X^{21\,\b\d}\,.
\qqq
and the base component of the principal connection transforms as
\qq\nn
\widehat\txA\bigl(\widehat\txm{}^{(7)}_2\bigl(\widehat\d_7^1,\widehat m_7^1\bigr),\widehat\txm{}^{(7)}_2\bigl(\widehat\d_7^2,\widehat m_7^2\bigr)\bigr)=\widehat\txA\bigl(\widehat m_7^1,\widehat m_7^2\bigr)+\sfd\widehat\la\bigl(\widehat\d_7^1,\widehat\d_7^2,\widehat m_7^1,\widehat m_7^2\bigr)\,,
\qqq
with
\qq\nn
\widehat\la\bigl(\widehat\d_7^1,\widehat\d_7^2,\widehat m_7^1,\widehat m_7^2\bigr)&=&\tfrac{1}{30}\,\bigl[W^{21\,ab}\,\z^{21}_{ab}+V^{21\,a\a}\,\bigl(\psi^{21}_{a\a}-\vep^\b\,\ovl\G{}^b_{\a\b}\,\z^{21}_{ab}\bigr)-U^{21\,\a\b}\,\bigl(4\upsilon^{21}_{\a\b}-\vep^\g\,\bigl(\ovl\G{}^a_{\a\b}\,\psi^{21}_{a\g}+8\ovl\G{}^a_{\a\g}\,\psi^{21}_{a\b}\bigr)\cr\cr
&&+2\bigl(y^a\,\ovl\G{}^b_{\a\b}+2\vep^\g\,\vep^\d\,\ovl\G{}^a_{\a\g}\,\ovl\G{}^b_{\b\d}\bigr)\,\z^{21}_{ab}\bigr)\bigr]\,,
\qqq
which we may -- once again -- rewrite conveniently as
\qq\nn
\widehat\la\bigl(\widehat\d_7^1,\widehat\d_7^2,\widehat m_7^1,\widehat m_7^2\bigr)=\widehat{\unl\la}\bigl(\vep,y,U^2,V^2,W^2;\z^{21},\psi^{21},\upsilon^{21}\bigr)-\widehat{\unl\la}\bigl(\vep,y,U^1,V^1,W^1;\z^{21},\psi^{21},\upsilon^{21}\bigr)
\qqq
in terms of the functions
\qq
\widehat{\unl\la}\bigl(\vep,y,U,V,W;\z^{21},\psi^{21},\upsilon^{21}\bigr)&:=&\tfrac{1}{30}\,\bigl[W^{ab}\,\z^{21}_{ab}+V^{a\a}\,\bigl(\psi^{21}_{a\a}-\vep^\b\,\ovl\G{}^b_{\a\b}\,\z^{21}_{ab}\bigr)\cr\cr
&&-U^{\a\b}\,\bigl(4\upsilon^{21}_{\a\b}-\vep^\g\,\bigl(\ovl\G{}^a_{\a\b}\,\psi^{21}_{a\g}+8\ovl\G{}^a_{\a\g}\,\psi^{21}_{a\b}\bigr)+2\bigl(y^a\,\ovl\G{}^b_{\a\b}+2\vep^\g\,\vep^\d\,\ovl\G{}^a_{\a\g}\,\ovl\G{}^b_{\b\d}\bigr)\,\z^{21}_{ab}\bigr)\bigr]\,. \label{eq:hatlamunl}
\qqq
Accordingly, we may take the lift of the supersymmetry to $\,\widehat\xcL\,$ to be as stated in
\berop\label{prop:smembndlie}
The principal $\bC^\x$-bundle $\,\widehat\xcL\,$ of \Reqref{eq:smembndl} equipped with the binary operation
\qq
\widehat\txm^{(8)}_2\ &:&\ \widehat\xcL\x\widehat\xcL\too\widehat\xcL\cr\cr 
&:&\ \bigl(\bigl(\widehat m_7^1,\widehat m_7^2,\widehat z_1\bigr),\bigl(\widehat n_7^1,\widehat n_7^2,\widehat z_2\bigr)\bigr)\longmapsto\bigl(\widehat\txm{}^{(7)}_2\bigl(\widehat m_7^1,\widehat n_7^1\bigr),\widehat\txm{}^{(7)}_2\bigl(\widehat m_7^2,\widehat n_7^2\bigr),\ee^{\sfi\,\widehat\la(\widehat m_7^1,\widehat m_7^2,\widehat n_7^1,\widehat n_7^2)}\cdot\widehat z_1\cdot\widehat z_2\bigr)\label{eq:smembLonLfix}
\qqq
with the inverse
\qq\nn
\widehat\Inv^{(8)}_2\ :\ \widehat\xcL\too\widehat\xcL\ :\ \bigl(\widehat m_7^1,\widehat m_7^2,\widehat z\bigr)\longmapsto\bigl(\widehat\Inv{}^{(7)}_2\bigl(\widehat m_7^1\bigr),\widehat\Inv{}^{(7)}_2\bigl(\widehat m_7^2\bigr),\ee^{\sfi\,(Z^{21\,ab}\,\z^{21}_{ab}+Y^{21\,a\a}\,\psi^{21}_{a\a}-4X^{21\,\a\b}\,\upsilon^{21}_{\a\b})}\cdot\widehat z^{-1}\bigr)
\qqq
and the neutral element
\qq\nn
\widehat e^{(8)}_2=(0,0,1)
\qqq
is a Lie supergroup. It is a supercentral extension
\qq\nn
\bd1\too\bC^\x\too\widehat\sfY^{[2]}\sfY_2^{[2]}\cM^{(1)}\x\bC^\x\xrightarrow{\ \pi_{\widehat\xcL}\ }\widehat\sfY^{[2]}\sfY_2^{[2]}\cM^{(1)}\too\bd1
\qqq
of the Lie supergroup $\,\sfY^{[2]}\sfY_2^{[2]}\cM^{(1)}\,$ determined by the CE super-2-cocycles corresponding to the CaE super-2-cocycle $\,(\pr_2^*-\pr_1^*)\widehat{\underset{\tx{\ciut{(2)}}}{\b}}$.
\eerop
\beroof
Straightforward, through inspection.
\eroof
At this stage, we may pass to the fibred cube 
\qq\nn
\widehat\sfY^{[3]}\sfY_2^{[2]}\cM^{(1)}\equiv\widehat{\sfY^{[2]}_2\cM^{(1)}}{}^{(7)}\x_{\sfY_2^{[2]}\cM^{(1)}}\widehat{\sfY^{[2]}_2\cM^{(1)}}{}^{(7)}\x_{\sfY_2^{[2]}\cM^{(1)}}\widehat{\sfY^{[2]}_2\cM^{(1)}}{}^{(7)}\,,
\qqq
equipped with the natural Lie-supergroup structure (with a binary operation $\,\widehat\txm{}^{(7)\,[3]}_2$) induced (through restriction) from the product structure on $\,(\widehat\sfY\sfY_2^{[2]}\cM^{(1)})^{\x 3}$,\ and look for a suitable connection-preserving isomorphism
\qq\nn
\mu_{\widehat\xcL}\ :\ \pr_{1,2}^*\widehat\xcL\ox\pr_{2,3}^*\widehat\xcL\xrightarrow{\ \cong\ }\pr_{1,3}^*\widehat\xcL\,.
\qqq
Comparison of the pullbacks of the connection 1-forms 
\qq\nn
(\pr_{1,2}^*+\pr_{2,3}^*-\pr_{1,3}^*)\widehat\txA\bigl(\widehat m_7^1,\widehat m_7^2,\widehat m_7^3\bigr)=0\,,
\qqq
in conjunction with Prop.\,\ref{prop:smembndlie} immediately suggest the natural choice
\qq
&&\mu_{\widehat\xcL}\left(\bigl(\bigl(\widehat m_7^1,\widehat m_7^2,\widehat m_7^3\bigr),\bigl(\widehat m_7^1,\widehat m_7^2,\widehat z_{1,2}\bigr)\bigr)\ox\bigl(\bigl(\widehat m_7^1,\widehat m_7^2,\widehat m_7^3\bigr),\bigl(\widehat m_7^2,\widehat m_7^3,\widehat z_{2,3}\bigr)\bigr)\right)\cr\cr
&:=&\bigl(\bigl(\widehat m_7^1,\widehat m_7^2,\widehat m_7^3\bigr),\bigl(\widehat m_7^1,\widehat m_7^3,\widehat z_{1,2}\cdot\widehat z_{2,3}\bigr)\bigr)\,. \label{eq:grpdstrwidehatL}
\qqq
A fibre map thus defined trivially satisfies the groupoid identity \eqref{eq:mugrpd} over $\,\widehat\sfY^{[3]}\sfY_2^{[2]}\cM^{(1)}$.\ Furthermore, it manifestly intertwines the Lie-supergroup structure on\qq\nn
\pr_{1,2}^*\widehat\xcL\ox\pr_{2,3}^*\widehat\xcL\equiv\bigl(\widehat\sfY^{[3]}\sfY_2^{[2]}\cM^{(1)}\x_{\widehat\sfY^{[2]}\sfY_2^{[2]}\cM^{(1)}}^{(1,2)}\widehat\xcL\bigr)\x_{\id_{\widehat\sfY^{[3]}\sfY_2^{[2]}\cM^{(1)}}\x\widehat\txm^{(8)}_2}\bigl(\widehat\sfY^{[3]}\sfY_2^{[2]}\cM^{(1)}\x_{\widehat\sfY^{[2]}\sfY_2^{[2]}\cM^{(1)}}^{(2,3)}\widehat\xcL\bigr)\,,
\qqq
where
\qq\nn
\alxydim{@C=1.5cm@R=1.5cm}{\widehat\sfY^{[3]}\sfY_2^{[2]}\cM^{(1)}\x^{(i,j)}_{\widehat\sfY^{[2]}\sfY_2^{[2]}\cM^{(1)}}\widehat\xcL \ar[r]^{\hspace{2cm}\pr_2} \ar[d]_{\pr_1} & \widehat\xcL \ar[d]^{\pi_{\widehat\xcL}} \\ \widehat\sfY^{[3]}\sfY_2^{[2]}\cM^{(1)} \ar[r]_{\pr_{i,j}} & \widehat\sfY^{[2]}\sfY_2^{[2]}\cM^{(1)}}\,,
\qqq
determined by the binary operation
\qq\nn
\bigl[\widehat\txm{}^{(7)\,[3]}_2\circ\pr_{1,5},\widehat\txm^{(8)}_2\circ\pr_{2,6},\widehat\txm{}^{(7)\,[3]}_2\circ\pr_{3,7},\widehat\txm^{(8)}_2\circ\pr_{4,8}\bigr]\ :\ \bigl(\pr_{1,2}^*\widehat\xcL\ox\pr_{2,3}^*\widehat\xcL\bigr)^{\x 2}\too\pr_{1,2}^*\widehat\xcL\ox\pr_{2,3}^*\widehat\xcL\,,
\qqq
with that on 
\qq\nn
\pr_{1,3}^*\widehat\xcL\equiv\widehat\sfY^{[3]}\sfY_2^{[2]}\cM^{(1)}\x_{\widehat\sfY^{[2]}\sfY_2^{[2]}\cM^{(1)}}^{(1,3)}\widehat\xcL\,,
\qqq
determined by
\qq\nn
\bigl(\widehat\txm{}^{(7)\,[3]}_2\circ\pr_{1,3},\widehat\txm^{(8)}_2\circ\pr_{2,4}\bigr)\ :\ \bigl(\pr_{1,3}^*\widehat\xcL\bigr)^{\x 2}\too\pr_{1,3}^*\widehat\xcL\,.
\qqq
This follows from the identity
\qq\nn
&&\widehat\la(\widehat\d_7^1,\widehat\d_7^2,\widehat m_7^1,\widehat m_7^2)+\widehat\la(\widehat n_7^2,\widehat n_7^3,\widehat m_7^2,\widehat m_7^3)\cr\cr
&=&\widehat{\unl\la}\bigl(\vep,y,U^2,V^2,W^2;\z^{21},\psi^{21},\upsilon^{21}\bigr)-\widehat{\unl\la}\bigl(\vep,y,U^1,V^1,W^1;\z^{21},\psi^{21},\upsilon^{21}\bigr)\cr\cr
&&+\widehat{\unl\la}\bigl(\vep,y,U^3,V^3,W^3;\z^{21},\psi^{21},\upsilon^{21}\bigr)-\widehat{\unl\la}\bigl(\vep,y,U^2,V^2,W^2;\z^{21},\psi^{21},\upsilon^{21}\bigr)\cr\cr
&=&\widehat{\unl\la}\bigl(\vep,y,U^3,V^3,W^3;\z^{21},\psi^{21},\upsilon^{21}\bigr)-\widehat{\unl\la}\bigl(\vep,y,U^1,V^1,W^1;\z^{21},\psi^{21},\upsilon^{21}\bigr)\cr\cr
&\equiv&\widehat\la(\widehat\d_7^1,\widehat\d_7^3,\widehat m_7^1,\widehat m_7^3)\,.
\qqq

Altogether, then, we establish the existence of a CaE super-1-gerbe
\qq\nn
\widehat\xcG=\bigl(\widehat\sfY\sfY_2^{[2]}\cM^{(1)},\pi_{\widehat\sfY\sfY_2^{[2]}\cM^{(1)}},\widehat{\underset{\tx{\ciut{(2)}}}{\b}},\widehat\xcL,\pi_{\widehat\xcL},\nabla_{\widehat\xcL},\mu_{\widehat\xcL}\bigr)
\qqq
over the fibred square $\,\sfY_2^{[2]}\cM^{(1)}\,$ of the supercentral extension $\,\sfY_2\cM^{(1)}\,$ of the support of the GS super-4-cocycle, in the sense of Def.\,\ref{def:s1gerbe}. We shall next construct a coherent product on the super-1-gerbe.

To this end, we consider the pullback surjective submersions
\qq\label{diag:pbsspro}
\alxydim{@C=2.cm@R=1.5cm}{\widehat\sfY^{i,j}\sfY_2^{[3]}\cM^{(1)}\equiv\sfY_2^{[3]}\cM^{(1)}\x_{\sfY_2^{[2]}\cM^{(1)}}\widehat\sfY\sfY_2^{[2]}\cM^{(1)} \ar[r]^{\hspace{2.25cm}\pr_2} \ar[d]_{\pi_{\widehat\sfY^{i,j}\sfY_2^{[3]}\cM^{(1)}}\equiv\pr_1} & \widehat\sfY\sfY_2^{[2]}\cM^{(1)} \ar[d]^{\pi_{\widehat\sfY\sfY_2^{[2]}\cM^{(1)}}} \\ \sfY_2^{[3]}\cM^{(1)} \ar[r]_{\pr_{i,j}} &  \sfY_2^{[2]}\cM^{(1)} }
\qqq
for $\,(i,j)\in\{(1,2),(2,3),(1,3)\}$,\ with (global) coordinates
\qq\nn
\widehat m^{(i,j)}=\bigl(\bigl(m_4^1,m_4^2,m_4^3\bigr),\bigl(m_4^i,m_4^j,X^{(i,j)},Y^{(i,j)},Z^{(i,j)}\bigr)\bigr)\in\widehat\sfY^{i,j}\sfY_2^{[3]}\cM^{(1)}\,,
\qqq
and equip them with the obvious product Lie-supergroup structure. Over these, we compare the (Deligne) tensor product of the pullback super-1-gerbes (we are dropping some obvious subscripts for the sake of transparency)
\qq\nn
\pr_{1,2}^*\widehat\xcG\ox\pr_{2,3}^*\widehat\xcG&=&\bigl(\widehat\sfY^{1,2;2,3}\sfY_2^{[3]}\cM^{(1)},\pi_{\widehat\sfY^{1,2}\sfY_2^{[3]}\cM^{(1)}}\circ\pr_1,\bigl(\pr_1^*\pr_2^*+\pr_2^*\pr_2^*\bigr)\widehat{\underset{\tx{\ciut{(2)}}}{\b}},\cr\cr
&&\pr_{1,3}^*\pr_{2,4}{}^*\widehat\xcL\ox\pr_{2,4}^*\pr_{2,4}^*\widehat\xcL,[\pi_{\pr_{1,3}^*\pr_{2,4}{}^*\widehat\xcL}\circ\pr_1],\cr\cr
&&\pr_{1,3}^*\pr_{2,4}^*\nabla_{\widehat\xcL}\ox\id+\id\ox\pr_{2,4}^*\pr_{2,4}^*\nabla_{\widehat\xcL},\pr_{1,3,5}^*\pr_{2,4,6}^*\mu_{\widehat\xcL}\ox\pr_{2,4,6}^*\pr_{2,4,6}^*\mu_{\widehat\xcL}\bigr)\,,
\qqq
written in terms of the obvious canonical projections (which will be made explicit below, wherever necessary) and the fibred product
\qq\nn
\alxydim{@C=2.cm@R=1.5cm}{\widehat\sfY^{1,2;2,3}\sfY_2^{[3]}\cM^{(1)}\equiv\widehat\sfY^{1,2}\sfY_2^{[3]}\cM^{(1)}\x^{\widehat 1\x\widehat 1}_{\sfY_2^{[3]}\cM^{(1)}}\widehat\sfY^{2,3}\sfY_2^{[3]}\cM^{(1)} \ar[r]^{\hspace{3.25cm}\pr_2} \ar[d]_{\pr_1} & \widehat\sfY^{2,3}\sfY_2^{[3]}\cM^{(1)} \ar[d]^{\pi_{\widehat\sfY^{2,3}\sfY_2^{[3]}\cM^{(1)}}} \\ \widehat\sfY^{1,2}\sfY_2^{[3]}\cM^{(1)} \ar[r]_{\pi_{\widehat\sfY^{1,2}\sfY_2^{[3]}\cM^{(1)}}} &  \sfY_2^{[3]}\cM^{(1)} }\,.
\qqq
with the pullback super-1-gerbe
\qq\nn
\pr_{1,3}^*\widehat\xcG=\bigl(\widehat\sfY^{1,3}\sfY_2^{[3]}\cM^{(1)},\pi_{\widehat\sfY^{1,3}\sfY_2^{[3]}\cM^{(1)}},\pr_2^*\widehat{\underset{\tx{\ciut{(2)}}}{\b}},\pr_{2,4}{}^*\widehat\xcL,\pr_{2,4}{}^*\nabla_{\widehat\xcL},\pr_{2,4,6}{}^*\mu_{\widehat\xcL}\bigr)\,.
\qqq
We perform the comparison over the fibred product
\qq\nn
\widehat\sfY^{1,2,3}\sfY_2^{[3]}\cM^{(1)}:=\widehat\sfY^{1,2;2,3}\sfY_2^{[3]}\cM^{(1)}\x^{\widehat 1\x\widehat 1}_{\sfY_2^{[3]}\cM^{(1)}}\widehat\sfY^{1,3}\sfY_2^{[3]}\cM^{(1)}
\qqq
surjectively submersed onto $\,\sfY_2^{[3]}\cM^{(1)}\,$ as {\it per} 
\qq\nn
\pi_{\widehat\sfY^{1,2,3}\sfY_2^{[3]}\cM^{(1)}}\equiv\pi_{\widehat\sfY^{1,2}\sfY_2^{[3]}\cM^{(1)}}\circ\pr_1\circ\pr_1\,.
\qqq
There, we find
\qq\nn
&&\bigl(\pr_3^*\pr_2^*-\pr_{1,2}^*\bigl(\pr_1^*\pr_2^*+\pr_2^*\pr_2^*\bigr)\bigr)\widehat{\underset{\tx{\ciut{(2)}}}{\b}}\bigl(\widehat m^{(1,2)},\widehat m^{(2,3)},\widehat m^{(1,3)}\bigr)\cr\cr
&=&\widehat{\underset{\tx{\ciut{(2)}}}{\b}}\bigl(m_4^1,m_4^3,X^{(1,3)},Y^{(1,3)},Z^{(1,3)}\bigr)-\widehat{\underset{\tx{\ciut{(2)}}}{\b}}\bigl(m_4^1,m_4^2,X^{(1,2)},Y^{(1,2)},Z^{(1,2)}\bigr)\cr\cr
&&-\widehat{\underset{\tx{\ciut{(2)}}}{\b}}\bigl(m_4^2,m_4^3,X^{(2,3)},Y^{(2,3)},Z^{(2,3)}\bigr)\cr\cr
&=&\tfrac{1}{30}\,\sum_{(i,j)\in\{(1,3),(1,2),(2,3)\}}\,(-1)^{i+j}\,\sfd\bigl(Z^{(i,j)\,ab}\,\sfd\z^{ji}_{ab}+Y^{(i,j)\,a\a}\,\sfd\psi^{ji}_{a\a}-4X^{(i,j)\,\a\b}\,\sfd\upsilon^{ji}_{\a\b}\bigr)\,,
\qqq
the derivation invoking the identities
\qq\nn
X\bigl(m_4^1,m_4^3\bigr)-X\bigl(m_4^1,m_4^2\bigr)-X\bigl(m_4^2,m_4^3\bigr)=0\,,\qquad X\in\{\xcX_{ab},\xcY_{c\a},\xcZ_{\b\g}\}\,.
\qqq
From the last result, we infer the existence of a trivial principal $\bC^\x$-bundle 
\qq
\pi_{\widehat\xcE}\equiv\pr_1\ &:&\ \widehat\xcE:=\widehat\sfY^{1,2,3}\sfY_2^{[3]}\cM^{(1)}\x\bC^\x\too\widehat\sfY^{1,2,3}\sfY_2^{[3]}\cM^{(1)}\cr\cr 
&:&\ \bigl(\widehat m^{(1,2)},\widehat m^{(2,3)},\widehat m^{(1,3)},\widehat\z^{1,2,3}\bigr)\longmapsto\bigl(\widehat m^{(1,2)},\widehat m^{(2,3)},\widehat m^{(1,3)}\bigr) \label{eq:smembndl2}
\qqq
with a principal $\bC^\x$-connection super-1-form
\qq\nn
\widehat\a\bigl(\widehat m^{(1,2)},\widehat m^{(2,3)},\widehat m^{(1,3)},\widehat\z^{1,2,3}\bigr)=\tfrac{\sfi\,\sfd\widehat\z^{1,2,3}}{\widehat\z^{1,2,3}}+\widehat\txa\bigl(\widehat m^{(1,2)},\widehat m^{(2,3)},\widehat m^{(1,3)}\bigr)
\qqq
with the base component
\qq\nn
\widehat\txa\bigl(\widehat m^{(1,2)},\widehat m^{(2,3)},\widehat m^{(1,3)}\bigr)&=&\widehat{\unl\txA}\bigl(\z^{31},\psi^{31},\upsilon^{31},X^{(1,3)},Y^{(1,3)},Z^{(1,3)}\bigr)-\widehat{\unl\txA}\bigl(\z^{21},\psi^{21},\upsilon^{21},X^{(1,2)},Y^{(1,2)},Z^{(1,2)}\bigr)\cr\cr
&&-\widehat{\unl\txA}\bigl(\z^{32},\psi^{32},\upsilon^{32},X^{(2,3)},Y^{(2,3)},Z^{(2,3)}\bigr)\,,
\qqq 
written in terms of the super-1-forms \eqref{eq:hatunlA}. The bundle is endowed with a Lie-supergroup structure determined by the requirement of (left-)invariance of its principal $\bC^\x$-connection 1-form, as stated in 
\berop\label{prop:smembndlie2}
The principal $\bC^\x$-bundle $\,\widehat\xcE\,$ of \Reqref{eq:smembndl2} equipped with the binary operation
\qq
\widehat\txm^{(9)}_{1,2,3}\ &:&\ \widehat\xcE\x\widehat\xcE\too\widehat\xcE\cr\cr 
&:&\ \bigl(\bigl(\widehat m^{(1,2)},\widehat m^{(2,3)},\widehat m^{(1,3)},\widehat\z^{1,2,3}_1\bigr),\bigl(\widehat n^{(1,2)},\widehat n^{(2,3)},\widehat n^{(1,3)},\widehat\z^{1,2,3}_2\bigr)\bigr)\cr\cr
&&\longmapsto\bigl(\bigl(\txm{}^{(4)\,[3]}_2\circ\pr_{1,3},\widehat\txm{}^{(7)}_2\circ\pr_{2,4}\bigr)^{\x 3}\bigl(\bigl(\widehat m^{(1,2)},\widehat n^{(1,2)}\bigr),\bigl(\widehat m^{(2,3)},\widehat n^{(2,3)}\bigr),\bigl(\widehat m^{(1,3)},\widehat n^{(1,3)}\bigr)\bigr),\cr\cr
&&\hspace{1cm}\ee^{\sfi\,\widehat\la^{1,2,3}(\widehat m^{(1,2)},\widehat m^{(2,3)},\widehat m^{(1,3)},\widehat n^{(1,2)},\widehat n^{(2,3)},\widehat n^{(1,3)})}\cdot\widehat\z^{1,2,3}_1\cdot\widehat\z^{1,2,3}_2\bigr)\,,\label{eq:smembLonLfix}
\qqq
where for 
\qq\nn
\widehat m^{(i,j)}&=&((m_4^1,m_4^2,m_4^3),(m_4^i,m_4^j,X^{(i,j)\,\a\b}_1,Y^{(i,j)\,a\g}_1,Z^{(i,j)\,bc}_1))\cr\cr
\widehat n^{(i,j)}&=&((n_4^1,n_4^2,n_4^3),(n_4^i,n_4^j,X^{(i,j)\,\a\b}_2,Y^{(i,j)\,a\g}_2,Z^{(i,j)\,bc}_2))\,,
\qqq 
with 
\qq\nn
m_4^i=(\theta^\a_1,x^a_1,\z^i_{1\,bc},\psi^i_{1\,d\b},\upsilon^i_{1\,\g\d})\,,\qquad\qquad n_4^i=(\theta^\a_2,x^a_2,\z^i_{2\,bc},\psi^i_{2\,d\b},\upsilon^i_{2\,\g\d}) 
\qqq
we have
\qq\nn
\widehat\la^{1,2,3}\bigl(\widehat m^{(1,2)},\widehat m^{(2,3)},\widehat m^{(1,3)},\widehat n^{(1,2)},\widehat n^{(2,3)},\widehat n^{(1,3)}\bigr):=\widehat{\unl\la}\bigl(\theta_1,x_1,X^{(1,3)}_1,Y^{(1,3)}_1,Z^{(1,3)}_1;\z^{31}_2,\psi^{31}_2,\upsilon^{31}_2\bigr)\cr\cr
-\widehat{\unl\la}\bigl(\theta_1,x_1,X^{(1,2)}_1,Y^{(1,2)}_1,Z^{(1,2)}_1;\z^{21}_2,\psi^{21}_2,\upsilon^{21}_2\bigr)-\widehat{\unl\la}\bigl(\theta_1,x_1,X^{(2,3)}_1,Y^{(2,3)}_1,Z^{(2,3)}_1;\z^{32}_2,\psi^{32}_2,\upsilon^{32}_2\bigr)
\qqq
is expressed in terms of the functions \eqref{eq:hatlamunl}, with the inverse
\qq\nn
\widehat\Inv^{(9)}_{1,2,3}\ &:&\ \widehat\xcE\too\widehat\xcE\ :\ \bigl(\widehat m_7^1,\widehat m_7^2,\widehat z\bigr)\longmapsto\bigl(\widehat\Inv{}^{(7)}_2\bigl(\widehat m_7^1\bigr),\widehat\Inv{}^{(7)}_2\bigl(\widehat m_7^2\bigr),\cr\cr
&&\hspace{3,7cm}\ee^{\sfi\,\sum_{(i,j)\in\{(1,2),(2,3),(1,3)\}}\,(-1)^{i+j}\,(Z^{(i,j)\,ab}\,\z^{ji}_{ab}+Y^{(i,j)\,a\a}\,\psi^{ji}_{a\a}-4X^{(i,j)\,\a\b}\,\upsilon^{ji}_{\a\b})}\cdot\widehat z^{-1}\bigr)
\qqq
and the neutral element
\qq\nn
\widehat e^{(8)}_2=(0,0,0,1)
\qqq
is a Lie supergroup. It is a supercentral extension
\qq\nn
\bd1\too\bC^\x\too\widehat\sfY^{1,2,3}\sfY_2^{[3]}\cM^{(1)}\x\bC^\x\xrightarrow{\ \pi_{\widehat\xcE}\ }\widehat\sfY^{1,2,3}\sfY_2^{[3]}\cM^{(1)}\too\bd1
\qqq
of the Lie supergroup $\,\widehat\sfY^{1,2,3}\sfY_2^{[3]}\cM^{(1)}\,$ determined by the CE super-2-cocycle corresponding to the CaE super-2-cocycle $\,(\pr_3^*\pr_2^*-\pr_{1,2}^*(\pr_1^*\pr_2^*+\pr_2^*\pr_2^*))\widehat{\underset{\tx{\ciut{(2)}}}{\b}}$.
\eerop
\beroof
Straightforward, through inspection.
\eroof

Next, we take the $\sfY^{[3]}_2\cM^{(1)}$-fibred square $\,\widehat\sfY^{1,2,3\,[2]}\sfY_2^{[3]}\cM^{(1)}$,\ with its canonical projections \qq\nn
&\pr_{1,2,3,4,7,8,9,10}\ :\ \widehat\sfY^{1,2,3\,[2]}\sfY^{[3]}_2\cM^{(1)}\too\widehat\sfY^{1,2;2,3}\sfY^{[3]}_2\cM^{(1)}\x_{\sfY^{[3]}_2\cM^{(1)}}\widehat\sfY^{1,2;2,3}\sfY^{[3]}_2\cM^{(1)}\,,&\cr\cr\cr
&\pr_{5,6,11,12}\ :\ \widehat\sfY^{1,2,3\,[2]}\sfY^{[3]}_2\cM^{(1)}\too\widehat\sfY^{1,3\,[2]}\sfY^{[3]}_2\cM^{(1)}\,,&
\qqq
alongside 
\qq\nn
\pr_{1,2,3,4,5,6}\,,\pr_{7,8,9,10,11,12}\ :\ \widehat\sfY^{1,2,3\,[2]}\sfY^{[3]}_2\cM^{(1)}\too\widehat\sfY^{1,2,3}\sfY^{[3]}_2\cM^{(1)}\,,
\qqq 
and compute (in the notation of Prop.\,\ref{prop:smembndlie2}, with $\,m_4^i=n_4^i$)
\qq\nn
&&\bigl(\pr_{1,2,3,4,7,8,9,10}^*\bigl(\pr_{2,6}^*+\pr_{4,8}^*\bigr)\widehat\txA+\pr_{7,8,9,10,11,12}^*\widehat\txa\bigr)\bigl(\bigl(\widehat m^{(1,2)},\widehat m^{(2,3)},\widehat m^{(1,3)}\bigr),\bigl(\widehat n^{(1,2)},\widehat n^{(2,3)},\widehat n^{(1,3)}\bigr)\bigr)\cr\cr
&=&\widehat{\unl\txA}\bigl(\z^{21},\psi^{21},\upsilon^{21},X^{(1,2)}_2,Y^{(1,2)}_2,Z^{(1,2)}_2\bigr)-\widehat{\unl\txA}\bigl(\z^{21},\psi^{21},\upsilon^{21},X^{(1,2)}_1,Y^{(1,2)}_1,Z^{(1,2)}_1\bigr)\cr\cr
&&+\widehat{\unl\txA}\bigl(\z^{32},\psi^{32},\upsilon^{32},X^{(2,3)}_2,Y^{(2,3)}_2,Z^{(2,3)}_2\bigr)-\widehat{\unl\txA}\bigl(\z^{32},\psi^{32},\upsilon^{32},X^{(2,3)}_1,Y^{(2,3)}_1,Z^{(2,3)}_1\bigr)\cr\cr
&&+\widehat{\unl\txA}\bigl(\z^{31},\psi^{31},\upsilon^{31},X^{(1,3)}_2,Y^{(1,3)}_2,Z^{(1,3)}_2\bigr)-\widehat{\unl\txA}\bigl(\z^{21},\psi^{21},\upsilon^{21},X^{(1,2)}_2,Y^{(1,2)}_2,Z^{(1,2)}_2\bigr)\cr\cr
&&-\widehat{\unl\txA}\bigl(\z^{32},\psi^{32},\upsilon^{32},X^{(2,3)}_2,Y^{(2,3)}_2,Z^{(2,3)}_2\bigr)\cr\cr
&=&\widehat{\unl\txA}\bigl(\z^{31},\psi^{31},\upsilon^{31},X^{(1,3)}_2,Y^{(1,3)}_2,Z^{(1,3)}_2\bigr)-\widehat{\unl\txA}\bigl(\z^{21},\psi^{21},\upsilon^{21},X^{(1,2)}_1,Y^{(1,2)}_1,Z^{(1,2)}_1\bigr)\cr\cr
&&-\widehat{\unl\txA}\bigl(\z^{32},\psi^{32},\upsilon^{32},X^{(2,3)}_1,Y^{(2,3)}_1,Z^{(2,3)}_1\bigr)\cr\cr
&\equiv&\widehat{\unl\txA}\bigl(\z^{31},\psi^{31},\upsilon^{31},X^{(1,3)}_1,Y^{(1,3)}_1,Z^{(1,3)}_1\bigr)-\widehat{\unl\txA}\bigl(\z^{21},\psi^{21},\upsilon^{21},X^{(1,2)}_1,Y^{(1,2)}_1,Z^{(1,2)}_1\bigr)\cr\cr
&&-\widehat{\unl\txA}\bigl(\z^{32},\psi^{32},\upsilon^{32},X^{(2,3)}_1,Y^{(2,3)}_1,Z^{(2,3)}_1\bigr)\cr\cr
&&+\widehat{\unl\txA}\bigl(\z^{31},\psi^{31},\upsilon^{31},X^{(1,3)}_2,Y^{(1,3)}_2,Z^{(1,3)}_2\bigr)-\widehat{\unl\txA}\bigl(\z^{31},\psi^{31},\upsilon^{31},X^{(1,3)}_1,Y^{(1,3)}_1,Z^{(1,3)}_1\bigr)\cr\cr
&\equiv&\bigl(\pr_{1,2,3,4,5,6}^*\widehat\txa+ \pr_{5,6,11,12}^*\pr_{2,4}^*\widehat\txA\bigr)\bigl(\bigl(\widehat m^{(1,2)},\widehat m^{(2,3)},\widehat m^{(1,3)}\bigr),\bigl(\widehat n^{(1,2)},\widehat n^{(2,3)},\widehat n^{(1,3)}\bigr)\bigr)\,,
\qqq
whereupon it becomes clear that we have a connection-preserving $\bC^\x$-bundle isomorphism
\qq\nn
\widehat\vep\ &:&\ \pr_{2,8}^*\widehat\xcL\ox\pr_{4,10}^*\widehat\xcL\ox\pr_{7,8,9,10,11,12}^*\widehat\xcE \xrightarrow{\ \cong\ }\pr_{1,2,3,4,5,6}^*\widehat\xcE\ox\widehat\pr_{6,12}^*\widehat\xcL\cr\cr
&:&\ \bigl(\widehat m^{1,2,3}_{12},\widehat m^{(1,2)},\widehat n^{(1,2)},\widehat z^{(1,2)}\bigr)\ox\bigl(\widehat m^{1,2,3}_{12},\widehat m^{(2,3)},\widehat n^{(2,3)},\widehat z^{(2,3)}\bigr)\ox\bigl(\widehat m^{1,2,3}_{12},\widehat n^{(1,2)},\widehat n^{(2,3)},\widehat n^{(1,3)},\widehat\z^{1,2,3}_2\bigr)\cr\cr
&&\hspace{0.5cm}\longmapsto\bigl(\widehat m^{1,2,3}_{12},\widehat m^{(1,2)},\widehat m^{(2,3)},\widehat m^{(1,3)},\widehat z^{(1,2)}\cdot\widehat z^{(2,3)}\cdot\widehat\z^{1,2,3}_2\bigr)\ox\bigl(\widehat m^{1,2,3}_{12},\widehat m^{(1,3)},\widehat n^{(1,3)},1\bigr)\,,
\qqq
written in an obvious shorthand notation using 
\qq\nn
\widehat m^{1,2,3}_{12}\equiv\bigl(\bigl(\widehat m^{(1,2)},\widehat m^{(2,3)},\widehat m^{(1,3)}\bigr),\bigl(\widehat n^{(1,2)},\widehat n^{(2,3)},\widehat n^{(1,3)}\bigr)\bigr)\,.
\qqq
The triviality of its form, in conjunction with that of the groupoid structure $\,\mu_{\widehat\xcL}\,$ on $\,\widehat\xcL\,$ established in \Reqref{eq:grpdstrwidehatL}, ensures that it satisfies the usual requirement of compatibility with the respective groupoid structures on $\,\pr_{1,2}^*\widehat\xcG\ox\pr_{2,3}^*\widehat\xcG\,$ and $\,\pr_{1,3}^*\widehat\xcG$.\ It is also in keeping with our definition of the super-0-gerbe isomorphism as $\,\widehat\vep\,$ homomorphically maps the Lie-supergroup structure on its domain into that on its codomain owing to the identity 
\qq\nn
&&\widehat\la\bigl(\widehat m^{(1,2)}_{7\,1},\widehat m^{(1,2)}_{7\,2},\widehat n^{(1,2)}_{7\,1},\widehat n^{(1,2)}_{7\,2}\bigr)+\widehat\la\bigl(\widehat m^{(2,3)}_{7\,1},\widehat m^{(2,3)}_{7\,2},\widehat n^{(2,3)}_{7\,1},\widehat n^{(2,3)}_{7\,2}\bigr)\cr\cr
&&+\widehat\la^{1,2,3}\bigl(\widehat m^{(1,2)}_2,\widehat m^{(2,3)}_2,\widehat m^{(1,3)}_2,\widehat n^{(1,2)}_2,\widehat n^{(2,3)}_2,\widehat n^{(1,3)}_2\bigr)\cr\cr
&=&\widehat{\unl\la}\bigl(\theta,x,X^{(1,2)}_2,Y^{(1,2)}_2,Z^{(1,2)}_2;\widetilde\z^{21},\widetilde\psi^{21},\widetilde\upsilon^{21}\bigr)-\widehat{\unl\la}\bigl(\theta,x,X^{(1,2)}_1,Y^{(1,2)}_1,Z^{(1,2)}_1;\widetilde\z^{21},\widetilde\psi^{21},\widetilde\upsilon^{21}\bigr)\cr\cr
&&+\widehat{\unl\la}\bigl(\theta,x,X^{(2,3)}_2,Y^{(2,3)}_2,Z^{(2,3)}_2;\widetilde\z^{32},\widetilde\psi^{32},\widetilde\upsilon^{32}\bigr)-\widehat{\unl\la}\bigl(\theta,x,X^{(2,3)}_1,Y^{(2,3)}_1,Z^{(2,3)}_1;\widetilde\z^{32},\widetilde\psi^{32},\widetilde\upsilon^{32}\bigr)\cr\cr
&&+\widehat{\unl\la}\bigl(\theta,x,X^{(1,3)}_2,Y^{(1,3)}_2,Z^{(1,3)}_2;\widetilde\z^{31},\widetilde\psi^{31},\widetilde\upsilon^{31}\bigr)-\widehat{\unl\la}\bigl(\theta,x,X^{(1,2)}_2,Y^{(1,2)}_2,Z^{(1,2)}_2;\widetilde\z^{21},\widetilde\psi^{21},\widetilde\upsilon^{21}\bigr)\cr\cr
&&-\widehat{\unl\la}\bigl(\theta,x,X^{(2,3)}_2,Y^{(2,3)}_2,Z^{(2,3)}_2;\widetilde\z^{32},\widetilde\psi^{32},\widetilde\upsilon^{32}\bigr)\cr\cr
&=&\widehat{\unl\la}\bigl(\theta,x,X^{(1,3)}_2,Y^{(1,3)}_2,Z^{(1,3)}_2;\widetilde\z^{31},\widetilde\psi^{31},\widetilde\upsilon^{31}\bigr)-\widehat{\unl\la}\bigl(\theta,x,X^{(1,2)}_1,Y^{(1,2)}_1,Z^{(1,2)}_1;\widetilde\z^{21},\widetilde\psi^{21},\widetilde\upsilon^{21}\bigr)\cr\cr
&&-\widehat{\unl\la}\bigl(\theta,x,X^{(2,3)}_1,Y^{(2,3)}_1,Z^{(2,3)}_1;\widetilde\z^{32},\widetilde\psi^{32},\widetilde\upsilon^{32}\bigr)\cr\cr
&\equiv&\widehat{\unl\la}\bigl(\theta,x,X^{(1,3)}_1,Y^{(1,3)}_1,Z^{(1,3)}_1;\widetilde\z^{31},\widetilde\psi^{31},\widetilde\upsilon^{31}\bigr)-\widehat{\unl\la}\bigl(\theta,x,X^{(1,2)}_1,Y^{(1,2)}_1,Z^{(1,2)}_1;\widetilde\z^{21},\widetilde\psi^{21},\widetilde\upsilon^{21}\bigr)\cr\cr
&&-\widehat{\unl\la}\bigl(\theta,x,X^{(2,3)}_1,Y^{(2,3)}_1,Z^{(2,3)}_1;\widetilde\z^{32},\widetilde\psi^{32},\widetilde\upsilon^{32}\bigr)\cr\cr
&&+\widehat{\unl\la}\bigl(\theta,x,X^{(1,3)}_2,Y^{(1,3)}_2,Z^{(1,3)}_2;\widetilde\z^{31},\widetilde\psi^{31},\widetilde\upsilon^{31}\bigr)-\widehat{\unl\la}\bigl(\theta,x,X^{(1,3)}_1,Y^{(1,3)}_1,Z^{(1,3)}_1;\widetilde\z^{31},\widetilde\psi^{31},\widetilde\upsilon^{31}\bigr)\cr\cr
&\equiv&\widehat\la^{1,2,3}\bigl(\widehat m^{(1,2)}_1,\widehat m^{(2,3)}_1,\widehat m^{(1,3)}_1,\widehat n^{(1,2)}_1,\widehat n^{(2,3)}_1,\widehat n^{(1,3)}_1\bigr)+\widehat\la\bigl(\widehat m^{(1,3)}_{7\,1},\widehat m^{(1,3)}_{7\,2},\widehat n^{(1,3)}_{7\,1},\widehat n^{(1,3)}_{7\,2}\bigr)\,,
\qqq
written for ($A\in\{1,2\}$)
\qq\nn
\widehat m^{(i,j)}_{7\,A}&=&\bigl(\bigl(\theta,x,\z^i,\psi^i,\upsilon^i\bigr),\bigl(\theta,x,\z^j,\psi^j,\upsilon^j\bigr),X^{(i,j)}_A,Y^{(i,j)}_A,Z^{(i,j)}_A\bigr)\,,\cr\cr
\widehat n^{(i,j)}_{7\,A}&=&\bigl(\bigl(\widetilde\theta,\widetilde x,\widetilde\z^i,\widetilde\psi^i,\widetilde\upsilon^i\bigr),\bigl(\widetilde\theta,\widetilde x,\widetilde\z^j,\widetilde\psi^j,\widetilde\upsilon^j\bigr),\widetilde X^{(i,j)}_A,\widetilde Y^{(i,j)}_A,\widetilde Z^{(i,j)}_A\bigr)\,.
\qqq
Thus, altogether, we have the desired product 1-isomorphism of super-1-gerbes
\qq\nn
\cM_{\widehat\xcG}\ :\ \pr_{1,2}^*\widehat\xcG\ox\pr_{2,3}^*\widehat\xcG\xrightarrow{\ \cong\ }\pr_{1,3}^*\widehat\xcG\,,
\qqq
given by
\qq\nn
\cM_{\widehat\xcG}=\bigl(\widehat\sfY^{1,2,3}\sfY_2^{[3]}\cM^{(1)},\id_{\widehat\sfY^{1,2,3}\sfY_2^{[3]}\cM^{(1)}},\widehat\xcE,\pi_{\widehat\xcE},\nabla_{\widehat\xcE},\widehat\vep\bigr)\,.
\qqq

Finally, we verify the existence of an associator 2-isomorphism 
\qq\label{diag:assoc}
\alxydim{@C=4.cm@R=2cm}{\pr_{1,2}^*\widehat\xcG\ox\pr_{2,3}^*\widehat\xcG\ox\pr_{3,4}^*\widehat\xcG \ar[r]^{\pr_{1,2,3}^*\cM_{\widehat\xcG}\ox\id_{\pr_{1,3}^*\widehat\xcG}} \ar[d]_{\id_{\pr_{1,2}^*\widehat\xcG}\ox\pr_{2,3,4}^*\cM_{\widehat\xcG}} & \pr_{1,3}^*\widehat\xcG\ox\pr_{3,4}^*\widehat\xcG \ar[d]^{\pr_{1,3,4}^*\cM_{\widehat\xcG}} \ar@{=>}[dl]|{\ \mu_{\widehat\xcG}\ } \\ \pr_{1,2}^*\widehat\xcG\ox\pr_{2,4}^*\widehat\xcG \ar[r]_{\pr_{1,2,4}^*\cM_{\widehat\xcG}} & \pr_{1,4}^*\widehat\xcG }
\qqq
For that purpose, we first consider surjective submersions of the pullback super-1-gerbes
\qq\nn
\pr_{i,j}^*\widehat\xcG\equiv\widehat\xcG_{i,j}\,,\qquad(i,j)\in\{(1,2),(2,3),(3,4),(1,3),(2,4),(1,4)\}
\qqq
and, subsequently, for their double tensor products
\qq\nn
\pr_{i,j}^*\widehat\xcG\ox\pr_{j,k}^*\widehat\xcG\equiv\widehat\xcG_{i,j}\ox\widehat\xcG_{j,k}\equiv\widehat\xcG_{i,j,k}\,,\qquad(i,j,k)\in\{(1,2,4),(1,3,4)\}\,,
\qqq
as well as for their triple tensor product
\qq\nn
\pr_{1,2}^*\widehat\xcG\ox\pr_{2,3}^*\widehat\xcG\ox\pr_{3,4}^*\widehat\xcG\equiv\widehat\xcG_{1,2}\ox\widehat\xcG_{2,3}\ox\widehat\xcG_{3,4}\equiv\widehat\xcG_{1,2,3,4}
\qqq
over $\,\sfY_2^{[4]}\cM^{(1)}$.\ For the first of them, we consider -- with hindsight -- two fibred products:
\qq\nn
\alxydim{@C=2.cm@R=1.5cm}{\widehat\sfY^{i,j}\sfY_2^{[4]}\cM^{(1)}\equiv\sfY_2^{[4]}\cM^{(1)}\x_{\sfY_2^{[2]}\cM^{(1)}}^{(i,j)}\widehat\sfY\sfY_2^{[2]}\cM^{(1)} \ar[r]^{\hspace{2.5cm}\pr_2} \ar[d]_{{\pi_{\widehat\sfY^{i,j}\sfY_2^{[4]}\cM^{(1)} }}\equiv\pr_1} & \widehat\sfY\sfY_2^{[2]}\cM^{(1)} \ar[d]^{\pi_{\widehat\sfY\sfY_2^{[2]}\cM^{(1)}}} \\ \sfY_2^{[4]}\cM^{(1)} \ar[r]_{\pr_{i,j}} &  \sfY_2^{[2]}\cM^{(1)} }
\qqq
and 
\qq\nn
\alxydim{@C=2.cm@R=1.5cm}{\widehat\sfY^{i(k)j}\sfY_2^{[4]}\cM^{(1)}\equiv\sfY_2^{[4]}\cM^{(1)}\x_{\sfY_2^{[3]}\cM^{(1)}}^{(i,k,j)}\widehat\sfY^{1,3}\sfY_2^{[3]}\cM^{(1)} \ar[r]^{\hspace{2.5cm}\pr_2} \ar[d]_{\pr_1} & \widehat\sfY^{1,3}\sfY_2^{[3]}\cM^{(1)} \ar[d]^{\pi_{\widehat\sfY^{1,3}\sfY_2^{[2]}\cM^{(1)}}} \\ \sfY_2^{[4]}\cM^{(1)} \ar[r]_{\pr_{i,j}} &  \sfY_2^{[3]}\cM^{(1)} }\,.
\qqq
These correspond to the two presentations of the pullback gerbe:
\qq\nn
\pr_{i,j}^*\widehat\cG\equiv\pr_{i,k,j}^*\pr_{1,3}^*\widehat\cG
\qqq
over $\,\sfY_2^{[4]}\cM^{(1)}$.\ The fibred products are naturally diffeomorphic (and Lie supergroup-isomorphic) and we denote the relevant diffeomorphism as
\qq\nn
\iota_{i(k)j}\ &:&\ \widehat\sfY^{i,j}\sfY_2^{[4]}\cM^{(1)}\xrightarrow{\ \cong\ }\widehat\sfY^{i(k)j}\sfY_2^{[4]}\cM^{(1)}\cr\cr 
&:&\ \bigl(\bigl(m_4^1,m_4^2,m_4^3,m_4^4\bigr),\bigl(m_4^i,m_4^j,X^{(i,j)},Y^{(i,j)},Z^{(i,j)}\bigr)\bigr)\cr\cr
&&\longmapsto\bigl(\bigl(m_4^1,m_4^2,m_4^3,m_4^4\bigr),\bigl(\bigl(m_4^i,m_4^k,m_4^j\bigr),\bigl(m_4^i,m_4^j,X^{(i,j)},Y^{(i,j)},Z^{(i,j)}\bigr)\bigr)\bigr)\,.
\qqq

For the double products, we choose another pair of surjective submersions:
\qq\nn
\alxydim{@C=1.5cm@R=1.5cm}{\widehat\sfY^{(i,j,k)}\sfY_2^{[4]}\cM^{(1)}\equiv\sfY_2^{[4]}\cM^{(1)}\x^{(i,j,k)}_{\sfY_2^{[3]}\cM^{(1)}}\widehat\sfY^{1,2;2,3}\sfY_2^{[3]}\cM^{(1)}  \ar[r]^{\hspace{1.25cm}\pr_2} \ar[d]_{\pi_{\widehat\sfY^{(i,j,k)}\sfY_2^{[4]}\cM^{(1)}}\equiv\pr_1} & \widehat\sfY^{1,2}\sfY_2^{[3]}\cM^{(1)}\x_{\sfY_2^{[3]}\cM^{(1)}}\widehat\sfY^{2,3}\sfY_2^{[3]}\cM^{(1)} \ar[d]^{\pi_{\widehat\sfY^{1,2}\sfY_2^{[2]}\cM^{(1)}}\circ\pr_1} \\ \sfY_2^{[4]}\cM^{(1)} \ar[r]_{\pr_{i,j,k}} &  \sfY_2^{[3]}\cM^{(1)} }
\qqq
and
\qq\nn
\alxydim{@C=1.5cm@R=1.5cm}{\widehat\sfY^{i,j;j,k}\sfY_2^{[4]}\cM^{(1)}\equiv\widehat\sfY^{i,j}\sfY_2^{[4]}\cM^{(1)}\x^{1\x 1}_{\sfY_2^{[4]}\cM^{(1)}}\widehat\sfY^{j,k}\sfY_2^{[4]}\cM^{(1)}  \ar[r]^{\hspace{3.25cm}\pr_2} \ar[d]_{\pr_1} & \widehat\sfY^{j,k}\sfY_2^{[4]}\cM^{(1)}  \ar[d]^{\pi_{\widehat\sfY^{j,k}\sfY_2^{[4]}\cM^{(1)} }} \\ \widehat\sfY^{i,j}\sfY_2^{[4]}\cM^{(1)} \ar[r]_{\pi_{\widehat\sfY^{i,j}\sfY_2^{[4]}\cM^{(1)} }} &  \sfY_2^{[4]}\cM^{(1)} }\,,
\qqq
the two being naturally identified by means of the diffeomorphism (and Lie supergroup-isomorphism)
\qq\nn
\iota_{i,j;j,k}\ &:&\ \widehat\sfY^{i,j;j,k}\sfY_2^{[4]}\cM^{(1)}\xrightarrow{\ \cong\ }\widehat\sfY^{(i,j,k)}\sfY_2^{[4]}\cM^{(1)}\cr\cr
&:&\ \bigl(\bigl(\bigl(m_4^1,m_4^2,m_4^3,m_4^4\bigr),\bigl(m_4^i,m_4^j,X^{(i,j)},Y^{(i,j)},Z^{(i,j)}\bigr)\bigr),\bigl(\bigl(m_4^1,m_4^2,m_4^3,m_4^4\bigr),\cr\cr
&&\hspace{.5cm}\bigl(m_4^j,m_4^k,X^{(j,k)},Y^{(j,k)},Z^{(j,k)}\bigr)\bigr)\bigr)\cr\cr
&&\longmapsto\bigl(\bigl(m_4^1,m_4^2,m_4^3,m_4^4\bigr),\bigl(\bigl(\bigl(m_4^i,m_4^j,m_4^k\bigr),\bigl(m_4^i,m_4^j,X^{(i,j)},Y^{(i,j)},Z^{(i,j)}\bigr)\bigr),\cr\cr
&&\hspace{.75cm}\bigl(\bigl(m_4^i,m_4^j,m_4^k\bigr),\bigl(m_4^j,m_4^k,X^{(j,k)},Y^{(j,k)},Z^{(j,k)}\bigr)\bigr)\bigr)\bigr)\,.
\qqq
They are to be thought as natural surjective submersions of the two presentations of the pullback gerbe:
\qq\nn
\pr_{i,j,k}^*\bigl(\pr_{1,2}^*\widehat\xcG\ox\pr_{2,3}^*\widehat\xcG\bigr)\equiv\pr_{i,j}^*\widehat\xcG\ox\pr_{j,k}^*\widehat\xcG\,.
\qqq
Finally, when it comes to the triple product, it is natural to work with two different surjective submersions:
\qq\nn
\alxydim{@C=1.5cm@R=1.5cm}{\widehat\sfY^{(1,2,3)}\sfY_2^{[4]}\cM^{(1)}\x^{1\x 1}_{\sfY_2^{[4]}\cM^{(1)}}\widehat\sfY^{3,4}\sfY_2^{[4]}\cM^{(1)} \ar[r]^{\hspace{2.cm}\pr_2} \ar[d]_{\pr_1} & \widehat\sfY^{3,4}\sfY_2^{[4]}\cM^{(1)} \ar[d]^{\pi_{\widehat\sfY^{3,4}\sfY_2^{[4]}\cM^{(1)} }} \\ \widehat\sfY^{(1,2,3)}\sfY_2^{[4]}\cM^{(1)} \ar[r]_{\pi_{\widehat\sfY^{(1,2,3)}\sfY_2^{[4]}\cM^{(1)}}} &  \sfY_2^{[4]}\cM^{(1)} }
\qqq
and 
\qq\nn
\alxydim{@C=1.5cm@R=1.5cm}{\widehat\sfY^{1,2}\sfY_2^{[4]}\cM^{(1)}\x^{1\x 1}_{\sfY_2^{[4]}\cM^{(1)}}\widehat\sfY^{(2,3,4)}\sfY_2^{[4]}\cM^{(1)} \ar[r]^{\hspace{2.cm}\pr_2} \ar[d]_{\pr_1} & \widehat\sfY^{(2,3,4)}\sfY_2^{[4]}\cM^{(1)} \ar[d]^{\pi_{\widehat\sfY^{(2,3,4)}\sfY_2^{[4]}\cM^{(1)}}} \\ \widehat\sfY^{1,2}\sfY_2^{[4]}\cM^{(1)} \ar[r]_{\pi_{\widehat\sfY^{1,2}\sfY_2^{[4]}\cM^{(1)}}} &  \sfY_2^{[4]}\cM^{(1)} }\,,
\qqq
into which the third one:
\qq\nn
\widehat\sfY^{1,2;2,3;3,4}\sfY_2^{[4]}\cM^{(1)}\equiv\widehat\sfY^{1,2}\sfY_2^{[4]}\cM^{(1)}\x^{1\x 1}_{\sfY_2^{[4]}\cM^{(1)}}\widehat\sfY^{2,3}\sfY_2^{[4]}\cM^{(1)}\x^{1\x 1}_{\sfY_2^{[4]}\cM^{(1)}}\widehat\sfY^{3,4}\sfY_2^{[4]}\cM^{(1)}\,,
\qqq
determined by the commutative diagram
\qq\nn
\alxydim{@C=2.5cm@R=1cm}{ & \widehat\sfY^{1,2;2,3;3,4}\sfY_2^{[4]}\cM^{(1)} \ar[rd]^{\pr_3} \ar[d]_{\pr_2} \ar[ld]_{\pr_1} & \\ \widehat\sfY^{1,2}\sfY_2^{[4]}\cM^{(1)} \ar[rd]_{\pi_{\widehat\sfY^{1,2}\sfY_2^{[4]}\cM^{(1)}}} & \widehat\sfY^{2,3}\sfY_2^{[4]}\cM^{(1)} \ar[d]_{\pi_{\widehat\sfY^{2,3}\sfY_2^{[4]}\cM^{(1)}}} & \widehat\sfY^{3,4}\sfY_2^{[4]}\cM^{(1)} \ar[ld]^{\pi_{\widehat\sfY^{3,4}\sfY_2^{[4]}\cM^{(1)}}} \\ & \sfY_2^{[4]}\cM^{(1)} & }\,,
\qqq
is readily mapped by the respective diffeomorphisms (and Lie-supergroup isomorphisms):
\qq\nn
\iota_{1,2;2,3}\x\id_{\widehat\sfY^{3,4}\sfY_2^{[4]}\cM^{(1)}}\ &:&\ \widehat\sfY^{1,2;2,3;3,4}\sfY_2^{[4]}\cM^{(1)}\xrightarrow{\ \cong\ }\widehat\sfY^{(1,2,3)}\sfY_2^{[4]}\cM^{(1)}\x^{1\x 1}_{\sfY_2^{[4]}\cM^{(1)}}\widehat\sfY^{3,4}\sfY_2^{[4]}\cM^{(1)}\cr\cr
\id_{\widehat\sfY^{1,2}\sfY_2^{[4]}\cM^{(1)}}\x\iota_{2,3;3,4}\ &:&\ \widehat\sfY^{1,2;2,3;3,4}\sfY_2^{[4]}\cM^{(1)}\xrightarrow{\ \cong\ }\widehat\sfY^{1,2}\sfY_2^{[4]}\cM^{(1)}\x^{1\x 1}_{\sfY_2^{[4]}\cM^{(1)}}\widehat\sfY^{(2,3,4)}\sfY_2^{[4]}\cM^{(1)}\,.
\qqq

At this stage, we may write out the principal $\bC^\x$-bundles of the composite 1-isomorphisms $\,\pr_{1,3,4}^*\cM_{\widehat\xcG}\circ(\pr_{1,2,3}^*\cM_{\widehat\xcG}\ox\id_{\widehat\xcG_{1,3}})\,$ and $\,\pr_{1,2,4}^*\cM_{\widehat\xcG}\circ(\id_{\pr_{1,2}^*\widehat\xcG}\ox\pr_{2,3,4}^*\cM_{\widehat\xcG})\,$ using all our hitherto findings. Thus, the former reads
\qq\nn
\alxydim{@C=1.cm@R=1.5cm}{\bC^\x \ar[r] & \pi_{2,3}^*\iota_{1,(3;3),4}^*\widetilde\pr_{2,3,5}^*\widehat\xcE\ox\pi_{1,2}^*\bigl(\widehat\pr_{1,2,4}^*\iota_{1,(2;2),3}^*\pr_{2,3,5}^*\widehat\xcE\ox\widehat\pr_{3,5}^*\pr_{2,4}^*\widehat\xcL\bigr) \ar[d] \\ & \widehat\sfY^{1,2;2,3;3,4}\sfY_2^{[4]}\cM^{(1)}\x^{1\x 1}_{\sfY_2^{[4]}\cM^{(1)}}\widehat\sfY^{1,3;3,4}\sfY_2^{[4]}\cM^{(1)}\x^{1\x 1}_{\sfY_2^{[4]}\cM^{(1)}}\widehat\sfY^{1,4}\sfY_2^{[4]}\cM^{(1)}}\,,
\qqq
where
\qq\nn
\pi_{2,3}\ &:&\ \widehat\sfY^{1,2;2,3;3,4}\sfY_2^{[4]}\cM^{(1)}\x^{1\x 1}_{\sfY_2^{[4]}\cM^{(1)}}\widehat\sfY^{1,3;3,4}\sfY_2^{[4]}\cM^{(1)}\x^{1\x 1}_{\sfY_2^{[4]}\cM^{(1)}}\widehat\sfY^{1,4}\sfY_2^{[4]}\cM^{(1)}\cr\cr
&&\too\widehat\sfY^{1,3;3,4}\sfY_2^{[4]}\cM^{(1)}\x^{1\x 1}_{\sfY_2^{[4]}\cM^{(1)}}\widehat\sfY^{1,4}\sfY_2^{[4]}\cM^{(1)}\,,\cr\cr\cr
\pi_{1,2}\ &:&\ \widehat\sfY^{1,2;2,3;3,4}\sfY_2^{[4]}\cM^{(1)}\x^{1\x 1}_{\sfY_2^{[4]}\cM^{(1)}}\widehat\sfY^{1,3;3,4}\sfY_2^{[4]}\cM^{(1)}\x^{1\x 1}_{\sfY_2^{[4]}\cM^{(1)}}\widehat\sfY^{1,4}\sfY_2^{[4]}\cM^{(1)}\cr\cr
&&\too\widehat\sfY^{1,2;2,3;3,4}\sfY_2^{[4]}\cM^{(1)}\x^{1\x 1}_{\sfY_2^{[4]}\cM^{(1)}}\widehat\sfY^{1,3;3,4}\sfY_2^{[4]}\cM^{(1)}\,,\cr\cr\cr
\widehat\pr_{1,2,4}\ &:&\ \widehat\sfY^{1,2;2,3;3,4}\sfY_2^{[4]}\cM^{(1)}\x^{1\x 1}_{\sfY_2^{[4]}\cM^{(1)}}\widehat\sfY^{1,3;3,4}\sfY_2^{[4]}\cM^{(1)}\cr\cr
&&\too\widehat\sfY^{1,2}\sfY_2^{[4]}\cM^{(1)}\x^{1\x 1}_{\sfY_2^{[4]}\cM^{(1)}}\widehat\sfY^{2,3}\sfY_2^{[4]}\cM^{(1)}\x^{1\x 1}_{\sfY_2^{[4]}\cM^{(1)}}\widehat\sfY^{1,3}\sfY_2^{[4]}\cM^{(1)}\,,\cr\cr\cr
\widehat\pr_{3,5}\ &:&\ \widehat\sfY^{1,2;2,3;3,4}\sfY_2^{[4]}\cM^{(1)}\x^{1\x 1}_{\sfY_2^{[4]}\cM^{(1)}}\widehat\sfY^{1,3;3,4}\sfY_2^{[4]}\cM^{(1)}\too\widehat\sfY^{3,4}\sfY_2^{[4]}\cM^{(1)}\x^{1\x 1}_{\sfY_2^{[4]}\cM^{(1)}}\widehat\sfY^{3,4}\sfY_2^{[4]}\cM^{(1)}\,,\cr\cr\cr
\widetilde\pr_{2,3,5}\ &:&\ \widehat\sfY^{(1,3,4)}\sfY_2^{[4]}\cM^{(1)}\x^{1\x 1}_{\sfY_2^{[4]}\cM^{(1)}}\widehat\sfY^{1(3)4}\sfY_2^{[4]}\cM^{(1)}\too\widehat\sfY^{1,2,3}\sfY_2^{[3]}\cM^{(1)}\,,\cr\cr\cr
\pr_{2,3,5}\ &:&\ \widehat\sfY^{(1,2,3)}\sfY_2^{[4]}\cM^{(1)}\x^{1\x 1}_{\sfY_2^{[4]}\cM^{(1)}}\widehat\sfY^{1(2)3}\sfY_2^{[4]}\cM^{(1)}\too\widehat\sfY^{1,2,3}\sfY_2^{[3]}\cM^{(1)}\,,\cr\cr\cr
\pr_{2,4}\ &:&\ \widehat\sfY^{3,4}\sfY_2^{[4]}\cM^{(1)}\x^{1\x 1}_{\sfY_2^{[4]}\cM^{(1)}}\widehat\sfY^{3,4}\sfY_2^{[4]}\cM^{(1)}\too\widehat\sfY^{[2]}\sfY_2^{[2]}\cM^{(1)}
\qqq
are the canonical projections and 
\qq\nn
\iota_{i,(j;j),k}\equiv\iota_{i,j;j,k}\x\iota_{i(j)k}\,,
\qqq
whereas the latter takes the form
\qq\nn
\alxydim{@C=1.cm@R=1.5cm}{\bC^\x \ar[r] & \unl\pi_{2,3}^*\iota_{1,(2;2),4}^*\widetilde{\unl\pr}_{2,3,5}^*\widehat\xcE\ox\unl\pi_{1,2}^*\bigl(\widehat{\unl\pr}_{1,4}^*\unl\pr_{2,4}^*\widehat\xcL\ox\widehat{\unl\pr}_{2,3,5}^*\iota_{2,(3;3),4}^*\unl\pr_{2,3,5}^*\widehat\xcE\bigr) \ar[d] \\ & \widehat\sfY^{1,2;2,3;3,4}\sfY_2^{[4]}\cM^{(1)}\x^{1\x 1}_{\sfY_2^{[4]}\cM^{(1)}}\widehat\sfY^{1,2;2,4}\sfY_2^{[4]}\cM^{(1)}\x^{1\x 1}_{\sfY_2^{[4]}\cM^{(1)}}\widehat\sfY^{1,4}\sfY_2^{[4]}\cM^{(1)}}\,,
\qqq
where
\qq\nn
\unl\pi_{2,3}\ &:&\ \widehat\sfY^{1,2;2,3;3,4}\sfY_2^{[4]}\cM^{(1)}\x^{1\x 1}_{\sfY_2^{[4]}\cM^{(1)}}\widehat\sfY^{1,2;2,4}\sfY_2^{[4]}\cM^{(1)}\x^{1\x 1}_{\sfY_2^{[4]}\cM^{(1)}}\widehat\sfY^{1,4}\sfY_2^{[4]}\cM^{(1)}\cr\cr
&&\too\widehat\sfY^{1,2;2,4}\sfY_2^{[4]}\cM^{(1)}\x^{1\x 1}_{\sfY_2^{[4]}\cM^{(1)}}\widehat\sfY^{1,4}\sfY_2^{[4]}\cM^{(1)}\,,\cr\cr\cr
\unl\pi_{1,2}\ &:&\ \widehat\sfY^{1,2;2,3;3,4}\sfY_2^{[4]}\cM^{(1)}\x^{1\x 1}_{\sfY_2^{[4]}\cM^{(1)}}\widehat\sfY^{1,2;2,4}\sfY_2^{[4]}\cM^{(1)}\x^{1\x 1}_{\sfY_2^{[4]}\cM^{(1)}}\widehat\sfY^{1,4}\sfY_2^{[4]}\cM^{(1)}\cr\cr
&&\too\widehat\sfY^{1,2;2,3;3,4}\sfY_2^{[4]}\cM^{(1)}\x^{1\x 1}_{\sfY_2^{[4]}\cM^{(1)}}\widehat\sfY^{1,2;2,4}\sfY_2^{[4]}\cM^{(1)}\,,\cr\cr\cr
\widehat{\unl\pr}_{1,4}\ &:&\ \widehat\sfY^{1,2;2,3;3,4}\sfY_2^{[4]}\cM^{(1)}\x^{1\x 1}_{\sfY_2^{[4]}\cM^{(1)}}\widehat\sfY^{1,2;2,4}\sfY_2^{[4]}\cM^{(1)}\too\widehat\sfY^{1,2}\sfY_2^{[4]}\cM^{(1)}\x^{1\x 1}_{\sfY_2^{[4]}\cM^{(1)}}\widehat\sfY^{1,2}\sfY_2^{[4]}\cM^{(1)}\,,\cr\cr\cr
\widehat{\unl\pr}_{2,3,5}\ &:&\ \widehat\sfY^{1,2;2,3;3,4}\sfY_2^{[4]}\cM^{(1)}\x^{1\x 1}_{\sfY_2^{[4]}\cM^{(1)}}\widehat\sfY^{1,2;2,4}\sfY_2^{[4]}\cM^{(1)}\cr\cr
&&\too\widehat\sfY^{2,3}\sfY_2^{[4]}\cM^{(1)}\x^{1\x 1}_{\sfY_2^{[4]}\cM^{(1)}}\widehat\sfY^{3,4}\sfY_2^{[4]}\cM^{(1)}\x^{1\x 1}_{\sfY_2^{[4]}\cM^{(1)}}\widehat\sfY^{2,4}\sfY_2^{[4]}\cM^{(1)}\,,\cr\cr\cr
\widetilde{\unl\pr}_{2,3,5}\ &:&\ \widehat\sfY^{(1,3,4)}\sfY_2^{[4]}\cM^{(1)}\x^{1\x 1}_{\sfY_2^{[4]}\cM^{(1)}}\widehat\sfY^{1(3)4}\sfY_2^{[4]}\cM^{(1)}\too\widehat\sfY^{1,2,3}\sfY_2^{[3]}\cM^{(1)}\,,\cr\cr\cr
\unl\pr_{2,4}\ &:&\ \widehat\sfY^{3,4}\sfY_2^{[4]}\cM^{(1)}\x^{1\x 1}_{\sfY_2^{[4]}\cM^{(1)}}\widehat\sfY^{3,4}\sfY_2^{[4]}\cM^{(1)}\too\widehat\sfY^{[2]}\sfY_2^{[2]}\cM^{(1)}\,,\cr\cr\cr
\unl\pr_{2,3,5}\ &:&\ \widehat\sfY^{(1,2,3)}\sfY_2^{[4]}\cM^{(1)}\x^{1\x 1}_{\sfY_2^{[4]}\cM^{(1)}}\widehat\sfY^{1(2)3}\sfY_2^{[4]}\cM^{(1)}\too\widehat\sfY^{1,2,3}\sfY_2^{[3]}\cM^{(1)}
\qqq
are the canonical projections.

Given the above choices, we conclude that the surjective submersion of the sought-after 2-isomorphism $\,\mu_{\widehat\xcG}\,$ may be chosen in the form 
\qq\nn
&&\sfY^{1,2;2,3;3,4}\sfY_2^{[4]}\cM^{(1)}\x^{1\x 1}_{\sfY_2^{[4]}\cM^{(1)}}\widehat\sfY^{1,2;2,4}\sfY_2^{[4]}\cM^{(1)}\x^{1\x 1}_{\sfY_2^{[4]}\cM^{(1)}}\widehat\sfY^{1,3;3,4}\sfY_2^{[4]}\cM^{(1)}\x^{1\x 1}_{\sfY_2^{[4]}\cM^{(1)}}\widehat\sfY^{1,4}\sfY_2^{[4]}\cM^{(1)}\cr\cr
&\equiv&\widetilde\sfY\sfY_2^{[4]}\cM^{(1)}\,,
\qqq
and, over it, we look for a principal $\bC^\x$-bundle and Lie-supergroup isomorphism 
\qq\nn
\widehat\g\ :\ \pi_{1,3,4}^*\bigl(\pi_{2,3}^*\iota_{1,(3;3),4}^*\widetilde\pr_{2,3,5}^*\widehat\xcE\ox\pi_{1,2}^*\bigl(\widehat\pr_{1,2,4}^*\iota_{1,(2;2),3}^*\pr_{2,3,5}^*\widehat\xcE\ox\widehat\pr_{3,5}^*\pr_{2,4}^*\widehat\xcL\bigr)\bigr)\cr\cr
\xrightarrow{\ \cong\ }\pi_{1,2,4}^*\bigl(\unl\pi_{2,3}^*\iota_{1,(2;2),4}^*\widetilde{\unl\pr}_{2,3,5}^*\widehat\xcE\ox\unl\pi_{1,2}^*\bigl(\widehat{\unl\pr}_{1,4}^*\unl\pr_{2,4}^*\widehat\xcL\ox\widehat{\unl\pr}_{2,3,5}^*\iota_{2,(3;3),4}^*\unl\pr_{2,3,5}^*\widehat\xcE\bigr)\bigr)\,,
\qqq
where
\qq\nn
\pi_{1,3,4}\ &:&\ \widetilde\sfY\sfY_2^{[4]}\cM^{(1)}\too\sfY^{1,2;2,3;3,4}\sfY_2^{[4]}\cM^{(1)}\x^{1\x 1}_{\sfY_2^{[4]}\cM^{(1)}}\widehat\sfY^{1,3;3,4}\sfY_2^{[4]}\cM^{(1)}\x^{1\x 1}_{\sfY_2^{[4]}\cM^{(1)}}\widehat\sfY^{1,4}\sfY_2^{[4]}\cM^{(1)}\,,\cr\cr
\pi_{1,2,4}\ &:&\ \widetilde\sfY\sfY_2^{[4]}\cM^{(1)}\too\sfY^{1,2;2,3;3,4}\sfY_2^{[4]}\cM^{(1)}\x^{1\x 1}_{\sfY_2^{[4]}\cM^{(1)}}\widehat\sfY^{1,2;2,4}\sfY_2^{[4]}\cM^{(1)}\x^{1\x 1}_{\sfY_2^{[4]}\cM^{(1)}}\widehat\sfY^{1,4}\sfY_2^{[4]}\cM^{(1)}\,.
\qqq
are the canonical projections. Equality of the connection super-1-forms on the common base of the principal $\bC^\x$-bundles under comparison,
\qq\nn
&&\pi_{1,3,4}^*\bigl(\pi_{2,3}^*\iota_{1,(3;3),4}^*\widetilde\pr_{2,3,5}^*\widehat\txa+\pi_{1,2}^*\bigl(\widehat\pr_{1,2,4}^*\iota_{1,(2;2),3}^*\pr_{2,3,5}^*\widehat\txa+\widehat\pr_{3,5}^*\pr_{2,4}^*\widehat\txA\bigr)\bigr)\cr\cr
&&\bigl(\check m^{(1,2)}_1,\check m^{(2,3)},\check m^{(3,4)}_1,\check m^{(1,2)}_2,\check m^{(2,4)},\check m^{(1,3)},\check m^{(3,4)}_2,\check m^{(1,4)}\bigr)\cr\cr
&=&\widehat\txa\bigl(\widehat m^{(1,3,(4))},\widehat m^{((1),3,4)}_2,\widehat m^{(1,(3),4)}\bigr)+\widehat\txa\bigl(\widehat m^{(1,2,(3))}_1,\widehat m^{((1),2,3)},\widehat m^{(1,(2),3)}\bigr)+\widehat\txA\bigl(\widehat{\unl m}^{(3,4)}_{7\,1},\widehat{\unl m}^{(3,4)}_{7\,2}\bigr)\cr\cr
&\equiv&\widehat{\unl\txA}\bigl(\z^{41},\psi^{41},\upsilon^{41},X^{(1,4)},Y^{(1,4)},Z^{(1,4)}\bigr)-\widehat{\unl\txA}\bigl(\z^{31},\psi^{31},\upsilon^{31},X^{(1,3)},Y^{(1,3)},Z^{(1,3)}\bigr)\cr\cr
&&-\widehat{\unl\txA}\bigl(\z^{43},\psi^{43},\upsilon^{43},X^{(3,4)}_2,Y^{(3,4)}_2,Z^{(3,4)}_2\bigr)\cr\cr
&&+\widehat{\unl\txA}\bigl(\z^{31},\psi^{31},\upsilon^{31},X^{(1,3)},Y^{(1,3)},Z^{(1,3)}\bigr)-\widehat{\unl\txA}\bigl(\z^{21},\psi^{21},\upsilon^{21},X^{(1,2)}_1,Y^{(1,2)}_1,Z^{(1,2)}_1\bigr)\cr\cr
&&-\widehat{\unl\txA}\bigl(\z^{32},\psi^{32},\upsilon^{32},X^{(2,3)},Y^{(2,3)},Z^{(2,3)}\bigr)\cr\cr
&&+\widehat{\unl\txA}\bigl(\z^{43},\psi^{43},\upsilon^{43},X^{(3,4)}_2,Y^{(3,4)}_2,Z^{(3,4)}_2\bigr)-\widehat{\unl\txA}\bigl(\z^{43},\psi^{43},\upsilon^{43},X^{(3,4)}_1,Y^{(3,4)}_1,Z^{(3,4)}_1\bigr)\cr\cr
&=&\widehat{\unl\txA}\bigl(\z^{41},\psi^{41},\upsilon^{41},X^{(1,4)}_2,Y^{(1,4)}_2,Z^{(1,4)}_2\bigr)-\widehat{\unl\txA}\bigl(\z^{21},\psi^{21},\upsilon^{21},X^{(1,2)}_1,Y^{(1,2)}_1,Z^{(1,2)}_1\bigr)\cr\cr
&&-\widehat{\unl\txA}\bigl(\z^{31},\psi^{31},\upsilon^{31},X^{(1,3)},Y^{(1,3)},Z^{(1,3)}\bigr)-\widehat{\unl\txA}\bigl(\z^{32},\psi^{32},\upsilon^{32},X^{(2,3)},Y^{(2,3)},Z^{(2,3)}\bigr)\cr\cr
&&-\widehat{\unl\txA}\bigl(\z^{43},\psi^{43},\upsilon^{43},X^{(3,4)}_1,Y^{(3,4)}_1,Z^{(3,4)}_1\bigr)\cr\cr
&\equiv&\widehat{\unl\txA}\bigl(\z^{41},\psi^{41},\upsilon^{41},X^{(1,4)},Y^{(1,4)},Z^{(1,4)}\bigr)-\widehat{\unl\txA}\bigl(\z^{21},\psi^{21},\upsilon^{21},X^{(1,2)}_2,Y^{(1,2)}_2,Z^{(1,2)}_2\bigr)\cr\cr
&&-\widehat{\unl\txA}\bigl(\z^{42},\psi^{42},\upsilon^{42},X^{(2,4)},Y^{(2,4)},Z^{(2,4)}\bigr)\cr\cr
&&+\widehat{\unl\txA}\bigl(\z^{21},\psi^{21},\upsilon^{21},X^{(1,2)}_2,Y^{(1,2)}_2,Z^{(1,2)}_2\bigr)-\widehat{\unl\txA}\bigl(\z^{21},\psi^{21},\upsilon^{21},X^{(1,2)}_1,Y^{(1,2)}_1,Z^{(1,2)}_1\bigr)\cr\cr
&&+\widehat{\unl\txA}\bigl(\z^{42},\psi^{42},\upsilon^{42},X^{(2,4)},Y^{(2,4)},Z^{(2,4)}\bigr)-\widehat{\unl\txA}\bigl(\z^{32},\psi^{32},\upsilon^{32},X^{(2,3)},Y^{(2,3)},Z^{(2,3)}\bigr)\cr\cr
&&-\widehat{\unl\txA}\bigl(\z^{43},\psi^{43},\upsilon^{43},X^{(3,4)}_1,Y^{(3,4)}_1,Z^{(3,4)}_1\bigr)\cr\cr
&\equiv&\widehat\txa\bigl(\widehat m^{(1,2,(4))}_2,\widehat m^{((1),2,4)},\widehat m^{(1,(2),4)}\bigr)+\widehat\txA\bigl(\widehat{\unl m}^{(1,2)}_{7\,1},\widehat{\unl m}^{(1,2)}_{7\,2}\bigr)+\widehat\txa\bigl(\widehat m^{(2,3,(4))},\widehat m^{((2),3,4)}_1,\widehat m^{(2,(3),4)}\bigr)\cr\cr
&=&\pi_{1,2,4}^*\bigl(\unl\pi_{2,3}^*\iota_{1,(2;2),4}^*\widetilde{\unl\pr}_{2,3,5}^*\widehat\txa+\unl\pi_{1,2}^*\bigl(\widehat{\unl\pr}_{1,4}^*\unl\pr_{2,4}^*\widehat\txA+\widehat{\unl\pr}_{2,3,5}^*\iota_{2,(3;3),4}^*\unl\pr_{2,3,5}^*\widehat\txa\bigr)\bigr)\cr\cr
&&\bigl(\check m^{(1,2)}_1,\check m^{(2,3)},\check m^{(3,4)}_1,\check m^{(1,2)}_2,\check m^{(2,4)},\check m^{(1,3)},\check m^{(3,4)}_2,\check m^{(1,4)}\bigr)\,,
\qqq
written in the shorthand notation ($A\in\{1,2\}$):
\qq\nn
\check m^{(i,j)}_{(A)}&:=&\bigl(\bigl(m_4^1,m_4^2,m_4^3,m_4^4\bigr),\bigl(m_4^i,m_4^j,X^{(i,j)}_{(A)},Y^{(i,j)}_{(A)},Z^{(i,j)}_{(A)}\bigr)\bigr)\,,\cr\cr\cr
\widehat m^{(i,j,(k))}_{(A)}&:=&\bigl(\bigl(m_4^i,m_4^j,m_4^k\bigr),\bigl(m_4^i,m_4^j,X^{(i,j)}_{(A)},Y^{(i,j)}_{(A)},Z^{(i,j)}_{(A)}\bigr)\bigr)\,,\cr\cr
\widehat m^{(i,(j),k)}_{(A)}&:=&\bigl(\bigl(m_4^i,m_4^j,m_4^k\bigr),\bigl(m_4^i,m_4^k,X^{(i,k)}_{(A)},Y^{(i,k)}_{(A)},Z^{(i,k)}_{(A)}\bigr)\bigr)\,,\cr\cr
\widehat m^{((i),j,k)}_{(A)}&:=&\bigl(\bigl(m_4^i,m_4^j,m_4^k\bigr),\bigl(m_4^j,m_4^k,X^{(j,k)}_{(A)},Y^{(j,k)}_{(A)},Z^{(j,k)}_{(A)}\bigr)\bigr)\,,\cr\cr\cr
\widehat{\unl m}^{(i,j)}_{7\,A}&:=&\bigl(m_4^i,m_4^j,X^{(i,j)}_{(A)},Y^{(i,j)}_{(A)},Z^{(i,j)}_{(A)}\bigr)\,,
\qqq
leads us to set
\qq\nn
&&\widehat\g\bigl(\bigl(\widehat M,\widehat m^{(1,3,(4))},\widehat m^{((1),3,4)}_2,\widehat m^{(1,(3),4)},\widehat\z_1\bigr)\ox\bigl(\widehat M,\widehat m^{(1,2,(3))}_1,\widehat m^{((1),2,3)},\widehat m^{(1,(2),3)},\widehat\z_2\bigr)\ox\bigl(\widehat M,\widehat{\unl m}^{(3,4)}_{7\,1},\widehat{\unl m}^{(3,4)}_{7\,2},\widehat z\bigr)\bigr)\cr\cr
&:=&\bigl(\widehat M,\widehat m^{(1,2,(4))}_2,\widehat m^{((1),2,4)},\widehat m^{(1,(2),4)},\widehat\z_1\cdot\widehat\z_2\cdot\widehat z\bigr)\ox\bigl(\widehat M,\widehat{\unl m}^{(1,2)}_{7\,1},\widehat{\unl m}^{(1,2)}_{7\,2},1\bigr)\ox\bigl(\widehat m^{(2,3,(4))},\widehat m^{((2),3,4)}_1,\widehat m^{(2,(3),4)},1\bigr)\,,
\qqq
where
\qq\nn
\widehat M\equiv\bigl(\widehat m^{(1,2)}_1,\widehat m^{(2,3)},\widehat m^{(3,4)}_1,\widehat m^{(1,2)}_2,\widehat m^{(2,4)},\widehat m^{(1,3)},\widehat m^{(3,4)}_2,\widehat m^{(1,4)}\bigr)\,.
\qqq
Clearly, for a 2-isomorphism thus defined, all coherence costraints, in particular those involving the trivial groupoid structure $\,\mu_{\widehat\xcL}\,$ on $\,\widehat\xcL\,$ and the product isomorphism $\,\widehat\vep\,$ (likewise trivial), are automatically satisfied. Hence, the very last property that remains to be checked is the homomorphic nature of the principal $\bC^\x$-bundle isomorphism $\,\widehat\g$.\ This follows readily from the identity
\qq\nn
&&\widehat\la^{1,2,3}\bigl(\widehat m^{(1,3,(4))},\widehat m^{((1),3,4)}_2,\widehat m^{(1,(3),4)},\widehat n^{(1,3,(4))},\widehat n^{((1),3,4)}_2,\widehat n^{(1,(3),4)}\bigr)\cr\cr
&&+\widehat\la^{1,2,3}\bigl(\widehat m^{(1,2,(3))}_1,\widehat m^{((1),2,3)},\widehat m^{(1,(2),3)},\widehat n^{(1,2,(3))}_1,\widehat n^{((1),2,3)},\widehat n^{(1,(2),3)}\bigr)+\widehat\la\bigl(\widehat{\unl m}^{(3,4)}_{7\,1},\widehat{\unl m}^{(3,4)}_{7\,2},\widehat{\unl n}^{(3,4)}_{7\,1},\widehat{\unl n}^{(3,4)}_{7\,2}\bigr)\cr\cr
&\equiv&\widehat{\unl\la}\bigl(\theta,x,X^{(1,4)},Y^{(1,4)},Z^{(1,4)};\widetilde\z^{41},\widetilde\psi^{41},\widetilde\upsilon^{41}\bigr)-\widehat{\unl\la}\bigl(\theta,x,X^{(1,3)},Y^{(1,3)},Z^{(1,3)};\widetilde\z^{31},\widetilde\psi^{31},\widetilde\upsilon^{31}\bigr)\cr\cr
&&-\widehat{\unl\la}\bigl(\theta,x,X^{(3,4)}_2,Y^{(3,4)}_2,Z^{(3,4)}_2;\widetilde\z^{43},\widetilde\psi^{43},\widetilde\upsilon^{43}\bigr)\cr\cr
&&+\widehat{\unl\la}\bigl(\theta,x,X^{(1,3)},Y^{(1,3)},Z^{(1,3)};\widetilde\z^{31},\widetilde\psi^{31},\widetilde\upsilon^{31}\bigr)-\widehat{\unl\la}\bigl(\theta,x,X^{(1,2)}_1,Y^{(1,2)}_1,Z^{(1,2)}_1;\widetilde\z^{21},\widetilde\psi^{21},\widetilde\upsilon^{21}\bigr)\cr\cr
&&-\widehat{\unl\la}\bigl(\theta,x,X^{(2,3)},Y^{(2,3)},Z^{(2,3)};\widetilde\z^{32},\widetilde\psi^{32},\widetilde\upsilon^{32}\bigr)\cr\cr
&&+\widehat{\unl\la}\bigl(\theta,x,X^{(3,4)}_2,Y^{(3,4)}_2,Z^{(3,4)}_2;\widetilde\z^{43},\widetilde\psi^{43},\widetilde\upsilon^{43}\bigr)-\widehat{\unl\la}\bigl(\theta,x,X^{(3,4)}_1,Y^{(3,4)}_1,Z^{(3,4)}_1;\widetilde\z^{43},\widetilde\psi^{43},\widetilde\upsilon^{43}\bigr)\cr\cr
&=&\widehat{\unl\la}\bigl(\theta,x,X^{(1,4)},Y^{(1,4)},Z^{(1,4)};\widetilde\z^{41},\widetilde\psi^{41},\widetilde\upsilon^{41}\bigr)-\widehat{\unl\la}\bigl(\theta,x,X^{(1,2)}_1,Y^{(1,2)}_1,Z^{(1,2)}_1;\widetilde\z^{21},\widetilde\psi^{21},\widetilde\upsilon^{21}\bigr)\cr\cr
&&-\widehat{\unl\la}\bigl(\theta,x,X^{(1,3)},Y^{(1,3)},Z^{(1,3)};\widetilde\z^{31},\widetilde\psi^{31},\widetilde\upsilon^{31}\bigr)-\widehat{\unl\la}\bigl(\theta,x,X^{(2,3)},Y^{(2,3)},Z^{(2,3)};\widetilde\z^{32},\widetilde\psi^{32},\widetilde\upsilon^{32}\bigr)\cr\cr
&&-\widehat{\unl\la}\bigl(\theta,x,X^{(3,4)}_2,Y^{(3,4)}_2,Z^{(3,4)}_2;\widetilde\z^{43},\widetilde\psi^{43},\widetilde\upsilon^{43}\bigr)\cr\cr
&\equiv&\widehat{\unl\la}\bigl(\theta,x,X^{(1,4)},Y^{(1,4)},Z^{(1,4)};\widetilde\z^{41},\widetilde\psi^{41},\widetilde\upsilon^{41}\bigr)-\widehat{\unl\la}\bigl(\theta,x,X^{(1,2)}_2,Y^{(1,2)}_2,Z^{(1,2)}_2;\widetilde\z^{21},\widetilde\psi^{21},\widetilde\upsilon^{21}\bigr)\cr\cr
&&-\widehat{\unl\la}\bigl(\theta,x,X^{(2,4)},Y^{(2,4)},Z^{(2,4)};\widetilde\z^{42},\widetilde\psi^{42},\widetilde\upsilon^{42}\bigr)\cr\cr
&&+\widehat{\unl\la}\bigl(\theta,x,X^{(1,2)}_2,Y^{(1,2)}_2,Z^{(1,2)}_2;\widetilde\z^{21},\widetilde\psi^{21},\widetilde\upsilon^{21}\bigr)-\widehat{\unl\la}\bigl(\theta,x,X^{(1,2)}_1,Y^{(1,2)}_1,Z^{(1,2)}_1;\widetilde\z^{21},\widetilde\psi^{21},\widetilde\upsilon^{21}\bigr)\cr\cr
&&+\widehat{\unl\la}\bigl(\theta,x,X^{(2,4)},Y^{(2,4)},Z^{(2,4)};\widetilde\z^{42},\widetilde\psi^{42},\widetilde\upsilon^{42}\bigr)-\widehat{\unl\la}\bigl(\theta,x,X^{(2,3)},Y^{(2,3)},Z^{(2,3)};\widetilde\z^{32},\widetilde\psi^{32},\widetilde\upsilon^{32}\bigr)\cr\cr
&&-\widehat{\unl\la}\bigl(\theta,x,X^{(3,4)}_1,Y^{(3,4)}_1,Z^{(3,4)}_1;\widetilde\z^{43},\widetilde\psi^{43},\widetilde\upsilon^{43}\bigr)\cr\cr
&\equiv&\widehat\la^{1,2,3}\bigl(\widehat m^{(1,2,(4))}_2,\widehat m^{((1),2,4)},\widehat m^{(1,(2),4)},\widehat n^{(1,2,(4))}_2,\widehat n^{((1),2,4)},\widehat n^{(1,(2),4)}\bigr)+\widehat\la\bigl(\widehat{\unl m}^{(1,2)}_{7\,1},\widehat{\unl m}^{(1,2)}_{7\,2},\widehat{\unl n}^{(1,2)}_{7\,1},\widehat{\unl n}^{(1,2)}_{7\,2}\bigr)\cr\cr
&&+\widehat\la^{1,2,3}\bigl(\widehat m^{(2,3,(4))},\widehat m^{((2),3,4)}_1,\widehat m^{(2,(3),4)},\widehat n^{(2,3,(4))},\widehat n^{((2),3,4)}_1,\widehat n^{(2,(3),4)}\bigr)\,,
\qqq
in which the symbols $\,\widehat n^{(i,j,(k))}_{(A)},\widehat n^{(i,(j),k)}_{(A)},\widehat n^{((i),j,k)}_{(A)},\widehat{\unl n}^{(i,j)}_{7\,A}\,$ are defined just as their counterparts with $\,n\,$ replaced by $\,m\,$ but with all coordinates tilded.

We conclude our analysis with 
\bedef\label{def:s2gerbe}
The \textbf{Green--Schwarz super-2-gerbe} of curvature $\,\underset{\tx{\ciut{(4)}}}{\txH}\,$ is the quintuple
\qq\nn
\sG^{(2)}_{\rm GS}:=\bigl(\sfY_2\cM^{(1)},\underset{\tx{\ciut{(3)}}}{\b}^{(4)},\widehat\xcG,\cM_{\widehat\xcG},\mu_{\widehat\xcG}\bigr)
\qqq
constructed in the preceding paragraphs.
\exdef

Our results are amenable to a straightforward abstraction in the spirit of Defs.\,\ref{def:CaEs0g} and \ref{def:CaEs1g}.  We leave it to the avid Reader to work out the obvious details of a definition of a Cartan--Eilenberg super-2-gerbe. Instead, we conclude the present work with the following postulative
\bedef\label{def:CaEspg}
Let $\,\txG\,$ be a Lie supergroup. A \textbf{Cartan--Eilenberg super-$p$-gerbe} over $\,\txG\,$ is a (bundle) $p$-gerbe, in the sense\footnote{Bundle $p$-gerbes were described as principal $B^p\bC^\x$-bundles with connection(s) geometrising (representatives of) classes in $\,H^{p+2}(M,\bZ)$.} of \Rcite{Gajer:1996}, with total spaces of all surjective submersions entering its definition endowed with the structure of a Lie supergroup and the submersions themselves -- Lie-supergroup homomorphisms, all connections invariant and all (connection-preserving) principal $\bC^\x$-bundle isomorphisms mapping homomorphically the respective Lie-supergroup structures into one another.
\exdef
With three explicit and physically relevant instantiations worked out in detail in the present paper, the above definition certainly awaits further elaboration and exemplification in topologically more complex supergeometric settings. Likewise, the definition of the Green--Schwarz super-$p$-gerbe on the super-Minkowski space begs for a generalisation that would place it in a category of supersymmetric (bundle) $p$-gerbes on homogeneous spaces of Lie supergroups. We leave these challenges to the future work.
\newpage

\section{Conclusions \& Outlook}\label{ref:CandO}

In the present paper, we put forward an essentially complete proposal of a novel geometrisation scheme for a family of super-$(p+2)$-cocyles representing classes in the Cartan--Eilenberg supersymmetry-invariant cohomology of the super-Minkowskian spacetime $\,{\rm sMink}(d,1\,\vert\,D_{d,1})\,$ (regarded as a Lie supergroup), of direct relevance to the construction of the Green--Schwarz super-$\si$-models of super-$p$-brane dynamics. The motivation for the geometrisation comes from the construction, due to Rabin and Crane \cite{Rabin:1984rm,Rabin:1985tv}, of an orbifold of the original supertarget $\,{\rm sMink}(d,1\,\vert\,D_{d,1})\,$ with respect to the natural geometric action of the discrete Kosteleck\'y--Rabin (lattice) supersymmetry group, the nontrivial topology of the orbifold being captured by the said Cartan--Eilenberg cohomology of the topologically trivial super-Minkowskian spacetime. The geometrisation scheme proposed hinges on the relation between the Cartan--Eilenberg cohomology of the Lie supergroup and the Chevalley--Eilenberg cohomology of its Lie superalgebra with values in the trivial module $\,\bR$,\ and on the correspondence between the second cohomology group in the latter cohomology and (equivalence classes) of supercentral extensions of the Lie superalgebra. It employs a family of Lie supergroups surjectively submersed over the original supertarget, of the type originally considered by de Azc\'arraga {\it et al.} \cite{Chryssomalakos:2000xd}, that arise from the supercentral extensions determined by distinguished super-2-cocycles methodically induced from the Green--Schwarz super-$(p+2)$-cocycles. These extended Lie supergroups were subsequently used as elementary ingredients in a construction of the super-$p$-gerbes, carried out explicitly for $\,p\in\{0,1,2\}$,\ along the lines of the standard bosonic geometrisation scheme for de Rham cocycles, due to Murray \cite{Murray:1994db}. \medskip 

The results reported in the present paper prompt a host of natural questions, and actually define a conrete formal context in which these may be formulated. Starting with those of the more fundamental nature, it is certainly tempting to seek an explicit relation between our construction and alternative approaches to supersymmetry in the context of superstring and related models, one such particularly attractive approach being at the heart of the proposal, originally conceived by Killingback \cite{Killingback:1986rd} and Witten \cite{Witten:1988dls}, elaborated by Freed \cite{Yau:1987}, recently revived by Freed and Moore \cite{Freed:2004yc}, and ultimately concretised in the higher-geometric language by Bunke \cite{Bunke:2009} ({\it cp}\ also \Rcite{Waldorf:2009uf} for an explicit construction), for a geometrisation of the Pfaffian bundle of the target-space Dirac operator, associated with fermionic contributions to the superstring path integral, in terms of a differential ${\rm String}$-structure on the target space. Another, and not entirely unrelated, idea that might -- given the r\^ole played by the algebra and (super)symmetry arguments in our construction -- lead to a deeper understanding of the geometrisation scheme proposed would be to look for an explicit and geometrically meaningful relation between the super-$p$-gerbes constructed in the present work, and in particular the towers of supercentral extensions of the Lie supergroups built over the super-Minkowski (resp.\ super-Poincar\'e) Lie supergroup, and the Lie-$n$-superalgebras and $L_\infty$-superalgebras of Baez {\it et al.} considered in Refs.\,\cite{Baez:2004hda6,Baez:2010ye,Huerta:2011ic}.

On the next level, we find directions in which the study initiated in the present paper could and should be completed. One such question follows directly from the identification of the super-$\si$-models under consideration as super-variants of the familiar WZW $\si$-models with standard Lie groups as targets -- the extensive knowledge of the symmetry structure of the latter raises concrete expectations concerning the amenability of various combinations of the rigid (or global) symmetries of the super-$\si$-models engendered by the left and right regular actions of the target Lie supergroup upon itself to gauging, and hence also the existence of a particular equivariant structure on the geometrisations constructed in this paper. These expectations will find a corroboration in the upcoming paper \cite{Suszek:2018sM}. Another one that complements the discourse developed herein in a manner absolutely critical from the physical point of view regards the actual (super)geometric and (super)algebraic content of the gauged supersymmetry, aka $\k$-symmetry, and its full-fledged (super-)gerbe-theoretic realisation in the (anticipated) form of an equivariant structure on the super-gerbe of the super-$\si$-model  -- this calls for a judicious reformulation of the super-$\si$-model and will also be discussed at length in \Rcite{Suszek:2018sM}. Finally, one is naturally led to launch an in-depth study of the (maximally) supersymmetric boundary conditions in the proposed higher-geometric formulation -- this points in the direction of a systematic study of super-$p$-gerbe (bi-)modules, or -- more generally -- the reconstruction of the associated higher categories of super-$p$-gerbes over $\,{\rm sMink}(d,1\,\vert\,D_{d,1})$.

Last but not least, our work paves the way to a variety of natural and interesting applications and extensions. One obvious line of development is the application of the formalism proposed to the super-$\si$-models on supertargets with the body of the general type $\,{\rm AdS}_{p+2}\x\bS^{d-p-2}\,$ whose exploration has led to remarkable progress in string theory, as seen from the phenomenological but also purely theoretical perspective. Here, the hope is that the ideas and constructions advanced in the present work prove sufficiently universal and technically robust to accomodate the extra complexity of these superbackgrounds whose super-$\si$-model description is -- after all -- structurally akin to that considered above. Another one that can be conceived is an explicit construction of a bosonisation/fermionisation defect (and the associated super-1-gerbe bi-brane) in the much tractable super-Minkowskian setting -- this promises to shed some light on the geometry behind the correspondence between worldsheet and target-space supersymmetry in superstring theory. We shall certainly return to these ideas in a future work.
\newpage

\appendix

\section{Conventions and Facts}\label{app:conv}

\becon\label{conv:asymm}
Fix natural numbers $\,m<n\in\bN\setminus\{0\}$.\ Given an arbitrary family $\,\{X_{i_1 i_2\ldots i_n}\}_{i_1,i_2,\ldots,i_n\in\xcI}\,$ of elements of an abelian group, indexed by a set $\,\xcI$,\ we define the (partial) symmetriser
\qq\nn
X_{(i_1 i_2\ldots i_m)i_{m+1}\ldots i_n}:=\tfrac{1}{m!}\,\sum_{\si\in\Sgt_m}\,X_{i_{\si(1)}i_{\si(2)}\ldots i_{\si(m)}i_{m+1}\ldots i_n}
\qqq
and the (partial) antisymmetriser
\qq\nn
X_{[i_1 i_2\ldots i_m]i_{m+1}\ldots i_n}:=\tfrac{1}{m!}\,\sum_{\si\in\Sgt_m}\,\sign(\si)\,X_{i_{\si(1)}i_{\si(2)}\ldots i_{\si(m)}i_{m+1}\ldots i_n}\,.
\qqq
\econ

\becon\label{conv:SignManifesto}
For the differential calculus on supermanifolds, we adopt the conventions of \Rcite{Deligne:1999sgn}. That is, given the standard coordinates $\,\{x^1,x^2,\ldots,x^d,\theta^1,\theta^2,\ldots,\theta^N\}\,$ on the superspace $\,\bR^{d\,\vert\,N}\,$ (parameterising locally a given supermanifold), we assign to a differential object $\,X\,$ (a superdifferential form, a supervector field \emph{etc.}) an additive bidegree composed of its Gra\ss mann and de Rham(-cohomology) degrees,
\qq\nn
\Deg(X):=\left(\widetilde{X},\deg_{\rm dR}(X)\right)\,. 
\qqq 
Thus, for the elementary objects, we have the assignments
\qq\nn
&\Deg(x^a)=(0,0)\,,\qquad\qquad\Deg(\theta^\a)=(1,0)\,,&\cr\cr
&\Deg(\sfd x^a)=(0,1)\,,\qquad\qquad\Deg(\sfd \theta^\a)=(1,1)\,,&\cr\cr
&\Deg(\tfrac{\p\ }{\p x^a})=(0,-1)\,,\qquad\qquad\Deg(\tfrac{\p\ }{\p\theta^\a})=(1,-1)\,.&
\qqq
Upon defining the product of bidegrees
\qq\nn
\corr{\Deg(X),\Deg(Y)}:=\widetilde{X}\cdot\widetilde{Y}+\deg_{\rm dR}(X)\cdot\deg_{\rm dR}(Y)\,,
\qqq
we have the bigraded commutativity relations
\qq\nn
XY=(-1)^{\corr{\Deg(X),\Deg(Y)}}\,YX\,.
\qqq
In particular, we find the following elementary supercommutation relations for the coordinate super-1-forms:
\qq\nn
& x^a\,x^b=x^b\,x^a\,,\qquad\qquad x^a\,\theta^\a=\theta^\a\,x^a\,,\qquad\qquad\theta^\a\,\theta^\b=-\theta^\b\,\theta^\a\,,&\cr\cr
& x^a\,\sfd x^b=\sfd x^b\,x^a\,,\qquad\qquad x^a\,\sfd\theta^\a=\sfd\theta^\a\,x^a\,,\qquad\qquad\theta^\a\,\sfd x^a=\sfd x^a\,\theta^\a\,,\qquad\qquad\theta^\a\,\sfd\theta^\b=-\sfd\theta^\b\,\theta^\a\,,&\cr\cr
&\sfd x^a\wedge\sfd x^b=-\sfd x^b\wedge\sfd x^a\,,\qquad\qquad\sfd x^a\wedge\sfd\theta^\a=-\sfd\theta^\a\wedge\sfd x^a\,,\qquad\qquad\sfd\theta^\a\wedge\sfd\theta^\b=\sfd\theta^\b\wedge\sfd\theta^\a\,.&
\qqq
These assignments are naturally ({\it i.e.}, additively) extended to object carrying multiple indices of the elementary type.
\econ

\becon\label{conv:Cliff}
For the Clifford algebra 
\qq\nn
\{\G_a,\G_b\}=2\eta_{ab}
\qqq
and its Majorana spin representations, we adopt the conventions of Refs.\,\cite{West:1998ey,Chryssomalakos:2000xd}, lowering resp.\ raising spacetime indices, wherever necessary, with the help of the Minkowskian metric $\,\eta\,$ resp.\ its inverse. Thus, in particular, the charge-conjugation matrix $\,C=(C_{\a\b})_{\a,\b\in\ovl{1,N}}\,$ has the properties
\qq
C\,\G^a\,C^{-1}=-\G^a{}^{\rm T}\,,\qquad\qquad C^{\rm T}=-\ep\,C\,,\quad\ep=\left\{ \barr{cl} -\sqrt{2}\,\cos\left(\tfrac{(d+2)\,\pi}{4}\right) & \tx{ if } d\in\{1,3,5,7,9\} \cr\cr -\sqrt{2}\,\cos\left(\tfrac{(d+1)\,\pi}{4}\right) & \tx{ if } d\in\{2,6,10\} \earr\right.\,,\cr\cr
\label{eq:Chargeprop}
\qqq
and we simply do not consider supertargets of dimensions and with properties other than those listed (although they are known to exist). Such restrictions ensure that -- \emph{for} $\,p>0\,$ -- all the matrices $\,C\,\G^{a_1 a_2\ldots a_p}\equiv C\,\G^{[a_1}\,\G^{a_2}\cdots\G^{a_p]}\,$ (and so also the matrices $\,C\,\G_{a_1 a_2\ldots a_p}=\eta_{a_1 b_1}\,\eta_{a_2 b_2}\cdots\eta_{a_p b_p}\,C\,\G^{b_1 b_2\ldots b_p}$) discussed in the main text are symmetric,  
\qq\label{eq:CGamSym}
\left(C\,\G^{a_1 a_2\ldots a_p}\right)^{\rm T}=C\,\G^{a_1 a_2\ldots a_p}\,,
\qqq
as long as the identity 
\qq
\ep=(-1)^{\frac{(p-1)\cdot(p-2)}{2}}
\qqq
is satisfied. As the basis of our concrete considerations is the standard Lie superalgebra \eqref{eq:sPoincalg}, we also want
\qq\label{eq:CGam2Sym}
\left(C\,\G^a\right)^{\rm T}=C\,\G^a\,,
\qqq
which further constrains the admissible pairs $\,(d,p)\,$ through the requirement
\qq\nn
\ep\must 1\,.
\qqq
For the sake of transparency of the formul\ae ~appearing in the article, we shall also use the shorthand notation
\qq\nn
\ovl\G{}^{a_1 a_2\ldots a_k}\equiv C\,\G^{a_1 a_2\ldots a_k}\,,\qquad\qquad\ovl\G_{a_1 a_2\ldots a_k}\equiv C\,\G_{a_1 a_2\ldots a_k}\,.
\qqq
Thus, in particular, the constructions discussed in the main text are well-defined for the following values of the dimensions: $\,d\in\{1,2,3,9,10\}\,$ for $\,p=1\,$ and $\,d=10\,$ ({\it i.a.}) for $\,p=2$.

The charge-conjugation matrix defines the fundamental bilinear form on spinors, 
\qq\nn
(\xi_1,\xi_2)\longmapsto\ovl\xi_1\,\xi_2\equiv\xi_1^\a\,C_{\a\b}\,\xi_2^\b\,,
\qqq
with the $\ep$-symmetry property
\qq\nn
\ovl\xi_2\,\xi_1=\ep\,\ovl\xi_1\,\xi_2\,.
\qqq
Note also the identity
\qq\label{eq:asymSpinVec}
\ovl\xi_2\,\G^a\,\xi_1=-\ovl\xi_1\,\G^a\,\xi_2
\qqq
that follows from \Reqref{eq:CGam2Sym} in the cases of interest.

In the distinguished case $\,(d,p)=(10,0)$,\ in which $\,\ep=1\,$ and hence \Reqref{eq:CGam2Sym} holds true, we also encounter the volume element of the corresponding Clifford algebra $\,\Cliff(\bR^{9,1})$,\ 
\qq\nn
\G_{11}:=\G^0\cdot\G^1\cdot\cdots\cdot\G^9\,.
\qqq 
It is readily seen to belong to the anticentre of $\,\Cliff(\bR^{1,9})$,
\qq\nn
\forall_{a\in\ovl{0,9}}\ :\ \{\G_{11},\G^a\}=0\,,
\qqq
and satisfy the elementary identites
\qq\nn
\G_{11}^2=\bd1_{D_{d,1}}\,,\qquad\qquad\ovl\G_{11}\equiv C\cdot\G_{11}=-\G^{\rm T}_{11}\cdot C\,.
\qqq
\econ

\section{A proof of Proposition \ref{prop:GSprim}}\label{app:GSprim}

In this section, we examine closed super-$(p+2)$-forms
\qq\nn
\underset{\tx{\ciut{(p+2)}}}{\txH}=\pr_1^*\bigl(\si\wedge\ovl\G_{a_1 a_2\ldots a_p}\,\si\bigr)\wedge e^{a_1 a_2\ldots a_p}\in Z^{p+2}_{\rm dR}\left({\rm sMink}^{d,1\,\vert\,ND_{d,1}}\right)\,.
\qqq
We have
\qq\nn
\underset{\tx{\ciut{(p+2)}}}{\txH}(\theta,x)=\sfd\left(\theta\,\ovl\G_{a_1 a_2\ldots a_p}\,\si(\theta)\wedge e^{a_1 a_2\ldots a_p}(\theta,x)\right)+\tfrac{p}{2}\,\theta\,\ovl\G_{a_1 a_2\ldots a_p}\,\si(\theta)\wedge\bigl(\si\wedge\ovl\G{}^{a_1}\,\si\bigr)(\theta)\wedge e^{a_2 a_3\ldots a_p}(\theta,x)\,,
\qqq
and so we may use the identity 
\qq\nn
(\ovl\G_{a_1 a_2\ldots a_p})_{\a(\b}\,\ovl\G{}^{a_1}_{\g\d)}=-\ovl\G{}^{a_1}_{\a(\b}\,(\ovl\G_{a_1 a_2\ldots a_p})_{\g\d)}\,,
\qqq
following directly from Eqs.\,\eqref{eq:ClifFierz} and \eqref{eq:CGamSym}, to rewrite the equality as
\qq\nn
\underset{\tx{\ciut{(p+2)}}}{\txH}(\theta,x)&=&\sfd\left(\theta\,\ovl\G_{a_1 a_2\ldots a_p}\,\si(\theta)\wedge e^{a_1 a_2\ldots a_p}(\theta,x)\right)-\tfrac{p}{2}\,\theta\,\ovl\G^{a_1}\,\si(\theta)\wedge\bigl(\si\wedge\ovl\G_{a_1 a_2\ldots a_p}\,\si\bigr)(\theta)\wedge e^{a_2 a_3\ldots a_p}(\theta,x)\cr\cr
&\equiv&\sfd\left(\theta\,\ovl\G_{a_1 a_2\ldots a_p}\,\si(\theta)\wedge e^{a_1 a_2\ldots a_p}(\theta,x)\right)+p\,\bigl(\si\wedge\ovl\G_{a_1 a_2\ldots a_p}\,\si\bigr)(\theta)\wedge\sfd x^{a_1}\wedge e^{a_2 a_3\ldots a_p}(\theta,x)\cr\cr
&&-p\,\underset{\tx{\ciut{(p+2)}}}{\txH}(\theta,x)\,.
\qqq
Thus,
\qq\nn
\underset{\tx{\ciut{(p+2)}}}{\txH}(\theta,x)&=&\sfd\left(\tfrac{1}{p+1}\,\theta\,\ovl\G_{a_1 a_2\ldots a_p}\,\si(\theta)\wedge e^{a_1 a_2\ldots a_p}(\theta,x)\right)+\tfrac{p}{p+1}\,\sfd x^{a_1}\wedge\underset{\tx{\ciut{(p+1)}}}{\chi}{}_{a_1}(\theta,x)
\qqq
with 
\qq\nn
\underset{\tx{\ciut{(p+1)}}}{\chi}{}_{a_1}:=\pr_1^*\bigl(\si\wedge\ovl\G_{a_1 a_2\ldots a_p}\,\si\bigr)\wedge e^{a_2 a_3\ldots a_p}\,,
\qqq
and we may next focus on the latter super-$(p+1)$-form. Reasoning as in the previous step, we find
\qq\nn
\underset{\tx{\ciut{(p+1)}}}{\chi}{}_{a_1}(\theta,x)&=&\sfd\left(\theta\,\ovl\G_{a_1 a_2\ldots a_p}\,\si(\theta)\wedge e^{a_2 a_3\ldots a_p}(\theta,x)\right)-\tfrac{p-1}{2}\,\theta\,\ovl\G{}^{a_2}\,\si(\theta)\wedge\bigl(\si\wedge\ovl\G_{a_1 a_2\ldots a_p}\,\si\bigr)(\theta)\wedge e^{a_3 a_4\ldots a_p}(\theta,x)\cr\cr
&=&\sfd\left(\theta\,\ovl\G_{a_1 a_2\ldots a_p}\,\si(\theta)\wedge e^{a_2 a_3\ldots a_p}(\theta,x)\right)-(p-1)\,\underset{\tx{\ciut{(p+1)}}}{\chi}{}_{a_1}(\theta,x)+(p-1)\,\sfd x^{a_2}\wedge\underset{\tx{\ciut{(p)}}}{\chi}{}_{a_1 a_2}(\theta,x)\,,
\qqq
or
\qq\nn
\underset{\tx{\ciut{(p+1)}}}{\chi}{}_{a_1}(\theta,x)=\sfd\left(\tfrac{1}{p}\,\theta\,\ovl\G_{a_1 a_2\ldots a_p}\,\si(\theta)\wedge e^{a_2 a_3\ldots a_p}(\theta,x)\right)+\tfrac{p-1}{p}\,\sfd x^{a_2}\wedge\underset{\tx{\ciut{(p)}}}{\chi}{}_{a_1 a_2}(\theta,x)
\qqq
with
\qq\nn
\underset{\tx{\ciut{(p)}}}{\chi}{}_{a_1 a_2}:=\pr_1^*\bigl(\si\wedge\ovl\G_{a_1 a_2\ldots a_p}\,\si\bigr)\wedge e^{a_3 a_4\ldots a_p}\,.
\qqq 
Repeating the above reduction procedure $p$ times, we eventually arrive at the equality
\qq\nn
\underset{\tx{\ciut{(p+1)}}}{\chi}{}_{a_1}=\sfd\underset{\tx{\ciut{(p)}}}{\b}{}_{a_1}\,,
\qqq
with 
\qq\label{eq:primba}
\underset{\tx{\ciut{(p)}}}{\b}{}_{a_1}(\theta,x)=\tfrac{1}{p}\,\sum_{k=1}^p\,\theta\,\ovl\G_{a_1 a_2\ldots a_p}\,\si(\theta)\wedge\sfd x^{a_2}\wedge\cdots\wedge\sfd x^{a_k}\wedge e^{a_{k+1} a_{k+2}\ldots a_p}(\theta,x)
\qqq
whence also
\qq\nn
\underset{\tx{\ciut{(p+1)}}}{\b}(\theta,x)=\tfrac{1}{p+1}\,\sum_{k=0}^p\,\theta\,\ovl\G_{a_1 a_2\ldots a_p}\,\si(\theta)\wedge\sfd x^{a_1}\wedge\sfd x^{a_2}\wedge\cdots\wedge\sfd x^{a_k}\wedge e^{a_{k+1} a_{k+2}\ldots a_p}(\theta,x)\,.
\qqq
\qed
\bigskip

\section{The Lie-superalgebra cohomology and its Chevalley--Eilenberg model}\label{app:LieAlgCohom}

In this appendix, we collect basic facts concerning the Lie-superalgebra cohomology that prove useful in an algebraic description of supertargets and of their differential geometry. In our exposition and discussion, we adopt the conventions of the original articles: \cite{Berezin:1970} by Berezin and Ka\v c, and \cite{Leites:1975} by Le{\"i}tes. 

We begin with the basic
\bedef\label{def:LSA}
A \textbf{Lie superalgebra} (to be abbreviated as \textbf{LSA}) over field $\,\bK\,$ is a pair $\,\left(\ggt,[\cdot,\cdot\}_\ggt\right)\,$ composed of a $\bK$-linear space $\,\ggt\,$ endowed with a $\bZ/2\bZ$-grading ({\it i.e.}, a \textbf{supervector space}, or a \textbf{$\bK$-linear superspace}) $\,|\cdot|_\ggt\,$ that induces a decomposition $\,\ggt=\ggt^{(0)}\oplus\ggt^{(1)}\,$ into a direct sum of homogeneous components, $\,|\cdot|_\ggt\rstr_{\ggt^{(n)}}=n$,\ and of a \textbf{Lie superbracket} (also termed a \textbf{supercommutator})
\qq\nn
[\cdot,\cdot\}_\ggt\ :\ \ggt\x\ggt\too\ggt\ :\ (X_1,X_2)\longmapsto[X_1,X_2\}_\ggt=-(-1)^{|X_1|_\ggt\cdot|X_2|_\ggt}[X_2,X_1\}_\ggt\,, 
\qqq
that preserves the grading,
\qq\nn
|[X,Y\}_\ggt|_\ggt\equiv|X|_\ggt+|Y|_\ggt\mod 2
\qqq
(written for arbitrary homogeneous $\,X,Y\in\ggt$), and has a vanishing \textbf{super-Jacobiator} (evaluated on arbitrary homogeneous elements $\,X_1,X_2,X_3\in\ggt$)
\qq\label{eq:sJac}
{\rm sJac}_\ggt(X_1,X_2,X_3)\\ \cr
:=(-1)^{|X_1|_\ggt\cdot|X_3|_\ggt}\,[[X_1,X_2\}_\ggt,X_3\}_\ggt+(-1)^{|X_3|_\ggt\cdot|X_2|_\ggt}\,[[X_3,X_1\}_\ggt,X_2\}_\ggt+(-1)^{|X_2|_\ggt\cdot|X_1|_\ggt}\,[[X_2,X_3\}_\ggt,X_1\}_\ggt=0\,.\nonumber
\qqq
Given two LSAs $\,\left(\ggt_A,[\cdot,\cdot\}_{\ggt_A}\right),\ A\in\{1,2\}$,\ an \textbf{LSA morphism} between them is a $\bK$-linear map $\,\chi\ :\ \ggt_1\too\ggt_2\,$ that preserves the $\bZ/2\bZ$-grading, $\,|\cdot|_{\ggt_2}\circ\chi=|\cdot|_{\ggt_1}$,\ and the Lie superbracket, 
\qq\nn
\chi\circ[\cdot,\cdot\}_{\ggt_1}=[\cdot,\cdot\}_{\ggt_2}\circ(\chi\x\chi)\,.
\qqq

A (\textbf{left}) \textbf{$\ggt$-module} is a pair $\,(V,\ell_\cdot)\,$ composed of a $\bK$-linear superspace with a decomposition $\,V=V^{(0)}\oplus V^{(1)}\,$ into homogeneous components induced by the $\bZ/2\bZ$-grading $\,|\cdot|_V$,\ and endowed with a left $\ggt$-action 
\qq\nn
\ell_\cdot\ :\ \ggt\x V\too V\ :\ (X,v)\longmapsto X\lact v
\qqq 
consistent with the $\bZ/2\bZ$-gradings, 
\qq\nn
|X\lact v|_V\equiv|X|_\ggt+|v|_V\mod 2\,,
\qqq 
and such that for any two homogeneous elements $\,X_1,X_2\in\ggt\,$ and $\,v\in V$, 
\qq\nn
[X_1,X_2\}_\ggt\lact v=X_1\lact(X_2\lact v)-(-1)^{|X_1|_\ggt\cdot|X_2|_\ggt}\,X_2\lact(X_1\lact v)\,.
\qqq
\exdef
\noindent The object of our main interest is introduced in
\bedef
Let $\,\left(\ggt,[\cdot,\cdot\}_\ggt\right)\,$ and $\,\left(\agt,[\cdot,\cdot\}_\agt\right)\,$ be two LSAs over a base field $\,\bK$.\ A \textbf{supercentral extension of $\,\ggt\,$ by $\,\agt\,$} is an LSA $\,\left(\widetilde\ggt,[\cdot,\cdot\}_{\widetilde\ggt}\right)\,$ over $\,\bK\,$ described by the short exact sequence of LSAs
\qq\label{eq:LSASES}
\brd0\too\agt\xrightarrow{\ \jmath_\agt\ }\widetilde\ggt\xrightarrow{\ \pi_\ggt\ }\ggt\too\brd0\,,
\qqq 
written in terms of an LSA monomorphism $\,\jmath_\agt\,$ and of an LSA epimorphism $\,\pi_\ggt$,\ and such that $\,\jmath_\agt(\agt)\subset\zgt(\widetilde\ggt)\,$ (the (super)centre of $\,\widetilde\ggt$). Hence, in particular, $\,\agt\,$ is necessarily supercommutative, that is $\,[\cdot,\cdot\}_\agt\equiv 0$.

Whenever $\,\pi_\ggt\,$ admits a \textbf{section} that is an LSA homomorphism, \textit{i.e.}, there exists
\qq\nn
\si\in\Hom_{\rm sLie}(\ggt,\widetilde\ggt)\,,\qquad\pi_\ggt\circ\si=\id_\ggt\,,
\qqq
the central extension is said to \textbf{split}.

An equivalence of central extensions $\,\widetilde\ggt_A, A\in\{1,2\}\,$ of $\,\ggt\,$ by $\,\agt\,$ is represented by a commutative diagram 
\qq\nn
\alxydim{@C=.75cm@R=.5cm}{ & & \widetilde\ggt_1 \ar[dd]^{\cong} \ar[dr] & & \\ \brd0 \ar[r] & \agt \ar[ur] \ar[dr] & & \ggt \ar[r] & \brd0 \\ & & \widetilde\ggt_2 \ar[ur] & & }\,,
\qqq
in which the vertical arrow is an LSA isomorphism.
\exdef
\noindent In close analogy with the purely Gra\ss mann-even case, equivalence classes of central extensions of LSAs are neatly captured by the cohomology of the latter. The relevant cohomology is specified in 
\bedef\label{def:LSAcohom}
Let $\,\left(\ggt,[\cdot,\cdot\}_\ggt\right)\,$ be an LSA over field $\,\bK\,$ and let $\,(V,\ell_\cdot)\,$ be a $\ggt$-module. A \textbf{$p$-cochain on $\,\ggt\,$ with values in} $\,V\,$ (also termed a \textbf{$p$-form on $\,\ggt\,$ with values in $\,V$}) is a $p$-linear map $\,\underset{\tx{\ciut{(p)}}}{\varphi}:\ggt^{\x p}\too V\,$ that is totally super-skewsymmetric, \textit{i.e.}, for any homogeneous elements $\,X_i\in\ggt,\ i\in\ovl{1,p}$,\ it satisfies
\qq\nn
\forall_{j\in\ovl{1,p-1}}\ :\ \underset{\tx{\ciut{(p)}}}{\varphi}(X_1,X_2,\ldots,X_{j-1},X_{j+1},X_j,X_{j+2},X_{j+3},\ldots,X_p)=-(-1)^{|X_j|_\ggt\cdot|X_{j+1}|_\ggt}\,\underset{\tx{\ciut{(p)}}}{\varphi}(X_1,X_2,\ldots,X_p)\,.
\qqq
Such maps form a $\bZ/2\bZ$-graded \textbf{group of $p$-cochains on $\,\ggt\,$ with values in} $\,V$,\ with the respective gradations $\,|\cdot|_p$,\ denoted by 
\qq\nn
C^p(\ggt,V)=C^p_0(\ggt,V)\oplus C^p_1(\ggt,V)\,,
\qqq 
with $\,\underset{\tx{\ciut{(p)}}}{\varphi}(X_1,X_2,\ldots,X_p)\in V_{\sum_{i=1}^p\,|X_i|_\ggt+n\mod 1}\,$ for $\,\underset{\tx{\ciut{(p)}}}{\varphi}\in C^p_n(\ggt,V)$,\ composed of \textbf{even} ($n=0$) and \textbf{odd} ($n=1$) $p$-cochains.

The family of these groups indexed by $\,p\in\bN\,$ forms a semi-bounded complex 
\qq\nn
C^\bullet(\ggt,V)\ :\ C^0(\ggt,V)\xrightarrow{\ \d_\ggt^{(0)}\ }C^1(\ggt,V)\xrightarrow{\ \d_\ggt^{(1)}\ }\cdots\xrightarrow{\ \d_\ggt^{(p-1)}\ }C^p(\ggt,V)\xrightarrow{\ \d_\ggt^{(p)}\ }\cdots
\qqq 
with the coboundary operators $\,\d_\ggt^{(p)}:C^p_n(\ggt,V)\too C^{p+1}_n(\ggt,V)\,$ determined by the formul\ae ~ (written for homogeneous elements $\,X,X_i\in\ggt,\ i\in\ovl{1,p+1}\,$ and $\,\underset{\tx{\ciut{(p)}}}{\varphi}\in C^p(\ggt,V)$)
\qq\nn
\bigl(\d_\ggt^{(0)}\underset{\tx{\ciut{(0)}}}{\varphi}\bigr)(X)&:=&(-1)^{|X|_\ggt\cdot|\underset{\tx{\ciut{(0)}}}{\varphi}|_0}\,X\lact\underset{\tx{\ciut{(0)}}}{\varphi}\,,\cr\cr
\bigl(\d_\ggt^{(p)}\underset{\tx{\ciut{(p)}}}{\varphi}\bigr)(X_1,X_2,\ldots,X_{p+1})&:=&\sum_{i=1}^{p+1}\,(-1)^{|X_i|_\ggt\,|\underset{\tx{\ciut{(p)}}}{\varphi}|_p+S(|X_i|_\ggt)}\,X_i\lact\underset{\tx{\ciut{(p)}}}{\varphi}(X_1,X_2,\underset{\widehat i}{\ldots},X_{p+1})\cr\cr
&&+\sum_{1\leq i<j\leq p+1}\,(-1)^{S(|X_i|_\ggt)+S(|X_j|_\ggt)+|X_i|_\ggt\cdot|X_j|_\ggt}\,\underset{\tx{\ciut{(p)}}}{\varphi}([X_i,X_j\},X_1,X_2,\underset{\widehat{i,j}}{\ldots},X_{p+1})\,,
\qqq
where
\qq\label{eq:Sumaltweights}
S(|X_i|_\ggt):=|X_i|_\ggt\cdot\sum_{j=1}^{i-1}\,|X_j|_\ggt+i-1\,.
\qqq
We distinguish the \textbf{group of $p$-cocycles}
\qq\nn
Z^p(\ggt,V):=\ker\,\d_\ggt^{(p)}\,,
\qqq
and the \textbf{group of $p$-coboundaries}
\qq\nn
B^p(\ggt,V):=\im\,\d_\ggt^{(p-1)}\,.
\qqq
The $\bZ/2\bZ$-graded homology groups of the complex $\,\left(C^\bullet(\ggt,V),\d_\ggt^{(\bullet)}\right)\,$ are called the \textbf{cohomology groups of $\,\ggt\,$ with values in} $\,V\,$ and denoted by
\qq\nn
H^p(\ggt,V):=H^p_0(\ggt,V)\oplus H^p_1(\ggt,V)\,,\qquad H^p_n(\ggt,V):=\frac{\ker\,\d_\ggt^{(p)}\rstr_{C^p_n(\ggt,V)}}{\im\,\d_\ggt^{(p-1)}\rstr_{C^{p-1}_n(\ggt,V)}}\,.
\qqq
\exdef
\noindent Let us write out -- with view to subsequent considerations and applications -- the relations defining a 2-cocycle and a 2-coboundary on an LSA $\,\ggt\,$ with values in a trivial $\ggt$-module $\,\agt$.\ Thus, for any homogeneous elements $\,X_1,X_2,X_3\in\ggt$,\ a 2-cocycle $\,\Th\in Z^2(\ggt,\agt)\,$ satisfies 
\qq\nn
(-1)^{|X_1|_\ggt\cdot|X_3|_\ggt}\,\Th\left([X_1,X_2\}_\ggt,X_3\right)+(-1)^{|X_3|_\ggt\cdot|X_2|_\ggt}\,\Th\left([X_3,X_1\}_\ggt,X_2\right)+(-1)^{|X_2|_\ggt\cdot|X_1|_\ggt}\,\Th\left([X_2,X_3\}_\ggt,X_1\right)=0\,,
\qqq
and a 2-coboundary obtained from a 1-cochain $\,\mu\in C^1(\ggt,\agt)\,$ evaluates as
\qq\nn
\bigl(\d_\ggt^{(1)}\mu\bigr)(X_1,X_2)=-\mu\left([X_1,X_2\}_\ggt\right)\,.
\qqq

\subsection{The algebraic meaning of $\,H^2_0(\ggt,\agt)$}

We shall now establish a natural correspondence between classes in $\,H^2(\ggt,\agt)\,$ and equivalence classes of supercentral extensions of $\,\ggt\,$ by a supercommutative LSA $\,\agt\,$ with a trivial $\ggt$-action. We begin our discussion with 
\berop\label{prop:ExtoCE}
Let $\,\left(\ggt,[\cdot,\cdot\}_\ggt\right)\,$ be an LSA, and let $\,\agt\,$ be a supercommutative LSA. An equivalence class of supercentral extensions $\,\left(\widetilde\ggt,[\cdot,\cdot\}_{\widetilde\ggt}\right)\,$ of $\,\ggt\,$ by $\,\agt\,$ canonically determines a class in $\,H^2_0(\ggt,\agt)$.\ This class vanishes iff the short exact sequence determined by the extensions splits.
\eerop
\beroof
The short exact sequence \eqref{eq:LSASES} implies the existence of a $\bK$-linear map $\,\si:\ggt\too\widetilde\ggt\,$ which preserves the $\bZ/2\bZ$-grading and satisfies the relation $\,\pi_\ggt\circ\si=\id_\ggt$,\ whence the (canonical) isomorphism of $\bK$-linear superspaces
\qq\nn
\widetilde\iota\ :\ \widetilde\ggt\xrightarrow{\ \cong\ }\agt\oplus\ggt\ :\ \widetilde X\longmapsto\bigl(\jmath_\agt^{-1}\left(\widetilde X-\si\circ\pi_\ggt(\widetilde X)\right),\pi_\ggt(\widetilde X)\bigr)\,.
\qqq
Indeed, the above map is well-defined as $\,\widetilde X-\si\circ\pi_\ggt(\widetilde X)\in\ker\,\pi_\ggt=\im\,\jmath_\agt\,$ and $\,\jmath_\agt\,$ is an isomorphism onto its image. Its inverse is explicitly given by
\qq\nn
\widetilde\iota^{-1}\ :\ \agt\oplus\ggt\too\widetilde\ggt\ :\ (A,X)\longmapsto\jmath_\agt(A)+\si(X)\,.
\qqq
We may, subsequently, promote $\,\widetilde\iota\,$ to the rank of an LSA isomorphism by inducing a Lie superbracket on the vector superspace $\,\agt\oplus\ggt\,$ from those on $\,\widetilde\ggt\,$ and $\,\ggt\,$ as per
\qq\nn
[(A_1,X_1),(A_2,X_2)\}_{\agt\oplus\ggt}&:=&\widetilde\iota\left([\widetilde\iota^{-1}(A_1,X_1),\widetilde\iota^{-1}(A_2,X_2)\}_{\widetilde\ggt}\right)\equiv\widetilde\iota\left([\si(X_1),\si(X_2)\}_{\widetilde\ggt}\right)\cr\cr
&=&\bigl(\jmath_\agt^{-1}\left([\si(X_1),\si(X_2)\}_{\widetilde\ggt}-\si\left([X_1,X_2\}_\ggt\right)\right),[X_1,X_2\}_\ggt\bigr)\,.
\qqq
Consistency of this definition is ensured by the properties of the $p$-linear map
\qq\nn
\Th_\si\ :\ \ggt^{\x 2}\too\agt\ :\ (X_1,X_2)\longmapsto\jmath_\agt^{-1}\left([\si(X_1),\si(X_2)\}_{\widetilde\ggt}-\si\left([X_1,X_2\}_\ggt\right)\right)\,.
\qqq
When evaluated on a pair of homogeneous elements of $\,\ggt$,\ it satisfies (in view of preservation of the $\bZ/2\bZ$-grading by $\,\si\,$ and $\,\jmath_\agt$)
\qq\nn
\Th_\si(X_2,X_1)=-(-1)^{|X_1|_\ggt\cdot|X_2|_\ggt}\,\Th_\si(X_1,X_2)\,,
\qqq
and so it is an even 2-cochain on $\,\ggt\,$ with values in $\,\agt$,\ the latter being understood as a trivial $\ggt$-module. Its coboundary reads
\qq\nn
\d_\ggt^{(2)}\Th_\si(X_1,X_2,X_3)=(-1)^{|X_1|_\ggt\cdot|X_3|_\ggt+1}\,\bigl[(-1)^{|X_1|_\ggt\cdot|X_3|_\ggt}\,\Th_\si\left([X_1,X_2\}_\ggt,X_3\right)+(-1)^{|X_3|_\ggt\cdot|X_2|_\ggt}\,\Th_\si\left([X_3,X_1\}_\ggt,X_2\right)\cr\cr
+(-1)^{|X_2|_\ggt\cdot|X_1|_\ggt}\,\Th_\si\left([X_2,X_3\}_\ggt,X_1\right)\bigr]=(-1)^{|X_1|_\ggt\cdot|X_3|_\ggt+1}\,\jmath_\agt^{-1}\bigl((-1)^{|X_1|_\ggt\cdot|X_3|_\ggt}\,[\si\left([X_1,X_2\}_\ggt\right),\si(X_3)\}_{\widetilde\ggt}\cr\cr
+(-1)^{|X_3|_\ggt\cdot|X_2|_\ggt}\,[\si\left([X_3,X_1\}_\ggt\right),\si(X_2)\}_{\widetilde\ggt}+(-1)^{|X_2|_\ggt\cdot|X_1|_\ggt}\,[\si\left([X_2,X_3\}_\ggt\right),\si(X_1)\}_{\widetilde\ggt}-\si\circ{\rm sJac}_\ggt(X_1,X_2,X_3)\bigr)\cr\cr
=(-1)^{|X_1|_\ggt\cdot|X_3|_\ggt}\,\jmath_\agt^{-1}\bigl((-1)^{|X_1|_\ggt\cdot|X_3|_\ggt}\,[\jmath_\agt\circ\Th_\si(X_1,X_2),\si(X_3)\}_{\widetilde\ggt}+(-1)^{|X_3|_\ggt\cdot|X_2|_\ggt}\,[\jmath_\agt\circ\Th_\si(X_3,X_1),\si(X_2)\}_{\widetilde\ggt}\cr\cr
+(-1)^{|X_2|_\ggt\cdot|X_1|_\ggt}\,[\jmath_\agt\circ\Th_\si(X_2,X_3),\si(X_1)\}_{\widetilde\ggt}
-{\rm sJac}_{\widetilde\ggt}\left(\si(X_1),\si(X_2),\si(X_3)\right)+\si\circ{\rm sJac}_\ggt(X_1,X_2,X_3)\bigr)=0\,,
\qqq
where in the last step we invoked the inclusion $\,\im\jmath_\agt\subset\zgt(\widetilde\ggt)$.\ We now readily find, for the induced Lie superbracket, the desired result
\qq\nn
{\rm sJac}_{\agt\oplus\ggt}\bigl((A_1,X_1),(A_2,X_2),(A_3,X_3)\bigr)=\bigl((-1)^{|X_1|_\ggt\cdot|X_3|_\ggt+1}\,\d_\ggt^{(2)}\Th_\si(X_1,X_2,X_3),{\rm sJac}_\ggt(X_1,X_2,X_3)\bigr)=(0,0)\,,
\qqq
and conclude that the supercentral extension does, indeed, canonically determine a 2-cocycle on $\,\ggt\,$ with values in $\,\agt$.

Let us, next, examine how the 2-cocycle changes when we pass to an equivalent supercentral extension. We now have two LSA monomorphisms $\,\jmath_{\agt,A}\ :\ \agt\too\widetilde\ggt_A,\ A\in\{1,2\}\,$ and two LSA epimorphisms $\,\pi_{\ggt,A}:\widetilde\ggt_A\too\ggt\,$ with the corresponding $\bK$-linear superspace sections $\,\si_A\ :\ \ggt\too\widetilde\ggt_A$.\ Taking into account the commutativity of the diagram
\qq\nn
\alxydim{@C=1.25cm@R=1cm}{ & & \widetilde\ggt^{(1)} \ar[dd]_\vep \ar[dr]^{\pi_{\ggt,1}} & & \\ \brd0 \ar[r] & \agt \ar[ur]^{\jmath_{\agt,1}} \ar[dr]_{\jmath_{\agt,2}} & & \ggt \ar[r] \ar@/^1pc/@{-->}[ul]^{\hspace{50pt}\si_1} \ar@/_1pc/@{-->}[dl]_{\hspace{50pt}\si_2} & \brd0 \\ & & \widetilde\ggt_2 \ar[ur]_{\pi_{\ggt,2}} & & }\,,
\qqq 
alongside the identity
\qq\nn
\pi_{\ggt,1}\circ\left(\vep^{-1}\circ\si_2-\si_1\right)=\pi_{\ggt,2}\circ\si_2-\pi_{\ggt,1}\circ\si_1=0\,,
\qqq
the latter implying the existence of a $\bK$-linear superspace homomorphism $\,\mu_\vep\ :\ \ggt\too\agt\,$ such that 
\qq\nn
\vep^{-1}\circ\si_2-\si_1=\jmath_{\agt,1}\circ\mu_\vep\,,
\qqq
we readily establish, for any homogeneous elements $\,X_1,X_2\in\ggt$,
\qq\nn
\jmath_{\agt,1}\circ(\Th_{\si_2}-\Th_{\si_1})(X_1,X_2)=\left(\vep^{-1}\circ\jmath_{\agt,2}\circ\Th_{\si_2}-\jmath_{\agt,1}\circ\Th_{\si_1}\right)(X_1,X_2)\cr\cr
=[\vep^{-1}\circ\si_2(X_1),\vep^{-1}\circ\si_2(X_2)\}_{\widetilde\ggt^{(1)}}-[\si_1(X_1),\si_1(X_2)\}_{\widetilde\ggt^{(1)}}-\jmath_{\agt,1}\circ\mu_\vep([X_1,X_2\}_\ggt)\cr\cr
=[\jmath_{\agt,1}\circ\mu_\vep(X_1),\vep^{-1}\circ\si_2(X_2)\}_{\widetilde\ggt^{(1)}}+[\si_1(X_1),\vep^{-1}\circ\si_2(X_2)\}_{\widetilde\ggt^{(1)}}-[\si_1(X_1),\si_1(X_2)\}_{\widetilde\ggt^{(1)}}\cr\cr
-\jmath_{\agt,1}\circ\mu_\vep([X_1,X_2\}_\ggt)=[\si_1(X_1),\jmath_{\agt,1}\circ\mu_\vep(X_2)\}_{\widetilde\ggt^{(1)}}-\jmath_{\agt,1}\circ\mu_\vep([X_1,X_2\}_\ggt)\cr\cr
=-\jmath_{\agt,1}\circ\mu_\vep([X_1,X_2\}_\ggt)\,,
\qqq
so that, altogether,
\qq\nn
\Th_{\si_2}-\Th_{\si_1}=\d^{(1)}_\ggt\mu_\vep\,,\qquad\tx{\textit{i.e.}},\qquad[\Th_{\si_2}]_\ggt=[\Th_{\si_1}]_\ggt\,.
\qqq

Finally, we prove the last statement of the proposition. The vanishing of (the class of) the 2-cocycle $\,\Th_\si\,$ for $\,\si\,$ an LSA section is obvious, hence it remains to demonstrate that, conversely, the cohomological triviality of $\,\Th_\si\,$ implies the existence of an LSA section. The statement of triviality of the 2-cocycle $\,\Th_\si\,$ rewrites neatly as
\qq\nn
[\si(X_1),\si(X_2)\}_{\widetilde\ggt}=\si_\mu([X_1,X_2\}_\ggt)\,,\qquad\si_\mu:=\si-\jmath_\agt\circ\mu\in\Hom_\bK(\ggt,\widetilde\ggt)\,. 
\qqq 
In view of the supercommutativity of $\,\jmath_\agt(\agt)$,\ this yields
\qq\nn
[\si_\mu(X_1),\si_\mu(X_2)\}_{\widetilde\ggt}=\si_\mu\bigl([X_1,X_2\}_\ggt\bigr)\,,
\qqq
and so $\,\si_\mu\,$ can be promoted to the rank of an LSA homomorphism. Since, furthermore, it satisfies the relation
\qq\nn
\pi_\ggt\circ\si_\mu=\pi_\ggt\circ\si-\pi_\ggt\circ\jmath_\agt\circ\mu=\pi_\ggt\circ\si=\id_\ggt\,,
\qqq
we can identify it as the sought-after LSA section of $\,\pi_\ggt$.
\eroof

From the point of view of physical applications in the setting of the GS super-$\si$-model, it is of utmost significance that the assignment of classes in $\,H^2_0(\ggt,\agt)\,$ to supercentral extensions of $\,\ggt\,$ by a supercommutative LSA $\,\agt\,$ detailed above may, in fact, be inverted. This is stated in
\berop\label{prop:CEtoExt}
Let $\,\left(\ggt,[\cdot,\cdot\}_\ggt\right)\,$ be an LSA, and let $\,\agt\,$ be a supercommutative LSA, regarded as a trivial $\ggt$-module. A class in $\,H^2_0(\ggt,\agt)\,$ canonically induces an equivalence class of supercentral extensions $\,\left(\widetilde\ggt,[\cdot,\cdot\}_{\widetilde\ggt}\right)\,$ of $\,\ggt\,$ by $\,\agt$.\ The extensions split iff the former class vanishes.
\eerop
\beroof
Given an arbitrary even 2-cocycle $\,\Th\in Z^2_0(\ggt,\agt)$,\ endow the $\bK$-linear space $\,\agt\oplus\ggt=:\widetilde\ggt\,$ with a $\bZ/2\bZ$-grading induced from that of its direct summands with a manifestly super-skewsymmetric 2-linear map
\qq\nn
[\cdot,\cdot\}_\Th\ :\ \widetilde\ggt^{\x 2}\too\widetilde\ggt\ :\ \bigl((A_1,X_1),(A_2,X_2)\bigr)\longmapsto\bigl(\Th(X_1,X_2),[X_1,X_2\}_\ggt\bigr)\,.
\qqq
It is easily checked that the map is a Lie superbracket,
\qq\nn
{\rm sJac}_{\widetilde\ggt}\bigl((A_1,X_1),(A_2,X_2),(A_3,X_3)\bigr)=\bigl((-1)^{|X_1|_\ggt\cdot|X_3|_\ggt+1}\,\d^{(2)}_\ggt\Th(X_1,X_2,X_3),{\rm sJac}_\ggt(X_1,X_2,X_3)\bigr)=(0,0)\,,
\qqq
and so $\,\left(\widetilde\ggt,[\cdot,\cdot\}_\Th\right)\,$ is an LSA.

The LSA $\,\agt\,$ being supercommutative, the standard injection $\,\jmath_\agt\ :\ \agt\too\widetilde\ggt\ :\ A\longmapsto(A,0)\,$ is an LSA monomorphism. The $\bK$-linear canonical projection $\,\pi_\ggt\ :\ \widetilde\ggt\too\ggt\ :\ (A,X)\longmapsto X$,\ on the other hand, is readily seen to be an LSA epimorphism with the obvious property $\,\ker\,\pi_\ggt=\im\,\jmath_\agt$,\ and so we obtain a short exact sequence of LSAs
\qq\nn
\brd0\too\agt\xrightarrow{\ \jmath_\agt\ }\widetilde\ggt\xrightarrow{\ \pi_\ggt\ }\ggt\too\brd0
\qqq
that identifies $\,\widetilde\ggt\,$ as a supercentral extension of $\,\ggt\,$ by $\,\agt$.

For cohomologous 2-cocycles, $\,\Th_2=\Th_1+\d_\ggt^{(1)}\mu,\ \mu\in C^1_0(\ggt,\agt)$,\ the scheme laid out above produces two Lie superbrackets on the supercentral extension $\,\widetilde\ggt=\agt\oplus\ggt\,$ of $\,\ggt\,$ by $\,\agt$,\ and the $\bK$-linear superspace automorphism 
\qq\nn
\vep_\mu\ :\ \widetilde\ggt\too\widetilde\ggt\ :\ (A,X)\longmapsto\bigl(A-\mu(X),X\bigr)
\qqq 
is easily verified to isomorphically map $\,(\widetilde\ggt,[\cdot,\cdot\}_{\Th_1})\,$ onto $\,(\widetilde\ggt,[\cdot,\cdot\}_{\Th_2})$,
\qq\nn
[\vep_\mu(A_1,X_1),\vep_\mu(A_2,X_2)\}_{\Th_2}&=&\bigl(\Th_2(X_1,X_2),[X_1,X_2\}_\ggt\bigr)=\bigl(\Th_1(X_1,X_2)-\mu([X_1,X_2\}_\ggt),[X_1,X_2\}_\ggt\bigr)\cr\cr
&\equiv&\vep_\mu\bigl([(A_1,X_1),(A_2,X_2)\}_{\Th_1}\bigr)\,.
\qqq

Whenever $\,\Th\,$ is a 2-coboundary, $\,\Th=\d_\ggt^{(1)}\mu,\ \mu\in C^1_0(\ggt,\agt)$,\ we may inject $\,\ggt\,$ into $\,\widetilde\ggt\,$ by means of a $\bK$-linear map 
\qq\nn
\si_\mu\ :\ \ggt\too\widetilde\ggt\ :\ X\longmapsto\bigl(-\mu(X),X\bigr)
\qqq
that manifestly defines a $\bK$-linear superspace section of $\,\pi_\ggt\,$ and lifts to an LSA monomorphism,
\qq\nn
[\si_\mu(X_1),\si_\mu(X_2)\}_\Th&=&[(-\mu(X_1),X_1),(-\mu(X_2),X_2)\}_\Th=\bigl(\Th(X_1,X_2),[X_1,X_2\}_\ggt\bigr)\cr\cr
&=&\bigl(-\mu\left([X_1,X_2\}_\ggt\right),[X_1,X_2\}_\ggt\bigr)\equiv\si_\mu
\left([X_1,X_2\}_\ggt\right)\,.
\qqq
In consequence, the associated short exact sequence of LSAs splits.

Conversely, an arbitrary LSA section of $\,\pi_\ggt\,$ is necessarily of the form 
\qq\nn
\si_\mu\ :\ \ggt\too\widetilde\ggt\ :\ X\longmapsto\bigl(-\mu(X),X\bigr)
\qqq
for some $\,\mu\in\Hom_\bK(\ggt,\agt)\,$ that preserves the $\bZ/2\bZ$-grading, and such that 
\qq\nn
\bigl(\Th(X_1,X_2),[X_1,X_2\}_\ggt\bigr)=[\si_\mu(X_1),\si_\mu(X_2)\}_\Th\must\si_\mu\left([X_1,X_2\}_\ggt\right)=\bigl(-\mu\left([X_1,X_2\}_\ggt\right),[X_1,X_2\}_\ggt\bigr)\,,
\qqq
so that $\,\Th=\d_\ggt^{(1)}\mu$,\ as claimed.
\eroof
\noindent Let us conclude the purely algebraic part of our exposition with the following simple reinterpretation that translates beautifully into differential (super)geometry {\it via} the correspondence stated in the next section. 

\brem\label{rem:LSApulltriv}
The existence of an extension of $\,\ggt\,$ by $\,\agt\,$ determined by $\,\Th\,$ is tantamount to a trivialisation of the pullback 2-cocycle
\qq\nn
\widetilde\Th:=\pi_\ggt^*\Th\ :\ \widetilde\ggt^{\x 2}\too\agt\ :\ \bigl((A_1,X_1),(A_2,X_2)\bigr)\longmapsto\Th(X_1,X_2)
\qqq
given by
\qq\nn
\widetilde\Th=\d_{\widetilde\ggt}^{(1)}\widetilde\mu\,,\qquad\widetilde\mu:=-\pi_\agt\ :\ \widetilde\ggt\too\agt\ :\ (A,X)\longmapsto-A\,.
\qqq
\erem

\subsection{The supergeometry of the Chevalley--Eilenberg Lie-superalgebra cohomology}

The LSA cohomology introduced above after \Rcite{Leites:1975} admits various explicit realisations. From the point of view of applications of immediate interest, both physical and geometric, the Chevalley--Eilenberg model of \Rcite{Chevalley:1948} seems most convenient. It is based on the elementary observation: The exterior (super)derivative of a left-invariant (super-)$p$-form $\,\underset{\tx{\ciut{(p)}}}{\om}\,$ on a Lie (super)group $\,\txG\,$ evaluates on a collection $\,(L_1,L_2,\ldots,L_{p+1})\,$ of Gra\ss mann-homogeneous left-invariant vector fields on that (super)group as 
\qq\nn
\sfd\underset{\tx{\ciut{(p)}}}{\om}(L_1,L_2,\ldots,L_{p+1})=\sum_{1\leq i<j\leq p+1}\,(-1)^{S(|L_a|)+S(|L_b|)+|L_b|\cdot|L_b|}\,\underset{\tx{\ciut{(p)}}}{\om}([L_a,L_b\},L_1,L_2,\underset{\widehat{i,j}}{\ldots},L_{p+1})\,,
\qqq
with the $\,S(|L_a|)\,$ defined as in \Reqref{eq:Sumaltweights}, which in conjunction with the isomorphism (identity) of LSA's (over $\,\bR$) between the (abstract) LSA $\,\ggt\,$ of the Lie supergroup $\,\txG\,$ and the LSA $\,\Xgt(\txG)^{\rm L}\,$ of left-invariant vector fields on $\,\txG\,$ (at the group unit),
\qq\nn
\left(\ggt,[\cdot,\cdot\}_\ggt\right)\cong\left(\Xgt(\txG)^{\rm L},[\cdot,\cdot\}\right)\,,
\qqq
leads to the fundamental
\bethe\label{thm:sCEmodelIdR}
Let $\,\txG\,$ be a Lie supergroup, and let $\,\left(\ggt,[\cdot,\cdot\}_\ggt\right)\,$ be its LSA (over $\,\bR$), which we take to act trivially on $\,\bR$.\ Denote by $\,\Om^p(\txG)^{\rm L}\,$ the $\bR$-linear superspace of left-invariant $p$-forms on $\,\txG$,\ and by $\,\sfd^{(p)}\,$ the restriction of the de Rham differential to $\,\Om^p(\txG)^{\rm L}$.\ There exists a canonical bijective cochain map
\qq\nn
\g\ :\ \left(C^\bullet(\ggt,\bR),\d_\ggt^{(\bullet)}\right)\too\left(\Om^\bullet(\txG,\bR)^{\rm L},\sfd^{(\bullet)}\right)
\qqq
that induces an isomorphism in cohomology,
\qq\nn
[\g]\ :\ H^\bullet(\ggt,\bR)\xrightarrow{\ \cong\ }H^\bullet_{\rm dR}(\txG,\bR)^{\rm L}\equiv{\rm CaE}^\bullet(\txG)\,,
\qqq
the latter being termed the \textbf{Cartan--Eilenberg cohomology} of $\,\txG$.
\ethe

The general theory leads to a correspondence between left-invariant de Rham super-2-cocycles on a Lie supergroup and supercentral extensions of that supergroup on which the pullbacks of those super-2-cocycles along the canonical projection can be trivialised in the Cartan--Eilenberg invariant cohomology\footnote{That is to say, the pullbacks admit smooth left-invariant primitives on the extended Lie supergroup.}. This correspondence is used amply in the construction of (super)geometric realisations of the Green--Schwarz super-$(p+2)$-cocycles over $\,{\rm sMink}^{d,1\,\vert\,ND_{d,1}}\,$ proposed in Section \ref{sec:GSgerbe}.

\section{A proof of Proposition \ref{prop:homformsgroup}}\label{app:homformsgroup}

We compute
\qq\nn
&&-30\underset{\tx{\ciut{(2)}}}{\D}{}_{\a\b}(\theta,x,\z,\psi,\upsilon)=\bigl(\ovl\G{}^a_{\a\b}\,\si^{(3)}_{a\g}+4\ovl\G{}^a_{\a\g}\,\si^{(3)}_{a\b}+4\ovl\G{}^a_{\b\g}\,\si^{(3)}_{a\a}\bigr)(\theta,x,\z,\psi)\wedge\si^\g(\theta)\cr\cr
&&-2\ovl\G{}^a_{\a\b}\,e_{ab}^{(2)}(\theta,x,\z)\wedge e^b(\theta,x)-2\ovl\G_{ab\,\a\b}\,\bigl(e^a\wedge e^b\bigr)(\theta,x)\cr\cr
&=&\sfd\bigl[-\bigl(\ovl\G{}^a_{\a\b}\,\si^{(3)}_{a\g}+4\ovl\G{}^a_{\a\g}\,\si^{(3)}_{a\b}+4\ovl\G{}^a_{\b\g}\,\si^{(3)}_{a\a}\bigr)(\theta,x,\z,\psi)\,\theta^\g\bigr]-\bigl[\ovl\G{}^a_{\a\b}\,\bigl(\ovl\G{}^b_{\g\d}\,e^{(2)}_{ab}(\theta,x,\z)+\ovl\G_{ab\,\g\d}\,e^b(\theta,x)\bigr)\cr\cr
&&+4\ovl\G{}^a_{\a\g}\,\bigl(\ovl\G{}^b_{\b\d}\,e^{(2)}_{ab}(\theta,x,\z)+\ovl\G_{ab\,\b\d}\,e^b(\theta,x)\bigr)+4\ovl\G{}^a_{\b\g}\,\bigl(\ovl\G{}^b_{\a\d}\,e^{(2)}_{ab}(\theta,x,\z)+\ovl\G_{ab\,\a\d}\,e^b(\theta,x)\bigr)\bigr]\wedge \si^\d(\theta)\,\theta^\g\cr\cr
&&-2\ovl\G{}^a_{\a\b}\,e_{ab}^{(2)}(\theta,x,\z)\wedge\sfd x^b+\ovl\G{}^a_{\a\b}\,\ovl\G{}^b_{\g\d}\,e_{ab}^{(2)}(\theta,x,\z)\wedge\si^\d(\theta)\,\theta^\g-2\ovl\G_{ab\,\a\b}\,\sfd x^a\wedge\sfd x^b\cr\cr
&&+2\ovl\G_{ab\,\a\b}\,\ovl\G{}^a_{\g\d}\,\si^\d(\theta)\,\theta^\g\wedge\sfd x^b-\tfrac{1}{2}\,\ovl\G_{ab\,\a\b}\,\theta\,\ovl\G{}^a\,\si(\theta)\wedge\theta\,\ovl\G{}^b\,\si(\theta)\cr\cr
&=&\sfd\bigl[-\bigl(\ovl\G{}^a_{\a\b}\,\si^{(3)}_{a\g}+4\ovl\G{}^a_{\a\g}\,\si^{(3)}_{a\b}+4\ovl\G{}^a_{\b\g}\,\si^{(3)}_{a\a}\bigr)(\theta,x,\z,\psi)\,\theta^\g+2\ovl\G{}^a_{\a\b}\,e_{ab}^{(2)}(\theta,x,\z)\,x^b\cr\cr
&&\hspace{.3cm}-2\ovl\G_{ab\,\a\b}\,x^a\,\sfd x^b-2\ovl\G_{ab\,\a\b}\,\ovl\G{}^a_{\g\d}\,\si^\d(\theta)\,\theta^\g x^b\bigr]\cr\cr
&&+2\bigl(\ovl\G{}^a_{\a\d}\,\ovl\G{}^b_{\b\g}+\ovl\G{}^a_{\b\d}\,\ovl\G{}^b_{\a\g}\bigr)\,e_{ab}^{(2)}(\theta,x,\z)\wedge\bigl(\si^\d(\theta)\,\theta^\g+\si^\g(\theta)\,\theta^\d+\sfd(\theta^\d\,\theta^\g)\bigl)\cr\cr
&&-\tfrac{1}{2}\,\bigl(\ovl\G{}^a_{\a\b}\,\ovl\G_{ab\,\g\d}+4\ovl\G{}^a_{\a\g}\,\ovl\G_{ab\,\b\d}+4\ovl\G{}^a_{\b\g}\,\ovl\G_{ab\,\a\d}\bigr)\,e^b(\theta,x)\wedge\bigl(\si^\d(\theta)\,\theta^\g+\si^\g(\theta)\,\theta^\d+\sfd(\theta^\d\,\theta^\g)\bigl)\cr\cr
&&-\bigl(\ovl\G{}^a_{\a\b}\,\ovl\G_{ab\,\g\d}+2\ovl\G{}^a_{\g\d}\,\ovl\G_{ab\,\a\b}\bigr)\,x^b\,\si^\g\wedge\si^\d(\theta)-\tfrac{1}{2}\,\ovl\G_{ab\,\a\b}\,\theta\,\ovl\G{}^a\,\si(\theta)\wedge\theta\,\ovl\G{}^b\,\si(\theta)\cr\cr
&=&\sfd\bigl[-\bigl(\ovl\G{}^a_{\a\b}\,\si^{(3)}_{a\g}+4\ovl\G{}^a_{\a\g}\,\si^{(3)}_{a\b}+4\ovl\G{}^a_{\b\g}\, \si^{(3)}_{a\a}\bigr)(\theta,x,\z,\psi)\,\theta^\g+2\ovl\G{}^a_{\a\b}\,e_{ab}^{(2)}(\theta,x,\z)\,x^b\cr\cr
&&\hspace{.3cm}-2\ovl\G_{ab\,\a\b}\,x^a\,\sfd x^b-2\ovl\G_{ab\,\a\b}\,\ovl\G{}^a_{\g\d}\,\si^\d(\theta)\,\theta^\g x^b\bigr]\cr\cr
&&+2\bigl[\bigl(\ovl\G{}^a_{\a\d}\,\ovl\G{}^b_{\b\g}+\ovl\G{}^a_{\b\d}\,\ovl\G{}^b_{\a\g}\bigr)\,e_{ab}^{(2)}(\theta,x,\z)+\bigl(\ovl\G{}^a_{\a\d}\,\ovl\G_{ab\,\b\g}+\ovl\G{}^a_{\b\d}\,\ovl\G_{ab\,\a\g}\bigr)\,e^b(\theta,x)\bigr]\wedge\sfd(\theta^\d\,\theta^\g)\cr\cr
&&-\bigl(\ovl\G{}^a_{\a\b}\,\ovl\G_{ab\,\g\d}+2\ovl\G{}^a_{\g\d}\,\ovl\G_{ab\,\a\b}\bigr)\,\bigl(\theta^\g\,e^b(\theta,x)+x^b\,\si^\g(\theta)\bigr)\wedge\si^\d(\theta)\cr\cr
&&-\tfrac{1}{2}\,\ovl\G_{ab\,\a\b}\,\theta\,\ovl\G{}^a\,\si(\theta)\wedge\theta\,\ovl\G{}^b\,\si(\theta)\cr\cr
&=&\sfd\bigl[-\bigl(\ovl\G{}^a_{\a\b}\,\si^{(3)}_{a\g}+4\ovl\G{}^a_{\a\g}\,\si^{(3)}_{a\b}+4\ovl\G{}^a_{\b\g}\,\si^{(3)}_{a\a}\bigr)(\theta,x,\z,\psi)\,\theta^\g+2\ovl\G{}^a_{\a\b}\,e_{ab}^{(2)}(\theta,x,\z)\,x^b\cr\cr
&&\hspace{.3cm}-2\bigl(2\ovl\G{}^a_{\a\d}\,\ovl\G{}^b_{\b\g}\,e_{ab}^{(2)}(\theta,x,\z)+\bigl(\ovl\G{}^a_{\a\d}\,\ovl\G_{ab\,\b\g}+\ovl\G{}^a_{\b\d}\,\ovl\G_{ab\,\a\g}\bigr)\,e^b(\theta,x)\bigr)\,\theta^\d\,\theta^\g\cr\cr
&&\hspace{.3cm}-2\ovl\G_{ab\,\a\b}\,x^a\,\sfd x^b-\ovl\G{}^a_{\a\b}\,\ovl\G_{ab\,\g\d}\,x^b\,\theta^\g\,\si^\d(\theta)\bigr]\cr\cr
&&+\bigl(2\ovl\G{}^a_{\a\d}\,\ovl\G{}^b_{\b\g}\,\ovl\G_{ab\,\ep\z}+\bigl(\ovl\G{}^a_{\a\d}\,\ovl\G_{ab\,\b\g}+\ovl\G{}^a_{\b\d}\,\ovl\G_{ab\,\a\g}\bigr)\,\ovl\G{}^b_{\ep\z}\bigr)\,\theta^\d\,\theta^\g\,\bigl(\si^\ep\wedge\si^\z\bigr)(\theta)\cr\cr
&&+\tfrac{1}{2}\,\bigl(\ovl\G{}^a_{\a\b}\,\theta\,\ovl\G_{ab}\,\si(\theta)+\ovl\G_{ab\,\a\b}\,\theta\,\ovl\G{}^a\,\si(\theta)\bigr)\wedge\theta\,\ovl\G{}^b\,\si(\theta)\,.
\qqq
The last two lines of the above expression define -- by construction -- a closed super-1-form 
\qq\nn
\underset{\tx{\ciut{(1)}}}{\eta}_{\a\b}(\theta)&=&\bigl(2\ovl\G{}^a_{\a\d}\,\ovl\G{}^b_{\b\g}\,\ovl\G_{ab\,\ep\z}+\bigl(\ovl\G{}^a_{\a\d}\,\ovl\G_{ab\,\b\g}+\ovl\G{}^a_{\b\d}\,\ovl\G_{ab\,\a\g}\bigr)\,\ovl\G{}^b_{\ep\z}\bigr)\,\theta^\d\,\theta^\g\,\bigl(\si^\ep\wedge\si^\z\bigr)(\theta)\cr\cr
&&+\tfrac{1}{2}\,\bigl(\ovl\G{}^a_{\a\b}\,\theta\,\ovl\G_{ab}\,\si(\theta)+\ovl\G_{ab\,\a\b}\,\theta\,\ovl\G{}^a\,\si(\theta)\bigr)\wedge\theta\,\ovl\G{}^b\,\si(\theta)\,,
\qqq
and we derive its global primitive using the formerly advertised homotopy formula, {\it cp}\ \Reqref{eq:shomform}.


\begin{thebibliography}{CdAIPB00}

\bibitem[AdA85]{Aldaya:1984gt}
V.~Aldaya and J.A. de~Azc{\'a}rraga, \emph{{``A note on the meaning of
  covariant derivatives in supersymmetry''}}, J. Math. Phys. \textbf{26}
  (1985), 1818--1821.

\bibitem[AETW87]{Achucarro:1987nc}
A.~Ach{\'u}carro, J.M. Evans, P.K. Townsend, and D.L. Wiltshire, \emph{{``Super
  p-branes''}}, Phys. Lett. \textbf{B198} (1987), 441--446.

\bibitem[AF08]{Arutyunov:2008if}
G.~Arutyunov and S.~Frolov, \emph{{``Superstrings on $\,{\rm
  AdS}_4\times\mathbb{CP}^3\,$ as a coset sigma-model''}}, JHEP \textbf{09}
  (2008), 129.

\bibitem[Alv85]{Alvarez:1984es}
O.~Alvarez, \emph{{``Topological quantization and cohomology''}}, Commun. Math.
  Phys. \textbf{100} (1985), 279--309.

\bibitem[AV74]{Akulov:1974xz}
V.P. Akulov and D.V. Volkov, \emph{{``Goldstone fields with spin 1/2"}}, Theor.
  Math. Phys. \textbf{18} (1974), 28--35.

\bibitem[Bat79]{Batchelor:1979a}
M.~Batchelor, \emph{{``The structure of supermanifolds''}}, Trans. Amer. Math.
  Soc. \textbf{253} (1979), 329--338.

\bibitem[BC04]{Baez:2004hda6}
J.C. Baez and A.S. Crans, \emph{{``Higher-dimensional algebra VI: Lie
  2-algebras''}}, Theor. Appl. Categor. \textbf{12} (2004), 492--528.

\bibitem[BCM+02]{Bouwknegt:2001vu}
P.~Bouwknegt, A.L. Carey, V.~Mathai, M.K. Murray, and D.~Stevenson,
  \emph{{``Twisted K-theory and K-theory of bundle gerbes''}}, Commun. Math.
  Phys. \textbf{228} (2002), 17--49.

\bibitem[BH11]{Baez:2010ye}
J.C. Baez and J.~Huerta, \emph{{``Division algebras and supersymmetry II''}},
  Adv. Theor. Math. Phys. \textbf{15} (2011), no.~5, 1373--1410.

\bibitem[BK70]{Berezin:1970}
F.A. Berezin and G.I. Ka{\v c}, \emph{{``Lie groups with commuting and
  anticommuting parameters''}}, Math. USSR Sbornik \textbf{11} (1970),
  311--325.

\bibitem[BL75]{Berezin:1975}
F.~A. Berezin and D.A. Le{\"i}tes, \emph{{``Supermanifolds"}}, Dokl. Akad. Nauk
  SSSR \textbf{224} (1975), 505--508.

\bibitem[BLNPST97]{Bandos:1997ui}
I.A.~Bandos and K.~Lechner and A.~Nurmagambetov and P.~Pasti and D.P.~Sorokin and M.~Tonin, \emph{{``Covariant action for the superfive-brane of M theory"}}, Phys. Rev. Lett. \textbf{78} (1997) 4332--4334.

\bibitem[Bry93]{Brylinski:1993ab}
J.-L. Brylinski, \emph{{Loop spaces, characteristic classes and geometric
  quantization}}, Progress in Mathematics, vol. 107, Birkh{\"a}user, 1993.

\bibitem[BS81]{Brink:1981nb}
L.~Brink and J.H. Schwarz, \emph{{``Quantum superspace''}}, Phys. Lett.
  \textbf{B100} (1981), 310--312.

\bibitem[BST86]{Bergshoeff:1985su}
E.~Bergshoeff, E.~Sezgin, and P.K. Townsend, \emph{{``Superstring actions in
  $\,D=3, 4, 6, 10\,$ curved superspace''}}, Phys. Lett. \textbf{B169} (1986),
  191--196.

\bibitem[BST87]{Bergshoeff:1987cm}
E.~Bergshoeff, E.~Sezgin, and P.~K. Townsend, \emph{{``Supermembranes and
  eleven-dimensional supergravity}}, Phys. Lett. \textbf{B189} (1987), 75--78.

\bibitem[Bun11]{Bunke:2009}
U.~Bunke, \emph{{``String structures and trivialisations of a Pfaffian line
  bundle''}}, Comm. Math. Phys. \textbf{307} (2011), 675--712.

\bibitem[BW84]{Bagger:1983mv}
J.~Bagger and J.~Wess, \emph{{``Partial breaking of extended supersymmetry"}},
  Phys. Lett. \textbf{B138} (1984), 105--110.

\bibitem[CCF11]{Carmeli:2011}
C.~Carmeli, L.~Caston, and R.~Fioresi, \emph{{Mathematical Foundations of
  Supersymmetry}}, European Mathematical Society, 2011.

\bibitem[CCWZ69]{Callan:1969sn}
C.G. Callan, Jr., S.R. Coleman, J.~Wess, and B.~Zumino, \emph{{``Structure of
  phenomenological Lagrangians. II"}}, Phys. Rev. \textbf{177} (1969),
  2247--2250.

\bibitem[CdAIPB00]{Chryssomalakos:2000xd}
C.~Chryssomalakos, J.A. de~Azc{\'a}rraga, J.M. Izquierdo, and J.C.
  P{\'e}rez~Bueno, \emph{{``The geometry of branes and extended
  superspaces''}}, Nucl. Phys. \textbf{B567} (2000), 293--330.

\bibitem[CE48]{Chevalley:1948}
C.~Chevalley and S.~Eilenberg, \emph{{``Cohomology theory of Lie groups and Lie
  algebras''}}, Trans. Amer. Math. Soc. \textbf{63} (1948), 85--124.

\bibitem[CJM02]{Carey:2002}
A.L. Carey, S.~Johnson, and M.K. Murray, \emph{{``Holonomy on D-branes''}},
  arXiv preprint: hep-th/0204199.

\bibitem[CJM+05]{Carey:2004xt}
A.L. Carey, S.~Johnson, M.K. Murray, D.~Stevenson, and B.-L. Wang,
  \emph{{``Bundle gerbes for Chern--Simons and Wess--Zumino--Witten
  theories''}}, Commun. Math. Phys. \textbf{259} (2005), 577--613.

\bibitem[Cla99]{Claus:1998fh}
P.~Claus, \emph{{``Super M-brane actions in $\,{\rm AdS}_4\times\mathbb{S}^7\,$
  and $\,{\rm AdS}_7\times \mathbb{S}^4$''}}, Phys. Rev. \textbf{D59} (1999),
  066003.

\bibitem[CMR12]{Cattaneo:2012}
A.S. Cattaneo, P.~Mnev, and N.~Reshetikhin, \emph{{``Classical and quantum
  Lagrangian field theories with boundary''}}, {Proceedings of the 11th
  Hellenic School and Workshops on Elementary Particle Physics and Gravity
  (CORFU2011) : Corfu, Greece, September 4-18, 2011} (K.~Anagnostopoulos,
  I.~Antoniadis, D.~Bahns, N.~Irges, A.~Kehagias, G.~Lazarides, D.~Luest,
  H.~Steinacker, and G.~Zoupanos, eds.), Proceedings of Science, 2012, p.~044.

\bibitem[CWZ69]{Coleman:1969sm}
S.R. Coleman, J.~Wess, and B.~Zumino, \emph{{``Structure of phenomenological
  lagrangians. I''}}, Phys. Rev. \textbf{177} (1969), 2239--2247.

\bibitem[dAI95]{deAzcarraga:1995}
J.A. De~Azc{\'a}rraga and J.M. Izquierdo, \emph{{Lie groups, Lie algebras,
  cohomology and some applications in physics}}, Cambridge Monographs On
  Mathematical Physics, Cambridge University Press, 1995.

\bibitem[dAL83]{deAzcarraga:1982njd}
J.A. de~Azc{\'a}rraga and J.~Lukierski, \emph{{``Supersymmetric particles in
  $\,N=2\,$ superspace: phase space variables and Hamiltonian dynamics"}},
  Phys. Rev. \textbf{D28} (1983), 1337--1345.

\bibitem[dAT89]{deAzcarraga:1989vh}
J.A. De~Azc{\'a}rraga and P.K. Townsend, \emph{{``Superspace geometry and
  classification of supersymmetric extended objects"}}, Phys. Rev. Lett.
  \textbf{62} (1989), 2579--2582.

\bibitem[DeW99]{DeWitt:1992}
B.~DeWitt, \emph{{Supermanifolds}}, Cambridge Monographs On Mathematical
  Physics, Cambridge University Press, 1999.

\bibitem[DF99]{Deligne:1999sgn}
P.~Deligne and D.W. Freed, \emph{{``Sign Manifesto''}}, vol.~1, pp.~357--363,
  American Mathematical Society, 1999.

\bibitem[DF05]{Douglas:1999hq}
M.R. Douglas and B.~Fiol, \emph{{``D-branes and discrete torsion. 2."}}, JHEP
  \textbf{09} (2005), 053.

\bibitem[DFGT09]{DAuria:2008vov}
R.~D'Auria, P.~Fre, P.A. Grassi, and M.~Trigiante, \emph{{``Superstrings on
  $\,{\rm AdS}(4)\x\bC P^3\,$ from supergravity"}}, Phys. Rev. \textbf{D79}
  (2009), 086001.

\bibitem[DHVW85]{Dixon:1985jw}
L.J. Dixon, J.A. Harvey, C.~Vafa, and E.~Witten, \emph{{``Strings on
  orbifolds''}}, Nucl. Phys. \textbf{B261} (1985), 678--686.

\bibitem[DHVW86]{Dixon:1986jc}
\bysame, \emph{{``Strings on orbifolds (II)''}}, Nucl. Phys. \textbf{B261}
  (1986), 285--314.

\bibitem[Dou98]{Douglas:1998xa}
M.R. Douglas, \emph{{``D-branes and discrete torsion"}}, arXiv preprint:
  hep-th/9807235.

\bibitem[dWPPS98]{deWit:1998yu}
B.~de~Wit, K.~Peeters, J.~Plefka, and A.~Sevrin, \emph{{``The M-theory
  two-brane in $\,{\rm AdS}_4\times\mathbb{S}^7\,$ and $\,{\rm AdS}_7\times
  \mathbb{S}^4$''}}, Phys. Lett. \textbf{B443} (1998), 153--158.

\bibitem[FFRS09]{Frohlich:2009gb}
J.~Fr\"ohlich, J.A. Fuchs, I.~Runkel, and C.~Schweigert, \emph{{``Defect lines,
  dualities, and generalised orbifolds''}}, Proceedings of the XVIth
  International Congress on Mathematical Physics: Prague, Czech Republic,
  3-–8 August 2009 (P.~Exner, ed.), World Scientific, 2009, pp.~608--613.

\bibitem[FG94]{Frohlich:1993es}
J.~Fr{\"o}hlich and K.~Gaw\c{e}dzki, \emph{{``Conformal Field Theory and
  Geometry of Strings''}}, Vancouver 1993, Proceedings, Mathematical Quantum
  Theory I: Field Theory and Many-Body Theory (J.~Feldman, R.~Froese, and L.M.
  Rosen, eds.), CRM Proceedings {\&} Lecture Notes, vol.~7, American
  Mathematical Society, 1994, pp.~57--97.

\bibitem[FG12]{Fre:2008qc}
P.~Fre and P.A. Grassi, \emph{{``Pure spinor formalism for $\,{\rm
  Osp}(N\,|\,4)\,$ backgrounds"}}, Int. J. Mod. Phys. \textbf{A27} (2012),
  1250185.

\bibitem[FGK88]{Felder:1988sd}
G.~Felder, K.~Gaw\c{e}dzki, and A.~Kupiainen, \emph{{``Spectra of
  Wess--Zumino--Witten models with arbitrary simple groups''}}, Commun. Math.
  Phys. \textbf{117} (1988), 127--158.

\bibitem[FM06]{Freed:2004yc}
D.S. Freed and G.W. Moore, \emph{{``Setting the quantum integrand of
  M-theory"}}, Commun. Math. Phys. \textbf{263} (2006), 89--132.

\bibitem[FMW83]{Ferrara:1983fi}
S.~Ferrara, L.~Maiani, and P.C. West, \emph{{``Non-linear representations of
  extended supersymmetry with central charges"}}, Z. Phys. \textbf{C19} (1983),
  267--273.

\bibitem[Fre87]{Yau:1987}
D.S. Freed, \emph{{``On determinant line bundles''}}, {Mathematical Aspects of
  String Theory, Proceedings of the Conference held at University of
  California, San Diego, July 21 -- August 1, 1986} (S.T. Yau, ed.), Advanced
  Series in Mathematical Physics, vol.~1, World Scientific, 1987.

\bibitem[Fre99]{Freed:1999}
\bysame, \emph{{Five Lectures on Supersymmetry}}, American Mathematical
  Society, 1999.

\bibitem[FSS14]{Fiorenza:2013nha}
D.~Fiorenza, H.~Sati, and U.~Schreiber, \emph{{``Super-Lie $n$-algebra
  extensions, higher WZW models, and super-$p$-branes with tensor multiplet
  fields"}}, Int. J. Geom. Meth. Mod. Phys. \textbf{12} (2014), 1550018.

\bibitem[FSW08]{Fuchs:2007fw}
J.A. Fuchs, C.~Schweigert, and K.~Waldorf, \emph{{``Bi-branes: Target space
  geometry for world sheet topological defects''}}, J. Geom. Phys. \textbf{58}
  (2008), 576--598.

\bibitem[FVP12]{Freedman:2012zz}
D.Z. Freedman and A.~Van~Proeyen, \emph{{``Supergravity''}}, Cambridge
  University Press, 2012.

\bibitem[Gaj96]{Gajer:1996}
 P.~Gajer, \emph{{``Geometry of Deligne cohomology''}}, Invent. Math. \textbf{127} (1997), 155--207.

\bibitem[Gaw72]{Gawedzki:1972ms}
K.~Gaw\c{e}dzki, \emph{{``On the geometrization of the canonical formalism in
  the classical field theory''}}, Rep. Math. Phys. \textbf{3} (1972), 307--326.

\bibitem[Gaw77]{Gawedzki:1977pb}
\bysame, \emph{{``Supersymmetries-mathematics of supergeometry''}}, Ann. Henri
  Poincare, Phys. Theor. \textbf{27} (1977), 335--366.

\bibitem[Gaw88]{Gawedzki:1987ak}
\bysame, \emph{{``Topological Actions in Two-Dimensional Quantum Field
  Theory''}}, {Nonperturbative Quantum Field Theory} (G.~{'t}~Hooft, A.~Jaffe,
  G.~Mack, P.~Mitter, and R.~Stora, eds.), Plenum Press, 1988, pp.~101--141.

\bibitem[Gaw91]{Gawedzki:1990jc}
\bysame, \emph{{``Classical origin of quantum group symmetries in
  Wess--Zumino--Witten conformal field theory''}}, Commun. Math. Phys.
  \textbf{139} (1991), 201--214.

\bibitem[Gaw99]{Gawedzki:1999bq}
\bysame, \emph{{``Conformal field theory: A case study''}}, arXiv preprint:
  hep-th/9904145.

\bibitem[Gaw02]{Gawedzki:2001ye}
\bysame, \emph{``boundary wzw, $\,\txg/\txh, \txg/\txg\,$ and cs theories''},
  Ann. Henri Poincare \textbf{3} (2002), 847--881.

\bibitem[Gaw05]{Gawedzki:2004tu}
\bysame, \emph{{``Abelian and non-Abelian branes in WZW models and gerbes''}},
  Commun. Math. Phys. \textbf{258} (2005), 23--73.

\bibitem[GK89a]{Gawedzki:1988nj}
K.~Gaw\c{e}dzki and A.~Kupiainen, \emph{{``Coset construction from functional
  integrals''}}, Nucl. Phys. \textbf{B320} (1989), 625--668.

\bibitem[GK89b]{Gawedzki:1988hq}
\bysame, \emph{{``$\txG/\txH\,$ conformal field theory from gauged WZW
  model''}}, Phys. Lett. \textbf{B215} (1989), 119--123.

\bibitem[GKO85]{Goddard:1984vk}
P.~Goddard, A.~Kent, and D.I. Olive, \emph{{``Virasoro algebras and coset space
  models''}}, Phys. Lett. \textbf{B152} (1985), 88--92.

\bibitem[GKW06]{Gomis:2006wu}
J.~Gomis, K.~Kamimura, and P.C. West, \emph{{``Diffeomorphism, kappa
  transformations and the theory of non-linear realisations''}}, JHEP
  \textbf{0610} (2006), 015.

\bibitem[GL71]{Golfand:1971iw}
Yu.~A. Golfand and E.P. Likhtman, \emph{{``Extension of the algebra of
  Poincar{\'e} group generators and violation of p invariance"}}, Pisma Zh.
  Eksp. Teor. Fiz. \textbf{13} (1971), 323--326.

\bibitem[GR02]{Gawedzki:2002se}
K.~Gaw\c{e}dzki and N.~Reis, \emph{{``WZW branes and gerbes''}}, Rev. Math.
  Phys. \textbf{14} (2002), 1281--1334.

\bibitem[GR03]{Gawedzki:2003pm}
\bysame, \emph{{``Basic gerbe over non simply connected compact groups''}}, J.
  Geom. Phys. \textbf{50} (2003), 28--55.

\bibitem[GS71]{Gervais:1971ji}
J.-L. Gervais and B.~Sakita, \emph{{``Field theory interpretation of
  supergauges in dual models"}}, Nucl. Phys. \textbf{B34} (1971), 632--639.

\bibitem[GS84a]{Green:1983wt}
M.B. Green and J.H. Schwarz, \emph{{``Covariant description of
  superstrings''}}, Phys. Lett. \textbf{B136} (1984), 367–--370.

\bibitem[GS84b]{Green:1983sg}
\bysame, \emph{{``Properties of the covariant formulation of superstring
  theories"}}, Nucl. Phys. \textbf{B243} (1984), 285--306.

\bibitem[GSoW08]{Gomis:2008jt}
J.~Gomis, D.~Sorokin, D. and L.~Wulff, \emph{{``The Complete $\,{\rm AdS}_4\x\bC P^3\,$ superspace for the type IIA superstring and D-branes"}}, JHEP \textbf{03} (2009), 015.

\bibitem[GSW08]{Gawedzki:2007uz}
K.~Gaw\c{e}dzki, R.R. Suszek, and K.~Waldorf, \emph{{``WZW orientifolds and
  finite group cohomology''}}, Commun. Math. Phys. \textbf{284} (2008), 1--49.

\bibitem[GSW10]{Gawedzki:2010rn}
\bysame, \emph{{``Global gauge anomalies in two-dimensional bosonic sigma
  models''}}, Comm. Math. Phys. \textbf{302} (2010), 513--580.

\bibitem[GSW11a]{Gawedzki:2008um}
\bysame, \emph{{``Bundle gerbes for orientifold sigma models''}}, Adv. Theor.
  Math. Phys. \textbf{15} (2011), 621--688.

\bibitem[GSW11b]{Gawedzki:2010G}
\bysame, \emph{{``Unoriented WZW-branes and gerbes''}}, in preparation.

\bibitem[GSW13]{Gawedzki:2012fu}
\bysame, \emph{{``The gauging of two-dimensional bosonic sigma models on
  world-sheets with defects''}}, Rev. Math. Phys. \textbf{25} (2013), 1350010.

\bibitem[GTTNB04]{Gawedzki:2001rm}
K.~Gaw\c{e}dzki, I.~Todorov, and P.~Tran-Ngoc-Bich, \emph{{``Canonical
  quantization of the boundary Wess--Zumino--Witten model''}}, Commun. Math.
  Phys. \textbf{248} (2004), 217--254.

\bibitem[HLP86]{Hughes:1986fa}
J.~Hughes, J.~Liu, and J.~Polchinski, \emph{{``Supermembranes''}}, Phys. Lett.
  \textbf{B180} (1986), 370--374.

\bibitem[HM85]{Henneaux:1984mh}
M.~Henneaux and L.~Mezincescu, \emph{{``A $\sigma$-model interpretation of
  Green--Schwarz covariant superstring action"}}, Phys. Lett. \textbf{B152}
  (1985), 340--342.

\bibitem[Hor96]{Hori:1994nc}
K.~Hori, \emph{{``Global aspects of gauged Wess--Zumino--Witten models''}},
  Commun. Math. Phys. \textbf{182} (1996), 1--32.

\bibitem[HP86]{Hughes:1986dn}
J.~Hughes and J.~Polchinski, \emph{{``Partially broken global supersymmetry and
  the superstring"}}, Nucl. Phys. \textbf{B278} (1986), 147--169.

\bibitem[HS02]{Hatsuda:2002hz}
M.~Hatsuda and M.~Sakaguchi, \emph{{``Wess--Zumino term for AdS
  superstring''}}, Phys. Rev. \textbf{D66} (2002), 045020.

\bibitem[HS03]{Hatsuda:2001pp}
\bysame, \emph{{``Wess--Zumino term for the AdS superstring and generalized
  In{\''o}n{\''u}--Wigner contraction"}}, Prog. Theor. Phys. \textbf{109}
  (2003), 853--867.

\bibitem[Hue11]{Huerta:2011ic}
J.C. Huerta, \emph{{``Division Algebras, Supersymmetry and Higher Gauge
  Theory"}}, Ph.D. thesis, UC, Riverside, 2011, arXiv preprint: 1106.3385.

\bibitem[Hul05]{Hull:2004in}
C.M. Hull, \emph{{``A geometry for non-geometric string backgrounds''}}, JHEP
  \textbf{10} (2005), 065.

\bibitem[Hul07]{Hull:2006qs}
C.~M. Hull, \emph{{``Global aspects of T-duality, gauged sigma models and
  T-folds"}}, JHEP \textbf{10} (2007), 057.

\bibitem[IK78]{Ivanov:1978mx}
E.~A. Ivanov and A.~A. Kapustnikov, \emph{{``General relationship between
  linear and nonlinear realizations of supersymmetry"}}, J. Phys. \textbf{A11}
  (1978), 2375--2384.

\bibitem[IK82]{Ivanov:1982bpa}
E.A. Ivanov and A.A. Kapustnikov, \emph{{``The nonlinear realization structure
  of models with spontaneously broken supersymmetry"}}, J. Phys. \textbf{G8}
  (1982), 167--191.

\bibitem[ISS71]{Isham:1971dv}
C.~J. Isham, A.~Salam, and J.~A. Strathdee, \emph{{``Nonlinear realizations of
  space-time symmetries. Scalar and tensor gravity''}}, Annals Phys.
  \textbf{62} (1971), 98--119.

\bibitem[Joh02]{Johnson:2003}
S.~Johnson, \emph{{``Constructions with Bundle Gerbes''}}, Ph.D. thesis, The
  University of Adelaide, Australia, 2002.

\bibitem[Kij73]{Kijowski:1973gi}
J.~Kijowski, \emph{{``A finite-dimensional canonical formalism in the classical
  field theory''}}, Commun. Math. Phys. \textbf{30} (1973), 99--128.

\bibitem[Kij74]{Kijowski:1974mp}
\bysame, \emph{{``Multiphase spaces and gauge in the calculus of
  variations''}}, Bull. Acad. Sc. Polon. \textbf{22} (1974), 1219--1225.

\bibitem[Kil87]{Killingback:1986rd}
T.P. Killingback, \emph{{``World-sheet anomalies and loop geometry''}}, Nucl.
  Phys. \textbf{B288} (1987), 578.

\bibitem[Kos77]{Kostant:1975}
B.~Kostant, \emph{{``Graded manifolds, graded Lie theory, and
  prequantization''}}, {Differential Geometrical Methods in Mathematical
  Physics, Proceedings of the Symposium Held at the University of Bonn, July
  1--4, 1975} (K.~Bleuler and A.~Reetz, eds.), Lecture Notes in Mathematics,
  vol. 570, Springer, 1977, pp.~177--306.

\bibitem[Kosz82]{Koszul:1982}
J.-L.~Koszul, \emph{``Graded manifolds and graded Lie algebras''},
{Proceedings of the International Meeting on Geometry and Physics (Florence, 1982)},
Pitagora, 1982, pp.~71--84.

\bibitem[KPSY89]{Karabali:1988au}
D.~Karabali, Q.-H. Park, H.J. Schnitzer, and Z.~Yang, \emph{{``A GKO
  construction based on a path integral formulation of gauged
  Wess-Zumino-Witten actions''}}, Phys. Lett. \textbf{B216} (1989), 307--312.

\bibitem[KR84]{Kostelecky:1983qu}
V.A. Kosteleck{\'y} and J.M. Rabin, \emph{{``Supersymmetry on a
  superlattice"}}, J. Math. Phys. \textbf{25} (1984), 2744--2748.

\bibitem[KS76]{Kijowski:1976ze}
J.~Kijowski and W.~Szczyrba, \emph{{``A canonical structure for classical field
  theories''}}, Commun. Math. Phys. \textbf{46} (1976), 183--206.

\bibitem[KT79]{Kijowski:1979dj}
J.~Kijowski and W.M. Tulczyjew, \emph{{A Symplectic Framework For Field
  Theories}}, Lecture Notes in Physics, vol. 107, Springer, 1979.

\bibitem[Lei75]{Leites:1975}
D.A. Leites, \emph{{``Cohomologies of Lie superalgebras''}}, Funct. Anal. Appl.
  \textbf{9} (1975), 340--341.

\bibitem[LR79]{Lindstrom:1979kq}
U.~Lindstr{\"o}m and M.~Ro{\v c}ek, \emph{{``Constrained local superfields"}},
  Phys. Rev. \textbf{D19} (1979), 2300--2303.

\bibitem[Mar97]{Martin:1997ns}
S.P. Martin, \emph{{``A Supersymmetry primer"}}, arXiv preprint: hep-ph/9709356
  [Adv. Ser. Direct. High Energy Phys. 21 (2010) 1-153].

\bibitem[McA00]{McArthur:1999dy}
I.N. McArthur, \emph{{``Kappa symmetry of Green--Schwarz actions in coset
  superspaces''}}, Nucl. Phys. \textbf{B573} (2000), 811--829.

\bibitem[Miy66]{Miyazawa:1966mfa}
H.~Miyazawa, \emph{{``Baryon number changing currents''}}, Prog. Theor. Phys.
  \textbf{36} (1966), no.~6, 1266--1276.

\bibitem[MS00]{Murray:1999ew}
M.K. Murray and D.~Stevenson, \emph{{``Bundle gerbes: stable isomorphism and
  local theory''}}, J. Lond. Math. Soc. \textbf{62} (2000), 925--937.

\bibitem[MT98]{Metsaev:1998it}
R.~R. Metsaev and A.A. Tseytlin, \emph{{``Type IIB superstring action in
  $\,{\rm AdS}_5\times\mathbb{S}^5\,$ background''}}, Nucl. Phys. \textbf{B533}
  (1998), 109--126.

\bibitem[Mur96]{Murray:1994db}
M.K. Murray, \emph{{``Bundle gerbes''}}, J. Lond. Math. Soc. \textbf{54}
  (1996), 403--416.

\bibitem[Rab87]{Rabin:1985tv}
J.M. Rabin, \emph{{``Supermanifold cohomology and the Wess--Zumino term of the
  covariant superstring action''}}, Commun. Math. Phys. \textbf{108} (1987),
  375--389.

\bibitem[RC85]{Rabin:1984rm}
J.M. Rabin and L.~Crane, \emph{{``Global properties of supermanifolds''}},
  Commun. Math. Phys. \textbf{100} (1985), 141--160.

\bibitem[Rog07]{Rogers:2007}
A.~Rogers, \emph{{Supermanifolds: Theory and Applications}}, World Scientific,
  2007.

\bibitem[RS08]{Recknagel:2006hp}
A.~Recknagel and R.R. Suszek, \emph{{``Non-commutative Weitzenb\"ock geometry,
  gerbe modules, and WZW branes''}}, JHEP \textbf{0802} (2008), 089.

\bibitem[RS09]{Runkel:2008gr}
I.~Runkel and R.R. Suszek, \emph{{``Gerbe-holonomy for surfaces with defect
  networks''}}, Adv. Theor. Math. Phys. \textbf{13} (2009), 1137--1219.

\bibitem[Sak00]{Sakaguchi:1999fm}
M.~Sakaguchi, \emph{{``IIB Branes and new space-time superalgebras''}}, JHEP
  \textbf{04} (2000), 019.

\bibitem[Sau89]{Saunders:1989jet}
D.J. Saunders, \emph{{``The Geometry of Jet Bundles''}}, London Mathematical
  Society Lecture Note Series, vol. 142, Cambridge University Press, 1989.

\bibitem[Sch67]{Schwinger:1967tc}
J.S. Schwinger, \emph{{``Chiral dynamics"}}, Phys. Lett. \textbf{B24} (1967),
  473--476.

\bibitem[Sch84]{Schwarz:1984}
A.S. Schwarz, \emph{{``On the definition of superspace"}}, Teoret. Mat. Fiz.
  \textbf{60} (1984), 37--42.

\bibitem[Seg04]{Segal:2002}
G.B. Segal, \emph{{``The Definition of Conformal Field Theory''}}, London
  Mathematical Society Lecture Note Series, vol. 308, Cambridge University
  Press, 2004.

\bibitem[Sie83]{Siegel:1983hh}
W.~Siegel, \emph{{``Hidden local supersymmetry in the supersymmetric particle
  action''}}, Phys. Lett. \textbf{B128} (1983), 397--399.

\bibitem[Sie84]{Siegel:1983ke}
\bysame, \emph{{``Light cone analysis of covariant superstrings''}}, Nucl.
  Phys. \textbf{B236} (1984), 311--318.

\bibitem[SS69a]{Salam:1969rq}
A.~Salam and J.~A. Strathdee, \emph{{``Nonlinear realizations. 1. The role of
  Goldstone bosons"}}, Phys. Rev. \textbf{184} (1969), 1750--1759.

\bibitem[SS69b]{Salam:1970qk}
\bysame, \emph{{``Nonlinear realizations. 2. Conformal symmetry''}}, Phys. Rev.
  \textbf{184} (1969), 1760--1768.

\bibitem[SSW07]{Schreiber:2005mi}
U.~Schreiber, C.~Schweigert, and K.~Waldorf, \emph{{``Unoriented WZW models and
  holonomy of bundle gerbes''}}, Commun. Math. Phys. \textbf{274} (2007),
  31--64.

\bibitem[Ste00]{Stevenson:2000wj}
D.~Stevenson, \emph{{``The Geometry of Bundle Gerbes''}}, Ph.D. thesis, The
  University of Adelaide, Australia, 2000.

\bibitem[Ste01]{Stevenson:2001grb2}
\bysame, \emph{{``Bundle 2-gerbes"}}, ArXiv (2001).

\bibitem[Sus11a]{Suszek:2011hg}
R.R. Suszek, \emph{{``Defects, dualities and the geometry of strings via
  gerbes. I. Dualities and state fusion through defects''}}, Hamb. Beitr. Math.
  Nr.\,{\bf 360} (2011) [arXiv preprint: 1101.1126 [hep-th]].

\bibitem[Sus11b]{Suszek:2011}
\bysame, \emph{{``The Gerbe Theory of the Bosonic Sigma-Model: The Multi-phase
  CFT, Dualities, and the Gauge Principle''}}, Acta Phys. Pol. B Proc. Suppl.
  \textbf{4} (2011), 425--460.

\bibitem[Sus12]{Suszek:2012ddg}
\bysame, \emph{{``Defects, dualities and the geometry of strings via gerbes II.
  Generalised geometries with a twist, the gauge anomaly and the gauge-symmetry
  defect''}}, Hamb. Beitr. Math. Nr.\,{\bf 361} (2011) [arXiv preprint:
  1209.2334 [hep-th]].

\bibitem[Sus13]{Suszek:2013}
\bysame, \emph{{``Gauge Defect Networks in Two-Dimensional CFT''}}, {Symmetries
  and Groups in Contemporary Physics, Proceedings of The XXIXth International
  Colloquium on Group-Theoretical Methods in Physics, Chern Institute of
  Mathematics, August 20-26, 2012}, Nankai Series in Pure, Applied Mathematics
  and Theoretical Physics, World Scientific, 2013, pp.~411--416.

\bibitem[Sus19]{Suszek:2018sM}
\bysame, \emph{{``Equivariant Cartan--Eilenberg supergerbes for the
  Green--Schwarz superbranes II. $\Ad_\cdot$-equivariance and Siegel's
  $\k$-symmetry in the super-Minkowskian setting"}}, in preparation.

\bibitem[SW83]{Samuel:1982uh}
S.~Samuel and J.~Wess, \emph{{``A superfield formulation of the non-linear
  realization of supersymmetry and its coupling to supergravity"}}, Nucl. Phys.
  \textbf{B221} (1983), 153--177.

\bibitem[Szc76]{Szczyrba:1976}
W.~Szczyrba, \emph{{``A symplectic structure on the set of Einstein metrics. A
  canonical formalism for general relativity''}}, Commun. Math. Phys.
  \textbf{51} (1976), 163--182.

\bibitem[UZ82]{Uematsu:1981rj}
T.~Uematsu and C.K. Zachos, \emph{{``Structure of phenomenological lagrangians
  for broken supersymmetry"}}, Nucl. Phys. \textbf{B201} (1982), 250--268.

\bibitem[VA72]{Volkov:1972jx}
D.V. Volkov and V.P. Akulov, \emph{{``Possible universal neutrino
  interaction"}}, Pisma Zh. Eksp. Teor. Fiz. \textbf{16} (1972), 438--440.

\bibitem[VA73]{Volkov:1973ix}
\bysame, \emph{{Is the neutrino a Goldstone particle?"}}, Phys. Lett.
  \textbf{46B} (1973), 109--110.

\bibitem[Var04]{Varadarajan:2004}
V.S. Varadarajan, \emph{{Supersymmetry for Mathematicians: An Introduction}},
  Courant Lecture Notes in Mathematics, vol.~11, American Mathematical Society,
  2004.

\bibitem[Vor84]{Voronov:1984}
A.A. Voronov, \emph{{``Maps of supermanifolds"}}, Teoret. Mat. Fiz. \textbf{60}
  (1984), 43--48.

\bibitem[Wal07]{Waldorf:2007mm}
K.~Waldorf, \emph{{``More morphisms between bundle gerbes''}}, Theory Appl.
  Categories \textbf{18} (2007), 240--273.

\bibitem[Wal13]{Waldorf:2009uf}
\bysame, \emph{{``String connections and Chern--Simons theory''}}, Trans. Amer.
  Math. Soc. \textbf{365} (2013), 4393--4432.

\bibitem[WC84]{Williams:1984}
D.~Williams and J.F. Cornwell, \emph{{``The Haar integral for Lie
  supergroups''}}, J. Math. Phys. \textbf{25} (1984), 2922--2932.

\bibitem[Wei52]{Weil:goc1952}
A.~Weil, \emph{{``Sur les th{\'e}or{\`e}mes de de Rham''}}, Comment. Math.
  Helv. \textbf{26} (1952), 119--145.

\bibitem[Wei68]{Weinberg:1968de}
S.~Weinberg, \emph{{``Nonlinear realizations of chiral symmetry"}}, Phys. Rev.
  \textbf{166} (1968), 1568--1577.

\bibitem[Wei99]{Weinberg:1999}
\bysame, \emph{{The Quantum Theory of Fields, Volume 3: Supersymmetry}},
  Cambridge University Press, 1999.

\bibitem[Wes99]{West:1998ey}
P.C. West, \emph{{``Supergravity, Brane Dynamics and String Duality''}},
  {Duality and Supersymmetric Theories. Proceedings, Easter School, Newton
  Institute, Euroconference, Cambridge, UK, April 7-18, 1997} (D.I. Olive and
  P.C. West, eds.), Cambridge University Press, 1999, pp.~147--266.

\bibitem[Wes00]{West:2000hr}
\bysame, \emph{{``Automorphisms, nonlinear realizations and branes"}}, JHEP
  \textbf{02} (2000), 024.

\bibitem[Wit84]{Witten:1983ar}
E.~Witten, \emph{{``Non-abelian bosonization in two dimensions''}}, Commun.
  Math. Phys. \textbf{92} (1984), 455--472.

\bibitem[Wit88]{Witten:1988dls}
\bysame, \emph{{``The Index of the Dirac Operator in Loop Space''}},
  {``Elliptic Curves and Modular Forms in Algebraic Topology. Proceedings of a
  Conference held at the Institute for Advanced Study, Princeton, September
  15-17, 1986''} (P.S. Landweber, ed.), Lecture Notes in Mathematics, vol.
  1326, Springer, 1988, pp.~161--181.

\bibitem[WZ74a]{Wess:1973kz}
J.~Wess and B.~Zumino, \emph{{``A Lagrangian model invariant under supergauge
  transformations"}}, Phys. Lett. \textbf{49B} (1974), 52--54.

\bibitem[WZ74b]{Wess:1974tw}
\bysame, \emph{{Supergauge transformations in four dimensions}}, Nucl. Phys.
  \textbf{B70} (1974), 39--50.

\end{thebibliography}
\end{document}